\documentclass[11pt, a4paper]{report}

\usepackage[english]{babel}
\usepackage[T1]{fontenc}
\usepackage{ae}

\usepackage{aas_macros}

\usepackage{amsmath, amssymb}
\usepackage{revsymb}
\usepackage{slashed}
\usepackage{bbm}
\usepackage{bm}
\usepackage{xcolor}
\usepackage{multirow, booktabs, dcolumn}
\usepackage{graphicx}
\usepackage{tabularx}
\usepackage{afterpage}

\usepackage[toc,page]{appendix}

\usepackage{geometry}
\usepackage[square, comma, numbers]{natbib}

\renewcommand{\contentsname}{Contents}
\usepackage[ small, labelfont=bf , margin=20pt, tableposition=top]{caption}
\usepackage{fancyhdr}
\usepackage[ps2pdf=true,colorlinks, pdfpagelabels]{hyperref}
\usepackage[figure,table]{hypcap} 
\hypersetup{
   bookmarksnumbered,
   pdfpagemode={UseOutlines},
   pdfauthor={Josef Pradler},
   pdftitle={The long-lived stau as a thermal relic}
   pdfsubject={Dissertation}
   ,citecolor={blue},
   linkcolor={blue}
} 

\newcolumntype{.}{D{.}{.}{-1}}

\newcommand{\VEV}[1]{\ensuremath{\langle #1 \rangle}}
\newcommand{\Order}{\ensuremath{{\mathcal{O}}}}   %
\newcommand{\Orderof}[1]{{\ensuremath{\mathcal{O}\left(#1\right)}}}
\newcommand{\dex}{\ensuremath{\mathrm{dex}}}   %

\newcommand{\hef}{\ensuremath{{}^4\mathrm{He}}}
\newcommand{\het}{\ensuremath{{}^3\mathrm{He}}}
\newcommand{\lisx}{\ensuremath{{}^6\mathrm{Li}}}
\newcommand{\lisv}{\ensuremath{{}^7\mathrm{Li}}}
\newcommand{\bes}{\ensuremath{{}^7\mathrm{Be}}}
\newcommand{\beet}{\ensuremath{{}^8\mathrm{Be}}}
\newcommand{\ben}{\ensuremath{{}^9\mathrm{Be}}}
\newcommand{\hefxm}{\ensuremath{(\hef X^-)}}
\newcommand{\beetxm}{\ensuremath{(\beet X^-)}}
\newcommand{\Trit}{\ensuremath{\mathrm{T}}}
\newcommand{\trit}{\ensuremath{\mathrm{T}}}
\newcommand{\Deut}{\ensuremath{\mathrm{D}}}
\newcommand{\deut}{\ensuremath{\mathrm{D}}}
\newcommand{\Ox}{\ensuremath{\mathrm{O}}}
\newcommand{\Fe}{\ensuremath{\mathrm{Fe}}}
\newcommand{\Hyd}{\ensuremath{\mathrm{H}}}
\newcommand{\Be}{\ensuremath{\mathrm{Be}}}

\newcommand{\nuc}[1]{\ensuremath{{}^{#1}\mathrm{N}}}
\newcommand{\mnuc}{\ensuremath{m_{\mathrm{N}}}}
\newcommand{\mred}{\ensuremath{m_{\mathrm{red}}}}

\newcommand{\abohr}{\ensuremath{a_{\mathrm{b}}}}
\newcommand{\Rcharge}[1]{\ensuremath{\VEV{r^2_c}^{1/2}_{#1}}}
\newcommand{\Rchargesq}[1]{\ensuremath{\VEV{r^2_c}_{#1}}}
\newcommand{\Rrms}{\ensuremath{\VEV{r^2}^{1/2}}}

\newcommand{\Ebindvar}{\ensuremath{E_{\mathrm{b}}^{\mathrm{var}}}}
\newcommand{\Ebindcoul}{\ensuremath{E^{\mathrm{coul}}_{\mathrm{b}}}}
\newcommand{\Ebindcouln}[1]{\ensuremath{E^{\mathrm{coul}}_{\mathrm{b},#1}}}

\newcommand{\BSx}[1]{\ensuremath{(#1\champ)}}
\newcommand{\BSnucx}{\ensuremath{(\mathrm{N}\champ)}}
\newcommand{\BSpx}{\ensuremath{(\proton\champ)}}

\newcommand{\BSbeetx}{\ensuremath{(\beet\champ)}}
\newcommand{\BSlisxx}{\ensuremath{(\lisx\champ)}}
\newcommand{\BShefx}{\ensuremath{(\hef\champ)}}

\newcommand{\Gammaph}{\ensuremath{\Gamma_{\mathrm{ph}}}}
\newcommand{\Gammarec}{\ensuremath{\Gamma_{\mathrm{rec}}}}
\newcommand{\Gammaphof}[1]{\ensuremath{\Gamma_{\mathrm{ph},#1}}}

\newcommand{\etabaryon}{\ensuremath{\eta_{\mathrm{b}}}}
\newcommand{\etacmb}{\ensuremath{\eta_{\mathrm{b}}({\mathrm{CMB}})}}
\newcommand{\nb}{\ensuremath{n_{\mathrm{b}}}}
\newcommand{\primordial}{{\ensuremath{\mathrm{p}}}}

\newcommand{\regCoul}[1]{\ensuremath{F_{#1}}}
\newcommand{\irregCoul}[1]{\ensuremath{G_{#1}}}
\newcommand{\phaseCoul}[1]{\ensuremath{\delta_{#1}}}
\newcommand{\phaseNuc}[1]{\ensuremath{\sigma_{#1}}}

\newcommand{\Lisix}{\ensuremath{{}^6 \mathrm{Li}}}

\newcommand{\Hefournarrow}{\ensuremath{{}^4\! \mathrm{He}}}
\newcommand{\taustau}{\ensuremath{\tau_{\widetilde{\tau}_1}}}
\newcommand{\Ystaudec}{\ensuremath{Y_{\widetilde{\tau}^-}}}

\newcommand{\monetwo}{\ensuremath{m_{1/2}}}
\newcommand{\mzero}{\ensuremath{m_{0}}}
\newcommand{\tanb}{\ensuremath{\tan{\beta}}}
\newcommand{\mgut}{\ensuremath{M_\mathrm{GUT}}}
\newcommand{\Omegatp}{\ensuremath{\Omega_{\widetilde{G}}^{\mathrm{TP}}}}
\newcommand{\Omegantp}{\ensuremath{\Omega_{\widetilde{G}}^{\mathrm{NTP}}}}

\newcommand{\Ebind}{\ensuremath{E_{\mathrm{b}}}}
\newcommand{\px}{\ensuremath{(\mathrm{p}X^-)}}

\newcommand{\Vmax}{\ensuremath{V_{\mathrm{max}}}}

\newcommand{\champ}{\ensuremath{X^{\! -}}}
\newcommand{\X}{\ensuremath{X}}
\newcommand{\xm}{\ensuremath{X^-}}
\newcommand{\mx}{\ensuremath{m_{X}}}
\newcommand{\tauX}{\ensuremath{\tau_{X}}}
\newcommand{\YX}{\ensuremath{Y_{X^-}}}
\newcommand{\YXdec}{\ensuremath{Y^{\mathrm{dec}}_{X^-}}}
\newcommand{\Ysldec}{\ensuremath{Y^{\mathrm{dec}}_{{\stauone}}}}
\newcommand{\Ysldecm}{\ensuremath{Y^{\mathrm{dec}}_{{\stauone^-}}}}

\newcommand{\Navogadro}{\ensuremath{N_{\mathrm{A}}}}
\newcommand{\mol}{\ensuremath{\mathrm{mol}}}

\newcommand{\eV}{\ensuremath{\mathrm{eV}}}
\newcommand{\keV}{\ensuremath{\mathrm{keV}}}
\newcommand{\MeV}{\ensuremath{\mathrm{MeV}}}
\newcommand{\GeV}{\ensuremath{\mathrm{GeV}}}
\newcommand{\TeV}{\ensuremath{\mathrm{TeV}}}

\newcommand{\seconds}{\ensuremath{\mathrm{s}}}

\newcommand{\barn}{\ensuremath{\mathrm{b}}}
\newcommand{\fm}{\ensuremath{\mathrm{fm}}}
\newcommand{\cm}{\ensuremath{\mathrm{cm}}}
\newcommand{\Mpc}{\ensuremath{\mathrm{Mpc}}}
\newcommand{\km}{\ensuremath{\mathrm{km}}}

\newcommand{\MPl}{\ensuremath{M_{\mathrm{P}}}}
\newcommand{\MP}{\ensuremath{M_{\mathrm{P}}}}

\newcommand{\proton}{\ensuremath{\mathrm{p}}}
\newcommand{\neutron}{\ensuremath{\mathrm{n}}}
\newcommand{\mproton}{\ensuremath{m_{\mathrm{p}}}}

\newcommand{\bquark}{\ensuremath{\mathrm{b}}}
\newcommand{\antibquark}{\ensuremath{\bar{\mathrm{b}}}}
\newcommand{\electron}{\ensuremath{\mathrm{e}^-}}
\newcommand{\positron}{\ensuremath{\mathrm{e}^+}}

\newcommand{\gravitino}{\ensuremath{{\widetilde{G}}}}
\newcommand{\st}{\ensuremath{{\tilde{\tau}_1}}}

\newcommand{\Bino}{\ensuremath{{\widetilde B}}}

\newcommand{\gluino}{\ensuremath{{\widetilde g}}}
\newcommand{\neutralino}{\ensuremath{{\widetilde \chi}^{0}_{1}}}
\newcommand{\chargino}{\ensuremath{{\widetilde{\chi}_1^{\pm}}}}

\newcommand{\mgr}{\ensuremath{m_{\widetilde{G}}}}
\newcommand{\mgrav}{\ensuremath{\mgr}}
\newcommand{\mgravitino}{\ensuremath{\mgr}}
\newcommand{\mneu}{\ensuremath{m_{{\widetilde \chi}^{0}_{1}}}}

\newcommand{\NLSP}{\ensuremath{\mathrm{NLSP}}}

\newcommand{\NTP}{\ensuremath{\mathrm{NTP}}}
\newcommand{\TP}{\ensuremath{\mathrm{TP}}}

\newcommand{\equil}{\ensuremath{\mathrm{eq}}}

\newcommand{\freezeout}{\ensuremath{\mathrm{f}}}
\newcommand{\CDM}{\ensuremath{\mathrm{dm}}}

\newcommand{\EM}{\ensuremath{\mathrm{em}}}
\newcommand{\HAD}{\ensuremath{\mathrm{had}}}

\newcommand{\rad}{\ensuremath{\mathrm{rad}}}
\newcommand{\dec}{\ensuremath{\mathrm{dec}}}
\newcommand{\GUT}{\ensuremath{\mathrm{GUT}}}
\newcommand{\Reheating}{\ensuremath{\mathrm{R}}}
\newcommand{\TR}{\ensuremath{T_{\Reheating}}}
\newcommand{\TL}{\ensuremath{T_{\mathrm{low}}}}
\newcommand{\Color}{\ensuremath{\mathrm{c}}}
\newcommand{\Weak}{\ensuremath{\mathrm{L}}}
\newcommand{\Hypercharge}{\ensuremath{\mathrm{Y}}}

\newcommand{\Xtau}{\ensuremath{X_{\tau}}}

\newcommand{\Atau}{\ensuremath{A_{\tau}}}

\newcommand{\mtau}{\ensuremath{m_{\tau}}}
\newcommand{\mLL}{\ensuremath{m_{\mathrm{LL}}}}
\newcommand{\mRR}{\ensuremath{m_{\mathrm{RR}}}}
\newcommand{\mstauone}{\ensuremath{m_{\stauone}}}
\newcommand{\mstautwo}{\ensuremath{m_{\stautwo}}}
\newcommand{\mstauL}{\ensuremath{m_{\stauL}}}
\newcommand{\mstauR}{\ensuremath{m_{\stauR}}}
\newcommand{\thetastau}{\ensuremath{\theta_{\stau}}}

\newcommand{\sel}{{\widetilde{e}_1}}
\newcommand{\smu}{{\widetilde{\mu}_1}}

\newcommand{\DsevEM}{\mathrm{D}^{\mathrm{sev}}_{\mathrm{em}}}
\newcommand{\DconsEM}{\mathrm{D}^{\mathrm{cons}}_{\mathrm{em}}}

\newcommand{\HeD}{^3\mathrm{He/D}}

\newcommand{\stauone}{\ensuremath{{\widetilde{\tau}_{1}}}}
\newcommand{\stautwo}{\ensuremath{{\widetilde{\tau}_{2}}}}
\newcommand{\stau}{\ensuremath{{\widetilde{\tau}}}}
\newcommand{\stauR}{\ensuremath{{\widetilde{\tau}_{\mathrm{R}}}}}
\newcommand{\stauL}{\ensuremath{{\widetilde{\tau}_{\mathrm{L}}}}}

\newcommand{\hhiggs}{\ensuremath{{h}^0}}
\newcommand{\Hhiggs}{\ensuremath{{H}^0}}
\newcommand{\Ahiggs}{\ensuremath{{A}^0}}
\newcommand{\Hphiggs}{\ensuremath{H}^+}
\newcommand{\Hmhiggs}{\ensuremath{H}^-}
\newcommand{\Hpmhiggs}{\ensuremath{H^{\pm}}}

\newcommand{\mtop}{\ensuremath{m_{t}}}

\newcommand{\lepton}{\ensuremath{\mathrm{l}}}

\newcommand{\mh}{\ensuremath{m_{\hhiggs}}}
\newcommand{\mH}{\ensuremath{m_{\Hhiggs}}}
\newcommand{\mA}{\ensuremath{m_{\Ahiggs}}}
\newcommand{\mHpm}{\ensuremath{m_{\mathrm{H}^{\pm}}}}

\newcommand{\Ystau}{\ensuremath{Y_{\widetilde{\tau}}}}
\newcommand{\Ystaueq}{\ensuremath{Y^{\mathrm{eq}}_{\widetilde{\tau}}}}
\newcommand{\YstaumO}{\ensuremath{Y^{\mathrm{m}\Omega}_{\widetilde{\tau}}}}

\newcommand{\mZ}{\ensuremath{M_{\mathrm{Z}}}}
\newcommand{\mW}{\ensuremath{M_{\mathrm{W}}}}

\newcommand{\ssqw}{\ensuremath{s^{2}_{W}}}

\newcommand{\sa}{\ensuremath{s_{\alpha}}}

\newcommand{\sapb}{\ensuremath{s_{\alpha+\beta}}}
\newcommand{\sbma}{\ensuremath{s_{\beta-\alpha}}}

\newcommand{\ca}{\ensuremath{c_{\alpha}}}

\newcommand{\cb}{\ensuremath{c_{\beta}}}
\newcommand{\ctwob}{\ensuremath{c_{2\beta}}}

\newcommand{\stwoth}{\ensuremath{s_{2\thetastau}}}

\newcommand{\csqth}{\ensuremath{c^{2}_{\thetastau}}}
\newcommand{\ctwoth}{\ensuremath{c_{2\thetastau}}}

\newcommand{\Lagrangian}{\ensuremath{\mathcal{L}}}
\renewcommand{\L}{\mathrm{L}}
\newcommand{\R}{\ensuremath{R}}

\newcommand{\stauROT}{\ensuremath{R_{\stau}}}
\newcommand{\stauMAT}{\ensuremath{\mathcal{M}_{\stau}^2}}

\newcommand{\couptriLR}[3]{\ensuremath{\widetilde{C}[#1,#2,#3]}}
\newcommand{\couptri}[3]{\ensuremath{C[#1,#2,#3]}}
\newcommand{\couptriDL}[3]{\ensuremath{C^{\mathrm{DL}}[#1,#2,#3]}}

\newcommand{\sigmavof}[1]{\ensuremath{\langle \sigma_{{#1}}v \rangle}}
\newcommand{\sigmatot}{\ensuremath{\sigma_{\mathrm{tot}}}}
\newcommand{\sigmav}{\ensuremath{\langle {\boldsymbol\sigma} v \rangle}}
\newcommand{\peff}{\ensuremath {P_{\mathrm{eff}}}}
\newcommand{\Tf}{\ensuremath {T_{\mathrm{f}}}}
\newcommand{\Tfone}{\ensuremath {T_{\mathrm{f1}}}}

\newcommand{\geff}{\ensuremath {g_{\mathrm{eff}}}}
\newcommand{\heff}{\ensuremath {h_{\mathrm{eff}}}}

\newcommand{\OmegaDM}{\Omega_{\mathrm{dm}}}
\newcommand{\nelectron}{n_{\mathrm{e}^-}}

\newcommand{\ngamma}{n_{\gamma}}

\newcommand{\melectron}{m_{\mathrm{e}}}
\newcommand{\Ein}{E^{\mathrm{in}}_{\gamma}}
\newcommand{\Eout}{E^{\mathrm{out}}_{\gamma}}
\newcommand{\Egamma}{E_{\gamma}}
\newcommand{\Texch}{T_{\mathrm{ex}}}

\newcommand{\etabarion}{\eta_{\mathrm{b}}}
\newcommand{\sigmaThomson}{\sigma_{\mathrm{T}}}

\newcommand{\vmol}{\ensuremath{v_{\text{m\o l}}}}
\newcommand{\vrel}{\ensuremath{v_{\mathrm{rel}}}}
\newcommand{\Ej}{\ensuremath{E_j}}
\newcommand{\Ejf}{\ensuremath{E_{jf}}}
\newcommand{\Ecm}{\ensuremath{E_{\mathrm{cm}}}}
\newcommand{\Ejcm}{\ensuremath{E_{j}^{\mathrm{cm}}}}
\newcommand{\Eecm}{\ensuremath{E_{e}^{\mathrm{cm}}}}
\newcommand{\Ejfcm}{\ensuremath{E_{jf}^{\mathrm{cm}}}}
\newcommand{\bj}{\ensuremath{\beta_j}}
\newcommand{\bcm}{\ensuremath{\beta_{\mathrm{cm}}}}
\newcommand{\gb}{\ensuremath{\gamma_{\beta}}}
\newcommand{\gcm}{\ensuremath{\gamma_{\mathrm{cm}}}}
\newcommand{\pj}{\ensuremath{p_j}}
\newcommand{\pjcm}{\ensuremath{p_j^{\mathrm{cm}}}}
\newcommand{\pjfcm}{\ensuremath{p_{jf}^{\mathrm{cm}}}}
\newcommand{\pjf}{\ensuremath{p_{jf}}}
\newcommand{\me}{\ensuremath{m_{e}}}

\newcommand{\thmin}{\ensuremath{{\theta_{\mathrm{min}}}}}

\newcommand{\omegapl}{\ensuremath{\omega_{\mathrm{pl}}}}
\newcommand{\kd}{\ensuremath{k_{\mathrm{D}}}}

\newcommand{\Ydec}{Y^{\mathrm{dec}}_{\mathrm{NLSP}}}

\newcommand{\cf}[2]{\chi_{#1}^{#2}}   
\newcommand{\cfb}[2]{\overline{\chi}_{#1}^{#2}}   

\begin{document}

\title{The long-lived stau as a thermal relic}
\author{Josef Pradler}
\date{}

\thispagestyle{empty}

\begin{center}

\vspace*{-1cm}

\noindent{\Large Technische Universit\"at M\"unchen, Physik Department, T30d }\\[0.3cm]

{\Large Max Planck Institut f\"ur Physik (Werner Heisenberg Institut) }\\

\vspace{1.7cm} 

{\Large\sf\sc Dissertation} \vfill
  {\huge \bf The long-lived stau as a thermal relic}\\[0.3cm]
\vspace{3.2cm}

{\Large\sf\sc Josef Pradler} \vfill

\end{center} \vfill

\noindent Vollst\"andiger Abdruck der von der Fakult\"at f\"ur Physik der
Technischen Universit\"at M\"unchen zur Erlangung des akademischen
Grades eines
\begin{center}
{\bf \sf Doktors der Naturwissenschaften (Dr. rer. nat.)}
\end{center}
genehmigten Dissertation.

\begin{center}
\begin{tabular}{lll}
Vorsitzender:              &    &  Univ.-Prof.~Dr.~L.~Oberauer
\\[0.2cm]
Pr\"ufer der Dissertation: & 1. & Univ.-Prof.~Dr.~A.~Ibarra\\
                           & 2. & Univ.-Prof.~Dr.~W.\,F.\,L.~Hollik
\end{tabular}
\end{center}

\noindent Die Dissertation wurde am 23.06.2009 bei der Technischen
Universit\"at M\"unchen einge\-reicht und durch die Fakult\"at f\"ur
Physik am 20.07.2009 angenommen. \cleardoublepage

\chapter*{Summary}
\phantomsection
\addcontentsline{toc}{chapter}{Summary}
\label{cha:introduction}

The results presented in this thesis have in part already been
published in 
Refs.~\cite{Pradler:2006hh,Pradler:2007is,Pradler:2007ar,Pospelov:2008ta,Pradler:2008qc}
listed overleaf (page~\pageref{publications}).
We consider physics beyond the Standard Model
which implies the existence a of long-lived electromagnetically
charged massive particle species (CHAMP) which we denote by
$\X^{\pm}$.
We discuss in detail the unique sensitivity the early Universe
exhibits on the mere presence and on the decay of such a particle. A
CHAMP can be realized in supersymmetric (SUSY) extensions of the Standard
Model. We carry out a detailed study of gravitino (\gravitino) dark
matter scenarios in which the lighter scalar tau (stau, $\stauone$) is the
lightest Standard Model superpartner so that~$\stauone = \X$. We also
provide a thorough investigation of the thermal freeze-out process of
$\stauone$.

\noindent The thesis is divided into three parts:

\noindent{\bf Part~\ref{part:bbn}}: In this part we consider a generic but
weak-scale CHAMP. In Chapter~\ref{cha:sbbn-and-particle-decays} we set
the stage for the coming investigations by shortly reviewing the
framework of Big Bang Nucleosynthesis (BBN), by working out the
typical CHAMP freeze-out abundance, and by reviewing the stringent
constraints arising from such a decaying component during/after BBN.
We also take a critical look at the BBN constraints arising from the
hadronic decay modes of an arbitrary exotic. In particular, we develop on
a refined treatment of the Coulomb stopping mechanism of charged
hadrons.

In Chapter~\ref{cha:catalyzed-bbn} we discuss the physics which
emerges when the light elements fused in BBN are captured by \champ\
at the time of primordial nucleosynthesis. Since the associated, most
striking effects were only discovered recently, we provide a detailed
exposition of the topic. In particular, we explicitly show how to
obtain the rates for bound state formation which carry a finite charge
radius correction of the nucleus.  In the remainder of this chapter,
which is based on \cite{Pospelov:2008ta}, we focus on the catalytic
production of \lisx\ and \ben.  There, we also discuss the issue of a
potential late-time catalysis due to proton-CHAMP bound states.
Upon solution of the full set of Boltzmann equations we obtain
stringent constraints on the primordial presence of long-lived~\champ\
from overproduction of \lisx.
Moreover, setting an upper limit on the abundance of primordial \ben\
allows us to constrain this scenario also from catalytic \ben\
production.

\noindent{\bf Part~\ref{part:two}}: The second part is devoted to scenarios in
which \gravitino\ is the lightest supersymmetric particle (LSP) and
\stauone\ is the next-to-lightest SUSY particle~(NLSP).
In Chapter~\ref{cha:grav-stau-scenario} we focus on the gravitino LSP
as a dark matter candidate.  We recollect the results on thermal
gravitino production, consider explicitly the post-inflationary
reheating process, and obtain an update on the upper bound on the
reheating temperature of the Universe from thermal production.

In Chapter~\ref{cha:stau-nlsp} we then focus on gravitino dark matter
scenarios in which \stauone\ is the NLSP. This chapter resembles many
of the results of the research papers
\cite{Pradler:2006hh,Pradler:2007is,Pradler:2007ar,Pospelov:2008ta}. We
constrain the gravitino-stau scenario by incorporating the BBN bounds
from \stauone-decays previously obtained in the literature. In
addition, the concrete realization of the long-lived CHAMP scenario
allows us to employ our results on the catalytic production of \ben\
and \lisx.  In the framework of the constrained minimal supersymmetric
Standard Model (CMSSM) a \stauone\ NLSP can be naturally
accommodated. There, we show that the novel catalytic effects severely
constrain the reheating temperature of the Universe and potentially
imply very heavy superparticle mass spectra which will be hard to
probe at the upcoming Large Hadron Collider (LHC) experiments. We also
consider explicitly the possibility of a non-standard cosmological
evolution and check for the viability of thermal leptogenesis.

\noindent{\bf Part~\ref{part:three}}: Chapter~\ref{cha:therm-relic-abund}
constitutes the final part of this thesis and is based on
\cite{Pradler:2008qc}. There, we take an in-depth look into the
chemical decoupling process of the long-lived \stauone\ from the
primordial plasma.
The quantity of interest is the thermal freeze-out abundance of the
stau. We identify its dependence on the crucial SUSY parameters and
also show that it sensitively depends on the details of the Higgs
sector. Stau annihilation into final state Higgses as well as resonant
annihilation via the heavy CP even Higgs boson can substantially
deplete the decoupling yield. Remarkably, we find these features are
already realized in the CMSSM. In those regions of the parameter space
even the most restrictive bounds from the thermal catalysis of BBN
reactions can potentially be respected. We discuss the implications
for the gravitino-stau scenario.

\clearpage

\vspace*{1cm}
\noindent The results obtained in this thesis have in part already been
published in the following references:

\begin{tabularx}{0.98\textwidth}{@{}>{}X@{}>{\hsize=0.8cm}X@{}l@{}X@{}r}
\label{publications}
\\[-0.6cm]

&\cite{Pradler:2006hh}&{\bf Constraints on the Reheating Temperature in }&&\\
&&{\bf  Gravitino Dark Matter Scenarios }\\
&&  J.~Pradler and F.~D.~Steffen \\
&& Phys.\ Lett.\ B 648, 224 (2007) [arXiv:hep-ph/0612291] \\[0.4cm]

&\cite{Pradler:2007is}&{\bf Implications of Catalyzed BBN in the CMSSM with}&&\\
&& {\bf Gravitino Dark Matter } \\
&&  J.~Pradler and F.~D.~Steffen \\
&& Phys.\ Lett.\ B  666, 181 (2008) [arXiv:0710.2213] \\[0.4cm] 

&\cite{Pradler:2007ar}&{\bf CBBN in the CMSSM } &&\\
&& J.~Pradler and F.~D.~Steffen \\
&& Eur.\ Phys.\ J.\ C  56, 287 (2008) [arXiv:0710.4548]\\[0.4cm] 

&\cite{Pospelov:2008ta}&{\bf Constraints on Supersymmetric Models from Catalytic }\\
&& {\bf Primordial Nucleosynthesis of Beryllium } \\
&& M.~Pospelov, J.~Pradler, and F.~D.~Steffen \\
&& JCAP 0811, 020 (2008)  [arXiv:0807.4287] \\[0.4cm]

&\cite{Pradler:2008qc}&{\bf Thermal Relic Abundances of Long-Lived Staus} &&\\
&&  J.~Pradler and F.~D.~Steffen  \\
&&  Nucl.\ Phys.\ B 809, 318 (2009) [arXiv:0808.2462] \\[0.4cm]

\end{tabularx}\\[0.4cm]


\cleardoublepage
\chapter*{Acknowledgements}
\phantomsection
\addcontentsline{toc}{chapter}{Acknowledgements}

I would first like to express my gratitude towards my research advisor
Frank Daniel Steffen at the Max Planck Institute for Physics (MPI) for
his continuous support, col\-laboration, and for the many interesting
discussions we had. I further thank him for the coordination of the
International Max Planck Research School from where I also have
received my funding.  I am thankful to the MPI for providing an
optimal place to work, in particular, to the secretary Rosita
Jurgeleit for her friendly help and to Thomas Hahn for his computer
support.

\noindent I would like to thank Alejandro Ibarra for being my official advisor
at the Technical University Munich and thus for providing the academic
framework to my PhD studies.

\noindent For his support, advice, and for holding together the enjoyable
atmosphere in the MPI ``Astroparticle Group'' I am grateful to Georg
Raffelt.

\noindent Many insights of the first part I owe to Maxim Pospelov whom I also
would like to thank for his invitation to the University of Victoria.
I am grateful to Gary Steigman for his friendly explanations during
his visit in Munich.  Simon Eidelman, Tilman Plehn and Stefan Hofmann
I want thank for
their general advice.

\noindent Many thanks to the friends which I had the chance to meet at the MPI,
in particular, to Steve Blanchet, Koushik Dutta, Florian Hahn-Woernle,
Max Huber, and Felix Rust.

\noindent For their friendship I am also most grateful to Ulrich Matt and Erik
H\"ortnagl.

\noindent I am deeply indebted to my family, foremost to my parents,
for their unconditional love and support and to Irina Bavykina for all
her understanding, encouragement, and patience over the last three
years.

\vspace*{0.8cm} \noindent\textit{To the memory of Florian Kunz.}


\cleardoublepage
\pdfbookmark[0]{\contentsname}{toc} 
\tableofcontents 
\listoffigures
\listoftables

\cleardoublepage
\pagenumbering{arabic}
\part{BBN with a long-lived CHAMP}
\label{part:bbn}
\cleardoublepage
\chapter{BBN and particle decays}
\label{cha:sbbn-and-particle-decays}

We start this work with a brief introduction into the framework of Big
Bang Nucleosynthesis (BBN) reviewing abundance predictions of some of
the primordial light elements and discussing their current
observational status (Sects.~\ref{sec:bbn-post-WMAP} and
\ref{sec:bbn-as-probe}).
In Sec.~\ref{sec:typic-champ-abund} we then carry out a simplified
treatment of the chemical decoupling of a long-lived
CHAMP~($\X^{\pm}$). This frames the thermal X-abundance region and in
Sec.~\ref{sec:typic-champ-abund} we shall see that with it are
associated strong limits on the energy release from $X$ decays
during/after~BBN.

In Sec.~\ref{sec:stopping power} we investigate in some detail the
stopping mechanism of injected particles short after the main stage of
primordial nucleosynthesis.  There, we will recover existing results
of the literature but also develop on a refined treatment of Coulomb
stopping of injected charged hadrons.

\section{Primordial nucleosynthesis after WMAP}
\label{sec:bbn-post-WMAP}

The cumulative evidence from observations of the Hubble expansion as
well as of the cosmic microwave background (CMB) radiation has put the
hot Big Bang model on firm footing. In addition, one of the pillars on
which modern day cosmology rests on is the framework of BBN. Relying
solely on Standard Model physics and a Friedmann-Robertson-Walker
Universe, an \textit{overall} agreement between the BBN predictions
and the observationally inferred primordial abundances of the light
elements \deut, \het, \hef, and \lisv\ is found. This is truly
striking given that those elements span nine orders of magnitude in
number and that light element observations are performed in vastly
different astrophysical sites. It is this concordance which provides
direct evidence that the Universe must once have had a temperature $T
\gtrsim 1\ \MeV$.

Standard BBN (SBBN) has only one free parameter, the baryon-to-photon
ratio $\etabaryon = \nb / n_{\gamma}$.  It measures the nucleon
content of the primordial plasma and controls the rates of the
processes which eventually lead to the fusion of the light elements.
With the measurements of the Wilkinson Microwave Anisotropy Probe
(WMAP) satellite
experiment~\cite{Spergel:2003cb,Spergel:2006hy,Komatsu:2008hk}
unprecedented precision data on the multipoles of the CMB angular
power spectrum became available. Based on a $\Lambda$CDM model, i.e.,
a flat Universe filled with baryons, cold dark matter, neutrinos, and
a cosmological constant, this has allowed one to pin the baryon
density down to~\cite{Dunkley:2008ie} $\Omega_{\mathrm{b}} h^2 =
0.02273 \pm 0.00062 $
with $h = H_0 / (100~\km\, \Mpc^{-1}\seconds^{-1}) $ parameterizing
the Hubble constant $H_0$. The value implies a baryon-to-photon ratio%
\footnote{This follows from the WMAP 5-year data set. For comparison,
  the 3-year result implied $ \etacmb = 6.116^{+0.197}_{-0.249}$
  ~\cite{Spergel:2006hy}.
  The conversion from $\Omega_{\mathrm{b}} h^2$ to $\etabaryon$
  requires knowledge of the average mass per
  baryon~\cite{Steigman:2006nf}.}
of~\cite{Dunkley:2008ie}
\begin{align}
  \label{eq:etacmb}
 \etacmb = (6.225 \pm 0.170)\times 10^{-10}
\end{align}
so that we have knowledge of the baryon content of the Universe at the
time of photon decoupling to $\sim 3\%$ accuracy (at $68\%$
C.L.). Using~(\ref{eq:etacmb}) and/or other non-BBN determinations of
$\etabaryon$ as input
for primordial nucleosynthesis makes BBN a parameter-free theory. When
we talk about the SBBN light element predictions in the following we
shall mean the outlined minimal framework of primordial
nucleosynthesis with $\etabaryon$ fixed by the CMB measurements.

\section{BBN as a probe for New Physics}
\label{sec:bbn-as-probe}

The comparison of SBBN predictions with the observationally inferred
primordial light element abundances makes the theory of primordial
nucleosynthesis a powerful tool to test and to constrain models of New
Physics. 

A true success of the standard cosmological model is the emerging
concordance between the SBBN predicted deuterium abundance and the
measurements of $\deut/\Hyd$ (in number) in hydrogen-rich clouds
absorbing the light of background quasars at high redshifts.
Those astrophysical sites are believed to be most appropriate to yield
an estimate on the primordial fraction $\deut/\Hyd|_\primordial$.
Including the latest measurement~\cite{Pettini:2008mq} of this ratio,
the weighted mean of seven determinations reads~\cite{Simha:2008mt}
\begin{align}
  \left. {\deut}/{\Hyd} \right|_\primordial = \left(2.70^{+0.22}_{-0.20} \right)\times 10^{-5} .
\end{align}
Conversely, with an uncertainty which is comparable to that of weak
and nuclear rates used in BBN codes, the SBBN deuterium abundance can
be predicted in the $\etabaryon$-range of interest
as~\cite{Steigman:2007xt}%
\begin{align}
  \label{eq:det-prediction}
  \left. {\deut}/{\Hyd} \right|_\primordial = 2.67(1 \pm 0.03)\times
  10^{-5} \, \left( \frac{6\times 10^{-10}}{\etabaryon} \right)^{1.6}
  = ( 2.52  \pm 0.11)\times 10^{-5} \ .
\end{align}
In the last expression we have used the CMB inferred baryon-to-photon
ratio (\ref{eq:etacmb}) and added errors in quadrature. As can be
seen, both values agree within their~\mbox{$\sim 1\sigma$} range.
Indeed, despite the difficult observations, deuterium is the
baryometer of choice.
Because of its weak binding energy, \deut\ is only destroyed in
astrophysical environments so that its post-BBN evolution is
monotonic. Moreover, the SBBN prediction shows a strong sensitivity on
the baryon-to-photon ratio, $\deut/\Hyd|_\primordial\propto \etabaryon
^{-1.6}$.  Any physical process which is triggered by extending the
SBBN framework must not spoil the agreement between prediction and
observation.

Though the agreement in deuterium is impressive it may still only be a
coincidence. Let us consider \hef\ which is the most tightly bound
element among the SBBN products.
The primordial \hef\ mass fraction is defined as%
\footnote{The convention to call $ Y_\primordial$ the mass fraction is
  slightly misleading since $m_{\alpha} = 3.97 m_{\proton}$ with
  $m_{\proton}$ ($m_{\alpha}$) denoting the mass of the proton
  (alpha-particle). However, what is observed
  is $ Y_\primordial$ and it represents the \hef\
  abundance in mass within $1\%$ accuracy.}
\begin{align}
  \label{eq:def-hef-mass-fraction}
  Y_\primordial \equiv \frac{4 n_{\hef}/n_{\Hyd}}{1 +
    4n_{\hef}/n_{\Hyd} } \simeq 0.25 
\end{align}
and with $\sim 25\%$ this makes \hef\ the second most abundant element
after hydrogen. The estimate in the last relation already follows from
the observation that most neutrons available are finally bound in
\hef\ and that the neutron-to-proton ratio in number at the onset of
BBN is $\neutron/\proton \simeq 1/7$.
Observationally, \hef\ is inferred from helium and hydrogen
recombination lines measured by now in more than 80 extragalactic HII
regions of low-metallicity. Following~\cite{Steigman:2007xt}, the
estimate for primordial mass fraction reads
$Y_\primordial = 0.240 \pm 0.006$ where the large adopted error
reflects the fact that systematic uncertainties may well dominate;
cf.~\cite{Steigman:2007xt} and references therein. Though the value is
somewhat low there is currently no clear discrepancy with its SBBN
prediction, the latest one reading $Y_\primordial = 0.2486 \pm
0.0002$~\cite{Cyburt:2008zz}.  It should be noted, however, that \hef\
is a poor baryometer varying only logarithmically
with~\etabaryon. Contrariwise, being very sensitive to
$\neutron/\proton$ and thus to the Hubble rate, \hef\ acts as a
powerful discriminator between models predicting additional
relativistic degrees of freedom at the onset of BBN.

Among the most generic ways how physics beyond the Standard Model can
affect the output of BBN are, e.g., a change in timing of the
reactions caused by new contributions to the Hubble expansion rate,
non-thermal nuclear reactions from late decays and annihilation of
heavy particles, and the thermal catalysis of nuclear reactions caused
by electromagnetic or strongly interacting relics. In this regard, the
stable lithium isotopes have attracted much attention because they
turn out to be very sensitive
on the latter two effects.

Standard BBN has a long-standing lithium problem. Let us first discuss
the heavier and more stable isotope \lisv. At $\etacmb$ it is produced
mainly in form of \bes\ via $\het + \hef \to \bes +\gamma$ which then
beta decays via electron capture into \lisv\ after BBN. The cross
section for the \bes\ fusion also dominates the error on the SBBN
prediction.  With a recent update of the reaction cross
section~\cite{Cyburt:2008zz} the authors tighten the SBBN prediction
to
\begin{align}
  \label{eq:lisv-SBBN-prediction}
  \lisv / \Hyd |_{\primordial} = \left( 5.24^{+0.71}_{-0.67}
  \right)\times 10^{-10} .
\end{align}
Lithium is observed in absorption spectra in the atmospheres of
metal-poor stars in the galactic halo as well as in stars of galactic
globular clusters. A link between the measured \lisv\ with a
primordial origin was first promoted in~\cite{Spite:1982dd}. What has
become known as the ``Spite-plateau'' was an observed constant lithium
abundance of $A(\mathrm{Li}) = 2.05 \pm 0.15$ in halo dwarfs of
low metallicity%
\footnote{ $[\mathrm{Fe}/\Hyd] = \log_{10}{(\mathrm{Fe}/\Hyd)} -
  \log_{10}{(\mathrm{Fe}/\Hyd)}_\odot$ with the solar abundance
  $\log_{10}{(\mathrm{Fe}/\Hyd)}_\odot \simeq
  -4.55$~\cite{2005ASPC..336...25A}.}
$-2.4 \leq[\mathrm{Fe}/\Hyd] \leq -1.1$ and which corresponds to
$\lisv / \Hyd = (1.12^{+0.46}_{-0.33})\times 10^{-10}$ using
${A(\mathrm{Li})}\equiv \log_{10}{(\mathrm{Li}/\Hyd)} + 12$.
Ever since much work has been done and other groups found similar
values so that there seems to be a clear discrepancy with the SBBN
prediction~(\ref{eq:lisv-SBBN-prediction}) being a factor of a few too
high. Indeed, the indication of a correlation of \lisv\ with
$[\mathrm{Fe}/\Hyd]$~\cite{Ryan:1999jq,Asplund:2005yt} tilts the
plateau so when extrapolating to smallest metallicities values as low
as $\lisv / \Hyd = 6.3\times 10^{-11}$~\cite{2007A&A...462..851B} have
been inferred. Moreover, such an increasing discrepancy is not
alleviated by the most recent observation that the \lisv\ abundance in
extremely metal-poor stars with $[\mathrm{Fe}/\Hyd] < -3$ is on
average $0.2\ \dex$ lower than in those (plateau) stars of higher
metallicity~\cite{Aoki:2009ce}.%
\footnote{\dex\ denotes the decimal exponent. For example, from
  $A(\mathrm{Li}) = 2.0$ to $A(\mathrm{Li}) = 1.8$ is
  $0.2\ \dex$.}
In this work we will not touch the \lisv\ problem. Instead, we
concentrate much of our attention to the second stable lithium
isotope~\lisx.

The measurements of \lisx\ in the atmospheres of old stars of low
metallicity are extremely difficult with only one firm detection in
the
1990s~\cite{1993ApJ...408..262S,1994ApJ...428L..25H,1998ApJ...506..405S,1999A&A...343..923C}
whereas other measurements of \lisx\ have changed into upper limits;
cf. \cite{2007A&A...462..851B} and references therein. More recently
\lisx\ has been observed in 9 more halo dwarfs with $ -3
\leq[\mathrm{Fe}/\Hyd] <- 1$ showing a similar isotopic ratio of
$\lisx/\lisv$ of $\sim 5\%$~\cite{Asplund:2005yt}.
This is tantalizing because it suggests the existence of a \lisx\
plateau mirroring the one for~\lisv. At first glance, this points to a
primordial origin of \lisx\ at the level of $\lisx/\Hyd|_{\primordial}
\sim \mathrm{few}\times 10^{-12}$. However, the story is complicated
by the fact that lithium is produced in galactic cosmic rays and may
as well have undergone stellar depletion. Whereas in standard stellar
models \lisv\ depletion is negligible
\cite{1990ApJS...73...21D,1992ApJS...78..179P}, \lisx\ is more fragile
and particularly destruction in proton burning is more
efficient. Indeed, non-standard models leading to lithium destruction,
e.g., from inward diffusion or from rotationally induced mixing have
been considered, trying to reconcile \lisv\ observations with its SBBN
prediction; cf.~\cite{Asplund:2005yt} and references therein.  However,
the absence of significant scatter in the stars of the Spite-plateau
demands a uniform depletion thus putting strong constraints on any of
such mechanisms. When considering upper bounds on the primordial
\lisx\ abundance many papers adopt values in the range
\begin{align}
\label{eq:Lisx-obs-upperbound-range}
  \lisx/\Hyd |_{\primordial} \leq 10^{-11} \div  10^{-10} .
\end{align}
Comparing this with the SBBN output $\lisx/\Hyd |_{\primordial} \sim
10^{-14}$ (see Sec.~\ref{sec:catalys-lisx-prod}) this isotope shows a
gaping discrepancy between prediction and observation; we refer the
reader to~Sec.~\ref{sec:typical-stau-abundance} for a further
discussion.

The lithium problem(s) has (have) particularly inspired non-standard
BBN scenarios seeking their solution. Most notably in this regard are
the possibility of the late-decay of a massive particle species [see
Sec.~\ref{sec:particle decays during BBN}] and the catalysis of
nuclear reactions; see Chap.~\ref{cha:catalyzed-bbn}.
In this thesis we will exclusively consider physics beyond the
Standard Model with a weak-scale long-lived CHAMP which we call
$\X^\pm$. We shall see that $\X$-decays as well the catalysis of
nuclear reactions due to the presence of \champ\ during BBN will pose
strong constraints on the CHAMP abundance/lifetime parameter space.

\section{\texorpdfstring{Typical CHAMP abundances}{Typical champ abundances}}
\label{sec:typic-champ-abund}

Let us assume that the temperature of the primordial plasma was $T \gg
\mx / 20$ with $\mx\gtrsim\Orderof{100\ \GeV}$ denoting the mass of
\X.  Then, \X\ has once been tracking an equilibrium abundance. With
dropping temperature, \X\ cannot maintain thermal equilibrium so that
the species freezes-out.%
\footnote{For a low reheating temperature scenario where \X\ may not
  achieve thermal equilibrium see \cite{Takayama:2007du}.}
This happens approximately at the time when the rate of
\X-annihilation drops below the Hubble expansion rate $H(T)$.

The key to the freeze-out abundance of \X\ lies in considering the
Boltzmann equation for the total \X\ number density $n_\X = n_{\X^+}+
n_{\X^-}$,
\begin{align}
  \label{eq:boltzmann-champ-in-number}
  \frac{d n_\X }{d t } 
  + 3 H n_\X = - \sigmavof{\mathrm{ann}} \left[ n_\X^2
  - (n_\X^{\mathrm{eq}})^2\right] .
\end{align}
The Hubble rate is given by 
\begin{align}
  \label{eq:hubble-rate}
  H(T)= \sqrt{\frac{\pi^2 \geff(T)}{  90}}\, \frac{T^2 }{ \MPl}
\end{align}
with $\geff$ radiation degrees of freedom and $\MPl$ denoting
the (reduced) Planck mass $\MPl\simeq 2.4\times 10^{18}\ \GeV$. The
quantity $\sigmavof{\mathrm{ann}}$ is found upon a thermal average of
the total annihilation cross section~$\sigma_{\mathrm{ann}} $ times
the ``relative velocity'' $v$; for details on the exact definition of
$\sigma_{\mathrm{ann}} $, $\sigmavof{\mathrm{ann}}$, and $v$ we refer
the reader to Sects.~\ref{sec:prim-ann} and~\ref{sec:energy-transf-binary}.  The equilibrium number density is
denoted by~$n_\X^{\mathrm{eq}}$.

For a relic species it is customary to scale out the dilution of the
number density due to the expansion of the Universe.  We define the
yield variable $Y_\X $ by dividing $n_\X$ by the entropy density 
$s(T) = 2\pi^2\,\heff(T)\,T^3/45 $ where $\heff$ is an effective
degrees of freedom parameter~\cite{Gondolo:1990dk}. In absence of
\X-destroying or -producing events and as long as no entropy is
released, $ Y_\X \equiv n_\X/s$ remains
constant. From~(\ref{eq:boltzmann-champ-in-number}) one then finds
\begin{align}
  \label{eq:boltzmann-champ-in-Y}
  \frac{d Y_\X}{d T} = \sqrt{ \frac{8 \pi^2 g_*(T)}{45}} \MPl
  \sigmavof{\mathrm{ann}} \Big[ Y_\X^2 - (Y^\mathrm{eq}_{\X})^2 \Big]
\end{align}
where~\cite{Gondolo:1990dk}
\begin{align} 
  \label{eq:gstar}
  g_*^{1/2} = \frac{\heff}{\sqrt{\geff}}\left( 1 +
    \frac{1}{3}\frac{T}{\heff}\frac{d\heff}{dT} \right) .
\end{align}

The exact solution of~(\ref{eq:boltzmann-champ-in-Y}) can be rather
involved and for the case where $X$ is the lighter stau,
$\X=\stauone$, this is presented in great detail in
Part~\ref{part:three} of this thesis.
Nevertheless, in order to get a feeling for the expected abundances of
an electromagnetically charged relic we can employ a simplified
treatment of decoupling which is based on the
non-relativistic limit for \X.%
\footnote{We disregard here effects on $Y_\X$ such as coannihilation,
  annihilation on the threshold, or resonant
  annihilations~\cite{Griest:1990kh}; see, however,
  Part~\ref{part:three}.}
In this limit, the equilibrium number density is given by
\begin{align}
  \label{eq:non-rel-eq-numberdensX}
  n^{\mathrm{eq}}_{\X} = g_\X \left( \frac{\mx T}{2 \pi} \right)^{3/2}
  e^{-\mx / T} \ .
\end{align}
and the thermally averaged cross section may be written
as~\cite{Griest:1990kh,Drees:2004jm}
\begin{align}
  \label{eq:thavg-non-relat}
  \sigmavof{\mathrm{ann}}_{\mathrm{n.r.}} \simeq
  \frac{1}{2\sqrt{\pi}}\left(\frac{\mx}{T}\right)^{3/2}
  \int_{0}^{\infty} dv\, v^2 ( \sigma_{\mathrm{ann}} v )\, e^{-\mx v^2 /
    4 T}
\end{align}

To find the (approximate) decoupling temperature $\Tf$ we equate
$ \sigmavof{\mathrm{ann}}\, n^{\mathrm{eq}}_{\X}(\Tf) = H(\Tf) $. With
the notation $x_{\mathrm{f}} = \mx/\Tf$ this yields the standard
result
\begin{align}
  x_{\mathrm{f}} \simeq \ln{\left( \frac{\sqrt{45 } x_{\mathrm{f}}^{1/2}
        \sigmavof{\mathrm{ann}} \, g_\X\, \mx \MPl }{ 2\, \pi^{5/2}\,
        {\geff}^{1/2} } \right) } \ .
\end{align}

Let us assume an \X\ annihilation cross section expanded in powers of
$v$, $\sigma_{\mathrm{ann}} v \simeq a + b v^2$ so that
with~(\ref{eq:thavg-non-relat}) $\sigmavof{\mathrm{ann}}$ develops the
form $\sigmavof{\mathrm{ann}}_{\mathrm{n.r.}}\simeq a + 6
b\,T/\mx$. Choosing $a = \alpha^2 /\mx^2 $ on dimensional grounds%
\footnote{For example, the cross section for annihilation into two
  photons reads $\sigma_{\gamma\gamma} v \simeq 2 \pi \alpha^2 /
  \mx^2$~\cite{Asaka:2000zh}. When considering the total $\X^\pm$
  abundance this gives $\sigma_{\mathrm{ann}} =
  \sigma_{\gamma\gamma}/2$ and hence $\sigmavof{\mathrm{ann}} = \pi
  \alpha^2 / \mx^2 + \Orderof{\mx/T}$.}
and considering $s$--wave annihilation, \mbox{$(b \ll a)$} we find
numerically $ x_{\mathrm{f}} \simeq 26 \div 24$ for $\mx = 0.1\div 1\
\TeV $ and $g_\X = 2$. This gives the abundance at the time of
chemical decoupling, $ Y_\X(\Tf) \simeq n^{\mathrm{eq}}_{\X}(\Tf) /
s(\Tf) $. For $T < \Tf$ we can neglect $Y^\mathrm{eq}_{\X}$
in~(\ref{eq:boltzmann-champ-in-Y}) so that when accounting for
residual \X\ annihilation from $\Tf$ to $T_0$ one finds for the
inverse of the freeze-out yield
\begin{align}
  \label{eq:relic-yield-estimate}
  \frac{1}{Y_\X(T_0)} = \frac{1}{Y_\X(\Tf)} + \sqrt{ \frac{8\pi^2}{45}
  } \MPl \int^{\Tf}_{T_0} dT\, \sqrt{g_*(T)} \sigmavof{\mathrm{ann}} 
\end{align}
with $T_0 = 2.725\ \mathrm{K}$~\cite{Yao:2006px} denoting the present
day photon temperature. Whenever we write $Y_\X$ in the following,
$Y_\X(T_0)$ is understood, i.e., the yield of the species \X\ it would
have had today had it not decayed.

Taking into the account the temperature dependence of $g_{*}$ by
interpolating the tabulated values in~\cite{Gondolo:2004sc} and
integrating~(\ref{eq:relic-yield-estimate}) yields the following
estimate on the $\X$ abundance
\begin{align}
  \label{eq:upper-bound-YX}
  Y_\X \lesssim 10^{-12} \left(\frac{\mx}{100\ \GeV} \right)
\end{align}
with an approximate linear scaling in $\mx$.  Note that an increase in $
\sigmavof{\mathrm{ann}}$ contributes linearly to ${Y_\X(T_0)^{-1}}$,
provided $x_{\mathrm{f}} = \mathrm{const}$.
Therefore, we have indicated that $Y_\X$ in~(\ref{eq:upper-bound-YX})
is a value more towards the upper end, corresponding to a guaranteed
annihilation cross section of electromagnetic strength. A stronger
coupling will allow \X\ to stay longer in equilibrium, thus receiving
an additional Boltzmann suppression.

Conversely, we can constrain $ Y_\X $ from below by assuming the
maximum cross section of mutual $\X^\pm$ annihilation which is given
by the unitarity limit~\cite{nla.cat-vn2263194}, $ \sigma_{\mathrm{u}} = \pi \lambdabar^2 $
($s$-wave); $\lambdabar$~denotes the de~Broglie wavelength of the
relative motion. Using $\lambdabar = 1 / (\mred v) $ together with the
reduced mass $\mred =\mx/2 $ one finds with~(\ref{eq:thavg-non-relat})
\begin{align}
  \sigmavof{\mathrm{ann,u}}_{\mathrm{n.r.}}
  = \frac{4\sqrt{\pi}}{\mx^2} \sqrt{\frac{\mx}{ T}} .
\end{align}
Employing this cross section yields an estimate on the smallest
possible freeze out abundance for a weak scale electromagnetically
charged relic. Using $\mx=100\ \GeV$ gives $ x_{\mathrm{f}} \simeq 40 $
and
\begin{align}
  \label{eq:lower-bound-YX}
   Y_\X \gtrsim 10^{-18} .
\end{align}

Let us see how this lower limit on the decoupling yield compares with
experimental bounds on charged cosmological relics from (negative)
searches of anomalous heavy isotopes of ordinary nuclei. For example,
in~\cite{Hemmick:1989ns} severe limits on the concentration of $\X^+$
in form of heavy hydrogen as well as $\X^-$ in low $Z$--nuclei have
been obtained for a weak scale relic in the mass range $100\ \GeV \leq
\mx \leq 10\ \TeV $. It was found that present day abundances in
excess of
\begin{align}
\label{eq:exp-limit}
  Y_{\X^+\!,\X^-} < 10^{-25}
\end{align}
are firmly excluded.%
\footnote{We have obtained the constraint from the right end of the
  ${}^{14}$C-line in Fig.~7 of~\cite{Hemmick:1989ns} and changed the
  normalization of $n_\X$ from baryon number to entropy. For a recent
  compilation of other such limits along with a thorough investigation
  of the decoupling yield of a generic electromagnetically- or
  color-charged particle species confer~\cite{Berger:2008ti}.}
Note that individual limits on either $\X^-$ or $\X^+$ exist for
masses ranging from a few \GeV\ to multi-\TeV\ and they can be
stronger than~(\ref{eq:exp-limit}) by more than ten orders of
magnitude; cf.~\cite{Yao:2006px}.  Comparing (\ref{eq:lower-bound-YX})
with (\ref{eq:exp-limit}) forces us to conclude that, under our
assumption of a standard cosmological evolution, $\X^\pm$ cannot be
stable.  Considering a charged thermal relic of finite lifetime with
abundances in the range~(\ref{eq:lower-bound-YX})
to~(\ref{eq:upper-bound-YX}) sets the stage for our further
investigations.

In the next chapter we shall also quantify the \X\ abundance
normalized to baryon number~$\nb$ instead of entropy. We define
\begin{align}
  X \equiv \frac{n_\X}{\nb} = \frac{Y_\X}{\etabaryon}
  \frac{s(T_0)}{n_{\gamma}(T_0)} \simeq 1.13\times 10^{10}\, Y_\X ,
\end{align}
where $n_\gamma = 2\, \zeta(3)\, T^3 / \pi^2$ is the photon number
density with $\zeta(x)$ denoting the Riemann Zeta function.  We have
chosen this notation in order to clearly distinguish the two different
normalizations and it will be clear from the context whether \X\
denotes the particle itself or its abundance. The previous estimates
(\ref{eq:upper-bound-YX}) and (\ref{eq:lower-bound-YX}) then translate
into
\begin{align} 
  \label{eq:abundance-estimate-X-norm-to-baryons}
  10^{-8} \lesssim X \lesssim 10^{-2} \left(\frac{\mx}{100\ \GeV}
  \right) .
\end{align}

It is also instructive to compare the \X\ abundance with that of
\hef. This will be of some importance in the discussion of catalytic
BBN effects where bound states of \hef\ with \champ\ play a key role.
Since to a very good approximation it follows from
(\ref{eq:def-hef-mass-fraction}) that
\begin{align}
  \hef\equiv n_{\hef}/\nb = Y_\primordial /4 \simeq 0.06
\end{align}
we see from (\ref{eq:abundance-estimate-X-norm-to-baryons}) that we
can expect that the number density of \X\ is typically smaller than
that of \hef\ unless \X\ is rather heavy; here, $600\ \GeV$ but in
concrete particle physics models with \X\ annihilating via a number of
channels, a heavier \X\ is required.
Indeed, when focusing on the particle content of the minimal
supersymmetric Standard Model (MSSM) (plus a gravitino LSP) with
$\stauone = \X$, the \X-abundance is determined by the standard
chemical decoupling and $\X \lesssim \hef$ holds unless
$m_{\X}\gtrsim\mathcal{O}(4\ \TeV)$; see
Sec.~\ref{sec:typical-stau-abundance}.

\section{Particle decays during BBN}
\label{sec:particle decays during BBN}

Using primordial nucleosynthesis as a consistency check
for the existence of long-lived particles \X\ has a long-standing
history.%
\footnote{In this section \X\ stands for an arbitrary, not
  necessarily electromagnetically charged, long-lived species.}
When exotics decay during or after BBN electromagnetic and/or hadronic
energy is injected into the plasma. Depending on timing, energy
deposition, and abundance of the decaying species, the light element
output can be affected significantly. The comparison with the
observational bounds yields constraints on the parameter space of \X.
From early works, e.g., \cite{Reno:1987qw,Dimopoulos:1988ue}, to
elaborate studies~\cite{Cyburt:2002uv,Jedamzik:2004er,Kawasaki:2004qu}
this is still an active field of research of continuing refinement and
increasing sophistication; for a most recent work see,
e.g,~\cite{Bailly:2008yy}. Since we will also make use of BBN
constraints on electromagnetic and hadronic energy release, we provide
here a cursory overview pointing out important features. For a more
detailed exposition of the topic we refer the reader
to~\cite{Cyburt:2002uv,Jedamzik:2004er,Kawasaki:2004qu} and references therein.

\begin{description}
\item[Electromagnetic constraints] The emerging constraints can be
  classified with respect to the decay mode of the exotic
  particle. When \X\ decays radiatively into primary high-energy
  photon(s) and/or electrons (charged leptons) an electromagnetic
  cascade is induced. The important processes are $e^\pm$ pair
  creation $(\gamma + \gamma_{\mathrm{bg}} \to e^- + e^+$),
  photon--photon scattering $(\gamma + \gamma_{\mathrm{bg}}\to \gamma
  + \gamma)$, Compton scattering $(\gamma + e_{\mathrm{bg}} \to \gamma
  + e)$, inverse Compton scattering $( e +\gamma_{\mathrm{bg}} \to e +
  \gamma )$, and pair creation on nuclei $(\gamma +
  \nuc{}_{\mathrm{bg}} \to e^+ + e^- + \nuc{})$. The subscript 'bg'
  denotes the particles which are in equilibrium with the plasma.
Since $n_{\gamma}/n_{e}|_{\mathrm{bg}}\sim 10^{10}$ the scattering
on background photons is very frequent. This leads to an efficient
thermalization of the cascade so that destruction of light elements
does not happen frequently.
However, once energetic photons are degraded below $E_\gamma \lesssim
\melectron^2/22 T$ \cite{Kawasaki:1994sc} they loose their ability to
pair create $e^\pm$ on $\gamma_{\mathrm{bg}}$. The soft photons of the
associated 'break-out' spectrum are then capable to efficiently
destroy those light elements whose binding energy
$\Ebind^{\mathrm{nuc}}$ lies below the threshold of electron pair
creation. For \Deut\ ($\Ebind^{\mathrm{nuc}} = 2.22\
\MeV$~\cite{Audi:2002rp}) this happens at $T\lesssim 10\ \keV$ whereas
$\hef$ ($\Ebind^{\mathrm{nuc}} = 28.3\ \MeV$~\cite{Audi:2002rp}) is destroyed when the
photon temperature drops below $T\lesssim 1\ \keV$. This corresponds
to respective cosmic times of $t\gtrsim 10^4\ \seconds$ and $t\gtrsim
10^6\ \seconds$ when thermal nucleosynthesis reactions have long
frozen out.

\begin{figure}[t]
\centerline{
\includegraphics[viewport=0 0 480 485,clip=true,width=0.5\textwidth]{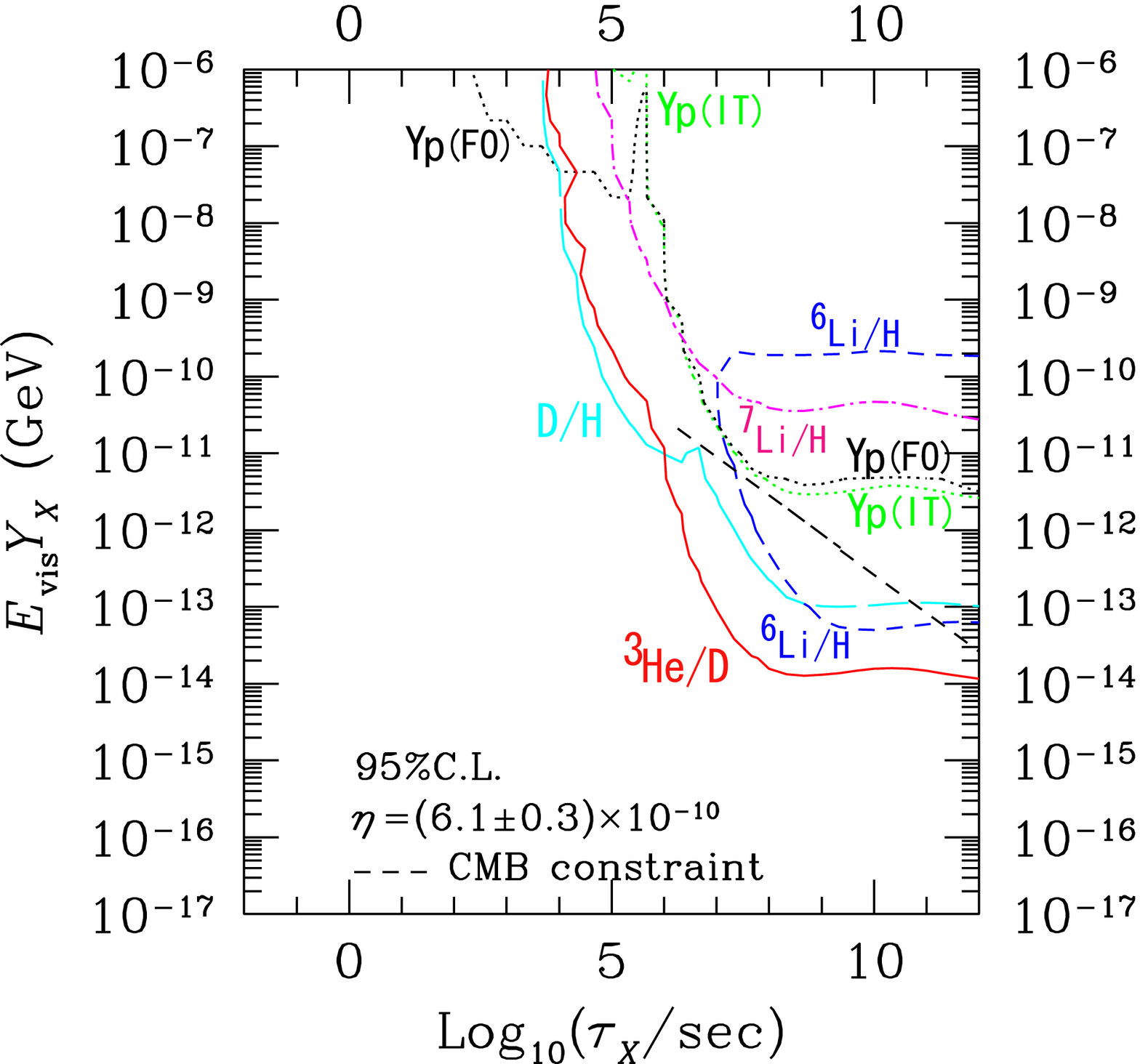}%
\includegraphics[viewport=0 0 480 485,clip=true,width=0.5\textwidth]{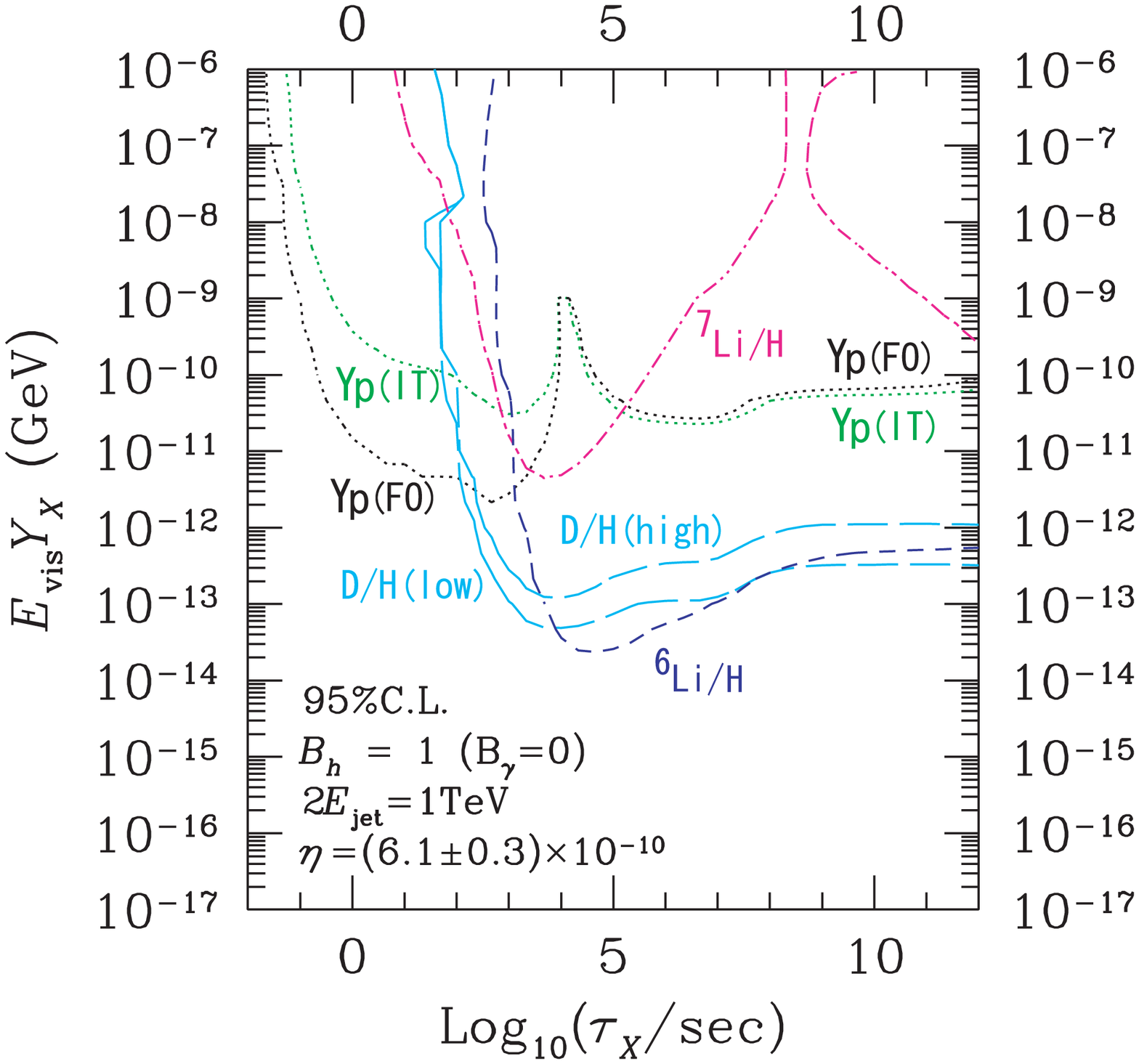}
}
\caption[Electromagnetic and hadronic BBN constraints]{
   Constraints on the electromagnetic (\textit{left figure}) and
  hadronic (\textit{right figure}) energy release from \X-decays for $\mx=1\
  \TeV$; the figures are taken from \cite{Kawasaki:2004qu}. The lines
  represent upper limits on the quantity $E_{\mathrm{vis}} Y_\X$,
  i.e., on the ``visible energy'' $E_{\mathrm{vis}} $ released per
  decay times their abundance prior to decay, $Y_\X$, and are plotted
  as a function of $\tauX$; see main text for discussion.}
\label{Fig:EM-HAD-BBN-constr} 
\end{figure}

Constraints on the electromagnetic energy release for $\mx=1\ \TeV$
are shown in the left Fig.~\ref{Fig:EM-HAD-BBN-constr} which is taken
from \cite{Kawasaki:2004qu}. The lines represent upper limits on the
quantity $E_{\mathrm{vis}} Y_\X$, i.e., on the ``visible energy''
$E_{\mathrm{vis}} $ released per decay times their abundance prior to
decay, $Y_\X$, and are plotted as a function of $\tauX$. Two dominant
constraints are visible.
For $\tauX \lesssim 10^6\ \seconds$ the most restrictive constraint,
labeled $\deut/\Hyd$, arises from the destruction of deuterium below
its observationally inferred primordial level. For larger $\tauX$,
however, \hef\ gets dissociated and $\deut$ (along with \het\ and
\trit) is also created. Since the \hef-target is very abundant,
\deut\ is indeed overproduced for $\tauX \gtrsim 10^6\ \seconds$.
Moreover, the combination $\het/\deut$ is then always produced
yielding the most stringent constraint on electromagnetic energy
release for $\tauX \gtrsim 10^6\ \seconds$.

\item[Hadronic Constraints]A second class of constraints on decaying
  \X\ during/after BBN comes from hadronic energy release into the
  plasma. For example, even if \X\ dominantly decays into photons, a
  non-vanishing hadronic branching ratio is expected from the
  conversion of a (virtual) photon into a quark-antiquark pair or from
  charged meson production on background photons (if kinematically
  allowed).
%
The partons emitted in the decay are quickly hadronized and the highly
energetic fragmentation products such as protons~(\proton),
neutrons~(\neutron), as well as their antiparticles are released. Also
long-lived mesons, namely, charged pions~$(\pi^\pm)$ and kaons
($K^\pm,\, K^0_{\mathrm{L}}$), with lifetimes $\Orderof{10^{-8}\
  \seconds}$ have a chance to interact with background nuclei before
decaying. 
Once an energetic hadron scatters on a background nucleus, essentially
$\proton$ or $\hef$, a hadronic shower is induced. In particular,
\hef\ may be destroyed with secondaries further participating in
interactions with the plasma constituents.

The energetic charged hadrons~$\nuc{}$ are downgraded in energy via
electromagnetic interactions, most importantly, by Coulomb scattering
$(\nuc{} + e^\pm_{\mathrm{bg}} \to \nuc{} + e^\pm)$, Compton
scattering $(\nuc{} + \gamma_{\mathrm{bg}} \to \nuc{} + \gamma )$, and
Bethe-Heitler scattering $( \nuc{} + \gamma_{\mathrm{bg}} \to \nuc{} +
e^+ + e^-)$. Injected neutrons loose their energy mainly by their
magnetic-moment interaction with $e^\pm_{\mathrm{bg}} $.
It is clear that emergent constraints on the hadronic energy release
will sensitively depend on the competition between the rate for
hadronic scattering and the rates for (electromagnetic) thermalization
and/or decay (of unstable particles).
Moreover, even after hadrons are stopped they may still induce
neutron-to-proton interconversion processes~\cite{Reno:1987qw}.

In the right Fig.~\ref{Fig:EM-HAD-BBN-constr} which is taken from
\cite{Kawasaki:2004qu} the constraints on $E_{\mathrm{vis}}Y_\X$ due
to hadro-dissociation as well as \neutron-\proton\ interconversion are
shown for a particle with $\mx = 1\ \TeV$ and hadronic branching ratio
$B_\mathrm{h} =1$. The effects from photo-dissociation are not
included. Note that this is an unrealistic situation since the hadron
stopping process itself as well as meson decays induce electromagnetic
showers.
For $\tauX\lesssim 100\ \seconds$, i.e., for $T\gtrsim 100\ \keV$, the
emitted hadrons essentially deposit all their kinetic energy
electromagnetically before interacting with the background
nuclei. However, interconversion processes which always lead to an
increase of $\neutron/\proton$ enhances the \hef\ output. The
associated constraints from \hef\ overproduction for two different
observationally adopted limits on the primordial mass fraction are
shown by the dotted lines labeled $Y_{\mathrm{p}}$. For larger
lifetimes, $\tauX\gtrsim 100\ \seconds$ ($T\lesssim 100\ \keV$),
mesons typically decay before interacting hadronically. However, the
stopping power for protons and neutrons rapidly decreases with
dropping temperature so that \hef\ is destined for being
destroyed. This yields the stringent $\deut/\Hyd$ constraint on
hadronic energy release for $\tauX\gtrsim 100\ \seconds$. Moreover, a
small fraction of the energetic spallation products $\trit$ and \het\
can scatter again on ambient \hef\ producing \lisx\
\cite{Jedamzik:1999di} (and \lisv). This non-thermally induced fusion
reaction gives the hadronic constraint labeled $\lisx/\Hyd$ in
Fig.~\ref{Fig:EM-HAD-BBN-constr}.  Since \lisx\ is efficiently
destroyed in (thermal) proton burning for temperatures $T\gtrsim 10\
\keV$ the constraint becomes the dominant one only for $\tauX\gtrsim
10^4\ \seconds$~\cite{Jedamzik:2004ip}.
\end{description}

So far, the discussion has been completely generic with our ignorance
on the nature of \X\ parameterized by $E_{\mathrm{vis}}$.
Constraining the particle's parameter space requires the specification
of the couplings of \X\ to Standard Model particles as well as its
mass~\mx. This allows for the determination of the decay modes of \X\
along with the computation of the associated average electromagnetic
and hadronic energy $E_{\mathrm{vis}}$ released per decay. Moreover,
the freeze-out abundance~$Y_\X$ can be calculated so that plots like
in Fig.~\ref{Fig:EM-HAD-BBN-constr} can be employed to constrain the
model.
In part~\ref{part:two} of the thesis we incorporate the most stringent
of the constraints for the case of a decaying stau in the gravitino
dark matter scenario. There, we also provide more details as soon the
problem of inclusion of such constraints becomes acute.


\section[A critical look at hadronic constraints for $T\lesssim 100\
\keV$]{\texorpdfstring{A critical look at hadronic constraints for
    \boldmath$T\lesssim 100\ \keV$ }{A critical look at hadronic BBN
    constraints}}
\label{sec:stopping power}

In the previous section we have noted that for $T\lesssim 100\ \keV$
($\tauX \gtrsim 100\ \seconds$) the BBN constraints on hadronic energy
release sensitively depend on the competition between the hadronic and
electromagnetic scattering rate (and potentially the lifetime). If
injected nucleons as well as their secondaries---either produced in
spallation reactions or ``up-scattered'' in elastic scatterings---are
predominantly thermalized by interactions on background nucleons or
nuclei, those constraints become stringent.
Underestimating the stopping power due to electromagnetic interactions
would lead to overly restrictive bounds on the hadronic energy
release.

For example, in the last section we have seen that the non-thermal
production of \lisx\ due to the energetic spallation debris \trit\ and
\het\ of destroyed \hef\ yields the dominant hadronic constraint for
$\tauX\gtrsim 10^4\ \seconds$. The reactions involved are $\trit +
\hef|_{\mathrm{bg}} \to \lisx + \neutron$ and $\het +
\hef|_{\mathrm{bg}} \to \lisx +\proton $~\cite{Jedamzik:1999di}. For
$T\lesssim 30 \keV$, i.e., for $t\gtrsim 10^3\ \seconds$, inverse
Compton scattering on background photons cannot prevent
\textit{low-energy} hadronic interactions~\cite{Reno:1987qw} above the
lithium formation threshold $[\Orderof{10\ \MeV}]$. The dominant
electromagnetic degradation mechanism is then Coulomb
scattering. However, the rapidly diminishing number of background
electrons (positrons) with dropping temperature also renders the
energy loss by Coulomb scattering increasingly inefficient.
Furthermore, in~\cite{Jedamzik:2004er} it is claimed that the non-thermal
\lisx\ output is boosted by a factor of ten because of a peculiarity
in the Coulomb process: Once the velocity $\beta$ of the energetic
mass-three nuclei drops below the thermal electron velocity
$\VEV{\beta_e}$, the stopping power seems to be strongly
suppressed.
This observation was first made in~\cite{Reno:1987qw}.

In light of these comments a critical look on the Coulomb stopping
process is warranted. We shall pay particular attention to the
velocity dependence of the cross sections, i.e., on $\beta$
and~$\VEV{\beta_e}$. In Sec.~\ref{sec:energy-transf-binary} (and
partly also in Sec.~\ref{sec:hadr-electr-scatt}) results from the
literature are reconciled. In Sec.~\ref{sec:cutoff-cons} we focus on
the stopping of charged hadrons and incorporate the proper
screening-prescription of the Coulomb interaction. In
Sec.~\ref{sec:discussion-coul-stop} we discuss the obtained results.

\subsection{Energy transfer in binary collisions}
\label{sec:energy-transf-binary}

A thorough investigation of the electromagnetic stopping of hadrons in
the context of primordial nucleosynthesis has first been presented
in~\cite{Reno:1987qw}. Indeed, the treatments in
\cite{Kawasaki:2004yh,Kawasaki:2004qu} (see
Fig.~\ref{Fig:EM-HAD-BBN-constr}) employ the results of that work.
Since the stopping power sensitively depends on the velocities of the
incident hadron and the target particles, we first reconcile the
general result on the energy transfer obtained
in~\cite{Reno:1987qw}. Though we encounter minor disagreements they
turn out to be without relevance.

Our starting point is the rate of energy loss due to binary
scatterings (A1) of Ref.~\cite{Reno:1987qw}
\begin{align}
  \label{eq:energy-loss-master}
  \frac{dE}{dt} = \sum_j \frac{g_j}{(2\pi)^3} \int  d\Omega\,
  d^3\mathbf{p}_j \, f_j(\mathbf{p}_j)
  \left[ 1\mp f_{jf}(\mathbf{p}_{jf}) \right] \Delta E_j
  \frac{d\sigma_j}{d\Omega} \vmol\, ,
\end{align}
where $\Delta \Ej$ denotes the energy transfer between the incident
hadron and a (background) particle species $j$ with three-momentum
$\mathbf{p}_j$ and $g_j$ internal degrees of freedom. The transfer is
weighted by the center-of-mass (CM) cross section
${d\sigma_j}/{d\Omega}$ and averaged over inital and final state
distribution functions $f_j$ and $f_{jf}$, respectively.
A subtle point is the appearance of the M\o ller
velocity~\cite{Gondolo:1990dk}
\begin{align}
  \label{eq:Moeller-velocity}
  \vmol \equiv \frac{F}{E \Ej} = [ (\bm{\beta} - \bm{\beta}_j)^2 - (\bm{\beta}
 \times \bm{\beta}_j) ]^{1/2} 
\end{align}
which is the relativistic generalization of the conventional relative
velocity $\vrel = |\bm\beta - \bm\beta_j|$.  The respective velocities
of the hadron and the target are given by $\bm{\beta} = \mathbf{p}/E$
and $\bm{\beta}_j = \mathbf{p}_j/E_j$ and $F = [(p\cdot\pj)^2 - m_j^2
M^2]^{1/2}$ denotes the Flux-factor.  Only in the CM frame or in the
rest frame of one of the incoming particles $\vmol$ coincides
with $\vrel$. We stress that $p= (E,\mathbf{p})^T$ and $\pj =
(\Ej,\mathbf{p}_j)^T$ denote the respective four-momenta of the
energetic nucleus and of the ambient target particle in the rest frame
of the thermal bath.  In that frame, and when $j$ is in thermal
equilibrium, the distribution functions $f_j$ take on their familiar
form: $f_{j(f)} = [\exp{(E_{j(f)}/T)\pm 1}]^{-1}$. The upper signs
in~(\ref{eq:energy-loss-master}) and in the last expression refer to
fermions, the lower to bosons. Finally, $E_{jf}$ is the energy of the
target after scattering and $m_j$ ($M$) is the mass of the target
(indicent nucleus).

The energy transfer $\Delta \Ej = \Ejf - \Ej$ can be obtained by a
series of Lorentz transformations: Since the scattering is elastic, in
the CM frame we have $\Ejcm = \Ejfcm$. Thus, we can obtain $\Ejf$ by a Lorentz
transformation $\Lambda=\Lambda_3\Lambda_2\Lambda_1$ of $\pj$ into the
CM frame followed by an inverse transformation of $\pjfcm$
back. $\Lambda$ is broken up as follows:%
\footnote{The explicit forms of $\Lambda_1$, $\Lambda_2$, and
  $\Lambda_3$ are given in the Appendix~\ref{sec:appendix-LT} at the
  end of this chapter.}
We choose $\bm{\beta}$ to lie along the $z$-axis and to have an angle
$\alpha$ with $\bm{\beta}_j\in yz$-plane
\begin{align}
  p = (E,0,0,\beta E)^T \quad \mathrm{and} \quad 
  p_j = (\Ej,0,\bj\Ej\sin{\alpha},\bj\Ej\cos{\alpha})^T
\end{align}
so that $\alpha = \pi$ corresponds to a ``head-on-head'' collision.
Boosting into the rest frame of the incident nucleus gives
\begin{align}
\pj'=\Lambda_1 \pj = 
\begin{pmatrix}
  \gb \Ej (1 - \beta \bj \cos{\alpha}) \\ 0 \\ \bj\Ej\sin{\alpha} \\
  \gb \Ej ( \bj \cos{\alpha} - \beta )
\end{pmatrix}
= ( \Ej', \mathbf{p}_j')^T 
\end{align}
where $\gb = (1-\beta^2)^{-1/2}$. Under Lorentz transformations the
M\o ller velocity changes as~\cite{Gondolo:1990dk}
\begin{align}
  \vmol' = \vmol
  \frac{1-\bm{\beta'}\cdot\bm{\beta'}_j}{1-\bm{\beta}\cdot\bm{\beta}_j}
\end{align}
which can be used to obtain the velocity of the ambient target in the
rest frame of the incident particle. Confirming the expression given
in \cite{Reno:1987qw} it reads
\begin{align}
  \beta_2 \equiv |\bm{\beta}_j'| =
  \frac{\vmol}{1-\beta\bj\cos{\alpha}} .
\end{align}
By the same token, the expression for the angle $\psi$ between
$\bm{\beta}_j'$ and the $z$-axis reads
\begin{align}
  \cos{\psi} = \bm{\beta}'_j \cdot \mathbf{e}'_z / \beta_2 =
  - \frac{\beta - \bj \cos{\alpha}}{\vmol}
\end{align}
which differs by a sign from~\cite{Reno:1987qw}.  Instead of
explicitly carrying out the rotation $\pj''=\Lambda_2 \pj'$ which
makes $\bm{\beta}_j''$ parallel to the $z'$-axis ($\vmol$ is lengthy)
we use our knowledge on the form of $\pj''$: $ \pj'' =
(\Ej',0,0,\beta_2 \Ej')^T$ since $\Ej' = \Ej''$. Boosting into the CM
frame using $\Lambda_3$ one finds
\begin{align}
   \pjcm=\Lambda_3 \pj'' = 
\begin{pmatrix}
  \gcm \Ej' (1 - \beta_2 \bcm) \\ 0 \\ 0 \\
  \gcm \Ej' (\beta_2 - \bcm )  
\end{pmatrix} ,  \quad
   p^{\mathrm{cm}} = 
\begin{pmatrix}
  \gcm M \\ 0 \\ 0 \\ - \gcm \bcm M  
\end{pmatrix} 
\end{align}
where $\bcm$ is obtained from $(p^{\mathrm{cm}}_{j})_z = -
(p^{\mathrm{cm}})_z$; $\gcm = (1-\bcm^2)^{-1/2}$.
We find
\begin{align}
  \bcm = \frac{\Ej' \beta_2}{ M + \Ej'} = \frac{\beta_2 \Ej \gb
    (1-\beta\bj \cos{\alpha})}{M + \gb \Ej (1-\beta\bj\cos{\alpha})}
\end{align}
In the CM frame, the scattered target three-momentum
$\mathbf{p}_{jf}^{\mathrm{cm}}$ has a scattering angle~$\theta$ with
$\mathbf{p}_{j}^{\mathrm{cm}}$ and both momenta span a plane with
azimuthal angle $\phi$. Thus, $\pjfcm =
(\Ejcm,\mathbf{p}_{jf}^{\mathrm{cm}} )^T$ is given by
\begin{align}
    \pjfcm=
\begin{pmatrix}
  \gcm \Ej' (1 - \beta_2 \bcm) \\ 
  \gcm \Ej' (\beta_2 - \bcm ) \sin{\theta} \cos{\phi} \\
  \gcm \Ej' (\beta_2 - \bcm ) \sin{\theta} \sin{\phi} \\
  \gcm \Ej' (\beta_2 - \bcm ) \cos{\theta} 
\end{pmatrix} 
\end{align}
and with $\pjf = \Lambda^{-1}\pjfcm =
\Lambda_1^{-1}\Lambda_2^{-1}\Lambda_3^{-1}\pjfcm$ we can transform
back into the rest frame of the thermal bath. This yields for the
energy of the scattered background particle
\begin{align}
\label{eq:Ejf}
  \Ejf &= \Ej \gb^2 \gcm (1 - \beta \bj \cos{\alpha}) \nonumber\\
       &\times \{ \gcm [ (1 - \bcm \beta_2) (1 + \beta\bcm
       \cos{\psi} )\nonumber\\
       & \hphantom{\times \{ \gcm [}
       - ( \bcm - \beta_2 )(\beta \cos{\psi} + \bcm ) \cos{\theta} ]
       \nonumber\\
       &  \hphantom{\times \{ } 
       + \beta(\bcm-\beta_2) \sin{\theta}\sin{\phi}\sin{\psi} \} 
\end{align}
from which $\Delta \Ej = \Ejf - \Ej$ is obtained.
We encounter some sign-differences and a different angular dependence
on $\phi$ in the last line with (A2) of \cite{Reno:1987qw}. This may
be due to a different definition of the coordinate system and turns
out to yield the same results.  Since the target medium is
unpolarized, $d\sigma_j/d\Omega$ is independent of $\phi$ and the
integration over the azimuthal angles in (\ref{eq:energy-loss-master})
can be performed upon which the last line of (\ref{eq:Ejf}) drops
out. Neglecting the Fermi blocking/Bose enhancement factors in
(\ref{eq:energy-loss-master}) we find (adopting the notation
of~\cite{Reno:1987qw}),
\begin{align}
  \label{eq:master-RS}
  \frac{dE}{dt} &= \sum_j \frac{g_j}{2\pi}
  \int_0^1  d\bj\,  m_j^4\bj^2\, \left(1-\bj^2\right)^{-3} f_j(\bj,T)\,
  I_j(\bj,\beta ,T) \, , \\
  \label{eq:I-RS}
  I_j(\bj,\beta ,T) &= \int d\theta\, d\alpha\, \sin{\theta}
  \sin{\alpha}\,
  \Delta_j \frac{d\sigma_j}{d\Omega} \ , \\
  \label{eq:Delta-RS}  
  \Delta_j & = \vmol \Big\{ \gb^2 \gcm^2 (1 - \beta \bj \cos{\alpha}) \nonumber\\
  &\hphantom{\vmol  \{}\times 
  \big[ (1 - \bcm \beta_2) (1 + \beta\bcm \cos{\psi} )  \nonumber\\
  &\hphantom{\vmol  \{ \times [ }
  - ( \bcm - \beta_2 )(\beta \cos{\psi} + \bcm \big) \cos{\theta} ] -
  1 \Big\}\ .
\end{align}

\subsection{Hadron-electron scattering}
\label{sec:hadr-electr-scatt}

Let us now focus on ``Coulomb scattering'' between an incident hadron
and background electrons (positrons) and compute
${d\sigma_j}/{d\Omega}$ for $j=e^\pm$.  Note that also neutral hadrons
scatter on $e^\pm$ via their magnetic moment interaction.

For spin-$1/2$ hadrons such as nucleons or \trit\ and \het\ nuclei the
hadron-photon vertex can be written as~\cite{Berestetsky:1982aq}
$ \Gamma_\mu = 2  M (G_e - G_m)   {P_\mu}/{P^2}  + G_m \gamma_\mu$
with $P = p^{\mathrm{cm}}+p_{f}^{\mathrm{cm}}$ and
$p_{f}^{\mathrm{cm}}$ denoting the (outgoing) four-momentum of the
nucleus. The respective electric and magnetic form factors $G_e$ and
$G_m$ depend on the (squared) four-momentum transfer $q^2 =
(p_{f}^{\mathrm{cm}} - p^{\mathrm{cm}})^2 $ and are normalized such
that $G_e(0) = Z$ is the charge in units of~$e$ and that $G_m(0) = \mu
$ is the magnetic moment in units of $e/2M$ of the ha\-dron.%
\footnote{The Sachs form factors $G_e$ and $G_m$ are convenient
  because no interference terms $\propto G_e G_m$ appear in the cross
  section~(\ref{eq:sigma-spin-half}); they are related to the Dirac
  and Pauli form factors $F_1$ and $F_2$ via $G_e= F_1 + F_2 q^2/4M^2$
  and $G_m= F_1 + F_2$~\cite{Sachs:1962zzc}.  With the definitions for
  $F_1$ and $F_2$ and for $E_e \gg \me$, in the laboratory frame, the
  Rosenbluth formula~\cite{Rosenbluth:1950yq} follows
  from~(\ref{eq:sigma-spin-half}).
  } 
  The (unpolarized) differential cross section for electron-hadron
  scattering is readily obtained.
  In accordance with~\cite{Berestetsky:1982aq} (typo in
  \cite{Reno:1987qw}) it reads using the Mandelstam variables $s,\ t$,
  and~$u$
\begin{align}
  \label{eq:sigma-spin-half}
  \frac{d\sigma_{1/2}}{dt} & = \frac{\pi \alpha^2}{[s-(M+\me)^2][s-(M-\me)^2]}
  \frac{1}{t^2(1-t/(4M^2))} \nonumber \\
  & \times \Big\{ G_e^2 [ (s-u)^2 + (4M^2-t) t ] \nonumber \\
  & \hphantom{\times \Big\{} 
  - \frac{t}{4M^2} G_m^2 [(s-u)^2 -  (4M^2 - t)(4\me^2+t)] \Big\} \ .
\end{align}

Owing to a different vertex structure for spin-0 hadrons such as pions
or \hef, $ \Gamma_\mu = F P_\mu $, where $F$ is the electromagnetic
form factor ($F(0)=Z$), one readily obtains~\cite{Berestetsky:1982aq}
\begin{align}
  \label{eq:sigma-spin-one}
  \frac{d\sigma_{0}}{dt} & = \frac{\pi \alpha^2 
    F^2 [ (s-u)^2 + (4M^2-t) t ]}{[s-(M+\me)^2][s-(M-\me)^2]t^2}.
\end{align}

We can expand (\ref{eq:sigma-spin-half}) and (\ref{eq:sigma-spin-one})
in terms of $x = \gb \Ej /M$ (typo in \cite{Reno:1987qw}). The
expansion is most likely to fail for scattering of (light)
ultra-relativistic nuclei at high temperatures of the thermal bath.
To see the validity of the expansion consider the typical energy of an
electron $\VEV{E_e} = \VEV{E_j}$ by using Maxwell-Boltzmann
statistics:
\begin{align}
  \VEV{E_e} = 3 T + \me \frac{K_1(\me/T)}{K_2(\me/T)}
\end{align}
Here, $K_{1/2}$ is the modified Bessel function of the first/second
kind. For example, with $T=0.1\ \MeV$, a kinetic energy
$T_{\mathrm{kin}} \equiv (\gamma - 1)M = 100\ \GeV\ (1\ \GeV) $ of the
nucleus, and $M=m_{\proton}$ one finds $\langle x \rangle \simeq 0.08\
(0.002)$. Thus for the cases of interest the expansion in $x$ is
fine. Since the cross-sections are independent of the azimuthal angle%
\footnote{An overall sign has been dropped since it can be fixed by
  the integration borders.}
\begin{align}
  \label{eq:dt-to-dOmega}
  \frac{d\sigma}{d\Omega} = \frac{p_*^2}{\pi} \frac{d\sigma}{dt} \quad
  \mathrm{with} \quad p_* = |\mathbf{p}^{\mathrm{cm}}| \simeq M \beta_2 (1
  - \beta \beta_e \cos{\alpha}) x 
\end{align}
where $\beta_e = \bj$. Neglecting $\me$ it follows
\begin{align*}
  s-M^2 &\simeq 2 M^2 x (1 - \beta \beta_e \cos{\alpha})  \\
  s-u & \simeq  4 M^2 x (1 - \beta \beta_e \cos{\alpha})  \\
  t & \simeq  - 2 M^2 \beta_2^2 x^2
  (1-\cos{\theta}) (1 - \beta \beta_e \cos{\alpha})^2  
\end{align*}
so that we find for the CM cross section for charged spin-1/2
\textit{and} spin-0 nuclei
\begin{align}
  \label{eq:sigma-ch}
  \frac{d\sigma_{\mathrm{ch}}}{d\Omega} & \simeq \frac{\alpha^2
    Z^2}{M^2 x^2} \ \frac{ 1 - \beta_2^2(1-\cos{\theta})/2} {\beta_2^2
    (1 - \beta \beta_e \cos{\alpha})^2 (1-\cos{\theta})^2}.
\end{align}
Note that we have made the approximation $F,G_e \simeq Z$ assuming
small $|q^2|$. Let us see if this is justified.  The maximum energy
transfer is realized in a back-to-back collision in the CM frame for
which $q^2 = t_{\mathrm{min}} = - 4 p_*^2 $. Considering the case that
$\beta\gg \bj$, i.e., the case when the electron is a stationary
target, it follows from~(\ref{eq:dt-to-dOmega}) that $p_* \simeq \gb
\beta \me$. For example, for a proton with $E = 100\ \GeV$ this gives
$Q^2 \equiv -q^2 \simeq 0.01\ \GeV^2$ for which $G_e(Q^2)$ is
practically unchanged from unity~\cite{Walker:1993vj}. Moreover, note
that setting $F,G_e = Z$ usually leads to an overestimation of the
cross section with $F$ and $G_e$ de\-creasing for increasing
$|q^2|$~\cite{Walker:1993vj,Schiavilla:2001qe}. Consequently, the
stopping power is overestimated leading to more conservative BBN
constraints.
For neutral hadrons we find%
\footnote{For example, $G_m(0) = -1.91$~\cite{Yao:2006px} for the
  neutron, being entirely anomalous. }
\begin{align}
  \label{eq:sigma-nc}
  \frac{d\sigma_{\mathrm{nc}}}{d\Omega} & \simeq \frac{\alpha^2
    G_m^2}{2 M^2} \ \frac{ 1 + \beta_2^2(1-\cos{\theta})/2}
  { 1-\cos{\theta}}.
\end{align}

We disagree in (\ref{eq:sigma-ch}) and (\ref{eq:sigma-nc}) with
\cite{Reno:1987qw} by a factor of $\beta_2^2$ in the denominator. The
disagreement arises as follows: Eq. (A5a) of \cite{Reno:1987qw} is
actually Eq. (139.5) of \cite{Berestetsky:1982aq}. In order to arrive
at the latter equation $ p_*^2 \simeq (\Ejcm)^2$ has been used
($\mathbf{p}_e^2\simeq \varepsilon_e^2$ in the notation of
\cite{Berestetsky:1982aq}). However, it is more accurate to use $
p_*^2 = (\gcm \bcm M)^2 = \beta_2^2 (\Eecm)^2 + \Orderof{x^2}$; recall
that $\beta_2$ is the velocity of the electron as seen from the rest
frame of the nucleus. We note in passing that up to corrections
$\Orderof{x^2}$ one has $\gcm \simeq 1$, $\bcm \simeq x \beta_2 (1 -
\beta \beta_e \cos{\alpha}) $, $\Ecm \simeq M $, and $\Eecm \simeq M x (1
- \beta \beta_e \cos{\alpha})$.

We can use the above expansion in $x$ to simplify $\Delta$~in
(\ref{eq:Delta-RS}) for the limiting cases of ultra-relativistic and
non-relativistic hadrons traversing the background plasma. Considering
$\beta \simeq 1$, i.e., an ultra-relativistic incident particle, and
therefore $\gb\gg 1$, $\beta_2\cos{\psi}\simeq -1$,
and $\vmol \simeq (1 - \beta \beta_e \cos{\alpha})$ we get to leading
order~$\Orderof{x^0}$
  \begin{align}
    \label{eq:delta-rel}
    \Delta_e^{\mathrm{rel}} \simeq \gb^2 (1 -  \beta_e \cos{\alpha})^2
    (1-\cos{\theta}) \ .
  \end{align}
 Conversely, for $\beta \ll 1$ and therefore
  $\gb^2\simeq 1+\beta^2$, $\beta_2 \simeq (\beta_e - \beta
  \cos{\alpha})/(1-\beta\beta_e \cos{\alpha})$, and $\vmol\simeq
  \beta_e -\beta\cos{\alpha}$ we find 
  \begin{align}
    \label{eq:delta-nrel}
    \Delta_e^{\mathrm{nrel}} \simeq (\beta^2 -
    \beta\beta_e \cos{\alpha})(\beta_e -\beta\cos{\alpha})(1-\cos{\theta})\ .
  \end{align}
  Both expressions agree with the ones obtained in~\cite{Reno:1987qw}
  with differing signs in~(\ref{eq:Delta-RS}) being compensated.

\subsection{Cutoff considerations for charged particles}
\label{sec:cutoff-cons}

\renewcommand{\bj}{\ensuremath{\beta_e}}

After having obtained the cross section for Coulomb and magnetic
moment scattering for hadrons on electrons we make the following
observation for charged particles:
Though the energy transfer in a collision is smallest in the forward
direction $\theta\to 0$ [as can be seen by the factor
$(1-\cos{\theta})$ in Eqs. (\ref{eq:delta-rel}) and
(\ref{eq:delta-nrel})], the divergence $(1-\cos{\theta})^{-2}$ in the
cross section for charged hadrons (\ref{eq:sigma-ch})---arising from
the long-range nature of the Coulomb interaction---is too strong to be
canceled. In this sense, the energy loss due to scatterings in the
forward direction gives the most efficient contribution. Cutting off
the angular integration in (\ref{eq:I-RS}) at~$\thmin$ leads to the
well known logarithmic dependence on~$\thmin$.
Of course, $\thmin$ has to be motivated.

In a plasma, i.e., in a gas of charged particles, correlation effects
lead to the screening of the long-range Coulomb interaction. In
Ref.~\cite{Kawasaki:2004qu} the authors determine the cutoff by
comparing the energy transfer to the electron with the plasma
frequency
\begin{align}
\label{eq:plasma-frequency}
 \omegapl^2= \frac{4\pi \alpha n_e}{\me}
\end{align}
where $n_e$ denotes the total electronic density
  \begin{align}
    \label{eq:total-electronic-density}
    n_e = n_{e^-} + n_{e^+} \simeq
    \begin{cases} 
      2\times({\me^2 T}/{\pi^2}) K_2(m/T) & \mathrm{for}\ T\gtrsim \me/26 \\[0.2cm]
      {7}/{8}\ \eta_{\mathrm{B}} n_\gamma & \mathrm{for}\ T\lesssim
      \me/26
    \end{cases}
  \end{align}
  In the first line we have neglected the electron chemical potential
  and in the second line we have imposed charge neutrality of the
  Universe. The upper relation in (\ref{eq:total-electronic-density})
  is derived by using Maxwell-Boltzmann statistics. For $T\gg \me$,
  i.e., for ultra-relativistic electrons/positrons, this implies an
  error by a factor of $3\zeta(3)/4 \simeq 0.9$. Note also that
  electrons freeze out in the temperature region of interest,
  $\me/26\simeq 20\ \keV$.

  The plasma frequency does, however, \emph{not} provide the correct
  scale~\cite{Raffelt:1985nk}. The screening of the electric field is
  a longitudinal phenomenon whereas the notion of the plasma frequency
  as an effective photon mass is associated with transverse plasma
  excitations. Electrons as well as the (ionized) light elements
  contribute to the screening with a scale~\cite{Raffelt:1996wa}
  \begin{align}
    k_{\mathrm{S}}^2 = \kd^2 + k_{\mathrm{I}}^2 = \frac{4\pi\alpha
      n_e}{T} + \frac{4\pi\alpha }{T} \sum_j Z_j^2 n_j ,
  \end{align}
  where $\kd$ denotes the Debye scale with Debye length
  $\lambda_{\mathrm{D}}=\kd^{-1}$; $n_j$ denotes the number density of
  nuclei with charge number $Z_j$. Note that $\kd$ and
  $\omega_{\mathrm{pl}}$ can be very different with
  $\omega_{\mathrm{pl}}/\kd = \sqrt{T/\me}$. However, in the
  temperature region of main interest $\omega_{pl}$ and
  $k_{\mathrm{D}}$ are within a factor of a few.  Moreover, since the
  screening scale will enter only logarithmically we neglect the
  contribution of the ions (in particular protons) in the following
  and set $k_{\mathrm{S}}\simeq \kd$.

  We shall distinguish two cases: For $\beta \gtrsim \langle\beta_e \rangle$ the
  electrons can be viewed as a stationary target. Thus, we follow the
  screening prescription obtained in \cite{Raffelt:1985nk} and replace
  (\ref{eq:sigma-spin-half}) via
  \begin{align}
    \label{eq:weak-screening}
    \frac{d\sigma}{dt} = \frac{f(s,t,u,M)}{\mathbf{q}^4} \quad
    \rightarrow \quad \left. \frac{d\sigma}{dt} \right|_{\mathrm{sc}} =
    \frac{f(s,t,u,M)}{\mathbf{q}^2(\mathbf{q}^2+\kd^2)}
    \quad \mathrm{for}\quad \beta \gtrsim \langle\beta_e \rangle
  \end{align}
  (for elastic scattering in the CM frame $t=-\mathbf{q}^2$) whereas
  for $\beta \lesssim \langle\beta_e \rangle$ electrons have time to
  rearrange so that the scattering resembles one on a Yukawa-like
  charge distribution with screening length $\lambda_{\mathrm{D}}$.  Then
  the correct prescription reads
  \begin{align}
    \label{eq:yukawa-screen-prescription}
    \frac{d\sigma}{dt} = \frac{f(s,t,u,M)}{\mathbf{q}^4} \quad
    \rightarrow \quad \left.\frac{d\sigma}{dt}\right|_{\mathrm{sc}} =
    \frac{f(s,t,u,M)}{(\mathbf{q}^2+\kd^2)^2}
    \quad \mathrm{for}\quad \beta \lesssim \langle\beta_e \rangle .
  \end{align}

  Given the above considerations we replace the scattering cross
  section~(\ref{eq:sigma-ch}) for charged particles employing the
  screening prescriptions (\ref{eq:weak-screening}) and
  (\ref{eq:yukawa-screen-prescription}) for the respective cases
  ${\beta\gtrsim\langle\bj\rangle}$ and
  ${\beta\lesssim\langle\bj\rangle}$. We find
\begin{align}
  \label{eq:screened-Sigma-fastNucleus}
  \left. \frac{d\sigma_{\mathrm{ch}}}{d\Omega} \right|_{\beta\gtrsim\langle\bj\rangle} \simeq
  \frac{\alpha^2 Z^2 }{x^2 M^2} \frac{ 1 - \beta_2^2(1-\cos{\theta})/2} 
  { \beta_2^2 (1 - \beta \bj \cos{\alpha})^2 (1-\cos{\theta})^2 
  + (1-\cos{\theta}) \kappa^2}\ ,
\end{align}
\begin{align}
  \label{eq:screened-Sigma-slowNucleus}
  \left. \frac{d\sigma_{\mathrm{ch}}}{d\Omega} \right|_{\beta\lesssim\langle\bj\rangle} \simeq
  \frac{\alpha^2 Z^2 \beta_2^2}{x^2 M^2} \frac{ 1 - \beta_2^2(1-\cos{\theta})/2} 
  { \left[ \beta_2^2 (1 - \beta \bj \cos{\alpha}) (1-\cos{\theta}) 
  +\kappa^2 /(1 - \beta \bj \cos{\alpha}) \right]^2 } 
\end{align}
with
\begin{align}
  \kappa^2 = \kd^2 / ( 2 M^2 x^2) = \kd^2/(2\gb^2\Ej^2).
\end{align}
In the region where $\kappa^2$ acts as a regulator, i.e., for
$\theta\to 0$, we make the immediate observation that
\begin{align*}
  \Delta^{\mathrm{rel/nrel}}_e
  \left. \frac{d\sigma_{\mathrm{ch}}}{d\Omega}
  \right|_{\beta\gtrsim\langle\bj\rangle} &\sim
  \frac{1}{\kappa^2} , \\
   \Delta^{\mathrm{rel/nrel}}_e
  \left. \frac{d\sigma_{\mathrm{ch}}}{d\Omega}
  \right|_{\beta\lesssim\langle\bj\rangle} &\sim
  \frac{1-\cos{\theta}}{\kappa^4}.
\end{align*}
For scattering in the forward direction the screening prescriptions
will yield a numerical difference only for
$(1-\cos{\theta})\lesssim\kappa^2$.  For our cases of interest
$\kappa^2$ is usually a very small quantity, e.g.,
$-\log{\kappa^2}\sim \Orderof{10\div 15}$ for $T=30\ \keV$ or
$\Orderof{3\div 8}$ for $T=300\ \keV$. Thereby, only in a very small
integration regime over $\theta$ both cross sections will be
significantly different---though the integrand is largest in this
area.

In order to decide which cross section is applicable for a given value
of the hadron velocity~$\beta$, it has to be compared with the average
electron/positron velocity
\begin{align}
  \langle \beta_e \rangle = \frac{2 T (\me + T)}{\me^2 K_2(\me/T)}
  e^{-\me/T}
\end{align}
which is obtained by using Maxwell-Boltzmann statistics; for $T\ll
\me$ the formula reduces to the standard result $ \langle \bj
\rangle_{\mathrm{n.r.}} = \sqrt{8 T / \pi\me} $.
Note that $\beta$ is related to the kinetic energy of the incident
hadron via $\beta = [ 1 - M^2/(T_{\mathrm{kin}}+M)^2 ]^{1/2}$. Thus,
for example, a proton with $T_{\mathrm{kin}} = 1\ \GeV\ (50\ \MeV)$
has $\beta = 0.88\ (0.31)$ so that $\beta$ drops below $ \langle \bj
\rangle $ for $T = 400\ \keV\ (20\ \keV)$.

\subsection{Discussion on Coulomb stopping}
\label{sec:discussion-coul-stop}

In this section we discuss in some detail the results on the energy
loss for charged particles due to Coulomb interactions with the
background electrons (positrons). We will also compare with a
treatment found in the literature.

To see the net effect of the different screening prescriptions on the
stopping power we perform a full numerical integration of
(\ref{eq:I-RS}) and (\ref{eq:master-RS}) using the Vegas
algorithm~\cite{Hahn:2004fe}. For the integration over the electron
(positron) velocity knowledge of the distribution function $f_e$ is
required. Though electrons are frozen out for $T\lesssim \me / 26
\simeq 20\ \keV$ they remain tightly coupled to the photon bath via
Thomson scattering. This ensures that electrons maintain kinetic
equilibrium so that we can make the approximation
\begin{align}
  \label{eq:electron-distro}
  f_e (E_e,T) \simeq   
  \begin{cases} 
      [\exp(E_e/T) + 1]^{-1} & \mathrm{for}\ T\gtrsim \me/26 \\[0.2cm]
      n_e  f_e^{\mathrm{eq}} /n^{\mathrm{eq}}_{e^-}
      & \mathrm{for}\ T\lesssim   \me/26
    \end{cases}
\end{align}
where $T$ denotes the photon temperature. For $T\lesssim \me/26$ we
use $f_e^{\mathrm{eq}} = \exp{(-E_e/T)} \simeq \exp[-
\mathbf{p}_e^2/(2\me T) -\me/T] $, i.e., we resort to
Maxwell-Boltzmann statistics in the non-relativistic limit.%
\footnote{It is shown in~\cite{Bernstein:1988bw} that $f_e = R^{-3}
  T_e^{-3/2} N_0 \exp{(-\mathbf{p}_e^2/2\me)}$ satisfies the Boltzmann
  equation in the non-relativistic limit with an elastic collision
  term due to Thomson scattering; $N_0 \propto n_\gamma R^3 =
  \mathrm{const}$ by comparison with
  (\ref{eq:total-electronic-density}).
  Defining the temperature of a non-relativistic particle species $j$
  with arbitrary distribution $f$ as $(3/2) T_j n_j \equiv g_j \int
  d^3\mathbf{p}_j/(2\pi)^3 \mathbf{p}_j^2/(2 m_j)
  f(\mathbf{p}_j)$~\cite{Bernstein:1988bw} it is found that the
  electron temperature $T_e$ tracks $T$ well until recombination,
  $T\simeq 0.3\ \eV$. There, $(T-T_e)/T =
  \Orderof{10^{-7}}$~\cite{Bringmann:2006mu}.}
From the definition $ n_e = g_e \int d^3\mathbf{p}_e/(2\pi)^3 f_e$ we
reproduce the second line of~(\ref{eq:total-electronic-density}) by
using $n_{e^-}^{\mathrm{eq}}$ in the form
of~(\ref{eq:non-rel-eq-numberdensX}) with $g_e = 2$; for the case
$T\gtrsim \me/26$ we use $g_e=4$.

\begin{figure}[t]
\includegraphics[width=\textwidth]{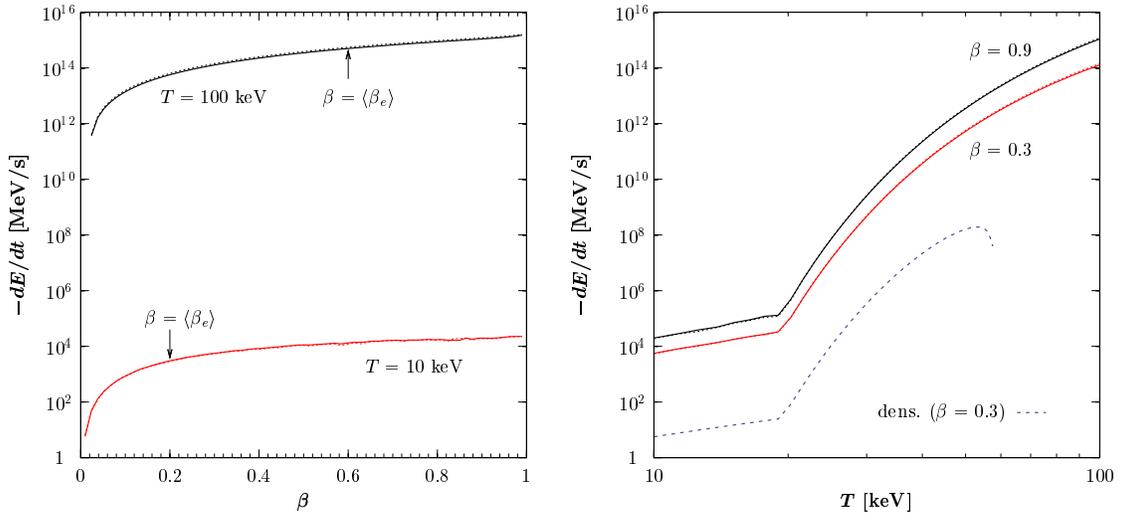}
\caption[Stopping power of a charged hadron]{We show the stopping
  power $dE/dt$ for an injected proton computed by numerical
  integration from~(\ref{eq:master-RS}) using the different screening
  prescriptions. Solid lines correspond to
  (\ref{eq:screened-Sigma-fastNucleus}) and dotted lines (hardly
  visible) are associated with~(\ref{eq:screened-Sigma-slowNucleus}).
  \textit{Left}: Shown is $-dE/dt$ as a function of the proton
  velocity $\beta$ at temperatures $T=100\ \keV$ and $10\ \keV$ as
  labeled. In addition, the points $\beta=\VEV{\beta_e}$ indicate
  which screening prescription should be used. \textit{Right}:
  $-dE/dt$ as a function of $T$ for a relativistic proton
  ($T_{\mathrm{kin}}= 1\ \GeV,\ \beta = 0.9$) and a non-relativistic
  proton ($T_{\mathrm{kin}}= 50\ \MeV,\ \beta = 0.3$) is plotted. In
  addition, the dashed line shows the energy loss due to the density
  effect in the non-relativistic case.}
\label{Fig:stopping} 
\end{figure}
In Fig.~\ref{Fig:stopping} we show the stopping power $dE/dt$ for an
injected proton computed by numerical integration
from~(\ref{eq:master-RS}) using the different screening
prescriptions. Solid lines correspond to
(\ref{eq:screened-Sigma-fastNucleus}) and dotted lines (hardly
visible) are associated with~(\ref{eq:screened-Sigma-slowNucleus}).
In the left panel we show $-dE/dt$ in units of $\MeV/\seconds$ as a
function of the proton velocity $\beta$ at temperatures $T=100\ \keV$
and $10\ \keV$ as labeled. In addition, the points
$\beta=\VEV{\beta_e}$ indicate which screening prescription should be
used. In the right panel we show $-dE/dt$ as a function of $T$ for a
relativistic proton ($T_{\mathrm{kin}}= 1\ \GeV,\ \beta = 0.9$) and a
non-relativistic proton ($T_{\mathrm{kin}}= 50\ \MeV,\ \beta =
0.3$).

From Fig.~\ref{Fig:stopping} we can make a number of observations. An
immediate one is that the stopping power is essentially insensitive to
the employed screening prescription. Concretely, we find that both
prescriptions yield a difference in $dE/dt$ by no more than $20\%$ for
the considered temperature range.
From the right panel we see that once the velocity $\beta$ of the
incident particle drops below the average electron velocity, the
stopping power starts to decrease rapidly. This confirms the
observation made in~\cite{Reno:1987qw} and it is also intuitive since
it becomes increasingly difficult to transfer momentum to the---on
average---faster electrons. Indeed, for $\beta<\bj$ the charged hadron
can even gain energy in a collision which is indicated by a sign-flip
of $I_e$ [Eq.~(\ref{eq:I-RS})] in the collinear region where $\alpha
\to 0$ ('head-on-back collision').
From the left panel we realize that the stopping power rapidly
decreases with dropping temperature. This is because for $T\lesssim
\me$ the number density of electrons and positrons is Boltzmann
suppressed. For $T\lesssim 20\ \keV$ the decrease is weaker because
the remaining electrons fail to track their exponentially decreasing
equilibrium abundance. More precisely, $dE/dt$ scales like $T^{3}$ for
such low temperatures because $n_e \sim a^{-3}$ and $ a \sim T^{-1}$
during radiation domination.

So far, we have only considered (screened) \textit{binary} collisions
of a fast charged particle traversing a QED plasma.  Such a treatment
gives an accurate description for those scattering events of the
particle with largest energy transfers, i.e., with smallest impact
parameters~$b$. Considering $b\gtrsim\lambda_{\mathrm{D}}$, the
nucleus scatters simultaneously on many electrons. (Note that in our
case the Debye length is much larger than the typical inter-particle
distance, $\lambda_{\mathrm{D}} \gg n_e^{-1/3}$.)
A sweeping external charge, i.e., a perturbation $\rho_{\mathrm{ext}}
= Z e \delta(\mathbf{r}-\boldsymbol{\beta}t)$, induces a macroscopic
electric field $\mathbf{E}$ in the medium which acts back on the
particle.
The resulting energy loss per unit path length can be found by
computing the work done on the particle.  It equals the force exerted onto
the charged hadron in direction opposite to its motion
\begin{align}
  \left. \frac{dE}{dx} \right|_{\kd b>1} = \frac{1}{\beta}
  \left. \frac{dE}{dt} \right|_{\kd b>1} = Z e\, \mathbf{e}_{\bm\beta}
  \cdot \mathbf{E}(\mathbf{r}=\bm\beta t) ,
\end{align}
where $\mathbf{e}_{\bm\beta}$ is a unit vector in $\bm\beta$
direction.
The electric field can be found by considering the macroscopic Maxwell
equations with dielectric permittivity $\varepsilon$.  In the
non-relativistic limit and using%
\footnote{Here, $\omega$ and $\mathbf{k}$ ($k=|\mathbf{k}|$) are the
  frequency and wave vector of the Fourier transformed fields and the
  expression is the limiting case for $\omega/(k\langle\bj\rangle)\gg 1$;
  see~\cite{1960ecm..book.....L}. }
$\varepsilon(\omega) \simeq 1- \omegapl^2/\omega^2 $ for a Maxwellian
plasma, the stopping power reads~\cite{1975clel.book.....J}
\begin{align}
  \label{eq:jackson-density-effect}
  -\left. \frac{dE}{dt}\right|_{\kd b>1} = \frac{Z^2\alpha}{\beta} \omegapl^2
  \ln{\left(\frac{1.123 \kd \beta}{\omegapl}
    \right)}
\end{align}

As can be seen by the dashed line in the left Fig.~\ref{Fig:stopping}
the contribution due to the 'density effect' is subleading.
Note that the formula was derived under the premise that the velocity
of the massive particle is large compared to the thermal speed of the
electrons.  Indeed, the argument of the logarithm in
(\ref{eq:jackson-density-effect}) is greater than unity only for
$\beta\gtrsim 0.6 \VEV{\bj}_{\mathrm{n.r.}}$ so that the line is
cut-off for $T\gtrsim 60\ \keV$. We remark that a full relativistic
treatment of the energy loss of a massive particle due to the
dielectric response of the medium for arbitrary velocities is complex
but will not affect significantly the above made conclusions; we refer
the reader to Landau's treatment in \cite{1960ecm..book.....L}.%
\footnote{The case of a hot QED plasma with $\me\ll e T$ has
  been treated within the framework of thermal field
  theory in~\cite{Braaten:1991jj}.}

We can compare the full numerical integration of (\ref{eq:master-RS})
with the treatment of Coulomb stopping presented
in~\cite{Kawasaki:2004qu}. The authors employ the results
of~\cite{Reno:1987qw} which also we have taken as a starting point.
Full numerical integration of~(\ref{eq:master-RS}) is not feasible
when scanning the $(\tauX,Y_\X)$ parameter space so that analytical
approximations based on (\ref{eq:delta-rel}) and (\ref{eq:delta-nrel})
have been used in~\cite{Kawasaki:2004qu}. Since we have observed that
the employed screening prescription affects $dE/dt$ only marginally
for $T\lesssim 100\ \keV$ and that $\kd$ and $\omegapl$---entering the
stopping power logarithmically---are not too different, it is not
surprising that we find overall agreement with~\cite{Kawasaki:2004qu}
on the energy degradation rate within a factor of a few.

We remark that obtaining constraints on the hadronic energy release of
decaying \X\ involves a fair amount of modelling and computation.
After calculation of the hadronic branching ratio in the decay of~\X,
each step involves uncertainties and approximations: Employing a
hadronization algorithm, computing the initial energy spectra of
secondaries, following the energy degradation and cascade formation
due to electromagnetic and hadronic processes, and finally obtaining
the yields of non-thermally produced light elements.
We have seen that already the seemlingly elementary process of Coulomb
stopping can become involved---especially when it is necessary to
apply it to a large range of incident particle energies and plasma
temperatures.
In the light of these comments we close this chapter by noting that we
have not found a radically different picture than that of previous
considerations which would strongly influence on the strength of the
hadronic constraints presented in~Fig.~\ref{Fig:EM-HAD-BBN-constr}.

\begin{subappendices}

\begin{section}{Lorentz transformations}
\label{sec:appendix-LT}
The explicit matrices for the Lorentz transformations performed
in~\ref{sec:energy-transf-binary} are given below. The matrix
$\Lambda_2$ describes an (active) rotation in counter-clockwise
direction around the $x'$-axis when looking towards the origin. The
inverse transformations $\Lambda_1^{-1}$, $\Lambda_2^{-1}$, and
$\Lambda_3^{-1}$ are obtained by the replacement $\beta\to -\beta$,
$\psi \to -\psi$, and $\bcm\to -\bcm$ in $\Lambda_1$, $\Lambda_2$, and
$\Lambda_3$, respectively.
\begin{align}
  \Lambda_1 & = 
  \begin{pmatrix}
    \gb & 0 & 0 & -\gb \beta \\
    0 & 1 & 0 & 0 \\
    0 & 0 & 1 & 0 \\
    -\gb \beta & 0 & 0 & \gb
  \end{pmatrix}
\quad
  \Lambda_2 = 
  \begin{pmatrix}
    1 & 0 & 0 & 0 \\
    0 & 1 & 0 & 0 \\
    0 & 0 & \cos{\psi} & -\sin{\psi} \\
    0 & 0 & \sin{\psi} & \cos{\psi}
  \end{pmatrix}
\\
  \Lambda_3 & = 
  \begin{pmatrix}
    \gcm & 0 & 0 & -\gcm \bcm \\
    0 & 1 & 0 & 0 \\
    0 & 0 & 1 & 0 \\
    -\gcm \bcm & 0 & 0 & \gcm
  \end{pmatrix}
\end{align}

\end{section}

\end{subappendices}


\cleardoublepage
\chapter{Bound states and catalysis of BBN}
\label{cha:catalyzed-bbn}

In this Chapter we now discuss (some of) the rich physics which
emerges when the light elements are captured by \champ\ during/after
the time of primordial nucleosynthesis. We start in
Sec.~\ref{sec:basic-prop-bound} by reviewing the basic properties of
such bound states. Section~\ref{sec:wave-functions} is devoted to the
calculation of the wave functions associated with the CHAMP-nucleus
system. This allows us in Sec.~\ref{sec:recomb-phot-rates} to obtain
recombination cross sections carrying a finite nuclear charge radius
correction. The detailed exposition in Sects.~\ref{sec:wave-functions}
and~\ref{sec:recomb-phot-rates} is of some value since (apart from
exceptions) those rates are not publicly available in the literature.

In Sec.~\ref{sec:catalyz-nucl-react} we then consider the catalysis of
BBN reactions. After a general review we employ the results from the
literature on the catalyzed production of \lisx\ and \ben\ and show
explicitly how to incorporate them into a Boltzmann network
calculation. In Sec.~\ref{Sec:pXcatalysis} we discuss the potential
impact of neutral proton-CHAMP bound states on the synthesized
elements. We close this chapter with Sec.~\ref{Sec:9BeConstraints} in
which we first infer an upper limit on primordial \ben\ and then
present the results of our CBBN calculation which heavily
constrains the \champ-abundance/lifetime parameter space.

\section{Basic bound state properties}
\label{sec:basic-prop-bound}

The presence of negatively charged massive particles \champ\
during/after primordial nucleosynthesis leads to the formation of
bound states $(\nuc{}\champ)$ with the ionized nuclei $\nuc{}$ of the
light elements. In this section we shall describe the basic properties
of such bound states.

In order to obtain a first estimate on the physical properties one can
immediately apply the standard formul\ae\ for the quantum mechanical
motion in a Coulomb field. The characteristic size of the
$(\nuc{}\champ)$ system is given by the Bohr radius of the system,
\begin{align}
  \label{eq:abohr}
  \abohr = \frac{1}{\mred Z \alpha} \sim \frac{29\ \fm}{A\, Z} ,
\end{align}
where $A$ and $Z$ are the atomic mass and charge number of the
nucleus, respectively. In the second relation we have used that the
reduced mass 
\begin{align}
  \mred =  \frac{\mnuc \mx}{\mnuc+\mx}
\end{align}
is given to good accuracy by the mass of nucleus, $\mred\simeq\mnuc$
($\mx\gg \mnuc$), and that roughly $\mnuc \sim A\,\mproton$ where
$\mproton = 938\ \MeV$~\cite{Yao:2006px} is the proton mass.
The binding energies of a \textit{point-like} nucleus orbiting
\champ\ are given by the well-known formula
\begin{align}
  \label{eq:ebindcoul}
  \Ebindcouln{n} = 
 - \frac{Z^2 \alpha^2 \mred}{2 n^2 } 
  \sim (- 25\ \keV) \, \frac{A\, Z^2}{ n^2} ,
\end{align}
where $n$ denotes the principal quantum number.

For the case of a proton bound state $\abohr^{\BSpx} \simeq 29\ \fm$
so that the system is a fac\-tor of $\mproton/\melectron\sim 1800 $
smaller than a hydrogen atom. Nevertheless, the \proton--\champ\
distance is still large when compared to the rms charge radius of the
proton, $\Rcharge{\proton} \simeq 0.88\
\fm$~\cite{2004ADNDT..87..185A}. The situation changes for heavier
nuclei. 
For example, considering \BSlisxx, one finds that $\abohr^{\BSlisxx}
\simeq 1.6\ \fm$ whereas the measured \lisx\ rms charge radius reads
$\Rcharge{\lisx}\simeq 2.54\ \fm$~\cite{2004ADNDT..87..185A}.  Thus,
we expect corrections to the na\"ive Bohr-like binding energies
(\ref{eq:ebindcoul}) once the finite size of the nucleus is taken into
account.

In order to obtain more realistic values for the ground state energy,
we need to make an assumption on the charge distribution of the
nucleus. A compilation thereof is presented in
\cite{1997ADNDT..67..207V}. We employ the Gaussian $\rho = e Z
(\xi/\pi)^{3/2} e^{-\xi r^2}$ with radial coordinate $r$
 from which the potential
 \begin{align}
   \label{eq:potential-for-gaussian-charge-distro}
   \phi(r) = - \frac{e Z}{4 \pi r}\, \mathrm{erf}\left({\sqrt{\xi} r}\right)
 \end{align}
 is obtained upon solution of Poisson's equation%
\footnote{We use Heaviside-Lorentz units with $e=\sqrt{4\pi \alpha}$.}
$\nabla^2 \phi = \rho$.  Requiring $ \VEV{r^2}_{\rho} =
\Rchargesq{\nuc{}}$ relates the parameter $\xi$ to the rms charge
radius; $ \VEV{r^2}_{\rho} =3/2\xi $. The error function is defined
by $\mathrm{erf}{\,(x)} = 2\pi^{-1/2}\,\int_0^x e^{-t^2}dt$.
\begin{table}
\label{tab:basic-prop-bound-states}
\caption[Basic properties of bound states]{Basic quantities for some selected
  light elements 
  and their bound states with \champ\ for $\mx\to\infty$. If necessary, 
  nuclear masses $\mnuc$ are derived$^\dagger$
  from~\cite{Audi:2002rp}. Rms charge radii $\Rcharge{N}$ are taken
  from~\cite{2004ADNDT..87..185A}, $\abohr$ denotes the Bohr radius
  (\ref{eq:abohr}), $\Ebindcouln{0}$ is the na\"ive Coulomb ground state
  energy~(\ref{eq:ebindcoul}), and $\Ebindvar$ provides realistic values
  of the binding  energy from the variational
  principle~(\ref{eq:variational-principle}).}
\begin{center}
\begin{tabularx}{0.92\textwidth}{rrcrrrr}
  \toprule bound state & $\mnuc\ [\MeV]$ & $\Rcharge{N}\ [\fm]$ &
  $\abohr\ [\fm]$ & $\Ebindcouln{0}\ [\keV]$
  & $\Ebindvar\ [\keV]$\\
  \midrule
  $(\proton\champ)$ & 938  & 0.88 & 28.8 & -25   & -25   &\\
  $(\deut\champ)$   & 1876 & 2.14 & 14.4 & -50   & -49   &\\
  $(\hef\champ)$    & 3727 & 1.67 & 3.6  & -397  & -347  &\\
  $(\lisx\champ)$   & 5601 & 2.54 & 1.6  & -1342 & -797  &\\
  $(\ben\champ)$    & 8393 & 2.52 & 0.8 & -3575 & -1469 &\\
  \bottomrule 
   \multicolumn{7}{l}{%
     \begin{minipage}{0.9\textwidth}
       \footnotesize ${}^\dagger$ Nuclear masses are obtained from
       atomic masses by subtracting $Z\melectron$ and correcting for
       the total binding energy of all electrons where we follow the
       prescription given in~\cite{Audi:2002rp}.
 \end{minipage}}\\[-0.1cm]%
\end{tabularx}
\end{center}
\end{table}

The above choice of the charge distribution is particularly convenient
because the electric
potential~(\ref{eq:potential-for-gaussian-charge-distro}) is given in
analytical form~(\ref{eq:potential-for-gaussian-charge-distro}).  This
makes an application of the Rayleigh-Ritz variational method
straightforward. Using the (unnormalized) trial wave function
\begin{align}
  \label{eq:trial-wf-bound-states}
  \psi(r; a, b) = e^{- a  r / \abohr} (1 + b r/\abohr )
\end{align}
with variational parameters $a$ and $b$ an \textit{upper bound} on the
true ground state energy $\Ebind$ can be obtained by minimizing the
right hand side of
\begin{align}
\label{eq:variational-principle}
  \Ebind \leq \frac{\int d^3\mathbf{r}\, \psi^* H \psi}{\int
    d^3\mathbf{r}\, \psi^* \psi } \ .
\end{align}
For the Hamiltonian $H$ of the $(\nuc{}\champ)$ system we use $H = -
(2\mnuc)^{-1}\nabla^2 + e\phi $, i.e., we take $\mx\to \infty$.

The results of minimization of (\ref{eq:variational-principle}) for
selected light elements along with some other basic quantities are
summarized in Table~\ref{tab:basic-prop-bound-states}. Note that the
Bohr radii of bound states with elements heavier than \hef\ lie
within the nuclear radii. Thereby, the true binding energy for those
systems is significantly reduced in magnitude as can be seen by comparing
$\Ebindcouln{0}$ with~$\Ebindvar$. We remark that the binding energy
is an important quantity since it directly influences on the bound
state fraction of the light nuclei.

\section{Wave functions of the relative motion}
\label{sec:wave-functions}

For the calculation of photo-dissociation and recombination cross
sections which include the finite charge radius correction, we are in
need of the actual wave functions of the $\nuc{}$--\champ\ system. In
the following we shall therefore obtain the wave functions for the
(\nuc{}\champ) bound states as well as for the $\nuc{}$--\champ\
continuum. It will also allow us to see how well our variationally
obtained upper bounds $\Ebindvar$ fit the actual value of the true
ground-state energy~$\Ebind$.

The Schr\"odinger equation for the radial part $R(r)= u(r)/r$ of the
wave function $\psi(r,\theta,\phi) = Y_{l,m}(\theta,\phi) R(r) $ of
the relative motion is given by
\begin{align}
  \label{eq:schroedinger-equation}
  \frac{d^2u}{dr^2} + \left[ 2 \mred (E - V) -
    \frac{l(l+1)}{r^2}\right]u = 0\, .
\end{align}
As usual, $Y_{l,m}(\theta,\phi)$ denotes the spherical harmonic with
orbital and magnetic quantum numbers $l$ and $m$. For the potential
$V$ we make the following choices
\begin{align}
  \label{eq:potentials-for-SE}
  V = \left\{
  \begin{array}{ll}
    - Z \alpha / r &\quad \mathrm{point}\\
    \hphantom{-} e\,\phi(r) &\quad \mathrm{gauss}\\
    -Z \alpha / (2 R_0)\: (3-r^2/R_0^2) &\quad \mathrm{h.sph} \quad (r\leq R_0)
  \end{array}\right.
\end{align}
where ``point'' stands for the Coulomb potential of a point-like
nucleus, ``gauss'' for a Gaussian charge distribution with $\phi$
defined in (\ref{eq:potential-for-gaussian-charge-distro}), and
``h.sph'' for a potential of a homogeneously charged sphere of squared
radius $R_0^2 = 5\Rchargesq{N}/3$~\cite{1997ADNDT..67..207V}. For
$r>R_0$, $V_{\mathrm{h.sph}}$ is to be continued by
$V_{\mathrm{point}}$.

\subsection{Discrete spectrum}
\label{sec:discrete-spectrum}

We solve (\ref{eq:schroedinger-equation}) for $E<0$ and the various
choices of V [Eq.~(\ref{eq:potentials-for-SE})] numerically. For fixed
$n = n_{\mathrm{r}} + l + 1$ we exploit the fact that the radial
function $R(r)$ and thus $u(r)$ vanishes $n_{\mathrm{r}}$ times;
$n_{\mathrm{r}}$ is the radial quantum number. The solution of
(\ref{eq:schroedinger-equation}) is fixed by imposing the standard
boundary conditions $u(\delta) = \delta^{l+1}$ and $u'(\delta) =(l +1)
\delta^l$ with $\delta \ll \Rcharge{N}$ and normalizing to unity,
$\int |u|^2dr = 1$.

\begin{figure}[h!tb]
\centerline{\includegraphics[totalheight=0.8\textheight]{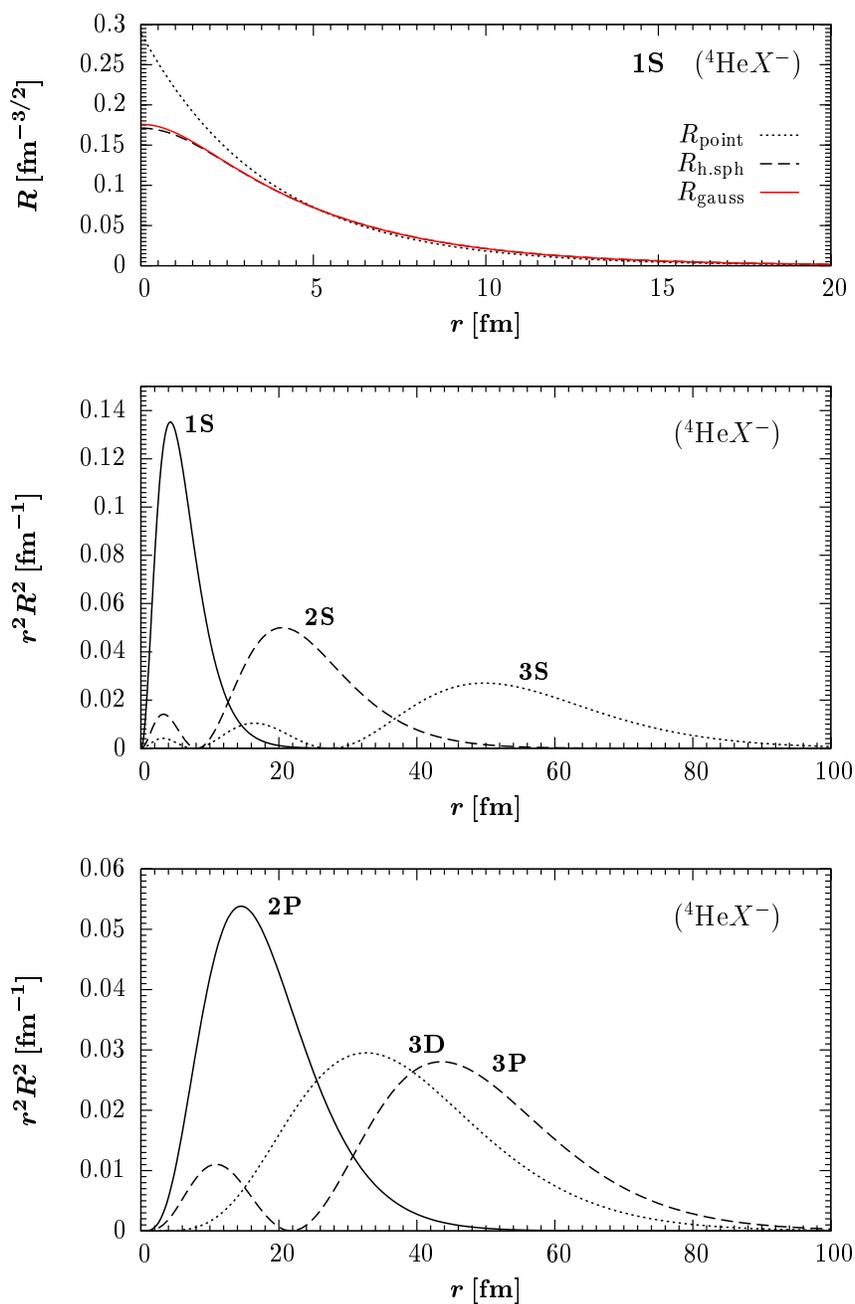}}
\caption[Radial wave functions for \BShefx]{ \textit{Top figure}:
  Radial wave functions $R$ of the \BShefx\ ground state for the
  choices (\ref{eq:potentials-for-SE}) of the potential as
  labeled. \textit{Lower figures}: probability densities $r^2 R^2$ of
  the \hef--\champ\ distance for eigenstates with $n\leq3$ and
  Gaussian charge distribution. All curves are obtained for
  $\mx\to\infty$.}
\label{Fig:Hef-waves} 
\end{figure}

At the top of Fig.~\ref{Fig:Hef-waves} we show the numerical solutions
of the normalized radial wave function $R_{nl} = R_{10}$ for the
\BShefx\ ground state (1S in the usual spectral notation) for the
different choices (\ref{eq:potentials-for-SE}) of the potential. An
attenuation of the wave functions with finite charge radius relative
to the Coulomb case can be seen at small $r$. At large radii, the wave
functions are Coulomb-like.%
\footnote{Of course, the Coulomb solution is simply given by $
  R_{\mathrm{point}} = \abohr^{-3/2} \exp{(-r/\abohr)}$}
It can further be seen that $R_{\mathrm{gauss}}\simeq
R_{\mathrm{h.sph}}$ for all $r$, i.e., the radial wave function for
$\BShefx$ is rather insensitive to the concrete choice of the charge
distribution. In the middle and at the bottom of
Fig.~\ref{Fig:Hef-waves} we plot $r^2 R^2$, i.e., the probability
density of the \hef--\champ\ distance, for $n\leq 3$ and
Gaussian charge distribution. Except for small radii, the curves
essentially resemble distributions obtained with Coulomb wave
functions. This is particularly true for the higher $l$ states as the
wave functions are pushed outwards due to the centrifugal term in
(\ref{eq:schroedinger-equation}).

\begin{table}[tb]
\begin{center}
  \caption[Spectrum and size of \BShefx]{Complete spectrum for
    \BShefx\ for $n\leq 3$. Binding energies are given for and $\mx\to
    \infty$ and $\mx = 100\ \GeV$ (bracketed values) for the
    potentials as labeled and given in
    (\ref{eq:potentials-for-SE}). Additionally, the expectation values
    $\VEV{r}$ and rms radius \Rrms\ for the ``gauss'' case are
    provided.}
\begin{tabularx}{0.80\textwidth}{@{\extracolsep{\fill}}Xr@{\:}r@{\qquad}r@{\:}r@{\qquad}rr}
\toprule
\multicolumn{3}{l}{${\hefxm}$} & \multicolumn{4}{r}{${\Rcharge{\hef}} = 1.67\ \fm$  }\\
\multicolumn{7}{r}{$\mx\to\infty\ (100\ \GeV)$, ${\abohr} = 3.63\ (3.76)\ \fm$}\\
\midrule
State & \multicolumn{2}{c}{$E^{\mathrm{point}}_{\mathrm{b}} [\keV]$} & \multicolumn{2}{c}{$E^{\mathrm{gauss}}_{\mathrm{b}}\ [\keV]$} & \multicolumn{1}{c}{$\VEV{r}\ [\abohr]$} & \multicolumn{1}{r}{$\Rrms\ [\abohr]$} \\
\midrule
 1S   &         $-$397         &  ($-$383)   & $-$348 &            ($-$338)             & 1.7                 & 2.0 \\
 2S   &         $-$99          &   ($-$96)   & $-$93  &             ($-$90)             & 6.4                 & 6.9 \\
 2P   &         $-$99          &   ($-$96)   & $-$99  &             ($-$96)             & 5.0                 & 5.5 \\
 3S   &         $-$44          &   ($-$43)   & $-$42  &             ($-$41)             & 14.1               & 15.0 \\
 3P   &         $-$44          &   ($-$43)   & $-$44  &             ($-$43)             & 12.5               & 13.4 \\
 3D   &         $-$44          &   ($-$43)   & $-$44  &             ($-$43)             & 10.5               & 11.2 \\
\bottomrule
\end{tabularx}
  \label{tab:hef-x-spectrum}
\end{center}
\end{table}

In Table~\ref{tab:hef-x-spectrum} the spectrum for the cases ``point''
and ``gauss'' [Eq.~(\ref{eq:potentials-for-SE})] is given for the
states plotted in Fig.~\ref{Fig:Hef-waves}. In addition, also the
expectation value $\VEV{r}$ as well as the rms radius
$\VEV{r^2}^{1/2}$ are given for the ``gauss'' case in units of
$\abohr= 3.63\ \fm$. We solve the Schr\"odinger
equation~(\ref{eq:schroedinger-equation}) for $\mx\to \infty$, i.e.,
for $\mred=\mnuc{}$, as well as for $\mx=100\ \GeV$ (bracketed values)
in order to study the influence of a finite $\champ$ mass on the
binding energies. The table shows that for $\mx=100\ \GeV$ this leads
to a shift of $10\ \keV$ for the ground state energy but the
correction quickly becomes marginal for the $n>1$ states. The same is
true when comparing the spectra for the different potentials. Whereas
the correction to the ground state energy is substantial, $49\, (45)\
\keV$, the higher states for the $\BShefx$ system essentially
coincide. It is, however, interesting to note that the Coulomb
degeneracy is broken. We refrain from showing the energies for the
case ``h.sph'' since they are the same as for the ``gauss'' case
(except for $n=1$ where a $1\ \keV$ shift is found.)  One can also see
that the variational determination of the ground state energy in
section \ref{sec:basic-prop-bound} gave an accurate result.

\begin{table}[tb]
\begin{center}
\caption[Spectrum and size of \BSbeetx]{As in
  Table~\ref{tab:hef-x-spectrum} but for \BSbeetx. In addition binding
  energies for the ``h.sph'' case are shown; see
  (\ref{eq:potentials-for-SE}). }
\begin{tabularx}{0.98\textwidth}{@{\extracolsep{\fill}}Xr@{\:}rr@{\:}r@{\quad}r@{\:}r@{\quad}rr}
\toprule
\multicolumn{5}{l}{${\beetxm}$}          & \multicolumn{4}{r}{${{\Rcharge{\beet}}} = 3.39\ \fm$  }\\
\multicolumn{9}{r}{$\mx\to\infty\ (100\ \GeV)$, ${\abohr} = 0.91\ (0.97)\ \fm$}\\
\midrule
State & \multicolumn{2}{c}{$\Ebindcoul\ [\keV]$} & \multicolumn{2}{c}{$E_{\mathrm{b}}^{\mathrm{h.sph}}\ [\keV]$}& \multicolumn{2}{c}{$\Ebind^{\mathrm{gauss}}\ [\keV]$} & \multicolumn{1}{c}{$\VEV{r}\ [\abohr]$} & \multicolumn{1}{r}{$\Rrms\ [\abohr]$} \\
\midrule
 1S   & $-$3176&  ($-$2956)  & $-$1118 &  ($-$1092)  & $-$1168 &  ($-$1138)   &    3.8    & 4.2  \\
 2S   & $-$794 &   ($-$739)  & $-$458  &   ($-$437)  & $-$475  &   ($-$453)   &    9.8    & 10.6 \\
 2P   & $-$794 &   ($-$739)  & $-$652  &   ($-$620)  & $-$650  &   ($-$618)   &    6.4    & 6.9 \\
 3S   & $-$353 &   ($-$328)  & $-$243  &   ($-$230)  & $-$249  &   ($-$236)   &   19.0    & 20.2 \\
 3P   & $-$353 &   ($-$328)  & $-$306  &   ($-$290)  & $-$307  &   ($-$290)   &   14.5    & 15.5 \\
 3D   & $-$353 &   ($-$328)  & $-$348  &   ($-$325)  & $-$346  &   ($-$323)   &   10.9    & 11.6 \\
\bottomrule
\end{tabularx}
\label{tab:beet-x-spectrum}
\end{center}
\end{table}

\begin{figure}[bt]
\centerline{\includegraphics[width=12cm]{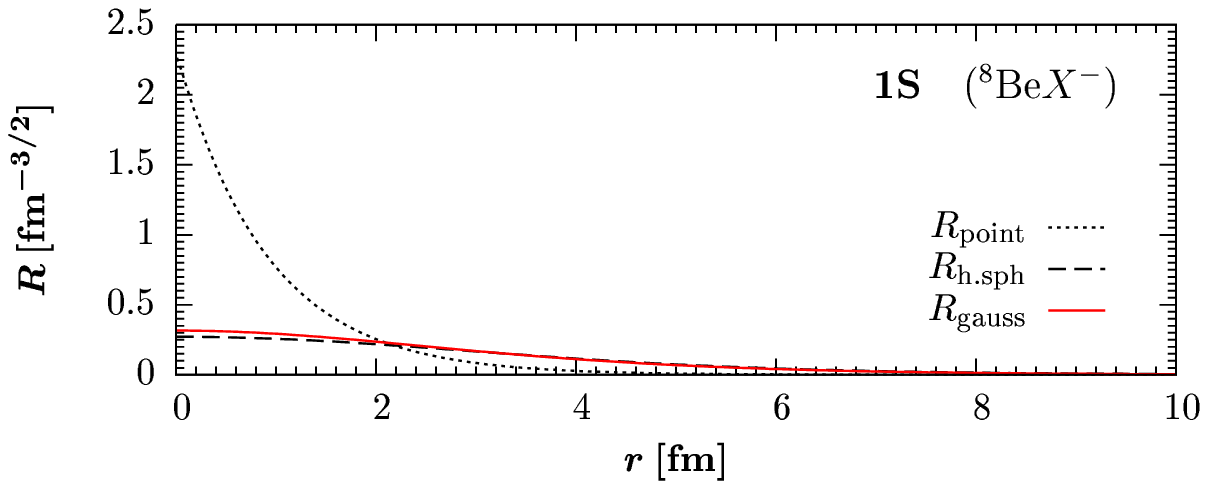}}
\caption[1S radial wave functions for \BSbeetx]{  1S
  radial wave functions for $\BSbeetx$ from solving
  (\ref{eq:schroedinger-equation}) with potentials
  (\ref{eq:potentials-for-SE}) as labeled and $\mx\to\infty$.}
\label{Fig:beet-waves} 
\end{figure}

For bound states of \champ\ with heavier nuclei than \hef, i.e., for
more compact systems, we expect a pronounced behaviour of the observed
effects above. Analogously to the case \BShefx\ we can analyze
\BSbeetx. This is an interesting system because \textit{free} \beet\
is unstable by $92\ \keV$ and decays into two alpha particles:
$\beet\to\hef + \hef$. Indeed, the stable \BSbeetx\ system is part of
a CBBN reaction chain which can open the path to primordial production
of \ben~\cite{Pospelov:2007js}; see Sec.~\ref{sec:catalyz-nucl-react}.
Since the lifetime of \beet\ is $\sim 10^{-16}\ \seconds$ no
experimental data on the charge radius of the isotope is available.
In this section we follow~\cite{Kamimura:2008fx} and adopt the value
$\Rcharge{\beet} = 3.39\ \fm$ which is based on a microscopic
$\hef+\hef$ model calculation~\cite{Hiyama:1997ub}.  Again, in
Fig.~\ref{Fig:beet-waves} we plot the 1S radial solutions of the
Schr\"odinger equation (\ref{eq:schroedinger-equation}) for the
various potentials~(\ref{eq:potentials-for-SE}).  The difference
between $R_{\mathrm{point}}$ and $R_{\mathrm{gauss}}$
($R_{\mathrm{h.sph}}$) is now substantial. Moreover, also a slight
difference between $R_{\mathrm{gauss}}$ and $R_{\mathrm{h.sph}}$ is
observable for smaller radii. We therefore expect a dependence of the
ground state energy on the adopted charge distribution.

In Table~\ref{tab:beet-x-spectrum} we provide the complete spectrum
for \BSbeetx\ with $n\leq 3$. Expectation values $\VEV{r}$ as well as
the rms radius $\VEV{r^2}^{1/2}$ are also computed for the ``gauss''
case in units of $\abohr= 0.91\ \fm$; $m_{\beet} =7.455\ \GeV$. Again,
we compare the energy eigenvalues for $\mx\to \infty$ with the ones
for $\mx=100\ \GeV$ (bracketed values). As can be seen, all states now
receive substantial corrections to the Coulomb values. Moreover, we
observe a $50\ \keV$ $(46\ \keV)$ shift in the 1S energy when changing the charge
distribution from Gaussian to square well (in $r$).

Finally, we have checked all variationally determined ground state
binding energies presented in Table~\ref{tab:basic-prop-bound-states}
of the last section by explicit computation of the wave function.  We
find that all $\Ebindvar$ given in
Table~\ref{tab:basic-prop-bound-states} are within $1\ \keV$ of the
numerically obtained result. Noteworthy may be the $1\ \keV$ shift for
\BSx{\hef}. 
Of course, not only the assumed distribution of charge influences on
$\Ebind$ but also the error on the measured or theoretically predicted
charge radius is a source of uncertainty. This is of pronounced
importance for the heavier nuclei because the bound state system is
more compact. For our purposes, however, it is not essential to pursue
this issue further; see Sec.~\ref{sec:catalys-ben-prod} for another
comment in the context of catalyzed \ben\ production.
In the following, we employ the binding energies determined
from the Gaussian charge distribution.

\subsection{Continuous spectrum}
\label{sec:continuous-spectrum}

We are also in need of solutions of the Schr\"odinger
equation~(\ref{eq:schroedinger-equation}) for $E>0$ if we want to
obtain charge-radius corrected bound-state formation cross sections.
The normalization of a numerically obtained solution is more involved
since the wave functions of the \mbox{$\nuc{}$--$\champ$}-continuum are not
bounded spatially.  However, a finite charge radius leads to a
modification of the Coulomb form of the potential only in the vicinity
of the origin.  Therefore, we can take the following approach: For
$r\gg\Rcharge{\nuc{}}$ the solution
of~(\ref{eq:schroedinger-equation}) has to be a linear combination of
the regular and irregular Coulomb wave functions \regCoul{kl}\ and
\irregCoul{kl}, respectively. They can be expressed as
\begin{subequations}
\label{eq:Coulomb-waves-Def}
\begin{align}
  \label{eq:regular-Coulomb-wave}
  \regCoul{kl}   &= \frac{1}{2} (Y_l+Y_l^*) \ , \\  
  \label{eq:irregular-Coulomb-wave}
  \irregCoul{kl} &=  \frac{1}{2 i} (Y_l^* - Y_l) \ ,
\end{align}
\end{subequations}
with~\cite{Yost:1936zz,1964coth.book.....G}
\begin{subequations}
  \label{eq:Yost-function}
\begin{align}
  Y_l & = + i  
  \frac{|\Gamma(l+1-i\eta)|}{\Gamma(l+1+i \eta)}
  e^{i \pi l /2} e^{\eta \pi / 2}\, W_{i\eta,\,l+1/2}(2 i k r) \ , \\
  Y_l^* & = -i 
  \frac{|\Gamma(l+1+i\eta)|}{\Gamma(l+1-i \eta)}
  e^{-i \pi l /2} e^{\eta \pi / 2}\, W_{-i\eta,\,l+1/2}(-2 i k r) \ . 
\end{align}
\end{subequations}

\begin{figure}[bt]
\centerline{\includegraphics[width=12cm]{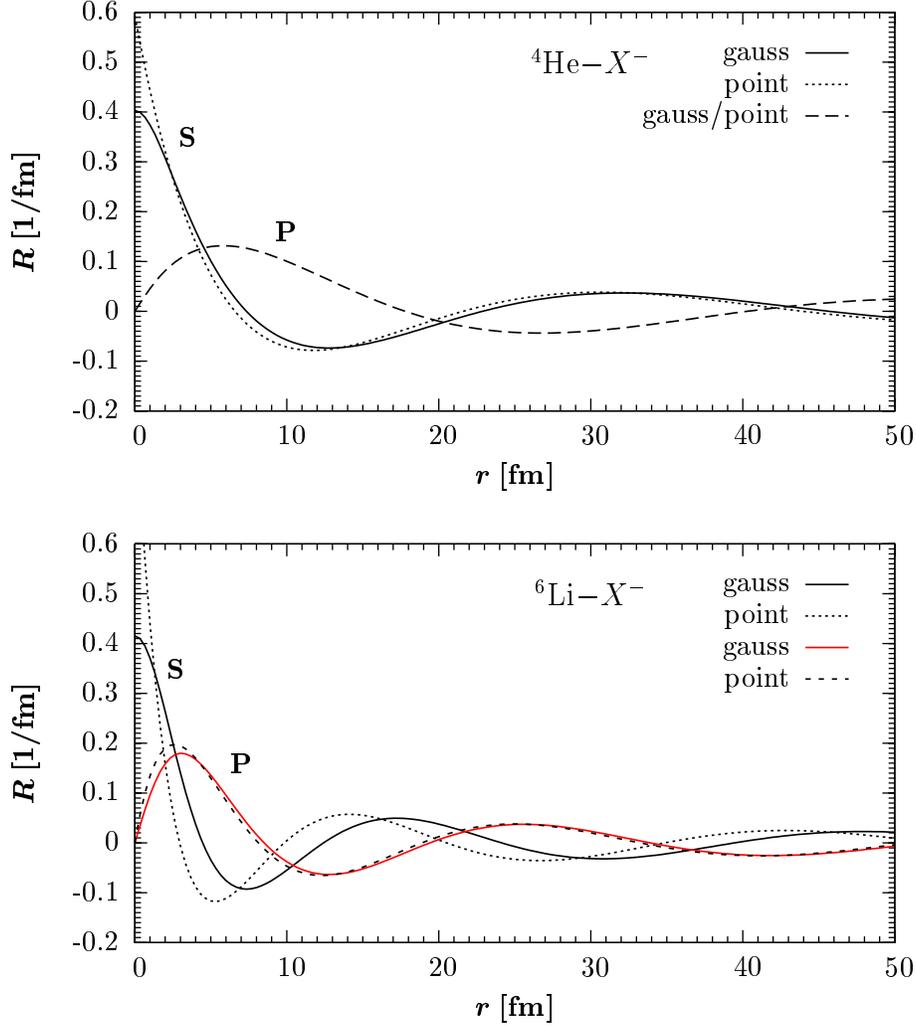}}
\caption[Continuum wave functions]{Continuum wave functions for
  \hef--\champ\ (top) and \lisx--\champ\ (bottom) with $k=10\ \MeV$
  and $\mx\to\infty$.}
\label{Fig:continuum-waves} 
\end{figure}

Here, $\eta = - 1/(k\abohr)$ denotes the Sommerfeld parameter for an
attractive Coulomb field where $\mathbf{k}$ is the wave vector of the
relative \nuc{}--\champ\ motion with $ |\mathbf{k}| =k= (2\mred
E)^{1/2}$; $\Gamma{(z)}$ is the Gamma
function~\cite{abramowitz+stegun} and $W_{a,\,b}(z)$ stands for
Whittaker's function~\cite{1963cma..book.....W}.

With $ W_{\pm a,b}(\pm z) = e^{\mp z/2} (\pm z)^{\pm a} \left[1 +
  \Orderof{z^{-1}} \right]$~\cite{1963cma..book.....W}
one finds that the asymptotic behavior of the wave
functions~(\ref{eq:Coulomb-waves-Def}) is given~by
\begin{subequations}
\label{eq:Coulomb-waves-ASYMPTOT}
\begin{align}
  \lim_{kr\to\infty} \regCoul{kl} &= +
  \sin{\big[ k r - \eta \ln{(2 k r) - \pi l /2 + \phaseCoul{l}}  \big]}  , \\
  \lim_{kr\to\infty} \irregCoul{kl} &= - \cos{\big[ k r - \eta \ln{(2 k
      r) - \pi l /2 + \phaseCoul{l}} \big]} ,
\end{align}
\end{subequations}
where the Coulomb phase is defined by $\phaseCoul{l} =
\arg{\,\Gamma{(l + 1 + i \eta)}}$.

Now, for $r\gg\Rcharge{\nuc{}}$ the radial solution of the
Schr\"odinger equation to a modified Coulomb potential can be written
as%
\footnote{This definition corresponds to normalization on the
  ``$k/2\pi$ scale'', $\int_0^\infty R_{k'l} R_{kl} r^2 dr = 2\pi
  \delta{(k'-k)}$. Note, however, that $\irregCoul{kl}$ is not regular
  at the origin. One has to introduce a cutoff factor if $R_{kl}$
  shall be an entire function; see~\cite{1960MNRAS.120..121B}.}
$ R^{\mathrm{out}}_{kl} = ({2}/{r})\, [a_l \regCoul{kl} - b_l
\irregCoul{kl}]$.
Retaining the asymptotic normalization
\begin{align}
  \lim_{kr\to\infty} R^{\mathrm{out}}_{kl} & = \frac{2}{r}\sin{\big[ k
    r - \eta \ln{(2 k r)} - \pi l /2 + \phaseCoul{l} + \phaseNuc{l}
    \big]} ,
\end{align}
it follows from comparison with~(\ref{eq:Coulomb-waves-ASYMPTOT}) that
the additional phase shift $\phaseNuc{l}(k)$ is given by $\phaseNuc{l}
= \tan{(b_l/a_l)}$. 

In the vicinity of the origin, i.e., for $r \lesssim \Rcharge{\nuc{}}$, the
numerically obtained wave function $R^{\mathrm{int}}_{kl}$ correctly
describes the solution to Schr\"odinger's equation. It can be
normalized by requiring a continuous transition at
$r=r_{\mathrm{fit}}$ to the outer solution
\begin{subequations}
  \begin{align}
    \label{eq:fitting-conditions}
    R^{\mathrm{int}}_{kl}(r_{\mathrm{fit}}) &=
    R^{\mathrm{out}}_{kl}(r_{\mathrm{fit}}) \ ,\\
    \left. \frac{d}{d r} R^{\mathrm{int}}_{kl}\right|_{r_{\mathrm{fit}}} &=
    \left. \frac{d}{d r}
      R^{\mathrm{out}}_{kl}\right|_{r_{\mathrm{fit}}}\ .
  \end{align}
\end{subequations}

Following the outlined approach, we
solve~(\ref{eq:schroedinger-equation}) for the relative motion of
various \nuc{}--\champ\ systems. As examples, we choose \hef--\champ\
and \lisx--\champ. At the times of BBN the relative velocity of the
\hef/\lisx--\champ\ system is Boltzmann distributed so that
$\VEV{k}\sim \sqrt{ \mred T}$ . Thus, a representative value is $k=10\
\MeV$ which corresponds to $T \simeq 20\, (30)\ \keV$ for
$\hef\,(\lisx)$ with $\mx\to\infty$. We join the inner solution with
the outer one at $r_{\mathrm{fit}} = 10\ \fm $. This determines the
phase shift~$\phaseNuc{l}$. We have checked that~$\phaseNuc{l}$ is
insensitive to the chosen value of $r_{\mathrm{fit}}$, provided
$r> \Rcharge{\nuc{}}$, and that $\phaseNuc{l} \to 0$ when
switching to a point-like nucleus. Using a Gaussian charge
distribution, we find for the S-wave $(l=0)$ of the
\hef(\lisx)--\champ\ system $\phaseNuc{0} = -0.21\,(-0.87) $ whereas
for the P-wave $(l=1)$ the phase shift is already significantly
reduced, $\phaseNuc{1} = -3.3\times 10^{-3}\,(-0.13)$.

In Fig.~\ref{Fig:continuum-waves} the corresponding wave functions for
\hef--\champ\ (top) and \lisx--\champ\ (bottom) are shown. As can be
seen, the wave functions $R_{k0}$ for the case ``gauss'' receive a
significant correction in comparison to the Coulomb case ``point''
which was already indicated by the size of the phase shifts
$\phaseNuc{0}$. Of course, the curves labeled ``point'' coincide with
the regular Coulomb functions
$(2/r)\regCoul{kl}$~[Eq.~(\ref{eq:regular-Coulomb-wave})]. Whereas for
the \hef--\champ\ system $\R_{k1}\simeq (2/r)\regCoul{k1}$, a
deviation from the regular Coulomb P-wave is visible in the
\lisx--\champ\ case.

When considering continuum wave functions for $k \to 0$ the numerical
evaluation of (\ref{eq:Yost-function}) is problematic. This case,
however, is the most important one in the computation of the
photo-dissociation cross section of \BSx{\nuc{}}
[Sec.~\ref{sec:phot-recomb-cross}]. Therefore, we need to consider the
Coulomb wave functions in a different form~\cite{1958MNRAS.118..504S},
\begin{subequations}
    \label{eq:coul-waves-QDM}
  \begin{align}
    \regCoul{kl} & = A\left({ i/k\abohr},l \right)^{1/2}
    \sqrt{\frac{\pi
        k \abohr}{2}} \, y_1\left({- i \eta} ,l ;  r/\abohr \right) , \\
    \irregCoul{kl} & = A\left({i/k\abohr },l \right)^{-1/2}
    \sqrt{\frac{\pi
        k \abohr}{2}} \, y_3\left({- i \eta} ,l ; r/\abohr \right) ,
  \end{align}
\end{subequations}
where $A\left({i}/{k \abohr},l \right) = \prod_{s=1}^{l}
\left[1+ (sk \abohr)^2\right] $ and $y_{1,3}$ are related to the
Whittaker functions; $A(x,0)=1$. Defined in this way, (\ref{eq:coul-waves-QDM})
satisfy the asymptotic behavior~(\ref{eq:Coulomb-waves-ASYMPTOT}). An
expansion in powers of $\kappa^{-2}$, i.e. in energy, for $y_1$
reads~\cite{1958MNRAS.118..504S} (see also~\cite{PhysRev.67.11})
\begin{subequations}
\label{eq:y1y3-expansion}
  \begin{align}
    y_1(\kappa,l;\rho) = \sum_{q=0}^{\infty} \kappa^{-2q}
    \sum_{p=2q}^{3q} a_{q,p}(l)(2\rho)^{(p+1)/2}
    J_{2l+1+p}\left(\sqrt{8\rho}\right) ,
  \end{align}
  whereas $y_3$ cannot be represented by a convergent expansion in
  powers of energy. However, an asymptotic expansion has been obtained
  in~\cite{1958MNRAS.118..504S},
  \begin{align}
    y_3(\kappa,l;\rho) = A(\kappa,l) \sum_{q=0}^{Q} \kappa^{-2q}
    \sum_{p=2q}^{3q}a_{q,p}(l)(2\rho)^{(p+1)/2}
    Y_{2l+1+p}\left(\sqrt{8\rho}\right) + \Orderof{\kappa^{-2Q-2}}.
  \end{align}
\end{subequations}
Here, $J_{2l+1+p}$ and $Y_{2l+1+p}$ are the respective Bessel functions of
the first and the second kind of order
$2l+1+p$~\cite{abramowitz+stegun} and the coefficients $a_{q,p}(l)$
satisfy recurrence relations ($a_{0,0}=1$); for details
see~\cite{1958MNRAS.118..504S}. With the expansion
\begin{align}
  A\left({i}/{k \abohr},l \right) = 1 + \frac{l(l+1)(2l+1)}{6}
  k^2\abohr^2 + \Orderof{k^4\abohr^4}
\end{align}
one obtains from (\ref{eq:y1y3-expansion})
\begin{subequations}
  \begin{align}
    \label{eq:zero-energy-coulomb-wf}
    \regCoul{kl} =& \sqrt{ \pi r k} \left\{ J_{2 l +1} \left(\sqrt{8 r
          / \abohr} \right) + \frac{(k\abohr)^2}{12}\left[
        l(l+1)(2l+1)\, J_{2 l +1} \left(\sqrt{8 r / \abohr} \right)
        \vphantom{\left(\frac{2r}{\abohr} \right)^{3/2}}
      \right.\right.  
    \nonumber \\ & 
    \left. \left.  -3(l+1)
        \left(\frac{2r}{\abohr}\right)\, J_{2 l +3} \left(\sqrt{8 r / \abohr} \right)
        + \left(\frac{2r}{\abohr} \right)^{3/2} J_{2 l +4}
        \left(\sqrt{8 r / \abohr} \right) + \Orderof{k^4\abohr^4}
      \right] \right\}
    \\
    \irregCoul{kl} =& \sqrt{ \pi r k} \left\{ Y_{2 l +1} \left(\sqrt{8 r
          / \abohr} \right) + \frac{(k\abohr)^2}{12}\left[
        l(l+1)(2l+1)\, Y_{2 l +1} \left(\sqrt{8 r / \abohr} \right)
        \vphantom{\left(\frac{2r}{\abohr} \right)^{3/2}}
      \right.\right.  
    \nonumber \\ & 
    \left. \left.  -3(l+1)
        \left(\frac{2r}{\abohr}\right)\, Y_{2 l +3} \left(\sqrt{8 r / \abohr} \right)
        + \left(\frac{2r}{\abohr} \right)^{3/2} Y_{2 l +4}
        \left(\sqrt{8 r / \abohr} \right) + \Orderof{k^4\abohr^4}
      \right] \right\}
  \end{align}
\end{subequations}
In the following section we employ those approximations in the
computation of the photo-dissociation cross section.

\section{Formation of bound states}
\label{sec:recomb-phot-rates}

The crucial quantity in the discussion of the catalysis of BBN
reactions is the bound state fraction $n_{\BSnucx}/n_{\nuc{}}$ of the
light elements \nuc{}. To this end we have to compute the cross
sections $\sigma_{\mathrm{rec}}$ for radiative recombination,
$\nuc{}+\champ \to \BSnucx +\gamma$, as well as $\sigma_{\mathrm{ph}}$
for the dissociation, $\BSnucx+\gamma_{\mathrm{bg}} \to \nuc{}+\champ$
due to background photons $\gamma_{\mathrm{bg}}$.

The rate (per nucleus \nuc{}) for \nuc{}--\champ\ recombination is
given by $\Gammarec = \sigmavof{\mathrm{rec}}\,
n_{\champ}$. Here, $\sigma_{\mathrm{rec}}$ has to be averaged over the
distribution of relative velocities $v$ between \nuc{}\ and \champ\
which is Maxwellian for a sub-\MeV\ plasma,
\begin{align}
  \label{eq:MB-velocity-distro}
  f_{v} = 4 \pi v^2 \left( \frac{\mred}{2 \pi T}\right)^{3/2}
  \exp{\left( \frac{- \mred v^2}{2T}  \right)}\ .
\end{align}
Hence,
\begin{align}
\label{eq:Th-Avg-Recombi}
\sigmavof{\mathrm{rec}} = \sqrt{\frac{8}{\pi \mred}} T^{-3/2}
\int_0^\infty d E_{\mathrm{k}}\, \sigma_{\mathrm{rec}} E_{\mathrm{k}}
e^{-E_{\mathrm{k}}/T} ,\
\end{align}
where $ E_{\mathrm{k}} = \mred v^2 / 2$ has been used. 

The rate of photo-dissociation $\Gammaph$ of $\BSnucx$ pairs depends
on the number of photons whose energy $E_\gamma$ exceed that of the ionization
potential $|\Ebind|$ of the bound state,
\begin{align}
\label{eq:number-of-ionizing-photons}
n_{\gamma}(E_{\gamma}>|\Ebind|) = \frac{1}{\pi^2} \int_{|\Ebind|}^{\infty}
 dE_\gamma\, \frac{E_\gamma^2}{e^{E_\gamma/T}-1}\ , 
\end{align}
and is given by $\Gammaph = \sigma_{\mathrm{ph}}\,
n_{\gamma}(E_\gamma>|\Ebind|) $.

The principle of detailed balance~\cite{nla.cat-vn2263194} relates the
cross sections via $\mathbf{p}^2 \sigma_{\mathrm{rec}} = 2
E_{\gamma}^2\sigma_{\mathrm{ph}} $ where $|\mathbf{p}|=\mred v$
denotes the momentum of the relative motion of the \nuc{}--\champ
system and the factor of two is a statistical factor accounting for
the two polarization degrees of freedom of the photon. From the
definition of the rates \Gammaph\ and \Gammarec\ together with
(\ref{eq:Th-Avg-Recombi}) and (\ref{eq:number-of-ionizing-photons}) it
follows that%
\footnote{It is used that $E_\gamma = E_{\mathrm{k}} + |\Ebind|$ and
  that $[\exp{(E_\gamma/T)}-1]^{-1} \simeq \exp{(-E_\gamma/T)}$ which
  holds well in the temperature regions of main interest.}
\begin{align}
  \label{eq:detailed-balance-rec-ph}
  \frac{ \Gammarec }{\Gammaph} = \left(
    \frac{2 \pi}{\mred T} \right)^{3/2} e^{|\Ebind|/T} n_{\champ} \ .
\end{align}

As long as $ \Gammaph(T),\,\Gammarec(T) \gtrsim H(T)$, i.e., as long
as recombination and break-up reactions happen frequently, the
concentrations of \nuc{},\ \champ, and \BSnucx\ have time to achieve
equilibrium values such that the reaction densities for recombination
and dissociation are equal, $ \Gammaph\, n_{\BSnucx} = \Gammarec\,
n_{\nuc{}}$. This yields the Saha equation for the bound state
fraction,%
\footnote{When used in this form one may need to impose that the
  number of bound states cannot be larger than the total number
  recombination partners available.}
\begin{align}
  \label{eq:stat-equilibrium-rec-ph}
  \frac{ n_{\BSnucx}}{ n_{\nuc{}} } = \frac{ \Gammarec }{\Gammaph} 
  \ .
\end{align}

\subsection{Photo-dissociation and recombination cross section}
\label{sec:phot-recomb-cross}

Since the early Universe is in a high-entropy state, bound states can
only form efficiently once $T\lesssim
|\Ebind|/40$~\cite{Kohri:2006cn}. At the relevant times, i.e., when
$\Gammaph \lesssim H$, only those photons in the high energy tail of
the spectrum with $E_\gamma \ge |\Ebind|$ are capable of destroying
\BSx{\nuc{}}; $\VEV{E_\gamma}\simeq 3 T$.
In addition, the binding energy is significantly smaller than the
associated light element mass so that a non-relativistic treatment of
the photoelectric effect is perfectly justified.

For the computation of the photo-dissociation cross section we can
employ Fermi's golden rule. The probability per unit time for a
nucleus bound to \champ\ to undergo a transition into the continuum
is given by 
\begin{align}
  \label{eq:fermis-golden-rule}
  d w_{\mathrm{ph}} = 2\pi |V_{fi}|^2 \delta{\left(-|\Ebind | + E_\gamma -
    E_{\mathrm{k}}\right)} d\rho \ . 
\end{align}
After the transition the nucleus has kinetic energy $ E_{\mathrm{k}}$
and momentum $\mathbf{p}$.%
\footnote{ Of course, strictly speaking, it is the energy and momentum
  of the relative motion. From the above explanations, however, it is
  clear that the \champ\ recoil is negligible.}
The density of final states is $d\rho = d^3 \mathbf{p}/(2\pi)^3 $
and the matrix element for absorption of a photon with momentum
$\mathbf{k}$ and energy $E_\gamma^2 = \mathbf{k}^2$ reads
\begin{align}
  V_{fi} = \frac{Z e}{\sqrt{2 E_\gamma}} \, e_\mu j^\mu_{fi} (-\mathbf{k})
\end{align}
where $j^\mu_{fi}(\mathbf{k}) = \int d^3x\,
e^{-i\mathbf{k}\cdot\mathbf{r}} j^\mu_{fi}(x)$ is the Fourier
transform of the transition current $j^\mu_{fi}(x) =
\overline{\psi}_{f}\gamma^\mu \psi_{i}$ and $e_\mu$ denotes the photon
polarization vector; see, e.g.,~\cite{Berestetsky:1982aq}. The cross
section is found by dividing~(\ref{eq:fermis-golden-rule}) by the
incident photon flux density. Averaging over photon polarizations [in
the gauge $(e^\mu) = (0,\mathbf{e})$], and integrating over
$E_{\mathrm{k}}$, the differential cross section for photo-dissociation
is given by
\begin{align}  
  \label{eq:differential-Photo-cross-section}
  \frac{d\sigma_{\mathrm{ph}}}{d\Omega_{\mathbf{p}}} & = 
  \frac{ Z^2 \alpha\, \mred |\mathbf{p}|}{4 \pi E_\gamma }
   |\mathbf{e_{k}} \times \mathbf{j}_{fi} |^2 \ ,
\end{align}
where $\mathbf{e_k}$ is a unit vector in $\mathbf{k}$-direction and
$\mathbf{j}_{fi}$ is the spatial part of $j^\mu_{fi} (-\mathbf{k})$.

We shall consider ionization from 1S as well as from 2S states. The
initial state wave function is $\psi_{i} = (4\pi)^{-1/2}
R_{1\mathrm{S} / 2 \mathrm{S} }$. The final state has to comprise a
plane wave in $\mathbf{p}$ direction together with an \textit{ingoing}
spherical wave~\cite{1965qume.book.....L}. In the partial wave
expansion,
\begin{align}
  \label{eq:parial-wave-ingoing}
  \psi_f = \frac{1}{2 |\mathbf{p}|} \sum_{l=0}^{\infty} i^l (2l+1)\,
  e^{-i(\phaseCoul{l}+\phaseNuc{l})} R_{|\mathbf{p}| l}
  P_{l}(\mathbf{e_p}\cdot \mathbf{e_r}) 
\end{align}
with unit vectors $\mathbf{e_x}$ in $\mathbf{x}$-direction; $ P_{l}$
are the Legendre polynomials~\cite{abramowitz+stegun}.  Note the
appearance of the additional phase shift~$\phaseNuc{l}$ coming from
the finite charge radius correction.

\begin{table}[tb]
\begin{center}
\caption[Photo-dissociation and recombination cross sections]{Listed
  below are the cross sections $\sigma_{\mathrm{ph}}$ for the
  threshold ($E_\gamma = |\Ebind|$) bound-free transition from 1S and
  2S states. From the definition~(\ref{eq:thavg-recombi-cs}) the
  averaged cross sections $\sigmavof{\mathrm{rec}}$ for recombination
  into 1S for $\px$ and into 1S+2S for \BSx{\hef}\ and  \BSx{\lisx}
  are obtained in the third column. Bracketed quantities refer to the
  ``point'' case.  In addition, a critical temperature of bound state
  formation~$T_\mathrm{rec}$ defined by
  $\Gammaph(T_{\mathrm{rec}})=H(T_{\mathrm{rec}})$ is given in the
  last column.}
\begin{tabularx}{0.89\textwidth}{@{\extracolsep{\fill}}Xr@{\:}rr@{\:}r@{\quad}r@{\:}r@{\quad}r@{.}l}
\toprule\\[-0.5cm]
bound state & 
\multicolumn{2}{c}{
\parbox[t]{1.8cm}{\centering
$\sigma_{\mathrm{ph}}^{\mathrm{1S}}$\\[0.2cm] $[\mathrm{m}\barn]$}
}
  &
\multicolumn{2}{c}{
\parbox[t]{2.1cm}{\centering
$\sigma_{\mathrm{ph}}^{\mathrm{2S}}$\\[0.2cm] $[\mathrm{m}\barn]  $}}& 
\multicolumn{2}{c}{
\parbox[t]{2.2cm}{\centering
$N_{\mathrm{A}}\sigmavof{\mathrm{rec}} T_9^{1/2}$\\[0.2cm] $[ \cm^3 \seconds^{-1}\mol^{-1}]$
}
}&
\multicolumn{2}{c}{
\parbox[t]{2.2cm}{\centering
$T_{\mathrm{rec}}$ \\[0.2cm] $[\keV]$
}
}
\\[0.9cm]
\midrule
 \BSx{\proton}   & 1870&          & 4380 &          & \: 3980 & {\tiny $(1\mathrm{S})$}\hspace*{1.1cm}      &  \qquad \quad 0&6    \\
 \BSx{\hef}      & 118 &          & 294  &   (278)  & 7260  &   (9230)  &    8&3       \\
 \BSx{\lisx}     & 34 &     (52)  & 103  &   (123)  & 6640  &   (25370) &    19&0     \\
\bottomrule
\end{tabularx}
\label{tab:photodiss-rec-scs}
\end{center}
\end{table}

Since the wavelength of the ionizing radiation ($\lambdabar =
1/|\Ebind|$ on the threshold) is much larger than the \BSx{\nuc{}}
dimensions, we can use the electric dipole approximation. The
associated selection rule implies $l=1$ for the continuum so that
\begin{align}
  \label{eq:psi-out-Photo}
  \psi_f = \frac{3 ie^{-i(\phaseCoul{1}+\phaseNuc{1})} }{2 |\mathbf{p}|
  } (\mathbf{e_p}\cdot \mathbf{e_r}) R_{|\mathbf{p}| 1} \ ,
\end{align}
and thus (in the dipole approximation)
\begin{align}
  \label{eq:transistion-current}
  \mathbf{j}_{fi} = -\frac{ 3 i}{\sqrt{16\pi} \mred |\mathbf{p}|} \int
  d^3\mathbf{r}\, (\mathbf{e_p}\cdot \mathbf{e_r}) R_{|\mathbf{p}| 1}
  \nabla R_{1\mathrm{S} / 2 \mathrm{S} } \ .
\end{align}

Performing all angular integrations in
(\ref{eq:differential-Photo-cross-section}) yields for the total
photo-dissociation cross section
\begin{align}
  \label{eq:Photo-cross-section}
  \sigma_{\mathrm{ph}} = \frac{2 \pi Z^2 \alpha }{3 \mred E_\gamma}
  \left[ \frac{1}{\sqrt{|\mathbf{p}|}} \int_0^\infty dr\, r^2
    R_{|\mathbf{p}| 1} \frac{\partial}{\partial r} R_{1\mathrm{S} / 2
      \mathrm{S} } \right] ^2 \ .
\end{align}
For $R_{|\mathbf{p}| 1}$ and $R_{1\mathrm{S} / 2\mathrm{S}}$ we employ
our numerically obtained solutions of the previous section which takes
into account the finite charge radius of the nucleus.
Note that on the ionization threshold $\sigma_{\mathrm{ph}} $ is
independent of $\mathbf{|p|}$. For a pure Coulomb field the momentum
dependence cancels analytically when using the leading term
in~(\ref{eq:zero-energy-coulomb-wf}). By the same token, numerically,
$[\dots]$ in~(\ref{eq:Photo-cross-section}) becomes insensitive to
$\mathbf{|p|}$. Thus, we find a constant cross section for $E_\gamma
\to |\Ebind|$. In this limit, using detailed balance, the averaged
recombination cross section reads
\begin{align}
  \label{eq:thavg-recombi-cs}
  \sigmavof{\mathrm{rec}} = \frac{4}{\sqrt{2\pi}} \left(
    \frac{\Ebind}{\mred} \right)^2
  \sqrt{\frac{\mred}{T}} \, \sigma_{\mathrm{ph}} \ .
\end{align}
from which $\Gammaph$ is readily obtained by
using~(\ref{eq:detailed-balance-rec-ph})
\begin{align}
  \Gammaph = \sigmavof{\mathrm{rec}} \left(\frac{\mred
      T}{2\pi}\right)^{3/2} e^{-|\Ebind|/T} .
\end{align}

In Table \ref{tab:photodiss-rec-scs} we present the results on the
threshold photo-dissociation cross sections $\sigma_{\mathrm{ph}}$ for
transistions from 1S and 2S states for the elements $\proton$, $\hef$,
and $\lisx$ and for $\mx\to \infty$. The bracketed values are for a
pure Coulomb potential whereas the other results are obtained by using
a Gaussian charge distribution. The respective values do not differ
very much. This is because the decrease of $|\Ebind|(=E_{\gamma})$ in
the denominator of (\ref{eq:Photo-cross-section}) when switching from
the ``point'' to the ``gauss'' case is counterbalanced by an increase
in the radial integral so that the net effect is small. However, the
reduction of the total (1S+2S) recombination cross section
$\sigmavof{\mathrm{rec}}$ from the hydrogen-like case is drastic. This
is due to the additional factor of $\Ebind^2$ in
(\ref{eq:thavg-recombi-cs}). In the last column we show the
temperature for which $\Gammaph(T_{\mathrm{rec}}) =
H(T_{\mathrm{rec}})$, i.e., the temperature when the formation of
bound-states can proceed efficiently---provided that~$\Gammarec\gtrsim
H$ and that the bound state is not destructed by another process.

Finally, we remark that for other (heavier) nuclei than the ones
presented in Table~\ref{tab:photodiss-rec-scs} the discussion of
recombination can become more involved. If the light element $\nuc{}$
possesses an excited state $N^*$ with a level splitting smaller than
the $\xm$ binding energy, then recombination may also proceed into
$(\mathrm{N}^*\champ)$ opening up the possibility of resonant
recombination. This was pointed out in~\cite{Bird:2007ge} where the
formation of $\BSx{\bes}$ was considered.

\section{Nuclear reactions with bound states and their catalysis}
\label{sec:catalyz-nucl-react}

After the freeze-out of weak interactions with the cease of n and p
interconversion processes,
light element fusion in SBBN proceeds via inelastic two-body nuclear
reactions%
\footnote{This does not include ``production'' processes like that of
  \lisv\ via electron capture by \bes\ or of \het\ by beta decay of
  \Trit, both of which, however, only happen at a much later time.}
\begin{subequations}
\begin{alignat*}{2}
B + C &\to D + E        &\qquad &    \mathbf{I}     \\
  B + C &\to F + \gamma &\qquad &  \mathbf{II} 
\end{alignat*}
\end{subequations}
with $B,\dots ,F$ denoting the nuclei of the light elements and the
arrow indicating the forward process, i.e., the exoergic direction
with positive $Q$ value. The reverse processes are typically
suppressed by $\exp{(-Q/T)}$ such as in
(\ref{eq:detailed-balance-rec-ph}) (which is an atomic process.) Only
elements with atomic mass number $A\leq 7$ are produced in relevant
quantities.

In presence of bound states of the light elements with \champ\ during
BBN the following additional types of inelastic reactions emerge as
particularly prominent,
\begin{subequations}
  \begin{alignat*}{2}
    \BSx{B} + C &\to D + E + \champ    &\qquad &  \mathbf{I}^{*}    \\
    \BSx{B} + C &\to \BSx{F} + \gamma  &\qquad &    \mathbf{II}^*    \\
    \BSx{B} + C &\to F + \champ &\qquad & \mathbf{III}  \\
    \BSx{B} + C &\to B + \BSx{C} &\qquad & \mathbf{IV}  \ .
  \end{alignat*}  
\end{subequations}
A first observation is that, in presence of bound states, the energy
gain of a nuclear reaction is altered. In the entrance channel, the
total available \textit{internal} energy is reduced due to the binding
of $B$ with \champ.  Thus, for example, $Q_{\mathbf{I^{*}}} =
Q_{\mathbf{I}} - |\Ebind^{\BSx{B}}|$ whereas additional energy becomes
available in the exit channel of $\mathbf{II^*}$ so that
$Q_{\mathbf{II^{*}}} = Q_{\mathbf{II}} - |\Ebind^{\BSx{B}}| +
|\Ebind^{\BSx{F}}|$.
Since $Q$ values of nuclear reactions are mainly in the \MeV\ to
multi-\MeV\ range, usually three-body break-up reactions
$\mathbf{I}^*$ rather than $\BSx{D} + E $ exit channels are realized.
The shift in energetics can also allow for resonances which are not
possible in SBBN. For example, type $\mathbf{II}^*$ can be realized in
resonant capture reactions whose intermediate excited state
$\BSx{F}^*$ decays into the $\BSx{F}$ ground state by $\gamma$
emission. If, instead, the nucleus is in an excited state $\BSx{F^*}$,
then also the $ F + \champ $ continuum acts as a concurrent
channel---provided that it is kinematically accessible. The latter is
an example of a reaction of type $\mathbf{III}$ which is of particular
interest since it has no SBBN counterpart.  \textit{Atomic} reactions
$\mathbf{IV}$ are called charge exchange reactions. They are also
important to consider because they can significantly affect the
relative concentrations of nuclei bound to \champ. They will be
discussed in Sec.~\ref{Sec:pXcatalysis}.

Reactions of the form $\BSx{B} + \BSx{C} \to \dots$ are only of
secondary importance. Their efficiency depends on the average relative
velocity between $\BSx{B}$ and $\BSx{C}$ which scales as $\mx^{-1/2}$.
Thus, for weak scale relics, the suppression of the average velocity
of \xm-containing bound states relative to the velocity of light
nuclei is from one to two orders of magnitude.

Another observation is that the screening of the charge of $B$ when in
bound state with \champ\ will lead to a modification of the SBBN cross
sections with charged ``projectiles'' $C$. It is customary to write
the cross sections of charged-particle induced reactions in the form
\begin{align}
  \label{eq:S-factor}
  \sigma(E_{\mathrm{k}}) = S(E_{\mathrm{k}})\, E_{\mathrm{k}}^{-1} e^{-2
    \pi \eta}
\end{align}
and which defines the astrophysical $S$-factor. The definition scales
out the ``geometrical'' cross section $\pi {\lambdabar}^2 \propto
E_{\mathrm{k}}^{-1}$ as well as the Coulomb penetration factor
$\exp{(-2 \pi \eta)}$. Note that during BBN ($T\lesssim 0.1\ \MeV$)
the thermal energy $\VEV{E_{\mathrm{k}}}\sim T$ of the reactants is
significantly smaller than the height of the Coulomb barrier $
E_{\mathrm{c}}\simeq\Orderof{\MeV}$; $\lambdabar = (\mred v)^{-1}$ is
the de~Broglie wavelength of the relative motion and $\eta =
Z_{B}Z_{C}\alpha /v>0$ denotes the earlier encountered Sommerfeld
parameter [below (\ref{eq:Yost-function})]. Using (\ref{eq:S-factor})
the definition of the thermally averaged cross
section~(\ref{eq:Th-Avg-Recombi}) becomes
\begin{align}
  \label{eq:gamow-factor}
  \sigmavof{\mathrm{rec}} = \sqrt{\frac{8}{\pi \mred}} T^{-3/2}
  \int_0^\infty d E_{\mathrm{k}}\, S(E_{\mathrm{k}})
  \exp{\left[-\frac{E_{\mathrm{k}} }{T}- \left(\frac{E_{\mathrm{G}}
      }{E_{\mathrm{k}} } \right)^{1/2} \right]} ,
\end{align}
where $E_{\mathrm{G}} = 2 \mred (\pi \alpha Z_B Z_C)^2$ is called the
Gamow energy. In absence of resonances the $S$-factor is only a slowly
varying function of $E_{\mathrm{k}}$ so that the integral is dominated
by the exponential which peaks at $E_0 = E_{\mathrm{G}}^{1/3}
(T/2)^{2/3}$ and which marks the energy range of most effective
nucleosynthesis (Gamow window). 

One may then attempt to account for the bound state in the entrance
channel by replacing $Z_B$ by $Z_{B} - 1$ and correcting for the
changed kinematics and energetics. Indeed, such a program has first
been carried out in a BBN network calculation
in~\cite{Kohri:2006cn}. When studying the effect on the charged
particle induced reactions, the authors find no significant changes in
the light element yields at the CMB inferred baryon asymmetry. Whereas
the compactness of the bound states with the heavier of the light
elements gives some justification to this procedure we will see in
Sec.~\ref{Sec:pXcatalysis} that, e.g., the large size of the
$\BSx{\proton}$ system plays a crucial role in obtaining a consistent
picture of BBN.

\subsection[Catalysis of \lisx\ production]{\texorpdfstring{Catalysis of
  $\mathbf{{}^6}$Li production}{Catalysis of 6Li production}}
\label{sec:catalys-lisx-prod}

The potential influence of bound states on the BBN paradigm was
already discussed almost twenty years ago
in~\cite{Dimopoulos:1989hk,DeRujula:1989fe,Rafelski:1989pz}.  However, only recently it has
been realized~\cite{Pospelov:2006sc} that the presence of \champ\ at
$T\lesssim 10\ \keV$ can lead to a tremendous enhancement of the
\lisx\ output.

In SBBN the freeze-out of \lisx\ from nucleosynthesis is dominated by
its production via radiative capture and its destruction via proton
burning,
\begin{subequations}
      \label{eq:SBBN-Li6-reactions}
  \begin{alignat}{2}
    \label{eq:SBBN-Li6-production}
    \hef + \deut &\to \lisx + \gamma &\quad  Q &= 1.47\ \MeV \ , \\
%
%
    \label{eq:proton-burning-Li6}
    \lisx + \proton &\to \het + \hef & Q &= 4.02\ \MeV \ ,
  \end{alignat}
\end{subequations}
respectively.
The cross section for the production
reaction~(\ref{eq:SBBN-Li6-production}) is very small with $ S \sim
10^{-8}\ \MeV\, \barn$~\cite{Angulo:1999zz} at the lowest energies.  For
example, the $S$-factor for $\het + \hef \to \bes +\gamma$ which is
the main source for $\lisv$ at $\etacmb$ reads $S(0)=5.8\times
10^{-4}\ \MeV\, \barn$~\cite{Cyburt:2008up}. The small SBBN output of
\lisx\ is attributed to the inefficiency of the production
process~(\ref{eq:SBBN-Li6-production}).%
\footnote{This is usually traced back to a weak quadrupole
  transition~(E2) in (\ref{eq:SBBN-Li6-production}). Theoretical
  calculations seem to suggest, however, that at the BBN relevant
  temperatures $(T\simeq 70\ \keV)$ the dipole transition (E1) is as
  important~\cite{Nollett:2000ch}.  Nevertheless, the dipole moment of
  the $\hef$--$\deut$ cluster is almost vanishing due to a similar
  charge-to-mass ratio which yields a very small cross section;
  cf.~\cite{Pospelov:2006sc}.  }
In this regard, also note that the destruction reaction
(\ref{eq:proton-burning-Li6}) has a large $S$-factor, $S(0) = 2.97\
\MeV\,\barn $~\cite{Angulo:1999zz}.

\begin{figure}[t]
\centerline{
\includegraphics[width=10cm]{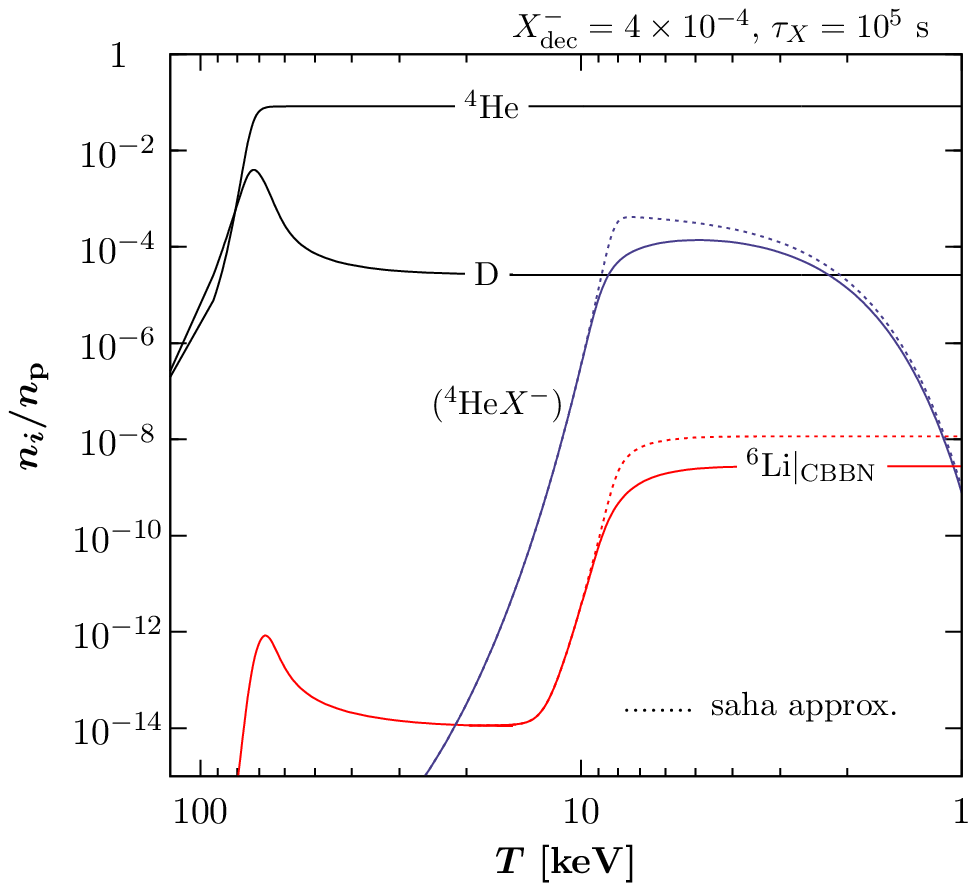}%
}
\caption[CBBN evolution of \lisx]{ Choosing $\champ_\dec = 4\times
  10^{-4}$ and $\tauX = 10^5\ \seconds$ the yields of \deut, \hef, and
  \lisx\ normalized to the proton number density~$n_\proton$ are
  shown. For $T\gtrsim 10\ \keV$ the light elements follow their SBBN
  evolution (the output thereof is produced using the BBN
  code~\cite{Pisanti:2007hk}.)  At lower temperatures bound
  states~\BSx{\hef} can form and catalyzed \lisx\ production proceeds
  via~(\ref{eq:CBBN-Li6-production}).  The dashed lines show the
  $\BSx{\hef}$ and \lisx\ abundance when the Saha
  approximation~(\ref{eq:stat-equilibrium-rec-ph}) is used. The
  resulting overestimation of \lisx\ production illustrates that a
  full numerical solution of~(\ref{eq:boltzmann-eqs}) is
  necessary. The CBBN variation of \deut\ and \hef\ is negligible.}
\label{Fig:CBBN-Li6-production}
\end{figure}

Let us briefly outline the evolution of the \lisx\ abundance in SBBN.
Once the temperature of the primordial plasma drops below $T\lesssim
0.1\ \MeV$ the deuterium bottleneck opens so that \lisx\ production
can proceed via (\ref{eq:SBBN-Li6-production}). A sharp drop in \deut\
below $T\lesssim 80\ \keV$ is accompanied by an associated decline in
\lisx; see left part of Fig.~\ref{Fig:CBBN-Li6-production}. Net
production of \lisx\ via (\ref{eq:SBBN-Li6-production}) soon freezes
out but proton burning (\ref{eq:proton-burning-Li6}) continues until
$\sim 10\ \keV$.

In the previous section we have seen that once the temperature drops
below $ 8\ \keV$ the photo-dissociation rate of $\BSx{\hef}$ freezes
out. Thus, provided that \tauX\ is large enough, the concentration of
$\BSx{\hef}$ can become substantial and fusion of \lisx\ is then
possible via the alternative path~\cite{Pospelov:2006sc}
\begin{alignat}{2}
  \label{eq:CBBN-Li6-production}
  \hefxm + \deut & \to \lisx + \champ & \quad Q & \simeq 1.13\ \MeV 
\end{alignat}
and which is a reaction of type~$\mathbf{III}$. Whereas the size of
the radiative capture cross section (\ref{eq:SBBN-Li6-production}) is
governed by the selection rules of the electromagnetic transition,
(\ref{eq:CBBN-Li6-production}) suggests a cross section which is
determined by the short distance behavior of the \hef--\deut\
cluster. Indeed, the original work~\cite{Pospelov:2006sc} estimates
$S_{\mathrm{CBBN}}/S_{\mathrm{SBBN}}\sim 10^8$ which points to a
cross section for~(\ref{eq:CBBN-Li6-production}) which is in the
ballpark of photonless SBBN reaction rates and implies the catalysis
of \lisx\ production.

Meanwhile, a dedicated quantum three-body calculation of the $\hef +
\deut + \champ$ system has become available~\cite{Hamaguchi:2007mp}
which confirms the catalytic picture. The authors find
$S(E_{\mathrm{G}}) = 0.038\ \MeV\barn$ at the Gamow peak position
$E_{\mathrm{G}} = 36.4\ \keV$. The thermally averaged cross section
reads~\cite{Hamaguchi:2007mp}
\begin{align}
\Navogadro \sigmavof{\mathrm{cat},\lisx}  = 
2.37 \times 10^8\, (1 - 0.34\, T_9)\, T_9^{-2/3}
  \exp{\left( -5.33\,  T_9^{-1/3} \right)}
\end{align}
and is given in the customary units of $ \cm^3\seconds^{-1}
\mol^{-1}$; $T_9$ denotes the temperature in units of $10^9\
\mathrm{K}$ and $\Navogadro$ is the Avogadro
constant~\cite{Mohr:2005zz}.

Since $\BSx{\hef}$ only forms for $T\lesssim 10\ \keV$, i.e., at a
time when the \hef\ and \deut\  abundances are essentially frozen out,
we can incorporate the effect of catalytic \lisx\ production in the
following way. At some low temperature $T < 20\ \keV$ we couple the
SBBN output into the network of Boltzmann equations
\begin{subequations}
  \label{eq:boltzmann-eqs}
\begin{align}
 \label{eq:boltzmann-BShef}
  - H T \frac{d }{dT}\, {\BSx{\hef}} & =
  \sigmavof{\mathrm{rec},{\hef}} \, \nb\, \hef\, \champ 
  - \Gammaphof{\hef}\,  {\BSx{\hef}} 
  \nonumber \\ 
  & \qquad  - \sigmavof{\mathrm{cat},\lisx} \, \nb\,\deut\, {\BSx{\hef}} 
  - \Gamma_{\X}\, {\BSx{\hef}}   
 \ ,  \\
 \label{eq:boltzmann-champ-free}
  - H T \frac{d }{dT}\, \champ & = 
  - \sigmavof{\mathrm{rec},{\hef}} \, \nb\,  \hef\, \champ 
  + \Gammaphof{\hef}\, {\BSx{\hef}}  
  \nonumber \\
  & \qquad  + \sigmavof{\mathrm{cat},\lisx} \, \nb\, \deut\, {\BSx{\hef}} 
  - \Gamma_{\X}\, {\BSx{\hef}}  
 \ ,   \\
\label{eq:boltzmann-hef}
    - H T \frac{d }{dT}\, \hef & = 
  - \sigmavof{\mathrm{rec},{\hef}}  \, \nb\,  \hef\, \champ 
  + \Gammaphof{\hef}\, {\BSx{\hef}} 
  \nonumber \\
  & \qquad + \Gamma_{\X}\, {\BSx{\hef}} 
 \ , \\
\label{eq:boltzmann-lisx}
 - H T \frac{d }{dT}\, \lisx & = 
 \sigmavof{\mathrm{cat},{\lisx}} \, \nb\, \deut \, \BSx{\hef} -
 \sigmavof{\mathrm{des},\lisx} \, \nb \,  \proton\, \lisx 
 \ ,\\
\label{eq:boltzmann-deut}
 - H T \frac{d }{dT}\, \deut & = 
 - \sigmavof{\mathrm{cat},{\lisx}} \, \nb\, \deut \, \BSx{\hef} \ . 
\end{align}
\end{subequations}
Light elements as well as bound state abundance are normalized to the
baryon number, $ \nuc{} \equiv n_{\nuc{}}/n_{\mathrm{b}}$, $\BSx{\hef}
= n_{\BSx{\hef}}/n_{\mathrm{b}} $, and $ \champ \equiv
n_{\champ}/n_{\mathrm{b}}$---better overview shall compensate for the
slight abuse of notation. 

The central input parameter for the catalytic production of \lisx\
(and \ben, see below) is the abundance of \xm\ at the time of its
recombination with \hef.  Above $10\ \keV$, we can track the resulting
(\hef\xm) abundance by using the
Saha-equation~(\ref{eq:stat-equilibrium-rec-ph}) since
photo-dissociation and recombination proceeds rapidly. Only at
$T\lesssim 10\ \keV$, (\hef\xm) starts to build up efficiently so that
we couple it into the full set of Boltzmann
equations~(\ref{eq:boltzmann-eqs}).
We parameterize \champ\ by its abundance prior to decay by introducing
$\champ_{\mathrm{dec}}$, where the superscript ``dec'' stands for
decoupling, and by the \xm\ lifetime $\tauX =\Gamma_{\X}^{-1} $,
so that the (total) \xm\ abundance at any moment during BBN is given
by $\champ(t)=\champ_\dec\times\exp(-t/\tau_X)$.
The SBBN cross section $\sigmavof{\mathrm{des},\lisx}$ for residual
\lisx\ destruction~(\ref{eq:proton-burning-Li6}) can be found
in~\cite{Caughlan:1987qf}. We solve~(\ref{eq:boltzmann-eqs}) using as
initial conditions the SBBN output values of the computer
code~\cite{Pisanti:2007hk}: $Y_\mathrm{p} \equiv 4 n_{\Hefournarrow} /
n_{\mathrm{b}} = 0.248 $, $\mathrm{D}/\mathrm{H} =2.6\times 10^{-5}$,
$\mathrm{\Lisix}/\mathrm{H} =1.14\times 10^{-14}$, and
$n_{\mathrm{p}}/n_{\mathrm{b}} = 0.75$; furthermore, $\geff = 3.36$
and $\heff = 3.91$.

In Fig.~\ref{Fig:CBBN-Li6-production} we show the evolution of the
\deut, \hef, and \lisx\ number densities normalized to~$n_\proton$ for
the exemplary choice $\champ_\dec = 4\times 10^{-4}$ and $\tauX =
10^5\ \seconds$. Rapid photo-dissociation delays bound-state formation
of $\BSx{\hef}$ until $T\sim 10\ \keV$. Once \BSx{\hef} forms,
catalyzed \lisx\ production proceeds efficiently
via~(\ref{eq:CBBN-Li6-production})  which leads to the steep rise
of \lisx\ at $T\sim 10\ \keV$.
As can be seen, the \deut\ and \hef\ reservoirs are essentially
unaffected by this. In addition, the dashed lines show the
$\BSx{\hef}$ and \lisx\ abundances when---instead of
solving~(\ref{eq:boltzmann-BShef})---the Saha
approximation~(\ref{eq:stat-equilibrium-rec-ph}) for the bound state
fraction is used. This results in an overestimation of the \lisx\
output. The reason is that \hef--\champ\ recombination itself is
efficient only for a short period after photo-dissociation freezes
out. Thus, in order to obtain a reasonable estimate on
$\lisx|_{\mathrm{CBBN}}$, a numerical solution of the Boltzmann
equations~(\ref{eq:boltzmann-eqs}) is necessary.

\subsection[Catalysis of \ben\ production]{\texorpdfstring{Catalysis of
  $\mathbf{{}^9}$Be production}{Catalysis of 9Be production}}
\label{sec:catalys-ben-prod}

Another dramatic catalytic enhancement is seen in the production of
\ben.  The yield of \ben\ in SBBN is tiny: Whereas the short lifetime
of \beet\ [see Sec.~\ref{sec:discrete-spectrum}] renders the neutron
capture reaction
$  \beet + n \to \ben + \gamma$
inefficient, \ben\ production via fusion on Li isotopes yields no more
than $\ben/\Hyd < 10^{-18}$ at \etacmb~\cite{Thomas:1992tq}.
The catalytic path to \ben\ is shown by the following
sequence~\cite{Pospelov:2007js}
\begin{eqnarray}
\label{eq:traf-Be9}
X^- \to (\hef X^-) \to (\beet X^-) \to  \ben 
\end{eqnarray}
which goes through the ``double bottleneck'' of \BSx{\hef} and
\BSx{\beet}. Bound states ($\beet$\xm) are formed by the radiative
fusion
\begin{subequations}
  \begin{align}
    \label{eq:radiative-alpha-fusion-beet}
    \hef + (\hef\xm) &\to (\beet\xm)+\gamma \ ,
    \intertext{and the catalysis of \ben\ production is triggered by
      the photonless recoil reaction}
    (\beet X^-) + n &\to  \ben +X^-
    \label{Eq:beet-neutron-capture}
    \ .
  \end{align}
\end{subequations}
The respective cross sections
for~(\ref{eq:radiative-alpha-fusion-beet})
and~(\ref{Eq:beet-neutron-capture}) have been obtained
in~\cite{Pospelov:2007js} and read
\begin{subequations}
  \label{eq:CS-A8-divide}
  \begin{align}
    \label{eq:alpha-fusion-beet-CS}
    \Navogadro \sigmavof{\mathrm{cat},{\beet}} &= 10^{5}\;
    T_9^{-3/2}\; [\ 0.95 \exp{(-1.02/T_9)} + 0.66 \exp{(-1.32/T_9)}\ ]
    ,
    \\
    \label{eq:ben-CS}    
    \Navogadro \sigmavof{\mathrm{cat},{\ben}} & \simeq 2\times 10^{9},
  \end{align}
\end{subequations}
in units of $\cm^3\seconds^{-1}\mol^{-1}$.  A comment is in order
here. Recently, the catalytic path~(\ref{eq:traf-Be9}) has been
questioned in~\cite{Kamimura:2008fx}. On theoretical grounds it is
argued that the charge radius of \beet , $\Rcharge{\beet} = 2.5\ \fm$,
adopted in~\cite{Pospelov:2007js} is too small; see
Sec.~\ref{sec:discrete-spectrum}. Since the neutron capture reaction
(\ref{Eq:beet-neutron-capture}) proceeds resonantly via the excited
state $\BSx{\ben_{1/2^+}} \to \ben_{3/2^-} + \champ$, the larger
charge radius $\Rcharge{\beet} = 3.39\ \fm$ as proposed
in~\cite{Kamimura:2008fx} decreases $|E_{\mathrm{b}}^{\BSx{\beet}}|$
and shifts the resonance $\Orderof{100~\keV}$ below threshold. A final
answer on the efficiency of (\ref{Eq:beet-neutron-capture}) can only
be obtained by a full quantum $\hef + \hef + \neutron + \champ$
four-body calculation and is announced in~\cite{Kamimura:2008fx} as
work in progress.

\begin{figure}[tb]
\centerline{\includegraphics[width=11cm]{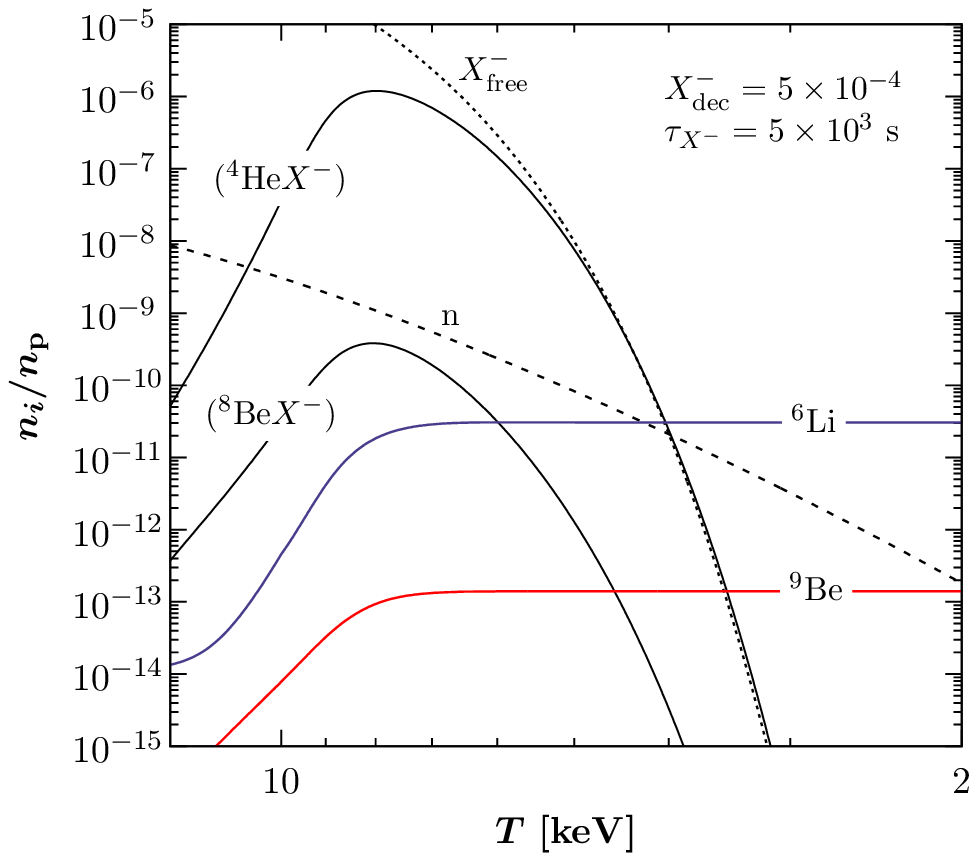}}
\caption[CBBN evolution of \ben]{Evolution of catalyzed \lisx\ and \ben\
  production shown together with the formation of the ``bottle-neck''
  abundances of (\hef\xm) and (\beet\xm) for $\champ_\dec=5\times 10^{-4}$
  and $\tauX=5\times10^3\ \mathrm{s}$. The dashed line gives the
  neutron abundance while the dotted line shows the abundance of free
  \xm.}
\label{Fig:BeEvol} 
\end{figure}

Given the unique sensitivity to physics beyond the Standard Model a
primordial origin of \ben\ offers, we choose to incorporate
(\ref{eq:traf-Be9}) into our reaction network. The following Boltzmann
equations describe the production of \ben ,
\begin{subequations}
  \label{eq:boltzmann-eqs-2}
\begin{align}
  - H T \frac{d }{dT}\, {\BSx{\beet}} & =
  \sigmavof{\mathrm{cat},{\beet}} \, \nb \, \hef\, \BSx{\hef}
 - \Gamma_{\X}\, {\BSx{\beet}}   
 \ ,  \\
  - H T \frac{d }{dT}\, {\ben} & =
  \sigmavof{\mathrm{cat},{\ben}} \, \nb  \, \neutron \, {\BSx{\beet}}
 \ ,  \\
  - H T \frac{d }{dT}\, \neutron & = 
  \frac{1}{2} \,  \sigmavof{\mathrm{fus},{\het}} \, \nb \, \deut \, \deut 
  + \sigmavof{\mathrm{fus},{\hef}} \, \nb \, \deut \, \trit 
  - \sigmavof{\mathrm{des},{\het}}  \, \nb\,\neutron\,  \het 
  \nonumber \\
  & \qquad
  - \sigmavof{\mathrm{cat},{\ben}} \, \nb \,
  \neutron \, {\BSx{\beet}} - \Gamma_{\neutron}\, \neutron
  \ .
\end{align}
\end{subequations}
Again, as in (\ref{eq:boltzmann-eqs}), the abundances are normalized
to $n_{\mathrm{b}}$ and written in an obvious notation; $\neutron \equiv
n_{\neutron}/\nb$.
For $T < 15\ \keV$, the SBBN neutron abundance can already be tracked
well by including the processes $\Deut + \Deut \to \neutron + \het$,
$\Trit + \Deut \to \neutron + \het$, and $\het + \neutron \to \proton
+ \Trit$ into the reaction network \cite{Mukhanov:2003xs}.  The
respective cross sections $ \sigmavof{\mathrm{fus},{\het}} $,
$\sigmavof{\mathrm{fus},{\hef}}$, and $\sigmavof{\mathrm{des},{\het}}$
are taken from~\cite{Caughlan:1987qf}.
Note that at the time of catalyzed \lisx\ fusion the $\deut$ reservoir
is essentially unaffected by those residual SBBN reactions so that we
can neglect the back-reaction on~(\ref{eq:boltzmann-deut}). The yields
of \trit\ and \het\ are taken from the output from an updated version
of the Kawano code~\cite{Kawano:1992ua}.
For the neutron lifetime we use $\tau_{\neutron} =
\Gamma_{\neutron}^{-1} = 885.7\ \seconds$~\cite{Yao:2006px}.

In addition, we supplement the right hand sides of
(\ref{eq:boltzmann-BShef}), (\ref{eq:boltzmann-champ-free}), and
(\ref{eq:boltzmann-hef}) with
\begin{subequations}
  \begin{align}
    \Delta^{(1)}_{\BSx{\hef}} & = - \sigmavof{\mathrm{cat},{\beet}} \,
    \nb \, \hef\,
    \BSx{\hef}   , \\
    \Delta^{(1)}_{\champ} & = \sigmavof{\mathrm{cat},{\ben}} \, \nb
    \, \neutron\,  \BSx{\beet} , \\
    \Delta^{(1)}_{\hef} & = \Delta^{(1)}_{\BSx{\hef}} ,
  \end{align}
\end{subequations}
respectively.
In our code we can neglect the formation of ($\beet$\xm) that proceeds
via molecular bound states (\hef$X^-_2$) \cite{Pospelov:2007js}. This
process becomes important only for a combination of large \YX\ and
large \tauX, i.e., a parameter region which is already excluded by
\lisx\ overproduction.  Also note that at the time when (\beet\xm)
form, their photo-dissociation is not important because of the high
binding energy $|E_{\mathrm{b}}^{(\beet\xm)}| \simeq 1170\ \keV$; see
Table~\ref{tab:beet-x-spectrum}.
It is important to note that we assume the SBBN value for the
deuterium abundance. The early decays of \xm\ may result in an
injection of nucleons into the system. This typically drives the
deuterium abundance upward, resulting in an enhanced number of
neutrons at later times and therefore in an increased output of \ben,
with the general scaling $\ben\sim {\rm const} \times ({\rm D/D_{\rm
    SBBN}})^2$. We choose to disregard this effect, noting its
model-dependent character. We are allowed to do so since its inclusion
can only make the \ben-derived bound on the \xm\ abundance stronger.

Figure~\ref{Fig:BeEvol} shows the evolution of catalyzed \lisx\ and
\ben\ production from the solution of the corresponding set of
Boltzmann equations below $T=10$ keV. The curves in
Fig.~\ref{Fig:BeEvol} are based on the values $\champ_\dec = 5\times
10^{-4}$ and $\tauX = 5\times10^3\ \mathrm{s}$. When the
``bottle-neck'' abundances of (\hef\xm) and (\beet\xm) form, the
catalytic paths (\ref{eq:CBBN-Li6-production}) and (\ref{eq:traf-Be9})
to \lisx\ and \ben\ open up, resulting in the asymptotic values
$\ben/\Hyd \simeq 10^{-13}$ and $\lisx/\Hyd \simeq 3\times 10^{-11}$.
The dashed line shows the neutron abundance and the dotted line the
free \xm\ abundance, which is dominated by its exponential decay.  We
remark in passing that residual recombinations of \hef\ with \xm\ lead
to the crossing of the (\hef\xm) and $X^-_{\mathrm{free}}$ lines at
late time.

\section{ Charge exchange reactions and late time catalysis}
\label{Sec:pXcatalysis}

In this Section we discuss the role of bound states of \champ\
with protons. The \BSx{\proton}\ system may have a large impact on the
BBN predictions because (i) the proton as a recombination
partner is the most abundant element and (ii) it is a neutral
system.  As can be seen from Table~\ref{tab:photodiss-rec-scs}, the
recombination of \px\ bound states becomes efficient only after the
temperature drops below 1~\keV\ which corresponds to a cosmic time of
$t\gtrsim 10^6\ \seconds$. Thus, the question
arises~\cite{Dimopoulos:1989hk,Kohri:2006cn,Jedamzik:2007cp} whether a
revival of fusion reactions, i.e., a late-time catalysis, can be
triggered by the (potential) high reactivity of ``neutron-like'' bound
states \px.

As we will see in Sec.~\ref{Sec:9BeConstraints} the presence of even a
modest number density of \champ\ during the recombination with helium
can lead to a production of \lisx\ and \ben\ at levels which are in
stark conflict with their observationally inferred primordial
values. In~\cite{Jedamzik:2007cp} it was claimed that large fractions
of the previously synthesized $\lisx$ at $T\simeq 8\ \keV$ can indeed
be reprocessed by \px\ via%
\footnote{At the time of publication of~\cite{Jedamzik:2007cp} the
  catalysis of \ben\ production~\cite{Pospelov:2007js} had not yet
  been realized.}
\begin{align}
  \label{eq:late-time-lisx-destruction}
  \px + \lisx \to \hef + \het +\champ
\end{align}
Thus, it was advocated that allowed ``islands'' reconcilable with
observations may well exist in the \champ\ abundance/lifetime
parameter space for $\tau_{\champ}\gtrsim 10^6\ \seconds$.

In the following we shall argue that this is not the case. One reason
is that any arising \px\ abundance is immediately intercepted by the
very efficient charge exchange reaction~\cite{Dimopoulos:1989hk}
\begin{equation} 
\label{pXHe-chExch-rct}
\BSx{\proton}+\hef\to \BSx{\hef} + \proton.
\end{equation}
This reaction may have a very large rate as its cross section is
determined by the actual size of the \px\ bound state that is of the
order of $\abohr\simeq 30\ \fm$ ($\Rrms{} = 50\ \fm$).

\begin{figure}[t]
\centerline{\includegraphics[width=10cm]{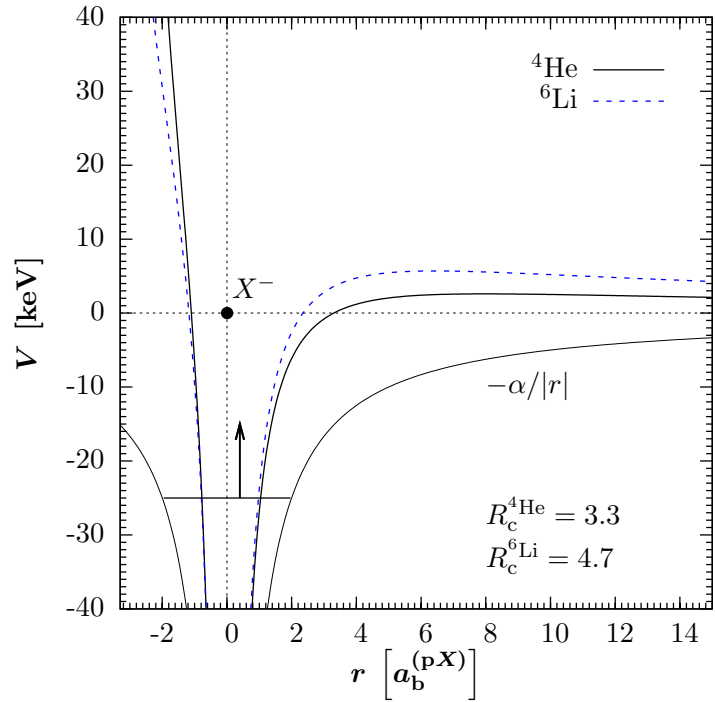}}
\caption[Proton potential for charge exchange]{ Potential energy of
  the proton in the field of \xm\ at $r=0$ and an incoming nucleus at
  $r=-R_{\mathrm{c}}$. The potential energy is plotted along the line
  connecting \xm\ with \hef\ (solid line) or \lisx\ (dashed line),
  respectively.  As the distance between the incoming nucleus and
  $X^-$ decreases, the potential well becomes more narrow, and the
  proton ground state energy level is pushed upward. The critical
  deconfinement distance
  $R_{\mathrm{c}}^{^4\mathrm{He},^6\mathrm{Li}}$ is defined as the
  distance at which the energy of the bound state found variationally
  becomes larger than the height of the barrier $V_{\mathrm{max}}$ to
  the right of~\xm.}
\label{Fig:Potential} 
\end{figure}

The charge exchange (\ref{pXHe-chExch-rct}) can best be understood by
employing a semi-classical picture.
Calling $R$ the separation between \hef\ and \xm\ (or, more generally,
the separation between \xm\ and the incoming nucleus of charge $Z$),
the one-dimensional slice of the proton potential energy in the field
of \xm\ and \hef\ is given by,
\begin{equation} 
\label{V}
V({\bf r}) = -\frac{\alpha}{r} + \frac{\alpha\,Z}{|{\bf r} - {\bf R}|},
\end{equation}
and is plotted in Fig.~\ref{Fig:Potential}. 
The limit of $R\to\infty$ corresponds to an unperturbed binding of the
proton to \xm\ with a binding energy of $\Ebind=-25~\keV$. For $Z>1$
and finite $R$, the curve has a maximum at positive values of $r$
referred to as $\Vmax$. As the \hef\ nucleus comes closer, $R$
decreases. At some point, the binding energy of the proton becomes
positive so that the tunneling of the proton to $r\to+\infty$ starts
to become viable.
For even smaller values of $R$, one can find the distance
$R_{\mathrm{c}}$ at which the probability for the deconfinement of the
proton approaches unity due to the fly-by of the \hef\ nucleus.
In order to estimate $R_{\mathrm{c}}$, we employ the variational
calculation of the proton energy in the potential (\ref{V}) by using
the (unnormalized) trial wave function for the ground state,
\begin{equation} 
\label{wf}
\psi(\mu, \nu) = \exp[-(\mu-\nu)R/(2 a \abohr)]\, (1+\nu R/b\abohr)^2,
\end{equation}
where $\mu$ and $\nu$ are elliptic coordinates and $a$ and $b$ the
minimization parameters. The coordinates are defined as 
$\mu=(r_1+r_2)/R$ 
and 
$\nu = (r_1-r_2)/R$, 
where $r_1$ and $r_2$ are the proton--nucleus and proton--\xm\
distances, respectively. We calculate the energy of the ground state
$\Ebindvar$ as a function of the distance~$R$.  This yields the
critical separation $R_{\mathrm{c}}$, i.e., the distance at which
$\Ebindvar(R_{\mathrm{c}})=\Vmax$, and which describes the situation
when even a metastable bound state simply cannot exist.

The cross section for the charge exchange reaction may then be
approximated by the geometric one with the impact parameter
$\rho=R_{\mathrm{c}} $, $\sigma = \pi R_{\mathrm{c}}^2$. 
The deconfining distances $R_{\mathrm{c}}$ together with the estimated
cross sections for charge exchange are presented in
Table~\ref{table-chex}. The associated thermally averaged cross
sections are given by
$ \sigmavof{\mathrm{ex}} = \sigma_{\mathrm{ex}} \VEV{v} =
\sigma_{\mathrm{ex}} \sqrt{8 T / (\pi \mred)} $ and are listed in the
Table~\ref{table-CBBN-rates} found at the end of this Chapter.
\begin{table}
  \caption[Proton deconfinement and charge exchange]{ 
    Deconfining distances $R_{\mathrm{c}}$ and charge exchange reaction
    cross sections on the \px\ target for incoming nuclei with
    different charges $Z$.}
\begin{center}
\begin{tabular}{rcccc}
\toprule
& Z = 1 & Z = 2 & Z = 3 & Z = 4 \\
\midrule
$R_{\mathrm{c}}\ [\fm]$ & 40  & 95 & 135 & 160  \\
$\sigma=\pi R_{\mathrm{c}}^2\ [\barn] $& 51   &280  & 580  & 850  \\
\bottomrule
\label{table-chex}  
\end{tabular}
\end{center}
\end{table}
As can be seen from Table~\ref{table-chex}, a \hef--\xm\ distance of
$\sim 95~\fm$ is sufficient to release the proton from the bound
state.
Consequently, the estimate points to a very large cross section of
almost $300$~bn for the charge exchange
reaction~(\ref{pXHe-chExch-rct}).

For the charge exchange on \hef\ [Eq.~(\ref{pXHe-chExch-rct})] an
exact solution of the three-body Schr\"o\-dinger equation has recently
become available~\cite{Kamimura:2008fx}. The authors confirm the
efficiency of the charge exchange and find
\begin{align}
  \label{eq:charge-exch-kamimura}
   \Navogadro \sigmavof{\mathrm{ex}, \hef} = 1.0 \times 10^{10} \
  \cm^3\seconds^{-1} \mol^{-1} .
\end{align}
This compares well with the estimate $\sim 4\times 10^9\
\cm^3\seconds^{-1} \mol^{-1} $ obtained from the semi-classical picture
when evaluated at the fiducial temperature of $1\ \keV$.%
\footnote{The different scaling $T^{-1/2}$ of the averaged cross
  sections obtained in the semi-classical approach stems from
  the fact that their energy dependence has not been resolved;
  cf.~\cite{Kamimura:2008fx}. }
Note that the charge exchange will mainly proceed into the $n=3$
level---preferentially with highest~$l$ (largest number of states);
cf.  Table~\ref{tab:hef-x-spectrum}. Capture into highly excited
orbits is indeed observed in charge exchange reactions of muons on
hydrogen which gives justification to the employed
semi-classical approach. Reference \cite{Jedamzik:2007cp} finds for the
charge exchange on \hef\ a rate (per particle pair) of $\sim
2\times10^7\ \cm^3\seconds^{-1} \mol^{-1}$ which is an underestimation
by two to three orders of magnitude.

We incorporate the charge exchange reaction on \hef\ in the network of
Boltzmann equations by solving
\begin{align}
  \label{eq:boltzmann-eq-chex}
  - H T \frac{d }{dT}\, {\BSx{\proton}} & =
  \sigmavof{\mathrm{rec},{\proton}} \, \nb \, \proton\, \champ
  - \Gammaphof{\proton}\,  {\BSx{\proton}}  \nonumber \\ & \qquad
  -  \sigmavof{\mathrm{ex},{\hef}} \, \nb \, \hef\, \px
 - \Gamma_{\X}\, {\BSx{\proton}}   
\end{align}
for the \px\ abundance. The back-reaction on free protons is
negligible so that we refrain here from writing an equation for
\proton. However, we need to supplement the right hand sides of
(\ref{eq:boltzmann-BShef}), (\ref{eq:boltzmann-champ-free}), and
(\ref{eq:boltzmann-hef}) by
\begin{subequations}
  \begin{align}
    \Delta^{(2)}_{\BSx{\hef}} &=  \sigmavof{\mathrm{ex},{\hef}} \,
    \nb \, \hef\, \BSx{\proton} , \\
    \Delta^{(2)}_{\champ} & = - \sigmavof{\mathrm{rec},{\proton}} \,
    \nb \, \proton\, \champ  + \Gammaphof{\proton}\,  {\BSx{\proton}}
    ,\\
    \Delta^{(2)}_{\hef} & = - \Delta^{(2)}_{\BSx{\hef}} ,
  \end{align}
\end{subequations}
respectively. Using the recent result~(\ref{eq:charge-exch-kamimura})
for the charge exchange cross section, we find that in the limit of
infinite lifetimes, $\tau_{X^-}\to\infty$,  the abundance of \px\
reaches its peak at around $T=0.7$ keV.  Its maximum abundance at
these temperatures can be well approximated as
\begin{equation} 
\frac{n^{\rm max}_{\BSx{\proton}}}{n_{\proton}} 
\simeq 
1.7 \times 10^{-7} \left(\frac{\champ}{10^{-2}}\right),
\label{pXmax}
\end{equation}
where the assumption $\champ \lesssim \hef$ has been made [see
Sec.~\ref{sec:typic-champ-abund}] ensuring the linear scaling
in~(\ref{pXmax}).%
\footnote{Using instead of (\ref{eq:charge-exch-kamimura}) the cross
  section $\sigmavof{\mathrm{ex},\hef}$ inferred from
  Table~\ref{table-chex} yields the coefficient~\mbox{$4\times
    10^{-7}$} in Eq.~(\ref{pXmax}); see~\cite{Pospelov:2008ta}.}
That such a small \px\ fraction only has a marginal impact on \lisx\
can be seen by comparing the destruction rate
for~(\ref{eq:late-time-lisx-destruction})~\cite{Kamimura:2008fx}
\begin{align}
  \label{eq:cat-dest-lisx}
  \Navogadro \sigmavof{\mathrm{cat,des},\lisx} = 1.6 \times 10^{8} \
  \cm^3\seconds^{-1} \mol^{-1} 
\end{align}
to the Hubble rate at the relevant temperature of $T = 0.7\
\keV$. Using $\champ=0.01$ we find
\begin{equation}
\label{comparetoH}
\left.\frac{ \sigmavof{\mathrm{cat,des},\lisx}
    \, n^{\rm max}_{(pX^-)}}{H}\right|_{T=0.7\ \keV} 
\simeq 10^{-3} 
\end{equation}
which tells us that the whole issue of \px\ mediated destruction of
\lisx\ (and accordingly of \ben ) is irrelevant so that the abundances
of \lisx\ and \ben\ fused at $T\simeq 8\ \keV$ remain unaffected at
$T\simeq 1\ \keV$.%
\footnote{The same conclusion---prior to the publication of
  \cite{Kamimura:2008fx}---has already been reached in
  \cite{Pospelov:2008ta} by assigning the maximal possible rate for
  the destruction process~(\ref{eq:late-time-lisx-destruction}) which
  is given by the unitarity bound.  }

We can further employ the results summarized in
Table~\ref{table-chex}. A successive chain of charge exchange
reactions with \px\ can lead to molecular states that are finally
destroyed in nuclear reactions with protons. In particular, $(\lisx
X^-_3)$ has a chance for a nuclear interaction with protons or helium
unsuppressed by a residual Coulomb barrier since it is a very compact
object.  We can solve for those molecular bound states by extending
our network of Boltzmann equations for $T$ smaller than a few \keV\
with
\begin{subequations}
    \label{eq:molecular-bs}
  \begin{align}
    - H T \frac{d }{dT}\, {\BSx{\lisx}} & =
    \sigmavof{\mathrm{ex},{\lisx}} \, \nb \, \px\, \lisx  
    -\Gamma_{\X}\, {\BSx{\lisx}} 
    \nonumber \\ & \qquad
    -\sigmavof{\mathrm{ex},{\BSx{\lisx}}} \, \nb \, \px\, \BSx{\lisx} ,
    \\
    - H T \frac{d }{dT}\, {(\lisx\champ_2)} & =
    \sigmavof{\mathrm{ex},{\BSx{\lisx}}} \, \nb \, \px\, \BSx{\lisx} 
    - 2\, \Gamma_{\X}\, (\lisx\champ_2) 
    \nonumber \\ & \qquad
    - \sigmavof{\mathrm{ex}, (\lisx\champ_2) } \, \nb \, \px\, (\lisx\champ_2) ,
    \\
    - H T \frac{d }{dT}\, {(\lisx\champ_3)} & =
    \sigmavof{\mathrm{ex},(\lisx\champ_2) } \, \nb \, \px\,(\lisx\champ_2) 
    - 3\, \Gamma_{\X}\, (\lisx\champ_3)
  \end{align}
\end{subequations}
and supplement the right hand sides of
Eq.~(\ref{eq:boltzmann-eq-chex}) with
\begin{subequations}
  \begin{align}
    \Delta^{(1)}_{\px}  & =  
    - \sigmavof{\mathrm{ex},{\lisx}} \, \nb \, \px\, \lisx ,
    -\sigmavof{\mathrm{ex},{(\lisx\champ)}} \, \nb \, \px\, (\lisx\champ) 
     \nonumber \\ & \qquad
    - \sigmavof{\mathrm{ex},(\lisx\champ_2) } \, \nb \, \px\,(\lisx\champ_2) ,
    \\
    \label{eq:feed-back-lisx}
    \Delta^{(1)}_{\lisx} & = 
    - \sigmavof{\mathrm{ex},{\lisx}} \, \nb \, \px\, \lisx +
    \Gamma_{\X} 
    \left[\BSx{\lisx} + 2  (\lisx\champ_2) + 3 (\lisx\champ_3) \right]
  \end{align}
\end{subequations}
A similar chain exists for \ben\, where the sequence of the charge
exchange reactions can proceed until $(\ben X^-_4)$. It is important
to note that the efficiency of this chain reaction depends very
sensitively on the concentration of the \px\ bound states and on the
mass of the \xm\ particle. The latter enters through the average
relative velocity of two heavy objects, e.g., $\px$ and $(\lisx X^-)$,
which in turn scales as $m_{X^-}^{-1/2}$.  Therefore, in the limit of
an infinitely heavy $X^-$, the chain will be cut off right at the
first step, terminating at~(\lisx\xm).
Also note that in (\ref{eq:feed-back-lisx}) we have made the
assumption that \lisx\ is not destroyed by the decay of
\champ. Clearly some of the recoiling \lisx\ nuclei will be destroyed
when released from the bound state. For the case of \BSx{\bes}\ this
has been investigated in~\cite{Bird:2007ge} where it was found that no
significant depletion of \bes\ takes place. On those grounds and
noting that only a small fraction of \lisx\ is locked in bound states
with \champ\ it is safe for us to disregard this effect.

\begin{figure}[t]
\centerline{\includegraphics[width=11cm]{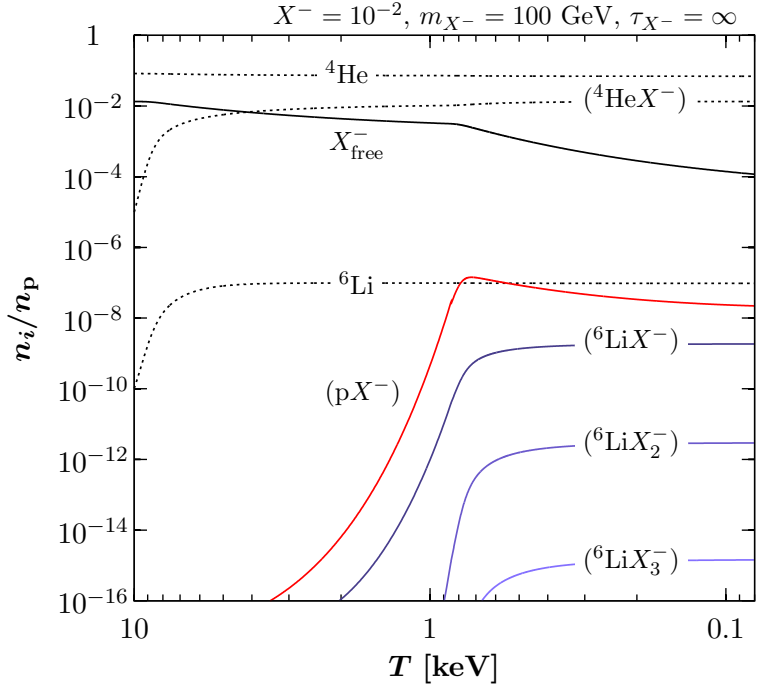}}
\caption[\px\ abundance after charge exchange]{ Evolution of
  primordial abundances as a function of time (or temperature $T_9$)
  from the input $\champ=0.01$, $m_{X^-}=100~\GeV$, and
$\tau_{X^-}\to\infty$. The \px\ abundance reaches its maximum of $\sim
1.7\times 10^{-7}$ at $T \simeq 0.7\ \keV$.  Around the same
temperatures, the abundance of unbounded CHAMPs,
$X^-_{\mathrm{free}}$, starts to decline more rapidly since it is
removed by the recombination with $\proton$ followed by the charge
exchange reaction on~\hef.}
\label{f2} 
\end{figure}

We run the full set of Boltzmann equations to determine the residual
concentrations of \px\ and of the molecular bound states of \lisx\
with \xm. The results are plotted in Fig.~\ref{f2}.  As one can see,
an initial concentration of per nucleon of $\champ = 10^{-2}$ results
in a \px\ abundance that never exceeds the maximum~(\ref{pXmax}),
leading to a progressively diminishing number of molecular states.

We remark that for the computation of Fig.~\ref{f2} we have not
included $\BSx{\lisx}$ formation via radiative recombination $\lisx
+\champ \to \BSx{\lisx} + \gamma$. This process is not important for
the present discussion because it does not affect~(\ref{pXmax}); \px\
is dominantly removed by charge exchange on \hef. We have checked this
by including the recombination process into our reaction network;
cf.~Table~\ref{tab:photodiss-rec-scs}. Whereas \BSx{\lisx}\ would form
around the same time as \BSx{\hef}\ it is important to note that
\BSx{\lisx}\ is also destroyed via $\BSx{\lisx}+\proton \to \hef +\het
+ \champ$ hindering its formation; the corresponding cross section is
given in~\cite{Kamimura:2008fx} and in Table~\ref{table-CBBN-rates}
with the name $\sigmavof{\mathrm{cat},\mathrm{des}2,\lisx}$.
Even with $\champ_\dec = 0.01$ we find that the final \BSx{\lisx}\
output is one order of magnitude below unbounded \lisx\
and~(\ref{pXmax}) remains unchanged. By the same token, we find that
the destruction process only has a minor impact on the total \lisx\
abundance (below 10\%).  For smaller values of $\champ_\dec$ the
effect is accordingly weaker.

To conclude this section, neither lithium nor beryllium synthesized in
CBBN processes at $8~\keV$ would be affected in any significant way by
the subsequent generation of \px\ bound states.
Thus---as we will show later---the part of the parameter space with a
typical freeze-out \xm\ abundance and a long \xm\ lifetime will be
confidently ruled out.

\subsection{Relaxation after charge exchange}
\label{sec:relaxation-after-chexch}

Above we have stated that the charge exchange on \hef\ will mainly
proceed into excited states of \BSx{\hef}. Though we have already
shown that \px\ bound states do not reach an abundance level where
late-time catalysis plays a role it is nevertheless amusing to see
whether $\px$ receives further depletion by the energetic photons
released from relaxation of \hef\ from its excited states $n=3$ into
the \BSx{\hef} ground state. Upon transition into the ground state
$\Ein \sim 300\ \keV$ will be injected into the plasma in form of
photons. This happens predominantly at $\Texch \simeq 0.7\ \keV$ and
the photons loose their ability to break up \px\ once they are
degraded below $\Eout \simeq 25\ \keV$.

The rate for photon-photon scattering with
$\gamma_{\mathrm{bg}}$---scaling with
$E_\gamma^3$~\cite{1990ApJ...349..415S}---is rapidly becoming
inefficient for $E_\gamma= \Orderof{100\ \keV}$ at the relevant
temperature so that we can neglect this photon-multiplying
process. Thus, the photons loose their energy mainly by Compton
scattering on background electrons. In the low-energy limit of Thomson
scattering the mean lifetime of a photon before scattering is
\begin{align}
\label{eq:tau-gamma}
  \tau_\gamma = \left.\frac{1}{\nelectron \sigmaThomson
    }\right|_{\Texch} \sim 10\ \seconds \quad \mathrm{with} \quad
  \sigmaThomson = \frac{8 \pi \alpha^2}{3 \melectron^2},
\end{align}
where we have used $\nelectron =  7/8\, \etabarion \ngamma$; see
(\ref{eq:total-electronic-density}). 

Since the energy loss of $\gamma$ particles in Compton scattering is
very small, we can write a differential equation for the systematic
energy transfer to the electrons
\begin{align}
\label{eq:ode-eloss-compton}
  \frac{d\Egamma}{dt}  = \langle\sigma \Delta\Egamma\rangle
  \nelectron .
\end{align}
Note that we have neglected a term $H\Egamma$ which would account for
the expansion of the Universe. We can do so since $\tau_\gamma\ll
H(\Texch)^{-1} \simeq 5\times 10^6\ \seconds$. The average energy loss per
scattering is given by
\begin{align}
  \label{eq:avg-eloss-compton}
  \langle\sigma \Delta\Egamma\rangle = \int \Delta\Egamma
  \frac{d\sigma}{d\Omega} d\Omega \ .
\end{align}
At $\Texch$ the rest frame of the electrons essentially coincides with
the frame of the thermal bath. Then ${d\sigma}/{d\Omega}$ is given by
the Klein-Nishina formula~\cite{1976tper.book.....J}
\begin{align}
  \frac{d\sigma}{d\Omega} = \frac{\alpha^2}{2 \melectron^2}
  \left( {\Egamma' \over \Egamma} \right)^2 
  \left( {\Egamma \over \Egamma'} + {\Egamma' \over \Egamma}
    -\sin^2{\theta} \right) 
\end{align}
with
\begin{align}
   {\Egamma' \over \Egamma}  = \frac{1}{1+(\Egamma/\melectron) (1-\cos{\theta})}.
\end{align}
From (\ref{eq:avg-eloss-compton}) together with $\Delta\Egamma =
\Egamma'-\Egamma$ and $\Egamma \ll \melectron$ it then follows that
\begin{align}
  \langle\sigma \Delta\Egamma\rangle \simeq
  -\sigmaThomson
  \frac{\Egamma^2}{\melectron}  .
\end{align}

Since the mean lifetime of the photon against Compton scattering is
much shorter compared to the Hubble time we can integrate
(\ref{eq:avg-eloss-compton}) from $\Ein$ to $\Eout$ by neglecting any
associated drop in temperature of the plasma. This defines a typical
escape time $\tau_{\mathrm{esc}}$, i.e. a thermalization time-scale
upon which the photon looses its ability to ionize \px. We find
\begin{align}
  \tau_{\mathrm{esc}} = \frac{8 \melectron}{7 \sigmaThomson \etabaryon
    n_\gamma(\Texch)}\left[ \frac{1}{\Eout} - \frac{1}{\Ein} \right]
  \simeq 160\ \seconds .
\end{align}

This has to be compared with the mean lifetime of an energetic photon
against ionization of \px,
\begin{align}
 \tau_{\mathrm{ph}} = \left.\frac{1}{n_{\px} \sigma_{\mathrm{ph}}
    }\right|_{\Texch} \sim 10^7\ \seconds 
\end{align}
and which is of the order of the Hubble time;
In the last step we have used $n_{\px} = n^{\mathrm{max}}_{\px} $ as
given in (\ref{pXmax}) with $\champ = 0.01$. The photo-dissociation cross
section has been obtained in Sec.~\ref{sec:phot-recomb-cross}; see
Table~\ref{tab:photodiss-rec-scs}. This tells us that the
thermalization of the injected photon happens very rapidly so that
those photons released in the relaxation process after charge exchange
are not capable of depleting \px\ any further.%
\footnote{The numerical values of the various times-scales sensitively
  depend on the exact value of fiducial temperature $\Texch$
  chosen. However, the argument is not affected by that.}
\section[Constraints on the \xm\ lifetime and
abundance]{\texorpdfstring{Constraints on the \boldmath$X^-$ lifetime and
    abundance}{Constraints on the champ lifetime and abundance}}
\label{Sec:9BeConstraints}
In order to constrain the (\tauX,\,\YX) parameter space from the
catalytic path (\ref{eq:traf-Be9}) to \ben~\cite{Pospelov:2007js}, we
need to set an \textit{upper limit} on its primordial abundance from existing observations.
It is generally accepted that the galactic evolution of the abundances
of Be, along with Li and B, are dominated by cosmic-ray
nucleosynthesis.  While Be is burned rapidly in stellar centers, it is
produced in cosmic rays by the spallation reactions of fast protons
and $\alpha$ particles hitting ambient CNO nuclei~\cite{1970Natur.226..727R,1971A&A....15..337M}.
As a consequence, the abundances of Be and O are linked, leading to a
secondary scaling, Be $\propto\, \Ox^2$~\cite{1990ApJ...364..568V}. On the other
hand, inverse spallation reactions of CNO nuclei, both produced and
accelerated in supernovae, will give a Be yield that is essentially
independent of the metallicity of the interstellar medium.  Such
primary processes, leading to Be $\propto\, \Ox$, are expected to play
a major role during the early galactic epochs~\cite{Fields:1999ib}.

The produced Be is subsequently supplemented in the outer layers of
stars. Thus, old stars which are far from the galactic center (and
thereby less affected by the galactic chemical evolution) bear the
potential to encode any pre-galactic origin of Be. Indeed, Be has been
observed in a number of Population II halo stars at very low
metallicities $[\Fe/\Hyd] \lesssim -2.5$.  Particularly noteworthy is
the detection in the star G~64--12 at $[\Fe/\Hyd] \simeq
-3.3$~\cite{Primas:2000gc}. The star's high Be value of
$\log_{10}(\Be/\Hyd) \simeq -13.05$ might suggest a possible
flattening in the Be trend during the early evolutionary phases of our
galaxy~\cite{Primas:2000gc}. Whether this really points to a
\textit{primordial} plateau or whether this indicates a Be dispersion
at lowest metallicities~\cite{Boesgaard:2005pf} is not clear at
present.

\begin{figure}[t]
\centerline{\includegraphics[width=0.49\textwidth]{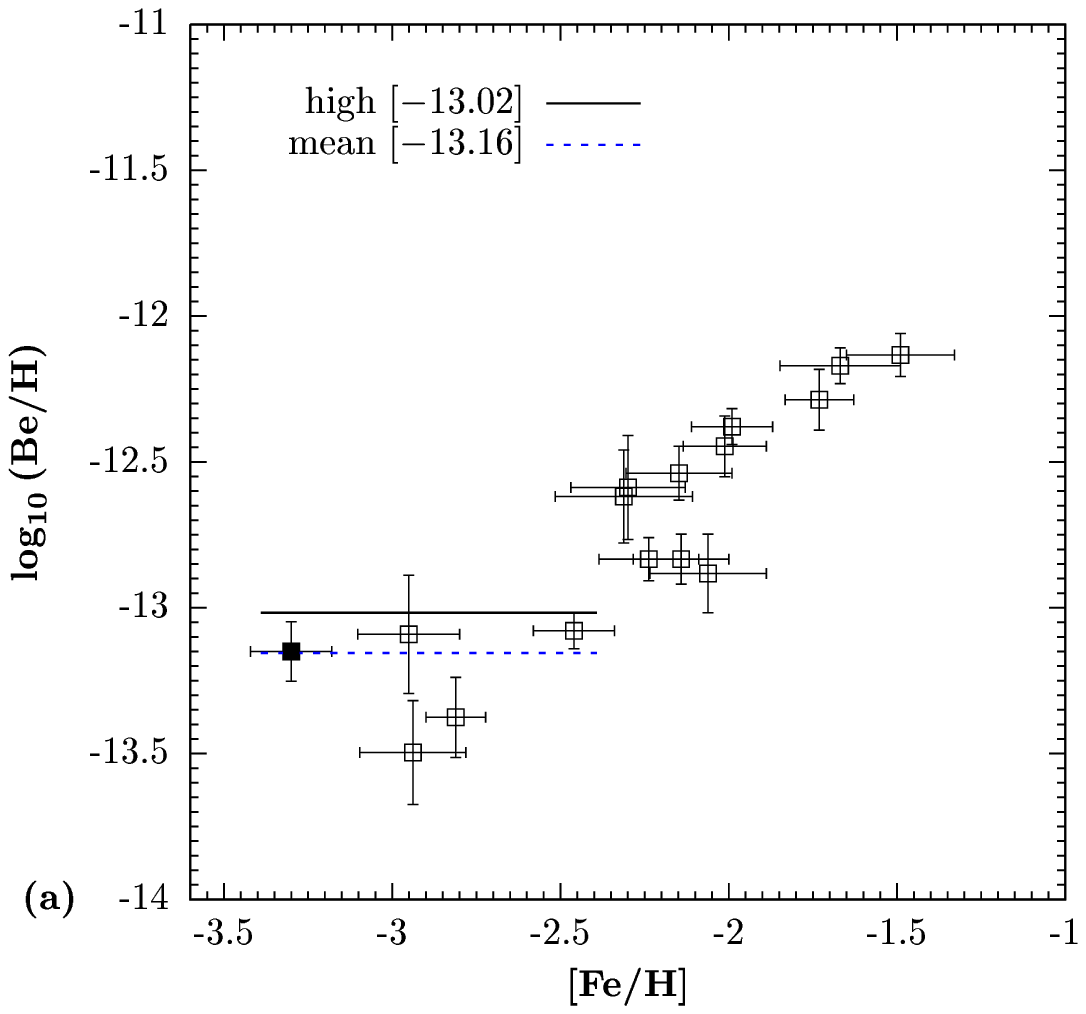}%
\hskip 0.3cm
\includegraphics[width=0.49\textwidth]{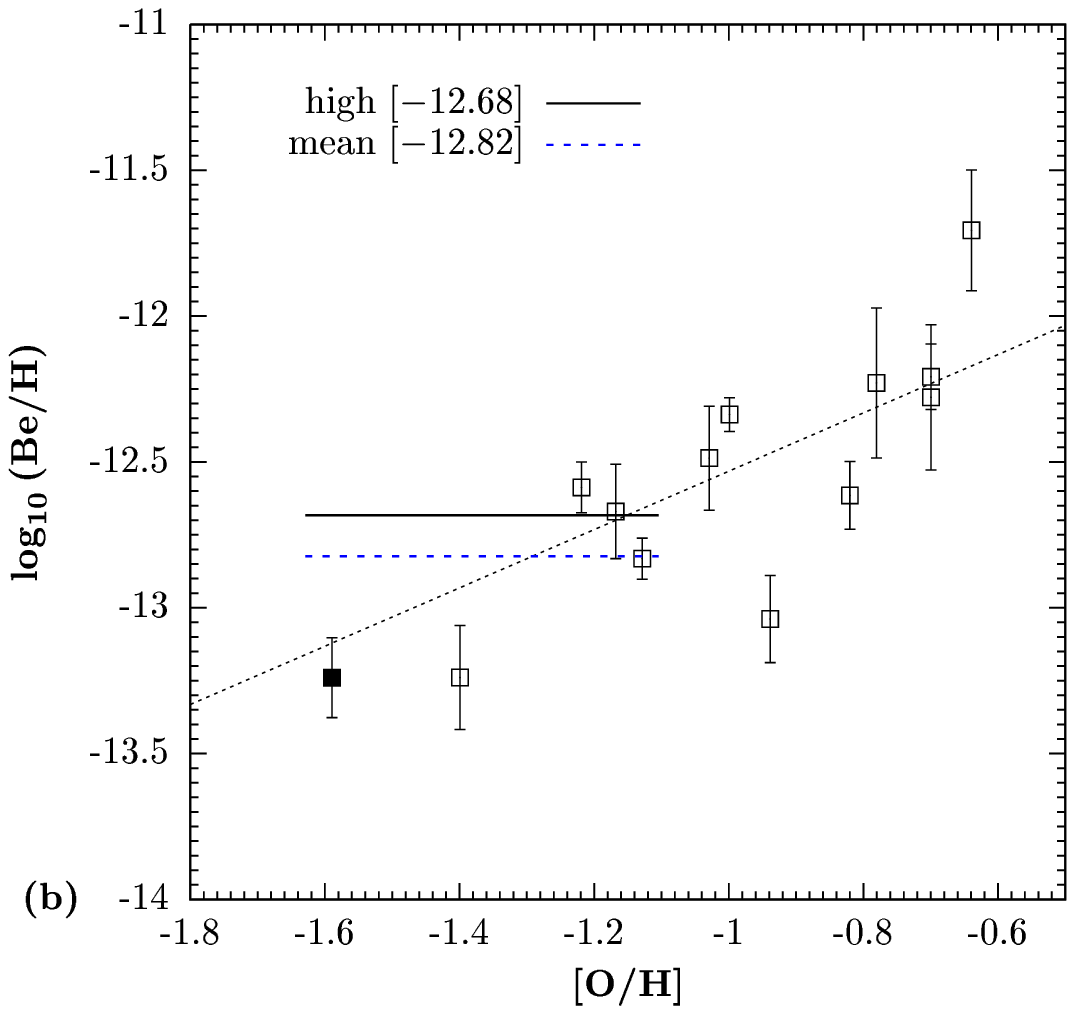}
}

\caption[Be observations in halo stars]{ Observations of Be in Pop II
  halo stars. In the left panel (a), the data is taken from Fig.~3a of
  Ref.~\cite{Primas:2000gc} and is plotted as a function of
  $[\Fe/\Hyd]$.  The right panel (b) shows the data from Fig.~6b of
  Ref.~\cite{Fields:2004ug} where $[\Ox/\Hyd]$ provides the
  metallicity indicator. The filled dots depict the data points
  associated with the star G~64--12. The solid lines give the inferred
  nominal upper limits on \ben\ from the weighted mean (dashed lines)
  of a sample of stars at lowest metallicity. Also shown in
  Fig.~\ref{Fig:Bedata}b is a fit of a primary scaling of Be; see main
  text.  }
\label{Fig:Bedata} 
\end{figure}
\begin{figure}[tb]
\centerline{\includegraphics[width=12cm]{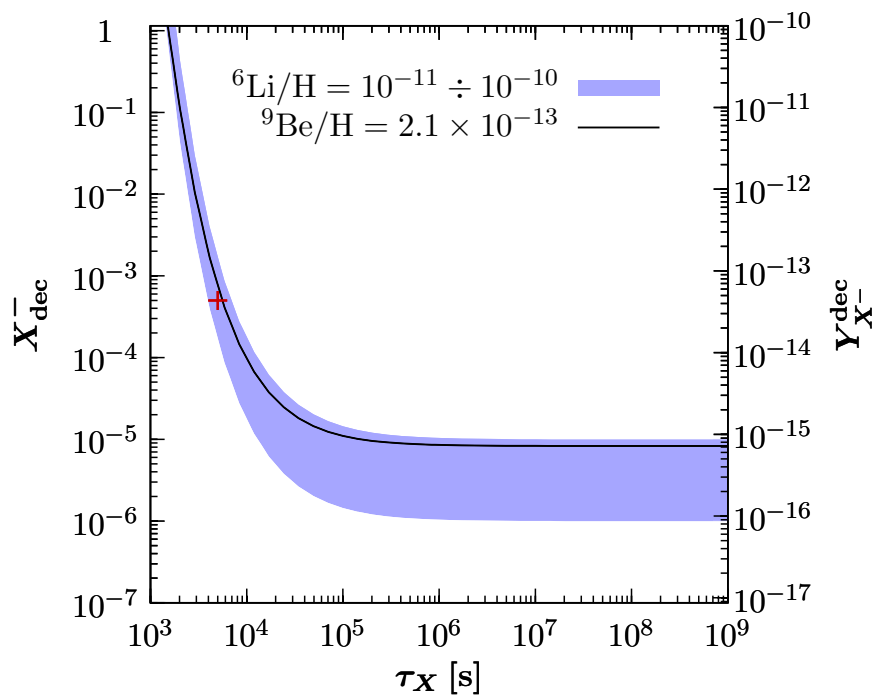}}
\caption[Contour plot of CBBN yields]{ Contour plot of CBBN abundance
  yields of \lisx\ and \ben\ in the $(\tau_{X},\champ_{\mathrm{dec}})$
  plane.  The solid line shows the limit
  (\ref{eq:upper-limit-Be9}). The region above this line is excluded
  by \ben\ overproduction.  The lower (upper) boundary of the band
  corresponds to $\lisx/\Hyd = 10^{-11}\ (10^{-10})$. The y-axis on
  the right-hand side indicates the \xm\ number density
  $n^{\mathrm{dec}}_{X^-}$ normalized to the entropy density,
  $\YXdec$. The cross shows the parameter point considered in
  Fig.~\ref{Fig:BeEvol}. }
\label{Fig:Ytau} 
\end{figure}

Figure~\ref{Fig:Bedata}a shows the original Be detection in
the star G~64--12%
\footnote{For consistency with the rest of the data points, the 1D LTE
  value has been plotted in Fig.~3a of the original
  reference~\cite{Primas:2000gc}.}
(filled dot) along with a subset of data points taken from Fig.~3a of
Ref.~\cite{Primas:2000gc}.  The data of Fig.~\ref{Fig:Bedata}b are
taken from Fig.~6b of the recent work \cite{Fields:2004ug} which also
uses $[\Ox/\Hyd]$ as a metallicity tracer. The latter paper discusses
the implications of a new temperature scale on the abundances of Li,
Be, and B. In principle, different assumed physical parameters which
characterize the stellar atmosphere may result in large systematic
shifts of the inferred abundances. In this regard, it is important to
note that Be is not overly sensitive to the assumed surface
temperature of the halo dwarfs \cite{Fields:2004ug}. In the following
we thus shall take a pragmatic approach: In both
Fig.~\ref{Fig:Bedata}a and Fig.~\ref{Fig:Bedata}b, we obtain the least
squares weighted mean (dashed lines) for a representative sample of
stars at lowest metallicities. From the variance of the fit, we can
extract a nominal $3\sigma$ upper limit (solid lines) on primordial
\ben. From Fig.~\ref{Fig:Bedata}b, we find%
\begin{equation}
  \label{eq:upper-limit-Be9}
  \log_{10} {\Be/\Hyd}|_{\mathrm{high}} = -12.68 
  \quad\Rightarrow\quad
  \ben/\Hyd \le 2.1\times 10^{-13}\ .
\end{equation}
Conversely, Fig.~\ref{Fig:Bedata}a yields \mbox{$\ben/\Hyd \lesssim
  10^{-13}$} while fitting only the last two data points with
$[\Ox/\Hyd] < -1.3$ in Fig.~\ref{Fig:Bedata}b would give $\ben/\Hyd
\lesssim 1.3\times 10^{-13}$. In our context, those values are less
conservative so that we use (\ref{eq:upper-limit-Be9}) in the
following. 
In Fig.~\ref{Fig:Bedata}b we have additionally fitted for a primordial
component, $\ben/\Hyd|_{\mathrm{p}}$, in combination with a primary
scaling, $\ben/\Hyd=\kappa\, (\Ox/\Hyd)\big/(\Ox/\Hyd)_{\odot}$. It
seems, however, that a purely primary mechanism with $\kappa\simeq
2.9\times 10^{-12}$ fits the data best since $\ben/\Hyd|_{\mathrm{p}}$ comes out
negligibly small.%
\footnote{For a proper comparison between different assumed surface
  temperature scales and corresponding fits of primary versus
  secondary scaling, see Ref.~\cite{Fields:2004ug}.}
Finally, we are aware that neither of the fitted mean values in
Fig.~\ref{Fig:Bedata} is very good in terms of $\chi^2$.  However, a
firm conjecture of a Be plateau is not the purpose of this work, and
indeed (\ref{eq:upper-limit-Be9}) does provide a sufficiently
conservative limit to work with.

We can now confront the constraint~(\ref{eq:upper-limit-Be9}) as well
(\ref{eq:Lisx-obs-upperbound-range}) with the CBBN yield of \ben\ and
\lisx\ obtained by solving the associated Boltzmann equations
presented in Sec.~\ref{sec:catalyz-nucl-react} for a wide variety of
(\tauX,\,$\champ_{\dec}$) combinations.  In Fig.~\ref{Fig:Ytau} we
obtain exclusion boundaries from catalyzed \ben\ and \lisx\ production
in the (\tauX,\,$\champ_\dec$) parameter space.  For convenience of
the reader, the \xm\ number density $n^{\mathrm{dec}}_{X^-}$
normalized to the entropy density, $\YXdec$ , is given on the
$y$-axis on the right-hand side. Above the solid line, \ben\ is in
excess with respect to (\ref{eq:upper-limit-Be9}) and thus
excluded. The shown band reflects the uncertainties in the
observational determination of \lisx. On the lower border, $\lisx/\Hyd
= 10^{-11}$ is fulfilled while $\lisx/\Hyd = 10^{-10}$ holds on the
upper border of the band. The cross indicates the exemplary parameter
point considered in Fig.~\ref{Fig:BeEvol}. At large lifetimes, the
linear scaling of \lisx\ with \champ\ can easily be seen from the
boundaries of the band.  Note that we find $\ben / \lisx $ in the
interval between $10^{-3}$ and $10^{-2}$, whenever CBBN is efficient,
which confirms the observation already made in
Ref.~\cite{Pospelov:2007js}.

\begin{table}
  \caption[Collection of CBBN rates]{Here we collect the key CBBN cross
    sections,
    i.e., the reaction rates per particle pair, $ N_{\mathrm{A}} \langle \sigma v
    \rangle$, used in the numerical solutions of the Boltzmann
    equations. They are given in units of
    $\mathrm{cm}^3\mathrm{s}^{-1}\mathrm{mol}^{-1}$ and $T_9 =
    T/10^9~\mathrm{K}$. The photo-dissociation cross sections
    $\sigmavof{\mathrm{ph}}$ are related to the rates $\Gammaph$ via
    $\Gammaph = n_{\gamma} \sigmavof{\mathrm{ph}}$. }
\begin{center}
\begin{tabular}{lcl}
\toprule\\[-0.5cm]
 process & name & rate
$[\cm^3\seconds^{-1}\mol^{-1}]$ \\[0.17cm]
\midrule\\[-0.40cm]
\multicolumn{3}{l}{
{\textit{Recombination and photo-dissociation:}}
}\\[0.15cm]
\: $p + \xm \to (p\xm) + \gamma$    & 
$\sigmavof{\mathrm{rec},\proton}$ &
$3980\; T_9^{-1/2}$
 \\[0.15cm]
\: $(p\xm) + \gamma_{\mathrm{bg}}\to  p + \xm $&
$\sigmavof{\mathrm{ph},\proton}$&
$1.18\times 10^{9}\; T_9^{-2}\; 
\exp{(-0.29/T_9)}$ 
 \\[0.15cm]
\: $\hef + \xm \to  (\hef\xm) + \gamma$  & 
$\sigmavof{\mathrm{rec},\hef}^{\dagger} $ &
$7900\; T_9^{-1/2}$
 \\[0.15cm]
\: $(\hef\xm) + \gamma_{\mathrm{bg}}\to \hef + \xm $ & 
$\sigmavof{\mathrm{ph},\hef}^{\dagger} $ &
$1.85\times 10^{10}\; T_9^{-2}\; 
\exp{(-4.03/T_9)} $\\[0.15cm]
\: $\lisx + \xm \to  (\lisx\xm) + \gamma $&
$\sigmavof{\mathrm{rec},\lisx}$ &
$6640\; T_9^{-1/2} $
 \\[0.15cm]
\: $(\lisx\xm) + \gamma_{\mathrm{bg}} \to \lisx + \xm $&
$\sigmavof{\mathrm{ph},\lisx} $&
$2.87\times 10^{10}\; T_9^{-2}\; 
\exp{(-9.25/T_9)} $
 \\[0.5cm]
\multicolumn{3}{l}{
{\textit{Charge exchange:}}
}\\[0.15cm]
\: $\px + \hef \to (\hef\xm) + \proton $ & 
$\sigmavof{\mathrm{ex},\hef}$&
$1.0\times 10^{10}$~\cite{Kamimura:2008fx}
\\[0.15cm] &  & $3.9\times 10^{10}\; T_9^{1/2}$ 
\\[0.15cm]
\: $\px + \lisx \to (\lisx\xm) + \proton $   & 
$\sigmavof{\mathrm{ex},\lisx} $&
$6.45\times 10^{10}\; T_9^{1/2} $
 \\[0.15cm]
\:$\px + (\lisx\xm) \to (\lisx\xm_2) + \proton $  & 
$\sigmavof{\mathrm{ex},\BSx\lisx} $ &
$3.37\times 10^{9}\; T_9^{1/2} \; 
 (1\ \TeV/m_{\xm})^{1/2} $
\\[0.15cm]
\:$\px + (\lisx\xm_2) \to (\lisx\xm_3) + \proton  $ & 
$\sigmavof{\mathrm{ex},(\lisx\champ_2)} $&
$5.25\times 10^{8}\; T_9^{1/2} \;  
(1\ \TeV/m_{\xm})^{1/2} $
 \\[0.5cm]
\multicolumn{3}{l}{
{\textit{$^6$Li destruction (from \cite{Kamimura:2008fx}):}}
}\\[0.15cm]
\: $(\proton\xm) + \lisx \to \hef + \het + \champ $  & 
$\sigmavof{\mathrm{cat,des},\lisx} $ &
$1.6\times 10^8 $ \\[0.15cm]
\: $(\lisx\xm) + \proton \to \hef + \het + \champ $  & $\sigmavof{\mathrm{cat},\mathrm{des}2,\lisx}$ &
$2.6 \times 10^{10}\; T_9^{-2/3} \exp{(-6.74\, T_9^{-1/3})} $ \\[0.5cm]
\multicolumn{3}{l}{
\textit{$^6$Li and $^9$Be catalysis (from \cite{Hamaguchi:2007mp} and
 \cite{Pospelov:2007js}):}
}\\[0.15cm]
\:  $(\hef\xm) + \Deut \to \lisx + \xm $&
$\sigmavof{\mathrm{\mathrm{cat},\lisx}} $&
$  2.37\times 10^8\; (1-0.34\, T_9)\,   T_9^{-2/3} $\\ &&\quad$\times\exp{(-5.33\,
    T_9^{-1/3})} $ 
\\[0.15cm]
\: $  \hef + (\hef\xm) \to (\beet\xm) + \gamma $&
$\sigmavof{\mathrm{\mathrm{cat},\beet}} $&
$ 10^{5}\; T_9^{-3/2}\; [\ 0.95 \exp{(-1.02/T_9)} $\\ &&\quad$+ 0.66
\exp{(-1.32/T_9)}\ ]  $
\\[0.15cm]
\:$  (\beet\xm) + n \to \ben + \xm    $&
$\sigmavof{\mathrm{\mathrm{cat},\ben}} $&
$ 2\times 10^{9}$ \\[0.15cm]
\bottomrule
   \multicolumn{3}{l}{%
     \begin{minipage}[t]{0.99\textwidth}
       $^\dagger$ \footnotesize For consistency, the rates employed
       for \BSx{\hef} match the ones
       from~\cite{Pospelov:2008ta}. From our numerical evaluation
       [Table~\ref{tab:photodiss-rec-scs}] the respective coefficients
       for $\sigmavof{\mathrm{rec},\hef}$ and
       $\sigmavof{\mathrm{ph},\hef}$ read $7260$ and $1.70\times 10^{10}$.
 \end{minipage}}\\[-0.1cm]%
\label{table-CBBN-rates}  
\end{tabular}
\end{center}
\end{table}

%
%

%


\cleardoublepage
 \part{The gravitino-stau scenario}
 \label{part:two}
\chapter{Gravitinos as a probe for the earliest epochs}
\label{cha:grav-stau-scenario}

\section{The gravitino-stau scenario}
\label{sec:gravitino-stau-intro}

In the first part of this thesis we have discussed the implications of
a generic electromagnetically charged massive particle species
$\X^\pm$ if it is present in the early Universe during/after the era
of BBN ($t\gtrsim 1\, \seconds$).
If \X\ is a weak scale thermal relic, it freezes out from the primordial
plasma at cosmic times $t\lesssim 10^{-7}\, \seconds$ so that the
question of the origin of its longevity arises.

Long-lived charged particles can naturally emerge in supersymmetric
(SUSY) extensions of the Standard Model.
In scenarios in which the gravitino $\gravitino$ is the lightest
supersymmetric particle (LSP), a long-lived $\X^\pm$ may be realized
if the lighter stau $\stauone$ is the next-to-lightest SUSY
particle (NLSP).
Assuming conserved \R-parity in this work the stau NLSP will be
typically long-lived because it can only decay into the \gravitino\
LSP with Planck-scale suppressed couplings.
Conserved \R-parity also implies that the gravitino is stable which
makes it a promising dark matter candidate.%
\footnote{The gravitino can also be dark matter if \R-parity is broken
  as long as it is ensured that the \gravitino-lifetime is of the
  order of the age of the Universe; see,
  e.g.,~\cite{Takayama:2000uz,Buchmuller:2007ui,Ibarra:2008qg,Hamaguchi:2009sz}
  and references therein.}

In this part of the thesis we consider gravitino dark matter scenarios
in which the \stauone\ is the NLSP. In the present chapter we briefly
introduce the gravitino and discuss immanent cosmological implications
which are independent of the nature of the NLSP.
In the next chapter we then work out the phenomenology of the
gravitino-stau scenario.

\section{Supergravity and basic properties of the gravitino}
\label{sec:basic-prop-grav}

In ordinary gauge theories the generators of the of the Poincar\'e
algebra commute with the generators of the internal (local) symmetry
such as, e.g., color $\mathrm{SU}(3)_{\mathrm{c}}$ in the Standard
Model. Indeed, it was shown~\cite{Coleman:1967ad} that any such
extension of the Poincar\'e algebra in a four-dimensional quantum
field theory (with non-zero scattering amplitudes) is necessarily
trivial in the sense that both algebras decouple.

Supersymmetry, however, is an extension of the space-time symmetry
which relates fermionic and bosonic degrees of freedom. The associated
particles form a supermultiplet. The spinorial supersymmetry generator
$Q$ obeys \textit{anti}-commu\-tation relations and the (simplest)
supersymmetry algebra reads%
\footnote{We follow the conventions used in~\cite{Haber:1984rc};
  $[\,\cdot\, ,\cdot\,]$ and $\{\cdot\,,\cdot\}$ denote the commutator and
  anti-commutator, respectively. }
\begin{subequations}
\begin{align}
  \{ Q_{\alpha} , \overline{Q}_{{\beta}} \} &= 2
  \gamma^\mu_{\alpha {\beta} } P_\mu ,
  \label{eq:fundamental-anticommutator-supercharge} \\
  \{ Q_{\alpha} , Q_{\beta} \} & = \{ \overline{Q}_{{\alpha}} ,
  \overline{Q}_{{\beta}} \} =0  , \\
  [ P_\mu , Q_{\alpha} ] &= [ P_\mu , \overline{Q}_{{\beta}} ]  = 0 ,
  \qquad [ P_\mu , P_\nu ] = 0   .
\end{align}
\end{subequations}
Note that (\ref{eq:fundamental-anticommutator-supercharge}) relates
$Q$ with the Poincar\'e algebra; $P_\mu$ generates
translations. Therefore, \textit{local} supersymmetry implies
\textit{local} Poincar\'e symmetry, i.e., invariance under general
coordinate transformations. This is exactly what we expect from a
theory of gravity so that local supersymmetry is also referred to as
supergravity. The gauge field of supergravity is the gravitino. It is
a spin-3/2 Majorana particle and can be written as a vector-spinor
$\psi_{\mu}$.

Particles within the same supermultiplet are degenerate in mass
because $ [ P^2 , Q_{\alpha} ] = 0$. Since we do not yet have
experimental evidence for supersymmetry, we know that it has to be a
broken symmetry if realized in nature.  Local supersymmetry offers the
appealing possibility to be broken spontaneously with a super-Higgs
mechanism operating.
The Goldstone fermion of supersymmetry breaking is absorbed by
\gravitino\ which thereby acquires its longitudinal, helicity~$\pm1/2$
degrees of freedom. After supersymmetry breaking the gravitino has
mass $\mgrav$. Depending on the underlying breaking mechanism, \mgrav\
can range from the eV scale up to scales beyond the TeV
region~\cite{Martin:1997ns}.

The phenomenology of the massive gravitino is then governed by the
following Lagrangian
\begin{align}
  \label{eq:schwinger-rarita-action}
  \mathcal{L} &= -\frac{1}{2} \varepsilon^{\mu\nu\rho\sigma}
  \overline{\psi}_{\mu} \gamma_5 \gamma_\nu
  \partial_\rho \psi_{\sigma} - \frac{1}{4} \mgrav 
   \overline\psi_{\mu} 
[ \gamma^\mu , \gamma^\nu ] \psi_{\nu} + \mathcal{L}_{\mathrm{int}} .
\end{align}
The first two terms describe a free massive spin-3/2
field~\cite{Rarita:1941mf} from which it can be shown that the free
gravitino field satisfies the Dirac equation $(i \slashed\partial -
\mgrav)\psi_{\mu}=0$ for each component $\mu$ and is subject to the
constraints $\gamma^\mu\psi_{\mu}=0$ and $\partial^\mu\psi_{\mu}=0$.
The interaction Lagrangian reads%
\footnote{For an explicit ``derivation'' of
  $\mathcal{L}_{\mathrm{int}}$ from the general supergravity
  Lagrangian see~\cite{Pradler:2007ne}. }
\begin{align}
  \label{eq:gravitino interaction lagrangian}
  \mathcal{L}_{\mathrm{int}} & =
  -\frac{i}{\sqrt{2}\MP} 
  \Big[
   ({D_{\mu}} \phi^{*i}) \overline{\psi}_\nu \gamma^\mu
  \gamma^\nu \cf{L}{i}
  - ({D_{\mu}} \phi^{i})\cfb{L}{i} \gamma^\nu \gamma^\mu  \psi_\nu
\Big]
 \nonumber \\  &\quad\,
 - \frac{i}{8\MP}\overline\psi_\mu [\gamma^\rho,\gamma^\sigma] \gamma^\mu
  {\lambda}^{(\alpha)a} {F}_{\rho\sigma}^{(\alpha)a} + \Orderof{\MP^{-2}}.
\end{align}
Focusing on a minimal particle content in the observable sector, the
fields $\phi$, $\chi_\mathrm{L}$, and~$\lambda$ denote the gauge
eigenstates of the scalars, chiral fermions, and gauginos of the MSSM;
${F}_{\rho\sigma}^{(\alpha)a}$ is the field strength tensor of the
gauge group $(\alpha)=
\left(\mathrm{SU}(3)_{\mathrm{c}},\mathrm{SU}(2)_{\mathrm{L}},\mathrm{U}(1)_{\mathrm{Y}}\right)$
with index $a$ of the associated adjoint representation and $D_{\mu}$
denotes the gauge-covariant derivative.
All matter fields are written in terms of left-handed four-spinors
$\chi_\mathrm{L}$ since they stem from left-chiral supermultiplets in
the general supergravity Lagrangian~\cite{wess:1992cp}. For example, a
right handed tau lepton $ {\tau}^{-}_R$ is written in terms of its
charge conjugate $({\tau}^{-}_R)^c$ which is a left-handed
spinor. Analogously the superpartner of $ {\tau}^{-}_R$ is written in
(\ref{eq:gravitino interaction lagrangian}) as~$({\widetilde
  \tau}_R)^*$.
Details aside, most important is the fact that the interactions of the
gravitino to the MSSM fields are fixed by the \mbox{(super)}symmetry and are
suppressed by inverse powers of~\MP\ which makes \gravitino\ an extremely weekly
interacting particle.%
\footnote{Couplings of a very light gravitino can be enhanced due to
  its longitudinal (goldstino) modes~\cite{Lee:1998aw}---a remnant of
  the super-Higgs mechanism.}

Being the gauge field of supergravity, the gravitino sits at the heart
of any locally supersymmetric theory. Despite its extremely weak
interaction gravitinos can be efficiently produced in the early
Universe. In the next section we discuss the case of thermal gravitino
production---a guaranteed source of potential relic gravitinos.

\section{Thermal gravitino production and reheating}
\label{sec:constr-rehe-temp}

The observed flatness, isotropy, and homogeneity of the Universe
suggest that its earliest moments were governed by
inflation~\cite{Linde:2005ht,Kolb:1990vq}. The inflationary expansion
is followed by a phase in which the Universe is reheated.  The
reheating process repopulates the Universe and provides the initial
conditions for the subsequent radiation-dominated epoch.  The
reheating temperature $T_{\Reheating}$ can be viewed as the initial
temperature of this early radiation-dominated epoch of our Universe.

The value of $T_{\Reheating}$ is an important prediction of inflation
models. While we do not have evidence for temperatures of the Universe
higher than $\Order(1~\MeV)$ (i.e., the temperature required by
primordial nucleosynthesis), inflation models can point to
$T_{\Reheating}$ well above
$10^{10}~\GeV$~\cite{Kolb:1990vq,Linde:1991km}.
While any initial population of gravitinos must be diluted away by the
exponential expansion during inflation~\cite{Khlopov:1984pf},
gravitinos are regenerated in scattering processes of particles that
are in thermal equilibrium with the hot primordial plasma.  The
efficiency of this thermal production of gravitinos during the
radiation-dominated epoch is sensitive to
$T_{\Reheating}$~\cite{Ellis:1984eq,Moroi:1993mb,Bolz:1998ek,Bolz:2000fu,Pradler:2007ne,Pradler:2006qh,Rychkov:2007uq}.

Gravitinos with $\mgr\gtrsim 1~\GeV$ have decoupling temperatures of
$T^{\gravitino}_{\freezeout}\gtrsim 10^{14}~\GeV$, as will be shown
below.
We consider thermal gravitino production in the radiation-dominated
epoch starting at $T_{\Reheating}<T^{\gravitino}_{\freezeout}$
assuming that inflation has diluted away any initial gravitino
population.%
\footnote{ In this work, we neglect gravitino production in
  inflaton-decays, cf.,
  e.g.,~\cite{Asaka:2006bv,Endo:2006tf,Endo:2007sz} and references
  therein. Though this non-thermal source can give a sizable
  contribution it is model-dependent and typically small when
  considering high reheating temperatures with the associated
  gravitino yield scaling as~$\TR^{-1}$~\cite{Endo:2007sz}. }
For $T_{\Reheating}<T^{\gravitino}_{\freezeout}$, gravitinos are not
in thermal equilibrium with the post-inflationary plasma. Accordingly,
the evolution of the gravitino number density $n_{\gravitino}$ with
cosmic time $t$ is described by the following Boltzmann
equation~\cite{Pradler:2007ne,Pradler:2006qh}
\begin{eqnarray}
&&
    \frac{dn_{\gravitino}}{dt} + 3 H n_{\gravitino} = C_{\gravitino}
\label{Eq:Boltzmann}\\
&&
        C_{\gravitino} 
        =
        \sum_{i=1}^{3} 
        \frac{3\zeta(3)T^6}{16\pi^3\MPl^2} 
        \left(1+\frac{M_i^2}{3\mgr^2}\right)
        c_i\, g_i^2
        \ln\left(\frac{k_i}{g_i}\right)
\label{Eq:CollisionTerm}        
\end{eqnarray}
The collision term $C_{\gravitino}$ involves the gaugino mass
parameters $M_i$, the gauge couplings $g_i$, and the constants $c_i$
and $k_i$ associated with the gauge groups U(1)$_\Hypercharge$,
SU(2)$_\Weak$, and SU(3)$_\Color$ as given in
Table~\ref{Tab:Constants}.
\begin{table}[t]
  \caption[Parameters of the gravitino collision term]{
    The gauge couplings $g_i$ and the constants $c_i$, $k_i$, $y_i$, 
    and $\beta_i^{(1)}$ associated with the gauge groups
    U(1)$_\Hypercharge$, SU(2)$_\Weak$, and SU(3)$_\Color$.}
  \label{Tab:Constants}
\begin{center}
\renewcommand{\arraystretch}{1.25}
\begin{tabular}{ccccccc}
\toprule
gauge group         & $i$ & $g_i$  & $c_i$ &  $k_i$ &  $(y_i/10^{-12})$ & $\beta_i^{(1)}$
\\ \midrule
U(1)$_\Hypercharge$ & 1 & $g'$       & 11    & 1.266  & 0.653 & 11
\\
SU(2)$_\Weak$       & 2 & $g$      & 27    & 1.312  & 1.604 & 1
\\
SU(3)$_\Color$ & 3 & $g_\mathrm{s}$  & 72   & 1.271  & 4.276 & -3
\\
\bottomrule
\end{tabular}
\end{center}
\end{table}
In expression~(\ref{Eq:CollisionTerm}) the temperature $T$ provides
the scale for the evaluation of $M_i$ and $g_i$. The given collision
term is valid for temperatures sufficiently below the gravitino
decoupling temperature, where gravitino disappearance processes can be
neglected. A primordial plasma with the particle content of the
MSSM in the high-temperature limit is
used in the derivation of~(\ref{Eq:CollisionTerm}).

The collision term~(\ref{Eq:CollisionTerm}) results from a consistent
gauge-invariant finite-temperature
calculation~\cite{Pradler:2007ne,Pradler:2006qh} following the
approach used in Ref.~\cite{Bolz:2000fu}.
Thus, in contrast to the previous estimates
in~\cite{Ellis:1984eq,Moroi:1993mb}, the expression for
$C_{\gravitino}$ is independent of arbitrary cutoffs. Note that the
field-theoretical methods of~\cite{Braaten:1991dd,Braaten:1989mz}
applied in its derivation require weak couplings, $g_i\ll 1$, and thus
high temperatures $T\gg 10^6~\GeV$.%
\footnote{For an alternative approach see \cite{Rychkov:2007uq}.}
Thus, in the following we focus on cosmological scenarios with
$\TR\gtrsim 10^6~\GeV$ which is also the most attractive temperature
range, e.g., for baryogenesis scenarios based on leptogenesis.

\begin{figure}[t]
\begin{center}
\includegraphics[width=0.6\textwidth]{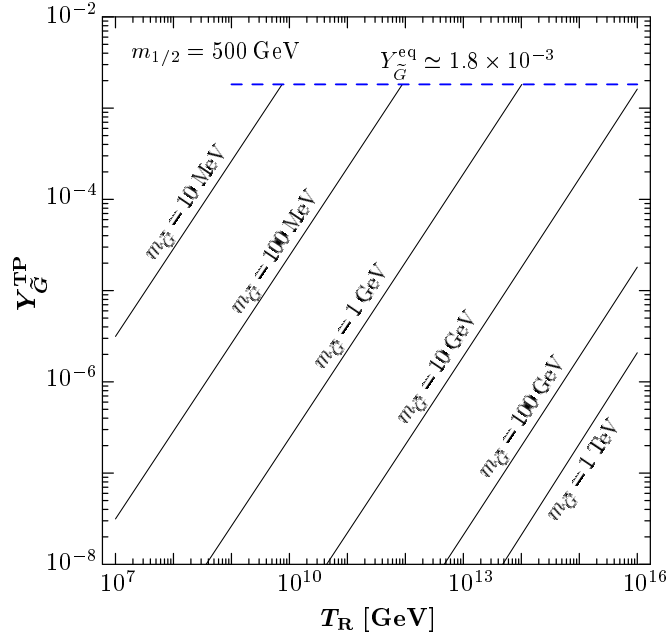} 
\caption[Gravitino yield from thermal production]{The thermally
  produced gravitino yield~(\ref{Eq:YgravitinoTP}) as a function of
  $T_R$ for $\mgr=10~\MeV$, $100~\MeV$, $1~\GeV$, $10~\GeV$,
  $100~\GeV$, and $1~\TeV$ (from left to right) and
  $M_{1,2,3}(M_{\GUT})=m_{1/2}=500~\GeV$. The dashed horizontal line
  indicates the equilibrium yield of a relativistic spin 1/2 Majorana
  fermion.}
\label{Fig:YgravitinoTP}
\end{center}
\end{figure}

Assuming conservation of entropy per comoving volume, the Boltzmann
equation~(\ref{Eq:Boltzmann}) can be solved to good approximation
analytically~\cite{Bolz:2000fu,Brandenburg:2004du}. At a temperature
$\TL\ll\TR$, the resulting gravitino yield from thermal production
reads
\begin{eqnarray}
        Y_{\gravitino}^{\TP}(\TL)
        &\equiv&
        \frac{n_{\gravitino}^{\TP}(\TL)}{s(\TL)}
        \,\,\simeq\,\,
        \frac{C_{\gravitino}(\TR)}{s(\TR)\,H(\TR)}
\nonumber\\
        &=&
        \sum_{i=1}^3
        y_i\, g_i^2(\TR)
        \left(1+\frac{M^2_{i}(\TR)}{3\mgr^2}\right) 
        \ln\left(\frac{k_i}{g_i(\TR)}\right)
        \left(\frac{\TR}{10^{10}\,\GeV} \right)
        \ ,
\label{Eq:YgravitinoTP}
\end{eqnarray}
where the constants $y_i$ are given in Table~\ref{Tab:Constants}.
These constants are obtained with an effective number of relativistic
degrees of freedom of
$\geff(\TR)=\heff(\TR)=228.75$
which follows from the fact that the entire MSSM particle content is
in thermal equilibrium and relativistic.
We evaluate $g_i(\TR)$ and $M_i(\TR)$ using the one-loop evolution
described by the renormalization group equation in the
MSSM~\cite{Martin:1997ns}:
\begin{align}
  g_i(T) 
  &=
  \left[
    g_i^{-2}(\mZ) - \frac{\beta_i^{(1)}}{8\pi^2}\ln\left(\frac{T}{\mZ}\right)
  \right]^{-1/2} ,
\label{Eq:running_coupling}\\
        M_i(T)
        &=
        \left(\frac{g_i(T)}{g_i(M_{\GUT})}\right)^2 M_i(M_{\GUT})
\label{Eq:running_gaugino_mass}
\end{align}
with the respective gauge coupling at the Z-boson mass, $g_i(\mZ)$,
and the $\beta_i^{(1)}$ coefficients listed in
Table~\ref{Tab:Constants}. For convenience we choose to parameterize
the gaugino masses in terms of their values at the scale of gauge
coupling unification $M_\GUT \simeq 2\times 10^{16}\ \GeV$.
We remark in passing that when considering only the SUSY-QCD
contribution in~(\ref{Eq:YgravitinoTP}) it is sometimes also
convenient to express the gravitino abundance in terms of the physical
gluino mass~\cite{Bolz:2000fu}. As has been shown
in~\cite{Buchmuller:2008vw} it is then important to employ a two-loop
running of the gluino mass since using~(\ref{Eq:running_gaugino_mass})
for the renormalization group evolution from the electroweak scale to
$\TR$ would underestimate the gravitino abundance by a factor of two.
In this work, however, we use the running gluino mass $M_3$ at $M_\GUT$ as
input where the effect is smaller (and working in the other
direction.)

For a standard cosmological history without release of entropy, the
gravitino yield from thermal production at the present temperature
$T_0$ is given by $Y_{\gravitino}^{\TP}(T_0) =
Y_{\gravitino}^{\TP}(\TL) $.
The resulting density parameter of thermally produced gravitinos reads
\begin{equation}
         \Omega_{\gravitino}^{\TP}h^2 
         =  \mgr\,Y_{\gravitino}^{\TP}(T_0)\,s(T_0)\,h^2/\rho_c
\label{Eq:OgravitinoTP}
\end{equation}
with $\rho_c/[s(T_0)h^2]=3.6\times
10^{-9}\,\GeV$~\cite{Yao:2006px}.

In Fig.~\ref{Fig:YgravitinoTP} the result~(\ref{Eq:YgravitinoTP}) for
the thermally produced gravitino yield $Y_{\gravitino}^{\TP}(\TL)$ is
shown as a function of $T_R$ for various values of $\mgr$ (solid
lines).  The curves are obtained with $m_{1/2}=500~\GeV$ for the case
of universal gaugino masses at $M_{\GUT}$:
$M_{1,2,3}(M_{\GUT})=m_{1/2}$. 
The dashed (blue) horizontal line indicates the equilibrium yield
\begin{equation}
  Y_{\gravitino}^{\equil} 
  \equiv \frac{n_{\gravitino}^{\equil}}{s}
  \approx 1.8 \times 10^{-3}
\label{Eq:Y_equil}
\end{equation}
which is given by the equilibrium number density of a relativistic
spin-1/2 Majorana fermion,
$n_{\gravitino}^{\equil} = 3\zeta(3)T^3/(2\pi^2)$.
For $T>T^{\gravitino}_{\freezeout}$,
$\geff(T)=\heff(T)=230.75$
since the spin-1/2 components of the gravitino are in thermal
equilibrium. In the region where the yield~(\ref{Eq:YgravitinoTP})
approaches the equilibrium value~(\ref{Eq:Y_equil}), gravitino
disappearance processes should be taken into account. This would then
lead to a smooth approach of the non-equilibrium yield to the
equilibrium abundance.  Without the back-reactions taken into account,
the kink position indicates a lower bound for
$T^{\gravitino}_{\freezeout}$.  Towards smaller $\mgr$,
$T^{\gravitino}_{\freezeout}$ decreases due to the increasing strength
of the gravitino couplings. For example, for $\mgr=1~\GeV$
($10~\MeV$), we find $T^{\gravitino}_{\freezeout}\gtrsim 10^{14}~\GeV$
($10^{10}~\GeV$).

\subsection{Reheating phase}
\label{sec:reheating-phase}

In the analytical expression~(\ref{Eq:YgravitinoTP}) we refer to $\TR$
as the initial temperature of the radiation-dominated epoch. 
So far we have not considered the phase in which the coherent
oscillations of a field $\phi$ dominates the energy budget of the
Universe and where one usually relates $\TR$ to the decay
width~$\Gamma_{\phi}$ of~$\phi$.
In the simplest models of inflation the decaying field $\phi$ which
reheats the Universe also drives the exponential expansion of the
Universe. In the following we shall simply refer to $\phi$ as the
inflaton field.

We can account for the (perturbative) reheating phase, by
considering~(\ref{Eq:Boltzmann}) together with the Boltzmann equations
for the energy densities of radiation and the inflaton field,
\begin{subequations}
  \label{eq:reheat-boltzmann}
  \begin{align}
    \frac{d\rho_{\rad}}{dt} + 4 H \rho_{\rad} &= \Gamma_{\phi}
    \rho_{\phi} \ ,
    \label{Eq:BEqRad}
    \\
    \frac{d\rho_{\phi}}{dt} + 3 H \rho_{\phi} &= -\Gamma_{\phi}
    \rho_{\phi} \ ,
    \label{Eq:BEqPhi}
  \end{align}
\end{subequations}
respectively. To relate $\Gamma_{\phi}$ with $\TR$ we first note that
the second term on the left hand side of~(\ref{Eq:BEqPhi}) indicates
that $\phi$ (when averaged over several oscillations)
scales like matter, $ \rho_{\phi} \propto a^{-3}$; $a$~denotes the
scale factor. Thus, in terms of the initial inflaton energy density
$\rho_{\phi,\mathrm{I}}$, the Hubble rate is given
by
\begin{align}
  H(a) = \sqrt{\frac{\rho_{\phi,\mathrm{I}}}{3 \MP^2}
    \left(\frac{a_{\mathrm{I}}}{a}\right)^3}
\end{align}
as long as $\rho_{\phi}$ dominates.  Assuming an instantaneous
conversion of $\rho_{\phi}$ into radiation when $\Gamma_{\phi} = \xi
H(\TR)$ with $\xi$ usually chosen to be a number between $1$ and $3$
then allows one to define a reheating temperature in terms of the
decay width of the inflaton field,
\begin{align}
  \TR^{\xi} \equiv \xi^{-1/2}  \left(\frac{90}{\geff(\TR)\pi^2}\right)^{1/4}
  \sqrt{\Gamma_{\phi}\,\MPl} .
\label{Eq:TRxi_definition}
\end{align}

\afterpage{\clearpage}
\begin{figure}[t]
\begin{center}
\includegraphics[width=0.70\textwidth]{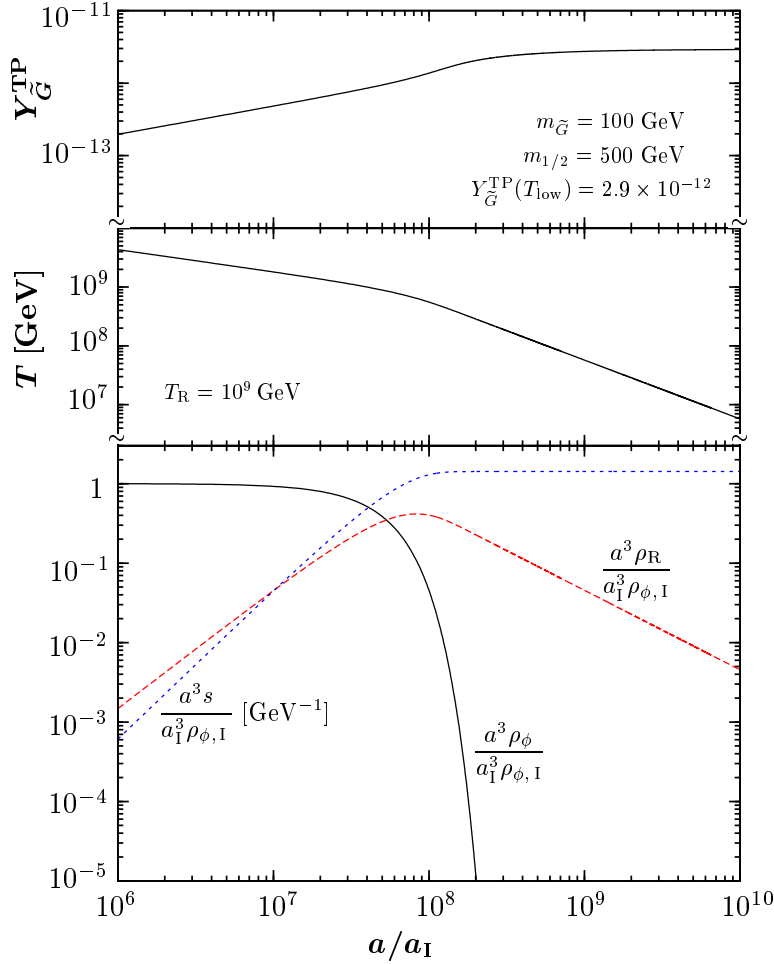} 
\caption[Reheating phase]{The results of a numerical integration of
  (\ref{Eq:Boltzmann}) and (\ref{eq:reheat-boltzmann}) is shown;
  $\Gamma_{\phi}$ is chosen such as to yield $\TR = 10^9\ \GeV$ for
  $\xi = 1.8$. In the lower figure comoving quantities as labeled and
  normalized to $a^3\rho_{\phi,\mathrm{I}}$ are plotted. The middle
  figure shows the evolution of the temperature $T$ of the thermal
  bath and in the top part of the figure the resulting gravitino
  abundance $Y_{\gravitino}^{\TP}$ is obtained for $\mgrav=100\
  \GeV$ and for $M_{1,2,3}=m_{1/2}$ at $M_{\GUT}$ with
  $m_{1/2}=500~\GeV$. This gives a final gravitino abundance of
  $Y_{\gravitino}^{\TP}(\TL) = 2.9\times 10^{-12}$; see main text for
  a discussion.}
\label{Fig:reheating}
\end{center}
\end{figure}
In Fig.~\ref{Fig:reheating} we show the results of a numerical
integration of (\ref{Eq:Boltzmann}) and (\ref{eq:reheat-boltzmann})
plotted against the scale factor with $\Gamma_{\phi}$ chosen such as
to yield $\TR = 10^9\ \GeV$ for $\xi = 1.8$.%
\footnote{For the actual integration we rewrite the Boltzmann
  equations (\ref{Eq:Boltzmann}) and (\ref{eq:reheat-boltzmann}) in
  terms of dimensionless quantities following~\cite{Kolb:2003ke}.  }
In the lower figure we plot comoving quantities normalized to
$a_{\mathrm{I}}^3\rho_{\phi,\mathrm{I}}$. At $a/a_{\mathrm{I}}\simeq
10^8$ the inflaton energy density decays exponentially and the
Universe enters the radiation dominated epoch which can be seen by the
turn-over of the dashed curve depicting the radiation density: there,
$\rho_{\rad}$ starts to scale as $a^{-4}$. This is also indicated by
the evolution of the temperature plotted in the middle part of the
figure. In the radiation dominated Universe $T$ scales with
$a^{-1}$. Also note that $\TR$ is not the maximum temperature of the
Universe. When the inflaton field decays entropy is produced (dotted
line in lower part) and the temperature scales as $T\propto
a^{-3/8}$. The production of entropy is also the reason why
$Y_{\gravitino}^{\TP}$ in the top part of the figure---despite the
large initial temperature---reaches its maximum value only in the
radiation dominated regime when $a^3 s$ is finally constant. Here,
$M_{1,2,3}=m_{1/2}$ at $M_{\GUT}$ with $m_{1/2}=500~\GeV$ and
$\mgrav=100\ \GeV$ has been chosen which gives a final gravitino
abundance of $Y_{\gravitino}^{\TP}(\TL) = 2.9\times 10^{-12}$.

With our result for the collision term~(\ref{Eq:CollisionTerm}), we
find that the gravitino yield obtained numerically is in good
agreement with the analytical expression~(\ref{Eq:YgravitinoTP}) for
$\xi = 1.8$. For an alternative $\TR$ definition with different $\xi$
the associated numerically obtained gravitino yield is described by
the analytical expression obtained after substituting $\TR$ with
$\sqrt{\xi/1.8}\,\TR^{\xi}$ in~(\ref{Eq:YgravitinoTP}).

A fitting formula on the gravitino yield which also includes the
effect of reheating and which was based on~\cite{Bolz:2000fu} has been
derived earlier in~\cite{Kawasaki:2004yh}. However, the production of
the helicity-1/2 component of the gravitino was neglected so that the
actual yield for $\mgr=100~\GeV$ was underestimated by about an order
of magnitude.  Accordingly, the $\TR$ bounds given
in~\cite{Kawasaki:2004yh,Kawasaki:2004qu,Kohri:2005wn} are
underestimated in the region $\mgr<1~\TeV$. Meanwhile---after
publication of~\cite{Pradler:2006hh} on which this section is based
on---an updated treatment~\cite{Kawasaki:2008qe} has become available
in which the authors now include the helicity-1/2 components as well
as the electroweak contributions~\cite{Pradler:2007ne,Pradler:2006qh}
of the thermal gravitino production. For a most recent discussion on
gravitino production during perturbative reheating see
also~\cite{Rangarajan:2008zb}.

\section[Constraints on $\TR$]{\texorpdfstring{Constraints on
    \boldmath$\TR$}{Constraints on T_R}}
\label{sec:TR-bounds-from-TP}

\begin{figure}[t]
\begin{center}
\includegraphics[width=0.6\textwidth]{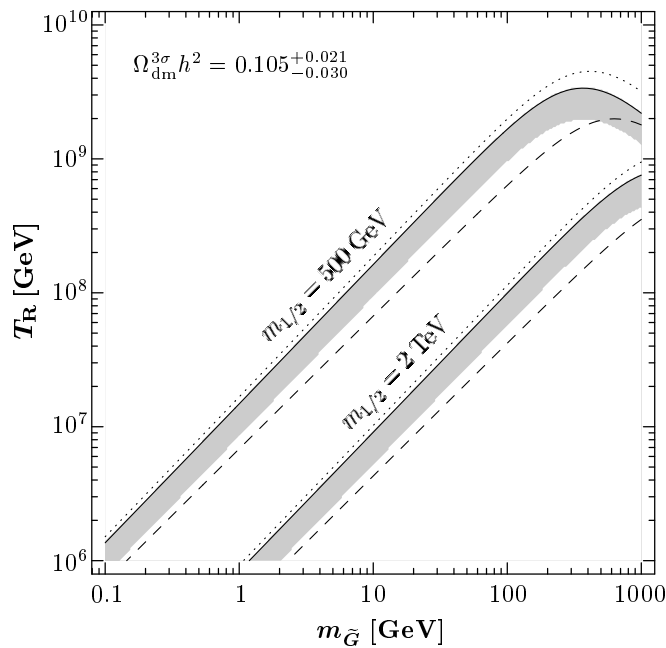} 
\caption[Upper bound on $\TR$]{Upper limits on the reheating
  temperature $T_{\Reheating}$. On the upper (lower) gray band,
  $\Omega_{\widetilde{G}}^{\TP}$ for $M_{1,2,3}=m_{1/2}=500~\GeV$
  ($2~\TeV$) at $M_{\GUT}$ agrees with $\Omega_{\CDM}^{3\sigma}$. The
  corresponding $\TR$ limits from the requirement
  $\Omega_{\widetilde{G}}^{\TP}h^2\leq 0.126$ shown by the dashed
  lines for $M_1/10=M_2/2=M_3=m_{1/2}$ at $M_{\GUT}$ and by the dotted
  lines for the SU(3)$_\mathrm{c}$
  contribution~\cite{Bolz:2000fu,Bolz2008336} with $M_3=m_{1/2}$ at
  $M_{\GUT}$. All lines are obtained with~(\ref{Eq:YgravitinoTP}).}
\label{Fig:UpperLimitTR}
\end{center}
\end{figure}
Since for the gravitino LSP the resulting density
$\Omega_{\gravitino}^{\TP}$ should not exceed the dark matter density
$\Omega_{\CDM}$, $T_{\Reheating}$ is bounded from
above~\cite{Moroi:1993mb}. Such a bound has to be compared with
predictions of the reheating temperature $T_{\Reheating}$ from
inflation models. Moreover, $T_{\Reheating}$ is important for our
understanding of the cosmic baryon asymmetry.  For example, successful
standard thermal leptogenesis~\cite{Fukugita:1986hr} can typically
require $T_{\Reheating}\gtrsim
10^9~\GeV$~\cite{Davidson:2002qv,Buchmuller:2004nz}.

We update the $T_{\Reheating}$ limits using the full gauge-invariant
result for the relic density of thermally produced gravitinos,
$\Omega_{\gravitino}^{\TP}$, to leading order in the Standard Model
gauge couplings~\cite{Pradler:2007ne,Pradler:2006qh}.%
\footnote{The computation of the electroweak contributions to thermal
  gravitino production was subject of the diploma
  thesis~\cite{Pradler:2007ne} of the author in which similar limits
  on \TR\ were already presented---however, without insight on the
  exact sensitivity of $\Omega_{\gravitino}^{\TP}$ on the reheating
  process; in this section $\xi=1.8$.}
In particular, this allows us to illustrate the dependence of the
bounds on the gaugino-mass relation at the scale of grand unification
$M_{\GUT}$.

The reheating temperature $T_{\Reheating}$ is limited from above in
the case of a stable gravitino LSP since $\Omega_{\gravitino}^{\TP}$
cannot exceed the dark matter density
$\Omega_{\CDM}$.
In the following we use~\cite{Spergel:2006hy,Yao:2006px} the WMAP
3-year result
\begin{equation}
        \Omega_{\CDM}^{3\sigma}h^2=0.105^{+0.021}_{-0.030} 
\label{Eq:OmegaDM}
\end{equation}

In Fig.~\ref{Fig:UpperLimitTR} we show the resulting upper limits on
$T_{\Reheating}$ using $\xi=1.8$ as a function of $\mgr$. On the gray
band, the thermally produced gravitino density~(\ref{Eq:OgravitinoTP})
is within the nominal $3\sigma$ range~(\ref{Eq:OmegaDM}). The upper
(lower) gray band is obtained for $M_{1,2,3}=m_{1/2}$ at $M_{\GUT}$
with $m_{1/2}=500~\GeV$ ($2~\TeV$).  From the requirement
$\Omega_{\widetilde{G}}^{\TP}h^2\leq 0.126$ the dashed lines show the
constraints for the exemplary non-universal
scenario~\cite{Anderson:1996bg} $M_1/10=M_2/2=M_3=m_{1/2}$ at
$M_{\GUT}$. Using the same requirement the dotted lines show the
SU(3)$_\mathrm{c}$ contribution~\cite{Bolz:2000fu,Bolz2008336} for
$M_3=m_{1/2}$ at $M_{\GUT}$.
As can be seen by comparing the dashed and dotted lines, the
electroweak contributions can be particularly important for the case
of non-universal gaugino masses at $M_{\GUT}$. On the other hand, for
gaugino masses which unify at $M_{\GUT}$, the dotted lines provide
already a very good estimate. In this regard, note that
Fig.~\ref{Fig:UpperLimitTR} is updated from the one presented
in~\cite{Pradler:2006hh} since the SU(3)$_{\mathrm{c}}$ contribution
to $\Omega_{\gravitino}^{\TP}$, originally obtained
in~\cite{Bolz:2000fu}, has meanwhile been corrected by the
authors~\cite{Bolz2008336}. For completeness, we also remark that,
after publication of \cite{Pradler:2006hh}, the upper limits on $\TR$
shown in Figs.~5 and~6 of Ref.~\cite{Cerdeno:2005eu} have also been
corrected. Previously, these figures underestimated the maximal value
of $\TR$ by a factor of four in the region in which
$\Omega_{\widetilde{G}}^{\TP}$ governs the limits.

The $\TR$ limits shown in Fig.~\ref{Fig:UpperLimitTR} are conservative
bounds that do only depend on $\mgr$ and the $M_i$ values at
$M_{\GUT}$. Taking into account contributions to
$\Omega_{\widetilde{G}}$ from NLSP decays will make those limits
stronger.  In the next chapter, we will account for this non-thermal
gravitino source in a systematic way.


\cleardoublepage
\chapter{The stau as the NLSP}
\label{cha:stau-nlsp}

In this chapter we now specialize on gravitino dark matter scenarios
in which the lighter stau $\stauone$ is the NLSP. 
Indeed, the appearance of $\stauone$ as the lightest Standard Model
superpartner is a commonplace occurrence even in models with
restrictive assumptions on the SUSY breaking sector such as the CMSSM.
The associated parameter region is usually not considered because of
severe upper limits on the abundance of massive stable charged
particles~(see Sec.~\ref{sec:typic-champ-abund}).  However, in
gravitino LSP scenarios \stauone\ is unstable and thereby a viable
option.

In this chapter, we first review the result on a frequently used range
of thermal freeze-out abundances of \stauone.
Employing such an estimate allows us to constrain the gravitino-stau
scenario from BBN limits on the electromagnetic and hadronic energy
release in the decay of \stauone\ in a rather model-independent
fashion. Moreover, we employ the results of
Chapter~\ref{cha:catalyzed-bbn} on the catalyzed light element
production of \lisx\ and \ben. We shall see that the associated
constraints pose the henceforth strongest limits on this
scenario. Specializing to the case of the CMSSM allows us to explore
concrete realizations of the gravitino-LSP stau-NLSP setting. In
particular, we shall find that the reheating temperature \TR\ is
heavily constrained by the novel CBBN bounds. We also explore the
possibility of a non-standard cosmological history to see whether one
can alleviate or even circumvent the strong restrictions on the
parameter space.

\section{Generic constraints on the gravitino-stau scenario}
\label{sec:typical-stau-abundance}

In order to set constraints on the outlined scenario we require
knowledge on the stau abundance in the early Universe.  In
Sec.~\ref{sec:typic-champ-abund} we have found that for a standard
cosmological history the decoupling temperature of a weak scale
charged particle satisfies $T_{\mathrm{f}}<\mstauone/20$. Thus, with a
post-inflationary reheating temperature $\TR$ above the decoupling
temperature, the $\stauone$ NLSP freezes out of the primordial plasma
as a cold thermal relic so that its yield after decoupling $\Ysldec$
is governed by its mass and its annihilation rate.  Thereby, $\Ysldec$
becomes sensitive to the mass spectrum and the couplings of the SUSY
model and representative values
\begin{equation}
  \Ysldec \simeq (0.4 \div 1.5)\times 10^{-13} 
  \left(\frac{\mstauone}{100~\GeV}\right)
\label{Eq:Yslepton}
\end{equation}
have been used to confront the gravitino-stau scenario with
cosmological
constraints~\cite{Asaka:2000zh,Fujii:2003nr,Steffen:2006hw,Pradler:2007is,Steffen:2008bt};
\mstauone\ denotes the mass of the lighter stau.
Equation~(\ref{Eq:Yslepton}) compares well with our upper
bound~(\ref{eq:upper-bound-YX}) derived from a purely dimensional
analysis. The yield~(\ref{Eq:Yslepton}) with a coefficient $0.7\times
10^{-13} $ is in good agreement with the curve in Fig.~1 of
Ref.~\cite{Asaka:2000zh} that has been derived for the case of a
purely `right-handed' $\stau\simeq\stauR$ NLSP with a mass that is
significantly below the masses of the lighter selectron and the
lighter smuon, $m_{\stau} \ll m_{\sel,\smu}$, and with a bino-like
lightest neutralino, $\neutralino\simeq\Bino$, that has a mass of
$m_{\Bino}=1.1\,m_{\stau}$.  In the case of an approximate slepton
mass degeneracy, $m_{\stau} \lesssim m_{\sel,\smu} \lesssim
1.1\,m_{\stau}$, the upper value in (\ref{Eq:Yslepton}) becomes
saturated due to slepton coannihilation processes~\cite{Asaka:2000zh}.
We shall see in Sec.~\ref{sec:exempl-param-scans} that approaching the
$\neutralino$--$\stau$ coannihilation region, $\mneu \approx
\mstauone$, even larger enhancement factors occur.  On the other hand,
a sizable left--right mixing of the stau NLSP is associated with an
increase of its MSSM couplings and thus with a reduction of
$Y_{\stauone}^{\mathrm{dec}}$. This will be discussed in
Part~\ref{part:three} where a systematic investigation of the stau
abundance and its sensitivity on the SUSY parameters will be given.
In this section, we shall focus on the more generic
$Y_{\stauone}^{\mathrm{dec}}$ values described by~(\ref{Eq:Yslepton}).

\begin{figure}[t!]
\centerline{\includegraphics[width=0.75\textwidth]{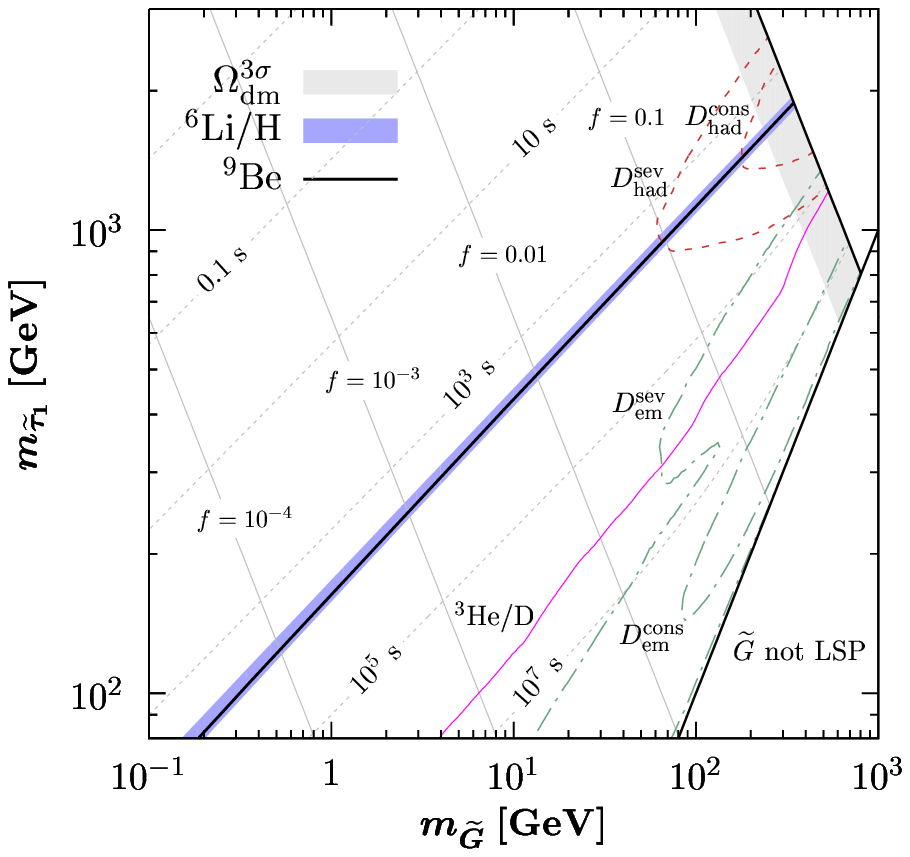}}
\caption[Generic constraints on the gravitino-stau scenario]{
  Constraints of the gravitino LSP stau NLSP scenario for $\Ysldec$
  given by~(\ref{eq:Ystau7}). In the light (gray) shaded region
  $\Omega_{\gravitino}^{\NTP}\in\OmegaDM^{3\sigma}$ holds with the
  region above being disfavored by $\Omega_{\gravitino}
  >\OmegaDM^{3\sigma}$. The thin gray straight lines show the contours
  on which $f\,\OmegaDM$ is provided by $\Omegantp$. The dotted gray
  lines show contours of $\tau_{\stauone}$ as labeled.  Catalyzed BBN
  production of \ben\ disfavors the the region below the thick solid
  line. In the dark shaded (blue) region $\lisx/\Hyd \in 10^{-11}\div
  10^{-10}$ holds. Below, \lisx\ is overproduced due to the bound
  state effects.  Hadronic energy release in stau decays disfavors the
  regions inside the dashed lines for different values of the adopted
  primordial D abundance (\ref{eq:adopted-D-abundances}). By the same
  token, the regions inside dash-dotted (green) curves are disfavored
  from the effect of electromagnetic energy release on D. Moreover,
  the region below the thin solid (pink) line is disfavored from
  overproduction of~$^3\mathrm{He}/\mathrm{D}$.}
\label{Fig:MstMgr} 
\end{figure}

In Fig.~\ref{Fig:MstMgr} we collect the cosmological constraints on
the gravitino-stau scenario by plotting $\mgrav$ versus
$\mstauone$. Let us go through the respective limits one by one:
\begin{description}
\item[Non-thermal gravitino production] Each $\stauone$ NLSP
  eventually decays into one $\gravitino$ LSP leading to a
  non-thermally produced (NTP) gravitino
  density~\cite{Asaka:2000zh,Feng:2004mt}:
\begin{equation}
  \Omegantp h^2
  = 
  \mgravitino\, \Ysldec\, s(T_0) h^2 / \rho_{\mathrm{c}} .
\label{Eq:GravitinoDensityNTP}
\end{equation}
This contributes to the relic gravitino density $\Omega_{\gravitino}$
which should not exceed the observationally inferred dark matter
density $\Omega_{\mathrm{dm}}$.  In Fig.~\ref{Fig:MstMgr} we choose as
representative value
\begin{align}
  \label{eq:Ystau7}
   \Ysldec = 0.7 \times 10^{-13} \left(\frac{\mstauone}{100~\GeV}\right) .
\end{align}
In the light shaded region in the upper right corner $\Omegantp h^2 $
agrees with $ \Omega_{\CDM}^{3\sigma}h^2$ of
Eq.~(\ref{Eq:OmegaDM}). The shading is limited from above by a solid
line which borders the above disfavored region in which $\Omegantp >
\OmegaDM$. Any additional contribution to $\OmegaDM$, such as a
thermally produced gravitino density $\Omegatp$
(Sec.~\ref{sec:constr-rehe-temp}), makes this constraint more
restrictive. This is indicated by the thin gray lines labeled with
$f=0.1$, $0.01$, $10^{-3}$, and $10^{-4}$, on which
(\ref{Eq:GravitinoDensityNTP}) obtained with~(\ref{eq:Ystau7})
satisfies $f\,\Omegantp=0.126$, respectively.
The timing of the \stauone\ NLSP decay into the gravitino LSP is
governed by the two-body decay mode $\stauone\to\gravitino\tau$ and
reads
\begin{equation}
        \tau_{\stauone} 
        \simeq
        \Gamma^{-1}(\stauone\to\gravitino\lepton)
        = 
        \frac{48 \pi \mgr^2 \MPl^2}{\mstauone^5} 
        \left(1-\frac{\mgr^2}{\mstauone^2}\right)^{-4}
\, .
\label{Eq:SleptonLifetime}
\end{equation}
in the limit $m_{\tau}\to 0$. Contour-lines thereof are shown in
Fig.~\ref{Fig:MstMgr} by the dotted gray lines.  Starting from the
upper left edge, they correspond to lifetimes of $10^{-3}\ \seconds$
and then---as labeled---to $0.1\ \seconds$, $10\ \seconds$, $10^3\ \seconds$, $10^5
\seconds$, and $10^7\ \seconds$.

\item[CBBN constraints] We can now incorporate the results of
  Chapter~\ref{cha:catalyzed-bbn} on the catalyzed fusion of \lisx\
  and \ben\ triggered by the bound state formation of
  $(\hef\stauone^-)$.  We use the CBBN constraints obtained in
  Fig.~\ref{Fig:Ytau} with $\YXdec = Y_{\stauone}^{\mathrm{\dec}}/2$, i.e.,
  we assume that there exists no asymmetry between positively and
  negatively charged staus [cf.~Sec.~\ref{sec:comm-stau-stau}]. For
  $Y_\stauone^{\mathrm{dec}}$ we use the
  estimate~(\ref{eq:Ystau7}). The shaded region in Fig.~\ref{Fig:Ytau}
  has corresponded to a \lisx\ output of $\lisx/\Hyd = 10^{-11} \div
  10^{-10}$ and which is now likewise associated with the dark (blue)
  shaded band in Fig.~\ref{Fig:MstMgr}.  On the upper border of the
  band a lithium abundance of $\lisx/\Hyd = 10^{-11}$ is attained
  whereas on the lower border $\lisx/\Hyd = 10^{-10}$ holds.  The
  region below the band is confidently ruled out by overproduction of
  \lisx.%
  \footnote{Using the initial estimate~\cite{Pospelov:2006sc} on the
    CBBN \lisx\ output, the associated constraint has first been shown
    in the $(\mgrav,\mstauone)$-plane in
    \cite{Steffen:2006wx}. Likewise, in this representation, the
    constraints from hadronic \stauone-decays have first been obtained
    in~\cite{Feng:2004mt}. However, both
    works~\cite{Feng:2004mt,Steffen:2006wx} are based on outdated
    light element yields so that we use our treatment of
    Chapter~\ref{cha:catalyzed-bbn} for \lisx\ and the update
    in~\cite{Steffen:2006hw} for the hadronic~\stauone-decays.}
  Moreover, we also show the CBBN constraint from primordial \ben\
  production which excludes a very similar region.
  
\item[Hadronic energy release] In Sec.~\ref{sec:particle decays during
    BBN} we have provided an overview over the physics of late
  decaying particles during/after BBN. In our concrete scenario we can
  now implement the stringent constraint on hadronic energy release
  from the observationally inferred primordial deuterium abundance.%
  \footnote{Additional constraints on hadronic energy release are
    imposed by the primordial abundances of $^4$He, $^3$He/D, $^7$Li,
    and
    $^6$Li/$^7$Li~\cite{Sigl:1995kk,Jedamzik:1999di,Jedamzik:2004er,Kawasaki:2004qu,Jedamzik:2006xz,Cyburt:2006uv}.
    However, in the region allowed by the $^9$Be and $^6$Li
    constraints from bound-state effects, i.e., $\tau_{\st}\lesssim
    \mathrm{few}\times 10^3~\seconds$, the considered D constraint on
    hadronic energy release is the dominant one as can be seen, e.g.,
    in Figs.~38--41 of~\cite{Kawasaki:2004qu} and in Figs.~6--8
    of~\cite{Jedamzik:2006xz}.}
  The limits are based on the severe and conservative upper bounds on
  the product $E_{\mathrm{vis}} Y_{\NLSP}$ [here, $Y_{\NLSP}=\Ystau$]
  obtained in Fig.~39 of~\cite{Kawasaki:2004qu} for (see references
  cited in~\cite{Kawasaki:2004qu}):
  \begin{subequations}
    \label{eq:adopted-D-abundances}
    \begin{align}
      \left(\mathrm{D}/\mathrm{H}\right)_{\mathrm{mean}} &=
      \left(2.78^{+0.44}_{-0.38}\right)\times 10^{-5} \quad
      \Rightarrow \quad \mathrm{severe~constraint},
      \label{Eq:Dmean}\\
      (\mathrm{D}/\mathrm{H})_{\mathrm{high}} &=
      \left(3.98^{+0.59}_{-0.67}\right)\times 10^{-5} \quad
      \Rightarrow \quad \mathrm{conservative~constraint}.
      \label{Eq:Dhigh}
    \end{align}
  \end{subequations}
  Recall from our discussion in Sec.~\ref{sec:bbn-as-probe} that
  (\ref{Eq:Dhigh}) is a rather high value
  on~$\deut/\Hyd|_{\primordial}$. Without trying to give extra
  credence to (\ref{Eq:Dhigh}), following~\cite{Kawasaki:2004qu}, we
  simply take it as a limiting value for D/H. The average injected
  hadronic energy $E_{\mathrm{vis}}$ has been obtained
  in~\cite{Steffen:2006hw} from computation of the 4-body decay of the
  stau NLSP into the gravitino, the tau, and a quark-antiquark pair
  for a purely right-handed $\stauone\simeq \stauR$ NLSP.  The effect
  of hadronic energy injection on primordial D disfavors the regions
  inside the dashed lines shown in Fig.~\ref{Fig:MstMgr}; see
  also Fig 16 in~\cite{Steffen:2006hw}.

\item[Electromagnetic energy release] The constraints resulting from
  the dissociation of light elements due to interaction with the
  electromagnetic cascades formed in stau decays are obtained for a
  ``visible'' electromagnetic energy of $E_{\mathrm{vis}}=
  0.3\,E_{\tau}$ of the tau energy
  \begin{align}
    E_{\tau}=\frac{\mstauone^2-\mgr^2+m_{\tau}^2}{2\mstauone}
  \end{align}
  released in $\stau\to\gravitino\tau$.  
  The $\DsevEM$ and $\HeD$ constraints result from the
  $E_{\mathrm{vis}}Y_{\NLSP}$ limits given in Fig.~42 of
  Ref.~\cite{Kawasaki:2004qu} and the $\DconsEM$ constraint from the
  $E_{\mathrm{vis}} Y_{\NLSP}$ limit given in Fig.~6 of
  Ref.~\cite{Cyburt:2002uv} .  It is the region to the right or
  inside of the dot-dashed (green) curves and the region to the right
  of the and thin solid (pink) line that are disfavored by the
  primordial abundances of D and $^3$He/D, respectively.
\end{description}

As can be seen in Fig.~\ref{Fig:MstMgr}, the constraints from the
catalytic production of \lisx\ and \ben\ are (essentially) the most
restrictive ones. Their coinciding position is due the fact that the
catalytic production of \lisx\ and \ben\ both depend on
$(\hef\stauone^-)$ (same timing) and that their output scales linearly
in \Ystau. Indeed, the associated constraints run parallel and in
vicinity of the $\taustau=10^3\ \seconds$ contour which corresponds to
the time at which $(\hef\stauone^-)$ formation starts to become
efficient. For the adopted stau abundance~(\ref{eq:Ystau7}) this
implies that lifetimes in the vicinity of $\taustau\lesssim 6\times
10^3\ \seconds$ are disfavored.

The electromagnetic $\Deut_{\EM}$ and $^3$He/D constraints are always
less restrictive than the CBBN constraints from $^9$Be and $^6$Li.
Only the hadronic constraint $\Deut_{\HAD}$ competes with the CBBN
constraints for $\mstauone\gtrsim 1\ \TeV$, i.e., in the lifetime
region $\taustau \simeq 10^3\ \seconds$ in which $\Ystau$ is largest.
Though for lifetimes shorter than about 100~s neutron-to-proton
interconversion processes affect \hef, the associated constraint is
about two orders of magnitude weaker (as can be seen in left panel of
Fig.~\ref{Fig:EM-HAD-BBN-constr}).
We also remark that the elevated content of D due to hadronic and
electromagnetic energy injection leads to an enhancement of
CBBN-produced \lisx\ and \ben. For example, if non-thermal processes
boost the deuterium abundance to the level of (\ref{Eq:Dhigh}), it
would lead to an enhancement of the \lisx\ output by a factor of $\sim
2$, while the corresponding enhancement factor in the case of \ben\ is
about~4.  We have not included this effect since it can make our
obtained limits only stronger.

Here we would like to emphasize that the $^9$Be and $^6$Li constraints
are the ones that are the least sensitive to the precise value of
$\Ysldec$ in the region $\Ysldecm\gtrsim 10^{-14}$. This results from
the fact that the limits are very steep in that region, as can be seen
in Fig.~\ref{Fig:Ytau}.  Indeed, a yield that is twice as large
as~(\ref{Eq:Yslepton}) will affect the position of the $^9$Be and
$^6$Li constraints only very mildly. In contrast, such an enhanced
yield---as encountered, e.g., in the case of slepton
coannihilations---leads to significant changes of the dark matter
constraint and the BBN constraints associated with
hadronic/electromagnetic energy injection, as can be seen explicitly
in Fig.~16 of Ref.~\cite{Steffen:2006hw}.

It should also be noted that an elevated slepton yield can lead to an
additional non-thermal output of \lisx\ for $\tau_{\stauone}\gtrsim
\mathrm{few}\times 10^2\ \seconds$. As discussed before, this is
because energetic spallation debris of destroyed \hef\ nuclei from
slepton decays can hit ambient \hef\ and thereby fuse
\lisx~\cite{Jedamzik:2004er,Kawasaki:2004qu,Jedamzik:2006xz}. This
mechanism depends sensitively on the hadronic branching ratio
$B_{\mathrm{h}}$ of the 4-body slepton decay into the gravitino, the
associated lepton, and a quark-antiquark pair for which typically
$B_{\mathrm{h}} \lesssim 3\times 10^{-3}$ for $\mstauone \lesssim 2\
\TeV$ (see Fig.~5 of Ref.~\cite{Steffen:2006hw}). Indeed, as discussed
in Ref.~\cite{Jedamzik:2007qk}, for those branching ratios, the effect
of CBBN on \lisx\ is the dominant one in the region which is not
already excluded by the D constraint.  Thus, for $\mstauone \lesssim
1.5\ \TeV$, our obtained limits on \lisx\ overproduction are only
marginally affected by the hadronic energy release of
$\stauone$-decays. However, for larger slepton masses, i.e., for
scenarios of large $\Ysldecm$ in conjunction with
$B_{\mathrm{h}}>10^{-3}$, the hadronic production of \lisx\ becomes
efficient so that only a simultaneous treatment of both effects can
decide on the accurate \lisx\ BBN output.%
\footnote{Using the catalysis of BBN reactions in order to seek for a
  simultaneous solution of both, the \lisx\ and \lisv\ problem [cf.
  Sec.~\ref{sec:bbn-post-WMAP}] has, e.g., been made in
  \cite{Cyburt:2006uv,Jedamzik:2007cp,Bird:2007ge}; for a most recent
  discussion which also includes the lithium output due to \X\ decays
  see \cite{Bailly:2008yy}.}
Note that this can make our presented limits on \lisx\ only
stronger. Thus, we are on the conservative side when neglecting such
additional contributions.
We note in passing that with a highly fine-tuned $\mstauone$-$\mgr$
degeneracy leading to $E_{\mathrm{vis}}\to 0$, any bound on energy
release can be evaded. However, the CBBN bounds remain.

Let us also comment on the reliability of the novel CBBN constraints
from \lisx\ and \ben\ overproduction and address the implications of
the associated restrictions on the $(\mgrav, \mstauone)$ parameter space:

\begin{description}
\item[Reliability of CBBN constraints] As already emphasized in
  Sec.~\ref{sec:bbn-post-WMAP}, observations of \lisx\ are extremely
  difficult. Whereas in the cold interstellar medium the lines of
  \lisx\ and \lisv\ are well resolved, measurements of the isotopic
  ratio $\lisx/\lisv$ in the outer layer of stars is complicated
  because the absorption lines of \lisx\ are not resolved
  spectroscopically with respect to the lines of \lisv\ due to thermal
  blending~\cite{Asplund:2005yt}. Indeed, the claim of a
  ``\lisx-plateau'' is being challenged in the recent
  papers~\cite{Cayrel:2007te,Cayrel:2008hk}. The presence of \lisx\ is
  inferred from an enhancement of the ``red wing'' of the \lisv\
  absorption line. It is argued that such a line asymmetry could also
  be mimicked by Doppler-shifts due to atmospheric convective motions.
  Accordingly, some of the observations may eventually turn out to
  provide only upper limits. Moreover, \lisx\ is more fragile than
  \lisv\ and would burn more efficiently at lower
  temperatures. Therefore, if there is a (yet unconfirmed) stellar
  mechanism (see, e.g.,~\cite{Korn:2006tv}) that resolves the lithium
  problem, i.e., that depletes \lisv\ by a factor of two or three,
  \lisx\ would have been depleted by a larger factor.

  Given those issues, we have adopted here a generous range on the
  observationally inferred upper limit on primordial \lisx\ with
  $\lisx/\Hyd = 10^{-10}$ being a very high value. In this regard it
  is also important to note that the catalytic effect on \lisx\ is
  very strong so that---for the purpose of setting constraints---we
  are not overly sensitive to the precise value of the upper
  bound. This can be seen by the fact that the dark shaded (blue) band
  in Fig.~\ref{Fig:MstMgr} is rather thin while spanning one order of
  magnitude in fused \lisx. In this respect,
  Sec.~\ref{Sec:pXcatalysis} becomes important in which we show that
  the destruction of large fractions of the previously synthesized
  \lisx\ by $(\proton\,\stauone^-)$ is not feasible. This rules out
  the possibility that allowed islands in the parameter region with
  large $Y_{\stauone}$/large $\tau_{\stauone}$---which was advocated
  to remain viable in Ref.~\cite{Jedamzik:2007cp}---exist.

  Unlike \lisx, \ben\ is firmly detected in a significant number of
  stars at low metallicity, and its observational status is not in
  doubt. (For the latest data on the \ben\ abundance in metal-poor
  stars, see, e.g.,
  \cite{Primas:2000gc,Primas:2000ee,Boesgaard:2005jf,Boesgaard:2005pf}.)
  Also note that stellar depletion would affect \ben\ less than either
  \lisv\ or \lisx\ since both \lisv\ and \lisx\ are more fragile than
  \ben. Moreover, the nuclear physics rates that enter in the
  calculation of \ben\ catalysis are dominated by resonances. Given
  the wealth of experimental information on the \ben\ resonances
  \cite{PhysRevC.63.018801,Sumiyoshi2002467}, this may eventually
  allow for very reliable calculations of the catalytic rates. Though
  it has recently been argued that the resonance in the final step
  (\ref{Eq:beet-neutron-capture}) in the fusion of \ben\ is shifted
  below threshold [cf. Sec.~\ref{sec:catalys-ben-prod}] it is a
  neutron induced reaction so that such a shift will affect the
  efficiency of the reaction but may not be fatal. Moreover, also
  note that we have adopted a very conservative upper limit on
  primordial \ben\ in (\ref{eq:upper-limit-Be9}) which can already be
  seen by mere optical inspection of Fig.~\ref{Fig:Bedata}b.
  Taking as a grain of salt that the final efficiency of the \ben\
  reaction is not fully established but noting the powerful physics
  potential a primordial origin of \ben\ offers we have chosen to
  incorporate this constraint. Moreover, given the fact that both,
  \lisx\ and \ben, show the same sensitivity on the gravitino-stau
  parameter space any conclusions drawn from \lisx\ are only
  corroborated and not altered by~\ben. We eagerly await further
  investigation of the critical catalyzed nuclear rates which shall
  give a final answer on the CBBN output of \ben.

\item[Implications of CBBN constraints]
  From Fig.~\ref{Fig:MstMgr} it is immediately clear that the new CBBN
  constraints imply a lower limit on $\mstauone$ given
  \mgrav. Clearly, a minimum value of the lightest Standard Model
  superpartner directly affects the testability of such SUSY scenarios
  at future colliders.

  For the gravitino dark matter scenario in which $\OmegaDM$ is
  exclusively provided by $\Omegantp$, i.e., which are situated in the
  light gray shaded band, formerly (marginal) allowed islands with
  $200\ \GeV\lesssim \mgrav\lesssim 400\ \GeV$ and $\mstauone\lesssim
  1.5\ \TeV$ are now confidently ruled out. Moreover, considering
  other values than $f=1$ the gravitino mass $\mgr$ is constrained to
  values well below 10\% of the slepton NLSP mass $m_{\stauone}$ for
  $\mstauone \lesssim\mathcal{O}(1\ \TeV)$. Thereby the kinematical
  determination of $\mgr$ proposed in~\cite{Buchmuller:2004rq} remains
  cosmologically disfavored at the next generation of particle
  accelerators.
  
  Of course, the CBBN constraints only emerge if $\taustau$ is large
  enough to allow for $(\hef\,\stauone^-)$ bound state formation.
  Thereby, scenarios with a gravitino mass of $\mgr\lesssim 200~\MeV$
  and $\mstauone\gtrsim 80~\GeV$---the latter of which is supported by
  the non-observation of long-lived charged sleptons at the Large
  Electron Positron Collider (LEP)~\cite{Yao:2006px}---are
  unconstrained from CBBN.
  Accordingly, for gauge-mediated SUSY breaking which typically
  predicts small values of $\mgr$, the CBBN constraint can be
  irrelevant.
  However, in gravity-mediated SUSY breaking the gravitino mass sets
  the scale for the soft breaking parameters so that $\mgrav\gtrsim
  10\ \GeV$ are the most natural values. Then, the CBBN constraints
  impose a lower limit of $\mstauone>400~\GeV$.
\end{description}

To extract further implications from the gravitino-stau scenario we
resort in the next section to concrete supersymmetric realizations by
full specification of the SUSY parameters. Thereby we obtain further
insight on the superparticle mass spectrum. Moreover, this allows us
obtain a stringent upper bound on the reheating temperature of the
Universe.

\section{The gravitino-stau scenario in the CMSSM}

We now consider gravitino dark matter scenarios in the framework of
the CMSSM where one assumes universal soft SUSY breaking parameters at
$M_{\GUT}$. The CMSSM yields phenomenologically acceptable spectra
with only four parameters and a sign: the gaugino mass parameter
$m_{1/2}$, the scalar mass parameter $m_0$, the trilinear coupling
$A_0$, the mixing angle $\tan\beta$ in the Higgs sector, and the sign
of the higgsino mass parameter~$\mu$.

In the CMSSM with the gravitino LSP, the next-to-lightest SUSY
particle is either the lightest neutralino $\neutralino$ or the
lighter stau $\stauone$.%
\footnote{A stop $\widetilde{t}_1$ NLSP is not feasible in the
  CMSSM~\cite{DiazCruz:2007fc}.}
The BBN constraints on electromagnetic and hadronic energy injection
disfavor the $\neutralino$ NLSP for $\mgr\gtrsim
100~\MeV$~\cite{Feng:2004mt,Roszkowski:2004jd,Cerdeno:2005eu}. For the
slepton NLSP case, the BBN constraints associated with
hadronic/electro\-magnetic energy injection have also been estimated
and found to be much weaker but still significant in much of the
parameter
space~\cite{Feng:2004mt,Roszkowski:2004jd,Cerdeno:2005eu,Steffen:2006hw}. In
the following, however, we shall see that the novel CBBN constraints
drastically change this picture.%
\footnote{Seeking a solution to the lithium problems, bound state
  effects within the framework of the CMSSM have first been considered
  in~\cite{Cyburt:2006uv}. We set a different focus: We constrain the
  CMSSM parameter space from which we mainly derive an upper bound on
  $\TR$.}

Let us give an overview of what is presented in the remainder of this
chapter. In Sec.~\ref{sec:lower-limit-monetwo} we consider the CBBN
restrictions on the \stauone\ lifetime. Employing the results on
$\mstauone$ from a renormalization group analysis this allows us to
relate the stau mass with the high-scale parameter $\monetwo$. The
obtained lower limit on the latter parameter can be translated into an
upper bound on the reheating temperature. This will be done in
Sec.~\ref{sec:upper-bound-TR}. Both limits will be based on the
estimate (\ref{eq:Ystau7}) of the stau decoupling yield.  We contrast
the obtained semi-analytical limits on $\monetwo$ and $\TR$ with
exemplary CMSSM parameter scans in
Sec.~\ref{sec:exempl-param-scans}. For those examples we will also
explicitly consider in Sec.~\ref{sec:late-time-entropy} the
possibility of a non-standard cosmological history. In this context,
we also check on the viability of thermal leptogenesis in
Sec.~\ref{sec:viab-therm-lept}.

\subsection[Lower limit on $\monetwo$]
{\texorpdfstring{Lower limit on $\mathbf{m}_{\mathbf{1/2}}$}
{Lower limit on m_1/2}}
\label{sec:lower-limit-monetwo}

In the previous section we have realized that the new bounds emerging
from the thermal catalysis of nuclear reactions yield the most
dominant restrictions on the gravitino-stau parameter space. We have
also noted that the discussion of those bounds is facilitated by the
fact that both, the constraint on \lisx\ as well as the one on \ben\
production, essentially disfavor the same region in the
$(\mgr,\mstauone)$ parameter space. In the following we focus on
\lisx\ and adopt as an upper limit on its primordial abundance
\begin{align}
  \label{eq:Li-obs}
  \Lisix/\mathrm{H} |_{\primordial} \lesssim 6\times 10^{-11}.
\end{align}
We thereby resort in this work to a more conservative point of view
than in our main discussions of the published
works~\cite{Pradler:2006hh,Pradler:2007is,Pradler:2007ar} which were
based on $ \Lisix/\mathrm{H} |_{\primordial} \lesssim 2\times
10^{-11}$~\cite{Cyburt:2002uv}. In this way, we give account to the
concerns cast in the previous section (without, however, going to the
very extreme using $10^{-10}$.)

\begin{figure}[t]
\centering
\includegraphics[width=0.60\textwidth,angle=0]{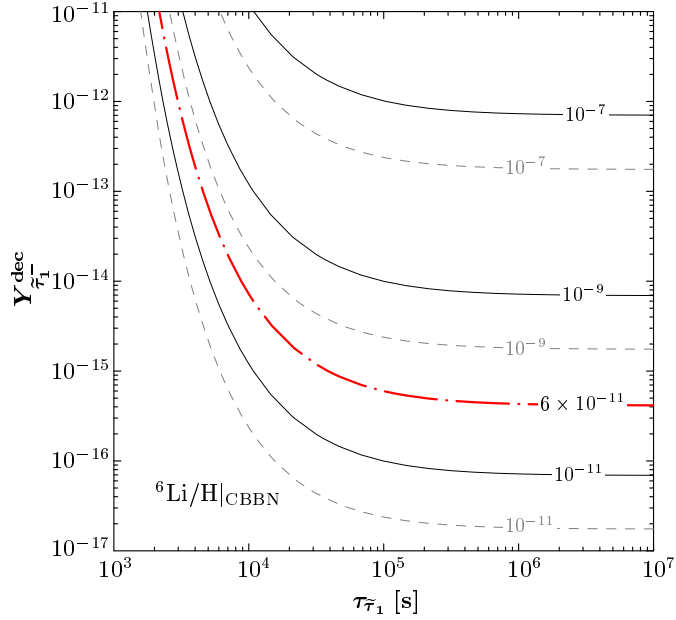}
\caption[Upper bound on \lisx]{The solid lines are contours of constant $\Lisix /
  \mathrm{H}$ as labeled and produced in CBBN. They are obtained by
  solving the full Boltzmann equations presented in
  Chapter~\ref{cha:catalyzed-bbn}. In addition, the dot-dashed (red)
  line shows the adopted limiting primordial
  abundance~(\ref{eq:Li-obs}) with the region above being disfavored
  from \lisx\ overproduction. The dashed lines show the \lisx\ output
  of the Boltzmann network when the Saha
  approximation~(\ref{eq:stat-equilibrium-rec-ph}) for the
  $(\hef\,\stauone^-)$ bound state fraction is used.  }
\label{fig:contours-Lisx}    
\end{figure}

The limit (\ref{eq:Li-obs}) is shown by the dash-dotted (red) line in
Fig.~\ref{fig:contours-Lisx} as a function of the yield of negatively
charged staus $\Ystaudec^{\mathrm{dec}}$ and $\taustau$. As in
Fig.~\ref{Fig:Ytau} the curve is obtained by numerical integration of
the Boltzmann equations presented in Chapter~\ref{cha:catalyzed-bbn}.
In addition, the solid lines show contours of constant \lisx\ output
for other values than~(\ref{eq:Li-obs}) as labeled. Moreover, the
dashed gray lines show the \lisx\ contours when the Saha approximation
(\ref{eq:stat-equilibrium-rec-ph}) is used for the bound-state
abundance. The associated overestimation of \lisx\ once more
demonstrates the importance of a full numerical solution of the
Boltzmann equations. We remark that Fig.~\ref{fig:contours-Lisx}
contains an improvement with respect to the corresponding figure
in~\cite{Pradler:2007is} in the sense that
proton-burning~(\ref{eq:proton-burning-Li6}) of \lisx\ has now been
included and which leads to some reduction of the final \lisx\
output. Moreover, we now use a $(\hef\,\stauone^-)$ recombination
cross section which includes the finite charge radius correction and
accounts for recombination into 1S and 2S states; see
Chapter~\ref{cha:catalyzed-bbn}.

Using the estimate (\ref{eq:Ystau7}) with $\Ystaudec^{\mathrm{dec}} =
Y^{\mathrm{dec}}_{\stau}/2$ we find from Fig.~\ref{fig:contours-Lisx}
that the amount of $\Lisix$ produced in CBBN can be in agreement with
(\ref{eq:Li-obs}) only for stau lifetimes of
\begin{align}
\label{eq:lifetimebound}
\taustau \lesssim 6\times 10^3\;\seconds .
\end{align}
As can be seen from the supergravity
prediction~(\ref{Eq:SleptonLifetime}) of $\taustau$, the requirement
(\ref{eq:lifetimebound}) implies a lower limit on the splitting
between $m_{\stau}$ and $\mgr$ provided $\mstauone \lesssim
\mathcal{O}(1\ \TeV)$.  Because of this hierarchy, the factor
$(1-\mgr^2/\mstauone^{2})^{-4}$ can be neglected in
Eq.~(\ref{Eq:SleptonLifetime}) in the following.

Let us now turn to the CMSSM. We employ the computer program
\texttt{SPheno~2.2.3}~\cite{Porod:2003um} to obtain the low-energy
supersymmetric particle spectrum from the high-scale input at
$M_\GUT$.  In the region in which~$\stauone$ is the NLSP, we find
\begin{align}
\label{eq:mstauone-estimate}
  \mstauone^2 \le 0.21 \monetwo^2 
\end{align}
by scanning over the following parameter range:
\begin{align*}
  \monetwo &= 0.1 \div 6\ \TeV,\\ 
  \tanb &= 2 \div 60, \\
 \mathrm{sgn\ \!}\mu &= \pm 1,\\
   -4 \mzero &< A_{0} < 4 \mzero,
\end{align*}
and with $\mzero$ as large as viable for a $\stauone$ NLSP.%
\footnote{We choose $m_{\mathrm{t}} = 172.5\ \GeV$ for the top quark
  mass. In addition, we use the Standard Model parameters
  $m_{\mathrm{b}}(m_{\mathrm{b}})^{\mathrm{\overline{MS}}} = 4.2\
  \GeV$,
  $\alpha_{\mathrm{s}}^{\mathrm{\overline{MS}}}(m_\mathrm{Z})=0.1172$,
  and $\alpha_{\mathrm{em}}^{-1\mathrm{\overline{MS}}}(m_{\mathrm{Z}})
  = 127.932 $.}

For small left-right mixing, $\stauone \simeq
\widetilde{\tau}_{\mathrm{R}}$, (\ref{eq:mstauone-estimate}) can be
understood qualitatively from the estimate for the mass of the
right-handed stau $m_{\widetilde{\tau}_{\mathrm{R}}}$ near the
electroweak scale \cite{Martin:1993ft}
\begin{align}
 \label{eq:mstauoneR-estimate}
  m_{\widetilde{\tau}_{\mathrm{R}}}^2 \simeq  0.15
  \monetwo^2 + \mzero^2
  -\sin^2{\theta_{W}}m_{\mathrm{Z}}^2\cos{2\beta}\ .
\end{align}
since $\mzero^2 \ll \monetwo^2$ in a large part of the $\stauone$ NLSP
region. In fact, (\ref{eq:mstauone-estimate}) tends to be saturated for
larger $\mzero$, i.e., in the stau-neutralino-coannihilation region
where the mass of the lightest neutralino $m_{\neutralino}\simeq
\mstauone$. This can be understood since the neutralino is bino-like in
this region and typically $m_{\neutralino}^2 \simeq 0.19\monetwo^2$.%
\footnote{This estimate is relatively  independent of $\tanb$ and
  valid in the $\monetwo$ region in which also the LEP bound on the
  Higgs mass~\cite{Yao:2006px}, $m_{\mathrm{h}}>114.4\ \GeV$, is respected.  }
In the remaining part of the stau NLSP
region, smaller values of $\mstauone$ 
satisfying, e.g., $\mstauone^2 = 0.15\monetwo^2$ can easily be
found.

To be on the conservative side, we set the stau NLSP mass $\mstauone$
to its maximum value at which (\ref{eq:mstauone-estimate}) is
saturated: $\mstauone^2 = 0.21\monetwo^2
$. Using~(\ref{Eq:SleptonLifetime}) this allows us to extract a lower
limit on the universal gaugino mass parameter from the
constraint~(\ref{eq:lifetimebound})
\begin{align}
  \label{eq:LowerLimitm12}
  \monetwo \ge 0.87\, \TeV \left( \frac{\mgr}{ 10\ \GeV}
  \right)^{2/5}  .
\end{align}

Since for a $\stauone$ NLSP typically $ \mzero^2 \ll \monetwo^2 $, it is
the gaugino mass parameter $\monetwo$ which sets the scale for the low
energy superparticle spectrum. Thus, depending on $\mgr$, the
bound~(\ref{eq:LowerLimitm12}) implies rather high values of the
superparticle masses. This is particularly true for the masses of the
squarks and the gluino since their renormalization group running from
$\mgut$ to $Q\simeq \mathcal{O}(1\ \TeV)$ is dominated by $M_3(Q)\simeq
\monetwo \alpha_{\mathrm{s}}(Q) /
\alpha_{\mathrm{s}}(\mgut)$. Therefore, for $\mgr \gtrsim 10~\GeV$,
the cosmologically favored region is associated with a mass range that
will be very difficult to probe at the Large Hadron Collider.

We stress that the scan over the entire natural CMSSM parameter space
has enabled us to set a bound on $\monetwo$ which depends on the
gravitino mass but is independent of the CMSSM parameters.%
\footnote{Similar limits have also been discussed in models in which
  the ratio $\mgr/\monetwo$ is bounded from
  below~\cite{Kersten:2007ab}.}
We, however, also remark that we have used the
estimate~(\ref{eq:Ystau7}) without accounting for the dependence of
$Y_\stauone^{\mathrm{dec}}$ on the SUSY parameters (other than
\mstauone). It will be thus important to reflect the
bound~(\ref{eq:LowerLimitm12}) on exemplary scenarios where
$Y_\stauone^{\mathrm{dec}}$ is computed in each point of the parameter
space. This will be done in
Sec.~\ref{sec:exempl-param-scans}. Moreover, in Part~\ref{part:three}
we will carry out a detailed study of the dependence of the stau
decoupling yield on the SUSY parameters. There, we will indeed find
that it is possible to evade (\ref{eq:LowerLimitm12}) in exceptional
cases.

Finally, we also note that in the derivation
of~(\ref{eq:LowerLimitm12}) we have only made use of the catalytic BBN
effects. In Sec.~\ref{sec:typical-stau-abundance} we have seen that
only the D constraint on hadronic energy release can compete with the
CBBN constraints. Accordingly, the D constraint can only tighten the
bounds on $\monetwo$ (and $\TR$ in the following). Thus, taking a
conservative point of view, we are allowed to neglect this
complication.

\subsection[Upper bound on \TR]
{\texorpdfstring{Upper bound on $\mathbf{T_R}$}
{Upper bound on T_R}}
\label{sec:upper-bound-TR}

The amount of gravitinos produced in thermal scattering is sensitive
to the reheating temperature $\TR$ and to the masses of the gauginos
and hence to $\monetwo$~\cite{Pradler:2006qh}. The associated
gravitino density can be approximated by%
\footnote{For a discussion on the definition of $\TR$ see the
  discussion in Sec.~\ref{sec:reheating-phase}; here, $\xi = 1.8$.}
\begin{align}
  \label{eq:omega-tp}
  \Omegatp h^2 \simeq 0.32 \Big( \frac{10\ \GeV}{\mgr} \Big) 
  \Big( \frac{\monetwo}{1\ \TeV} \Big)^2 \Big(
    \frac{\TR}{10^{8}\ \GeV} \Big) .
\end{align}
This follows from~(\ref{Eq:YgravitinoTP}). Here we use that the
running gaugino masses $M_i$ associated with the gauge groups
$\mathrm{SU}(3)_{\mathrm{c}}$, $\mathrm{SU}(2)_{\mathrm{L}}$, and
$\mathrm{U}(1)_{\mathrm{Y}}$ satisfy $M_3:M_2:M_1 \simeq 3:1.6:1 $ at
a representative scale of $10^8\,\GeV$ at which we also evaluate the
respective gauge couplings. Furthermore, we only need to take into
account the production of the spin-$1/2$ components of the gravitino
since (\ref{eq:LowerLimitm12}) implies $M_i^2/3\mgr^2 \gg 1$ for $\mgr
\gtrsim1\ \GeV$.

For a given $\monetwo$, the reheating temperature $\TR$ is limited
from above because $\Omegatp h^2$ cannot exceed the dark matter
density~(\ref{Eq:OmegaDM}). Using the derived lower bound
(\ref{eq:LowerLimitm12}) allows us to extract the upper limit:
\begin{align}
  \label{eq:UpperLimitTR}
  \TR \lesssim 5\times 10^7 \ \GeV \left( \frac{\mgr}{10\ \GeV}
  \right) ^{1/5} .
\end{align}
This constraint is a slowly varying function of $\mgr$:
$(\mgr/10\, \GeV)^{1/5} = 0.6 \div 2.5$ for $\mgr = 1\, \GeV \div 1\, \TeV$.
Therefore, (\ref{eq:UpperLimitTR}) poses a strong bound on
$\TR$ for the natural gravitino LSP mass range in gravity-mediated
supersymmetry breaking scenarios.%
\footnote{Similar, but less restrictive limits have been obtained
  in~\cite{Steffen:2008bt} by relaxing the CMSSM-specific
  splitting~(\ref{eq:mstauone-estimate}).}

Note that the constraint~(\ref{eq:UpperLimitTR}) relies on thermal
gravitino production only. In addition, gravitinos are produced in
stau NLSP decays with the respective
density~(\ref{Eq:GravitinoDensityNTP}). While the precise value of
$Y^{\mathrm{dec}}_{\stauone}$ depends on the concrete choice of the
CMSSM parameters, the upper limit (\ref{eq:UpperLimitTR})
can only become more stringent by taking $\Omegantp$ into
account (provided that (\ref{eq:Ystau7}) is not substantially
depleted.)

\subsection{Exemplary parameter scans in the CMSSM}
\label{sec:exempl-param-scans}

Taking into account gravitinos from thermal production and from late
decays of the lightest Standard Model superpartner we can confront our
numerical findings with the above derived semi-analytical limits on
$\monetwo$ and $\TR$ for various values of $\mgr$.
Considering concrete CMSSM scenarios allows us to compute the thermal
and non-thermal gravitino production in each point of the parameter
space without relying on typical values of the decoupling yield
$Y_\NLSP$ of the NLSP such as~(\ref{Eq:Yslepton}).

Earlier studies of $T_{\Reheating}$ constraints within the CMSSM used
the result of~\cite{Bolz:2000fu} to explore the viability of
$T_{\Reheating}\gtrsim 10^9\,\GeV$~\cite{Roszkowski:2004jd,Cerdeno:2005eu}.
Our study presents also scans for $T_{\Reheating}$ as low as
$10^7\,\GeV$ based on~(\ref{Eq:YgravitinoTP}) which includes
electroweak contributions to thermal gravitino
production~\cite{Pradler:2007ne}.%
\footnote{Meanwhile, after publication of~\cite{Pradler:2006hh},
  related works appeared~\cite{Bailly:2008yy,Bailly:2009pe}; cf. also
  Sec.~\ref{sec:TR-bounds-from-TP}.}

\begin{figure}[t]
\begin{center}
\includegraphics[width=0.62\textwidth]{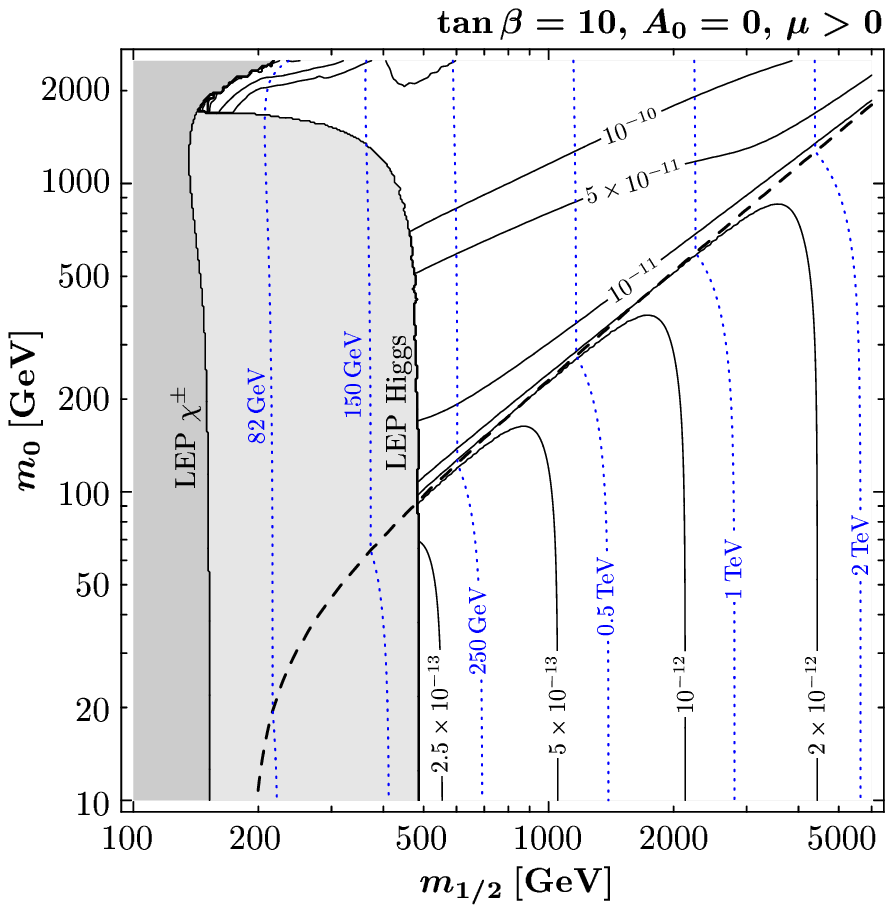}
\vskip 0.2cm
\includegraphics[width=0.62\textwidth]{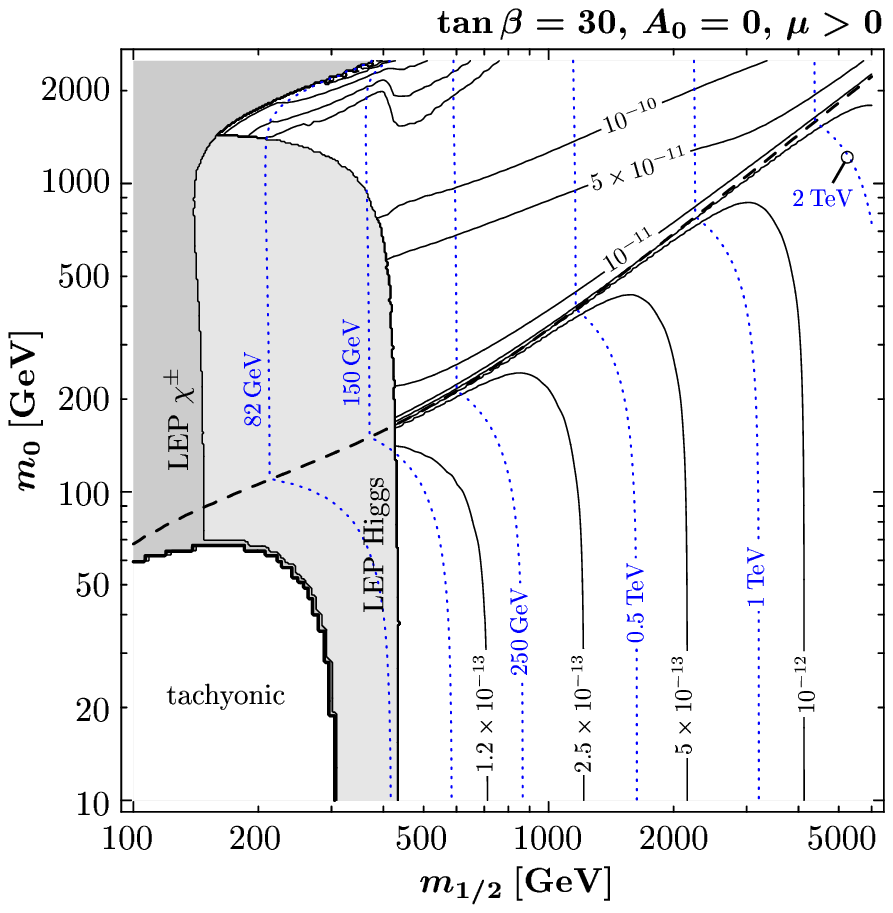} 
\caption[NLSP yield contours]{Contours of $Y_{\NLSP}(T_0)$
  (solid black lines) and $m_{\NLSP}$ (dotted blue lines) in the
  $(m_{1/2},m_0)$ plane for $A_0=0$, $\mu>0$, $\tan\beta=10$ (upper
  panel) and $\tan\beta=30$ (lower panel). Above (below) the dashed
  line, $m_{\neutralino}<m_{\stau}$ ($m_{\stau}<m_{\neutralino}$). The
  medium gray and the light gray regions at small $m_{1/2}$ show the
  mass bounds $m_{\chargino}>94~\GeV$ and $\mh>114.4~\GeV$ from
  chargino and Higgs searches at LEP~\cite{Yao:2006px}.}
\label{Fig:YNLSP}
\end{center}
\end{figure}

In Fig.~\ref{Fig:YNLSP} the solid (black) and dotted (blue) lines show
respectively contours of $Y_{\NLSP}(T_0)$ and $m_{\NLSP}$ in the
$(m_{1/2},m_0)$ plane for $A_0=0$, $\mu>0$, $\tan\beta=10$ (left
panel) and $\tan\beta=30$ (right panel). Above (below) the dashed
line, $m_{\neutralino}<m_{\stau}$ ($m_{\stau}<m_{\neutralino}$). The
medium gray and the light gray regions at small $m_{1/2}$ are excluded
respectively by the mass bounds $m_{\chargino}>94~\GeV$ and
$\mh>114.4~\GeV$ from chargino and Higgs searches at
LEP~\cite{Yao:2006px}. The leftmost dotted (blue) line indicates the
LEP bound $m_{\stau}>81.9~\GeV$~\cite{Yao:2006px}.
For $\tan\beta=30$, tachyonic sfermions occur in the low-energy
spectrum at points in the white corner labeled as ``tachyonic.''
For those scans we employ the computer program
\texttt{SuSpect~2.34}~\cite{Djouadi:2002ze} to calculate the
low-energy spectrum of the superparticles and the Higgs bosons.%
\footnote{In this section, we have used the following values:
  $m_{\mathrm{t}} = 172.5\ \GeV$,
  $m_{\mathrm{b}}(m_{\mathrm{b}})^{\mathrm{\overline{MS}}} = 4.23\
  \GeV$,
  $\alpha_{\mathrm{s}}^{\mathrm{\overline{MS}}}(m_\mathrm{Z})=0.1172$,
  and $\alpha_{\mathrm{em}}^{-1\mathrm{\overline{MS}}}(m_{\mathrm{Z}})
  = 127.90896$.}
Assuming a standard cosmological history, the yield $Y_{\NLSP}(T_0)$
is obtained from the $\Omega_{\NLSP}h^2$ values provided by the
computer program \texttt{micrOMEGAs~1.3.7}~\cite{Belanger:2001fz,
  Belanger:2004yn}.

The contours shown in Fig.~\ref{Fig:YNLSP} are independent of $\mgr$
and $\TR$. Therefore, they can be used to interpret the results shown
in the figures below. Note the sensitivity of both $Y_{\stau}(T_0)$
and $m_{\stau}$ on $\tan\beta$.  By going from $\tan\beta=10$ to
$\tan\beta=30$, $Y_{\stau}(T_0)$ decreases by about a factor of two at
points that are not in the vicinity of the dashed line, i.e., that are
outside of the $\stau$--$\neutralino$ coannihilation region. While
$m_{\stau}$ becomes somewhat smaller by increasing $\tan\beta$ to 30,
the $\tan\beta$ dependence of $m_{\neutralino}$ is negligible.

Let us now explore the parameter space in which the relic gravitino
density matches the observed dark matter density
$\OmegaDM^{3\sigma}$~(\ref{Eq:OmegaDM}),
\begin{align}
        0.075
        \leq 
        \Omega_{\gravitino}^{\TP}h^2+\Omega_{\gravitino}^{\NTP}h^2 
        \leq 
        0.126
        \ .
\label{Eq:OgravitinoConstraint}
\end{align} 
Now, $\TR$ and $\mgr$ appear in addition to the traditional CMSSM
parameters. We focus on $\mgr\gtrsim 1~\GeV$ since the soft SUSY
breaking parameters of the CMSSM are usually assumed to result from
gravity-mediated SUSY breaking. However, we do not restrict our study
to fixed relations between $\mgr$ and the soft SUSY breaking
parameters such as the ones suggested, for example, by the Polonyi
model.

\begin{figure*}[t]
\begin{center}
\includegraphics[width=0.5\textwidth]{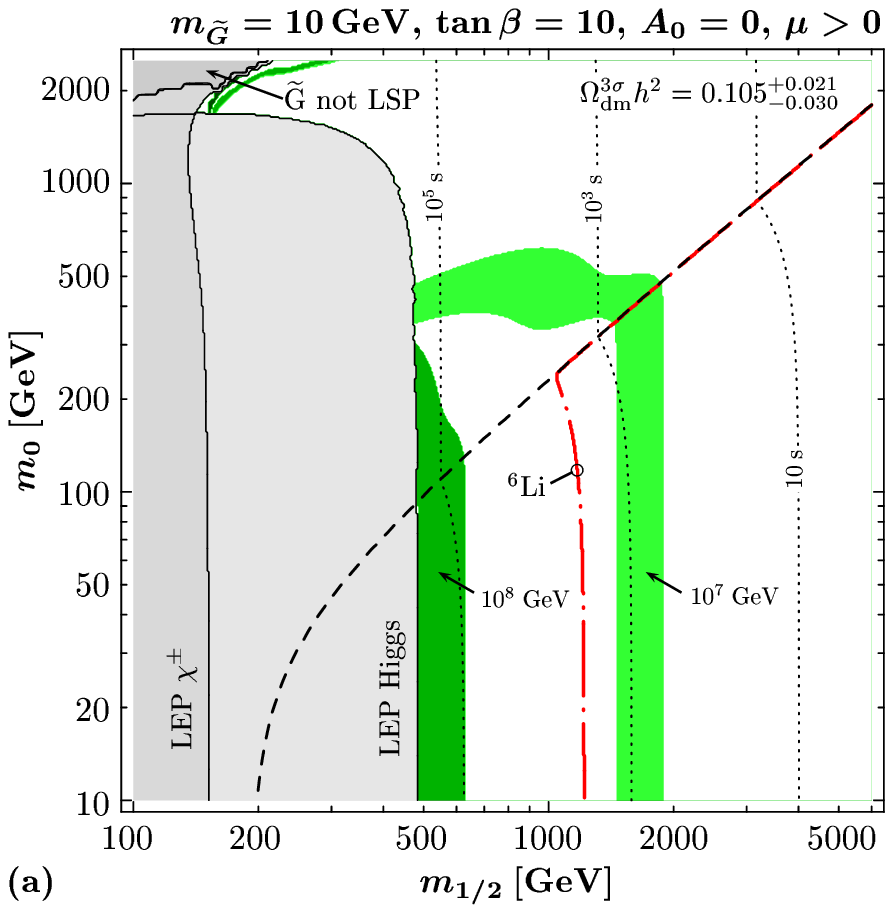}%
\includegraphics[width=0.5\textwidth]{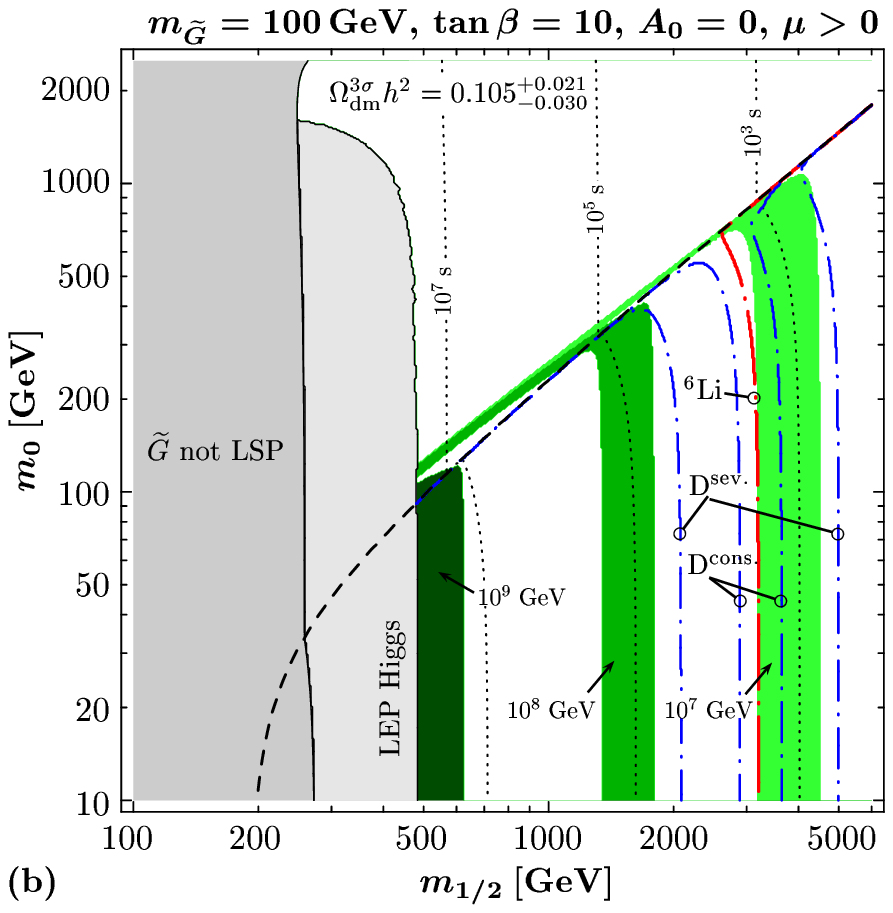}
\includegraphics[width=0.5\textwidth]{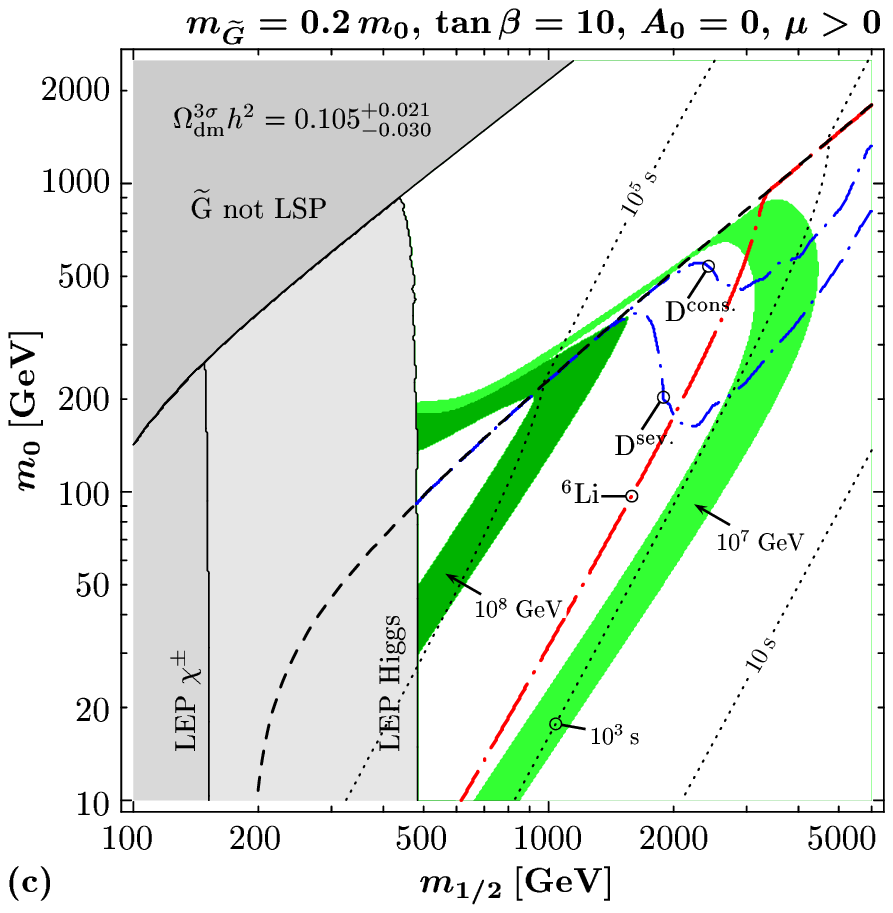}%
\includegraphics[width=0.5\textwidth]{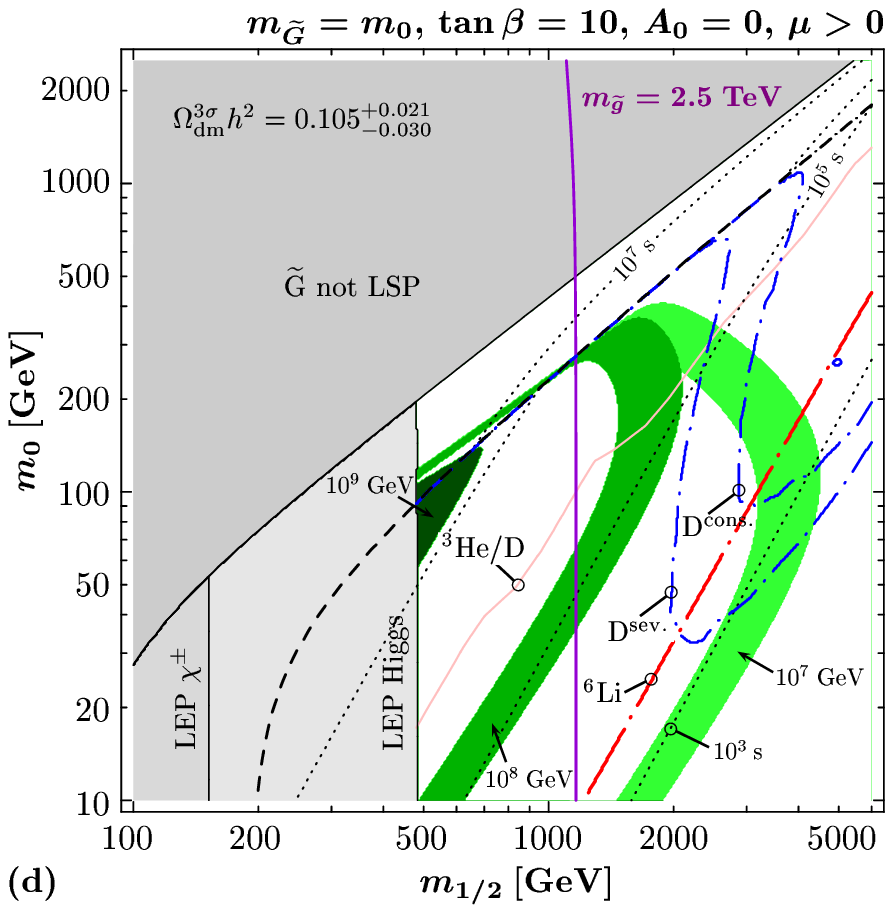}%
\caption[CMSSM planes for $\tan{\beta}=10$]{ The $(m_{1/2},m_0)$
  planes for $\tan\beta=10$, $A_0=0$, $\mu>0$, and the choices
  (a)~$\mgr=10~\GeV$, (b)~$\mgr=100~\GeV$, (c)~$\mgr=0.2\,m_0$, and
  (d)~$\mgr=m_0$.  In each panel, the light, medium, and dark shaded
  (green) bands indicate the regions in which
  $\Omega_{\gravitino}h^2\in \OmegaDM^{3\sigma}h^2$ for $\TR=10^7$,
  $10^8$, and $10^9~\GeV$, respectively.
  The medium gray and the light gray regions at small $m_{1/2}$ are
  excluded respectively by chargino and Higgs searches at LEP.  In the
  dark gray region, the gravitino is not the LSP. The dotted lines
  show contours of the NLSP lifetime. Below the dashed line,
  $m_{\stau}<m_{\neutralino}$. With the $\stau$ NLSP, the region to
  the left of the long-dash-dotted (red) line is
  cosmologically disfavored by bound-state effects on the primordial $^6$Li
  abundance~\cite{Pospelov:2006sc}. 
  The effects of late hadronic energy injection on the primordial D
  abundance~\cite{Steffen:2006hw} disfavor the $\stau$ NLSP region
  between the short-dash-dotted (blue) lines in
  panel~(b) and the one above the corresponding lines in panels~(c)
  and~(d). In addition, in (d) the electromagnetic $\het/\deut$
  constraint and the gluino mass contour $m_{\gluino}=2.5\ \TeV$ are
  shown as labeled.  }
\label{Fig:CMSSMtB10}
\end{center}
\end{figure*}
\afterpage{\clearpage}

In Fig.~\ref{Fig:CMSSMtB10} the light, medium, and dark shaded (green)
bands show the $(m_{1/2},m_0)$ regions that satisfy the upper
limit~(\ref{Eq:OgravitinoConstraint}) for $\TR=10^7$, $10^8$, and
$10^9~\GeV$, respectively, where $\tan\beta=10$, $A_0=0$, $\mu>0$. The
four panels are obtained for the choices (a)~$\mgr=10~\GeV$,
(b)~$\mgr=100~\GeV$, (c)~$\mgr=0.2\,m_0$, and (d)~$\mgr=m_0$. In the
dark-gray region, the gravitino is not the LSP. The regions excluded
by the chargino and Higgs mass bounds and the line indicating
$m_{\neutralino}=m_{\stau}$ are identical to the ones shown in the
upper panel of Fig.~\ref{Fig:YNLSP}.  The dotted lines show contours of
the NLSP lifetime~(\ref{Eq:SleptonLifetime}).  For the $\neutralino$
NLSP, we calculate $\tau_{\neutralino}$ from the expressions given in
Sec.~IIC of Ref.~\cite{Feng:2004mt}.

The $\tau_{\NLSP}$ contours in Fig.~\ref{Fig:CMSSMtB10} illustrate
that the NLSP decays during/after BBN. Successful BBN predictions
therefore imply cosmological constraints on $\mgr$, $m_{\NLSP}$, and
$Y_{\NLSP}$~\cite{Feng:2004mt,Roszkowski:2004jd,Cerdeno:2005eu,Steffen:2006hw}.
Indeed, as stressed before, it has been found that the considered
$\neutralino$ NLSP region is completely disfavored for $\mgr\gtrsim
100~\MeV$ by constraints from late electromagnetic and hadronic energy
injection~\cite{Feng:2004mt,Roszkowski:2004jd,Cerdeno:2005eu,Cyburt:2006uv}.
In the $\stauone$ NLSP region, the constraints from electromagnetic and
hadronic energy release are important but far less severe than in the
$\neutralino$ NLSP case.

Including the constraints from the bound-state effects this picture
changes. As we have already seen in the previous section, in most of
the $\stau$ NLSP parameter space, the bounds from the catalysis of
\lisx\ and \ben\ can be much more severe than the ones from late
energy injection.
We incorporate the $(\taustau,Y_{\stauone^-})$-dependent CBBN
constraint on \lisx\ from Fig.~\ref{fig:contours-Lisx} which is shown
in Fig.~\ref{Fig:CMSSMtB10} by the long dash-dotted (red) line. 
The \stauone-NLSP parameter space to the left of this line is
excluded.

We thereby update our figures presented in the published
work~\cite{Pradler:2006hh}. In those figures, the \lisx\ output was
taken from the work~\cite{Pospelov:2006sc} which provided the initial
estimate on the efficiency of the catalyzed production. Meanwhile, the
dedicated quantum-three-body calculation~\cite{Hamaguchi:2007mp}
became available which lead to a reduction of the $S$-factor of
(\ref{eq:CBBN-Li6-production}) by roughly one order in magnitude. As
discussed in Chapter~\ref{cha:catalyzed-bbn}, we have incorporated
this state-of-the-art result together with other improvements in our
Boltzmann network equation.

In the lifetime region $\taustau\lesssim \mathrm{few}\times 10^3\
\seconds$ which is unconstrained by CBBN only the constraint from
deuterium on the hadronic energy release becomes important.
Following the approach explained in
Sec.~\ref{sec:typical-stau-abundance}, we incorporate the constraint
on hadronic energy release for D.  In Fig.~\ref{Fig:CMSSMtB10} these
are shown by the short-dash-dotted (blue) lines. The D constraint
disfavors the region between the corresponding lines in panel~(b) and
the region above the corresponding lines in panels~(c) and~(d).  In
panel~(a) the D constraint does not appear. In addition, for
orientation, in panel~(d) we also include the electromagnetic
$\het/\deut$ constraint as a thin (pink) line and show the gluino mass
contour $m_{\gluino}= 2.5\ \TeV$ which is a near to vertical thick
(violet) line.

Indeed, one finds in each panel of Fig.~\ref{Fig:CMSSMtB10} that the
highest $\TR$ value allowed by the considered BBN constraints is about
$10^7~\GeV$. The bands obtained for $\TR\gtrsim 10^8~\GeV$ are located
completely within the region disfavored by the $^6$Li bound.  In
previous gravitino dark matter studies within the CMSSM
that did not take into account bound-state effects on the primordial
$^6$Li abundance,
much higher temperatures of up to about $10^9~\GeV$ were believed to
be allowed~\cite{Roszkowski:2004jd,Cerdeno:2005eu,Pradler:2006qh}.

\begin{figure}[t]
  \begin{center}
    \includegraphics[width=0.5\textwidth]
{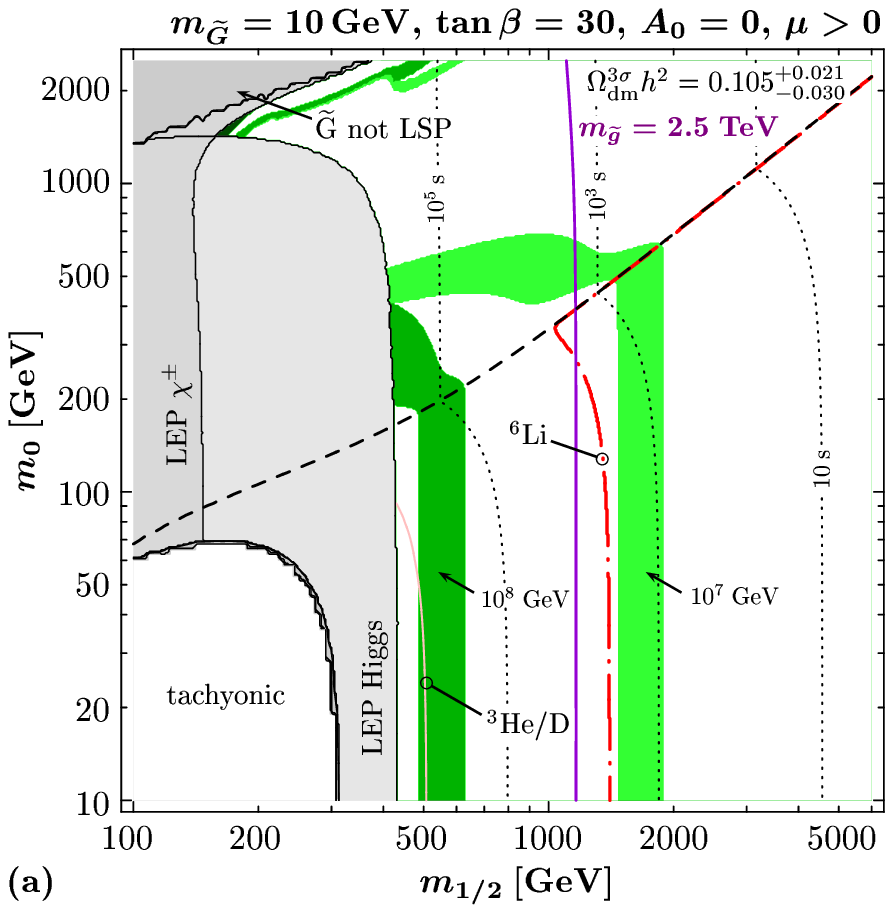}%
\includegraphics[width=0.5\textwidth]{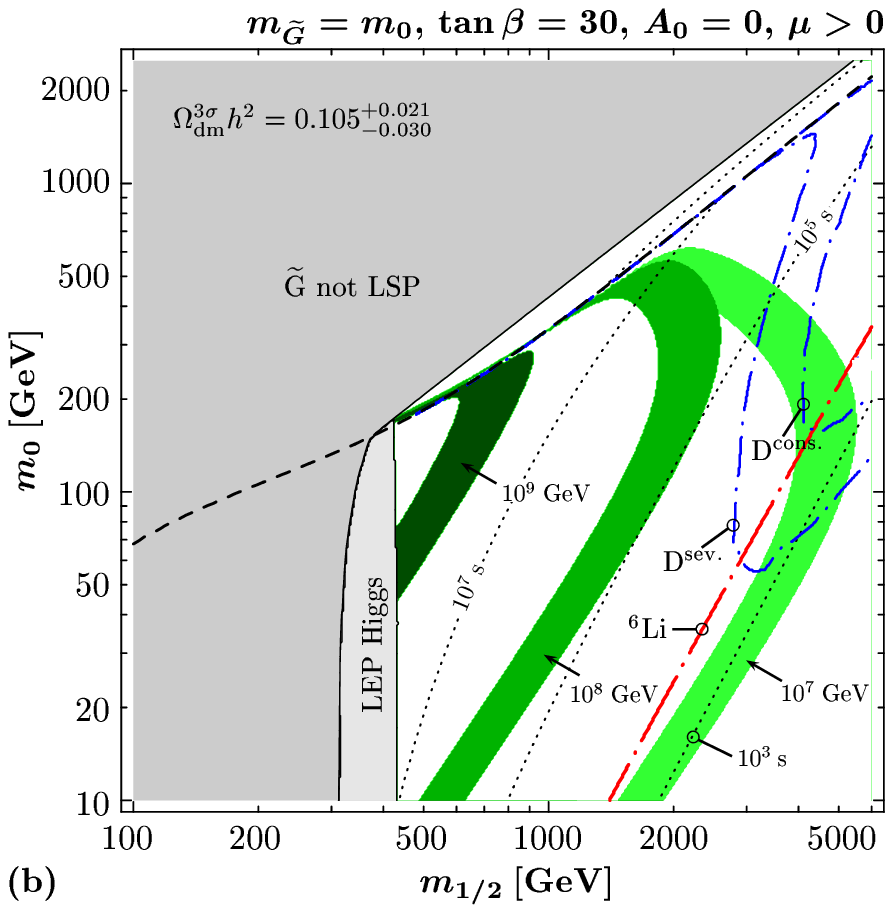} 
\caption[CMSSM planes for $\tan{\beta}=30$]{CMSSM planes as in
  Fig.~\ref{Fig:CMSSMtB10}, but for $\tan\beta=30$, $A_0=0$,
  $\mu>0$. In (a) $\mgr=10\ \GeV$ and in (b) $\mgr=\mzero$ has been
  chosen.}
\label{Fig:CMSSMtB30}
\end{center}
\end{figure}

The constraint $\TR\lesssim 10^7~\GeV$ remains if we consider larger
values of $\tan\beta$. This is demonstrated in
Fig.~\ref{Fig:CMSSMtB30} for $\tan\beta=30$, $A_0=0$, $\mu>0$, and~(a)
$\mgr=10\ \GeV$ and~(b)~$\mgr=m_0$.
The shadings (colors) and line styles are identical to the ones in
Fig.~\ref{Fig:CMSSMtB10}.

Let us comment on the dependence of the considered BBN constraints on
the assumed primordial abundances of D and $^6$Li. As can be seen in
Figs.~\ref{Fig:CMSSMtB10} and~\ref{Fig:CMSSMtB30}, the constraint from
late hadronic energy release is quite sensitive on the assumed
primordial D abundance. In contrast, even if we relax the restrictive
$^6$Li bound on $Y_{\NLSP}/2$ by two orders of magnitude, we still
find $\TR\lesssim 10^7~\GeV$. For example, the $^6$Li constraint
relaxed in this way would appear in Fig.~\ref{Fig:CMSSMtB10}~(b) as an
almost vertical line slightly above~$m_{1/2}=2.5~\TeV$.

\begin{description}
\item[Limit on \boldmath$\monetwo$] The limit (\ref{eq:LowerLimitm12})
  emerges since $\mstauone$ scales with $\monetwo$ [see
  Fig.~\ref{Fig:YNLSP}] and since $\taustau$ is fixed once $\mgr$ and
  $\mstauone$ are specified.  The choice $\mgr = 10 \ \GeV$ in
  Fig.~\ref{Fig:CMSSMtB10}~(a) and Fig.~\ref{Fig:CMSSMtB30}~(a) allows
  for an immediate comparison of the exemplary CMSSM scenarios with
  (\ref{eq:LowerLimitm12}). Only in the vicinity of the dashed line,
  i.e., in the $\stauone$--$\neutralino$ coannihilation region, the
  position of the $\Lisix$ constraint approaches its conservative
  lower limit on \monetwo.  This is because $\stauone$ becomes heavier
  for larger $\mzero$ which shortens $\taustau$ for fixed $\mgr$.
  Contrariwise, the splitting between the actual position of the
  $\Lisix$ constraint and (\ref{eq:LowerLimitm12}) is larger for
  smaller $\mzero$.
  This is slightly more pronounced in Fig.~\ref{Fig:CMSSMtB30}~(a)
  than in Fig.~\ref{Fig:CMSSMtB10}~(a) and results from the fact that
  the increase in $\tanb$ leads to a decrease in $\mstauone$ so that
  $\taustau$ becomes larger for fixed~$\mgr$.

  That the lower limit~(\ref{eq:LowerLimitm12}) can imply high values
  of the superparticle masses is illustrated by the vertical (violet)
  lines in Figs.~\ref{Fig:CMSSMtB10}~(d) and~\ref{Fig:CMSSMtB30}~(a)
  which show the gluino mass contour $m_{\widetilde{g}} = 2.5\
  \TeV$. In this regard, also note that the mass of the lighter stop
  is $ m_{\widetilde{t}_1} \simeq 0.7 m_{\widetilde{g}} $ in those
  $\stauone$ NLSP regions with $m_{\mathrm{h}}>114.4\ \GeV$.
  Since there the gaugino mass parameter sets the scale for the low
  energy superparticle spectrum, depending on $\mgr$, the
  bound~(\ref{eq:LowerLimitm12}) implies high values of the
  superparticle masses which can be associated with a mass range that
  will be difficult to probe at the LHC.

\item[Limit on \boldmath$\TR$] The limit (\ref{eq:UpperLimitTR})
  relies on thermal gravitino production only, $\Omegatp \sim
  \TR$. Thus the upper limit on $\TR$ becomes more stringent by taking
  $\Omegantp$ into account. In Figs.~\ref{Fig:CMSSMtB10}~(d) and
  \ref{Fig:CMSSMtB30}~(b) we have fixed $\mgr = \mzero$. Thereby, the
  non-ther\-mal production~(\ref{Eq:GravitinoDensityNTP}) becomes more
  important for larger values of $\mzero$.  In addition,
  $Y_{\stau}^{\mathrm{dec}}$ takes on its maximum at a given
  $\monetwo$ in the $\stau$--$\neutralino$ coannihilation region.
  This leads to the bending of the
  bands~(\ref{Eq:OgravitinoConstraint}) towards lower $\monetwo$.
  This indicates that~(\ref{eq:UpperLimitTR}) indeed seems to provide
  a good estimate.

  While the constraint $\TR\lesssim 10^7~\GeV$ is found for each of
  the considered $\mgr$ relations, one cannot use the $^6$Li bound to
  set bounds on $\mstauone$ without insights into $\mgr$. The $^6$Li
  (and similarly the \ben) bound disappears for
  $\tau_{\stau}\lesssim 10^3~\seconds$~\cite{Pospelov:2006sc}
  which is possible even for $\mstauone=\Order(100~\GeV)$ provided
  $\mgr$ is sufficiently small;
  see~(\ref{Eq:SleptonLifetime}). However, the constraints on $\TR$
  become more severe towards small $\mgr$ as is shown in
  Fig.~\ref{Fig:UpperLimitTR}. Thus, the constraint $\TR\lesssim
  10^7~\GeV$ cannot be evaded by lowering $\mgr$
  provided~$\TR<T_{\freezeout}^{\gravitino}$.
  An upper limit on $\TR$ of $10^7~\GeV$ can be problematic for
  inflation models and baryogenesis scenarios. This finding can thus
  be important for our understanding of the thermal history of the
  Universe.

\end{description}

\subsection{Late-time entropy production}
\label{sec:late-time-entropy}

The constraints shown above are applicable for a standard thermal
history during the radiation-dominated epoch.
Such a standard cosmological evolution may, e.g., be accomplished by
considering only a minimal framework such as the MSSM.%
\footnote{Even inflation seems to be feasible within the MSSM with the
  inflaton being a gauge invariant combination of squark and slepton
  fields~\cite{Allahverdi:2006iq}. 
  Strictly speaking, however, neutrinos do not obtain masses within
  this framework which are needed to explain the observed neutrino
  oscillations~\cite{GonzalezGarcia:2007ib}.}

On the other hand, focusing on a gravitino LSP we also explicitly
consider the gravity sector. For example, it is well known that
supergravity and string theories generically suffer from the
appearance of scalar fields which can give rise to ``cosmological
moduli problems''~\cite{Coughlan:1983ci,Banks:1993en,deCarlos:1993jw}.
Typically, the interaction of such an exotic (eventually) massive
field~$\phi$ to the MSSM sector is suppressed by a high-energy scale
such as $M_\GUT$ or $\MP$. The problem arises because $\phi$ easily
drops out (or never has reached) thermal equilibrium.

For a massive particle species which is in equilibrium, the energy
density $\rho_\phi$ becomes exponentially suppressed once the
temperature drops below its mass. However, if $\phi$ is frozen-out,
$\rho_{\phi}/\rho_{\mathrm{rad}}$ starts to scale as $T^{-1}$ so that,
eventually, the energy density in $\phi$ dominates over the one in
radiation. Thus, if $\phi$ lives sufficiently long, it is possible
that a substantial amount of entropy is released in its
out-of-equilibrium decay.%
\footnote{ Other entropy production events after inflation are, e.g.,
  Q-ball decays; cf.~\cite{Kasuya:2007cy} and references therein.
  Gravitino dark matter scenarios with late-time entropy production
  have been considered previously for gauge-mediated SUSY breaking
  where~$\TR>T^{\gravitino}_{\freezeout}$~\cite{Baltz:2001rq,Fujii:2002fv,Fujii:2003iw,Jedamzik:2005ir,Lemoine:2005hu}.
}
Since an entropy release tends to erase any pre-existing quantity
which itself has originated from a non-equilibrium process---such as
$\etabaryon$---this can be problematic. Moreover, additional unwanted
relics can be produced in the decay of $\phi$ leading, e.g., to too
much dark matter. However, as we shall see in the following, entropy
production can also have a positive effect~\cite{Buchmuller:2006tt}.

The change $d(a^3s)\equiv dS = dQ/T$ in entropy per comoving volume is
found from the ``heat added'' in the decay, $dQ = a^3 \Gamma_{\phi}
\rho_{\phi} dt$~\cite{Scherrer:1984fd}. The associated evolution of
the entropy per comoving volume is thus described by
\begin{equation}
        \frac{dS}{dt}
        =\frac{\Gamma_{\phi}\rho_{\phi} a^3}{T}
        =\left(\frac{2\pi^2}{45}\,\geff\right)^{1/3} 
        \Gamma_{\phi}\rho_{\phi} a^4 S^{-1/3} .
\label{Eq:EntropyEvolution}
\end{equation}
We have already indicated by the chosen notation that the situation is
somewhat similar to that at the end of inflation. Thus, we can solve
(\ref{Eq:EntropyEvolution}) by simultaneously considering the
Boltzmann equation for $\rho_{\phi}$~in the form~(\ref{Eq:BEqPhi})
together with the Friedmann equation which governs the evolution of
the scale factor $a$ of the Universe; $\Gamma_{\phi}$ denotes the
decay width of $\phi$.

From the discussion of Sec.~\ref{sec:reheating-phase} we know that the
temperature after the decay can be expressed in terms
of~$\Gamma_{\phi}$,
\begin{equation}
  T_{\mathrm{after}} 
  \equiv
  \left[\frac{10}{\geff(T_{\mathrm{after}})\pi^2}\right]^{1/4}
  \sqrt{\Gamma_{\phi}\,\MPl}
  \ ,
\label{Eq:Tafter_definition}
\end{equation}
which satisfies $\Gamma_{\phi}=3 H(T_{\mathrm{after}})$.
Note that primordial nucleosynthesis imposes a lower limit on this
temperature.  It mainly emerges from the fact that a re-thermalization
of all the three neutrino species after the reheating process does not
happen instantaneously. Insufficient thermalization affects the Hubble
rate, thereby the $\neutron/\proton$ freeze-out value, and thus the
\hef\ output in BBN [see Sec.~\ref{sec:typic-champ-abund}]. The bounds
derived
in~\cite{Kawasaki:1999na,Kawasaki:2000en,Hannestad:2004px,Ichikawa:2005vw}
are in the range
\begin{equation}
        T_{\mathrm{after}}\gtrsim 0.7\!-\!4~\MeV
        \ .
\label{Eq:Tafter_limit}
\end{equation}

\begin{figure}[t]
\begin{center}
\includegraphics[width=0.65\textwidth]{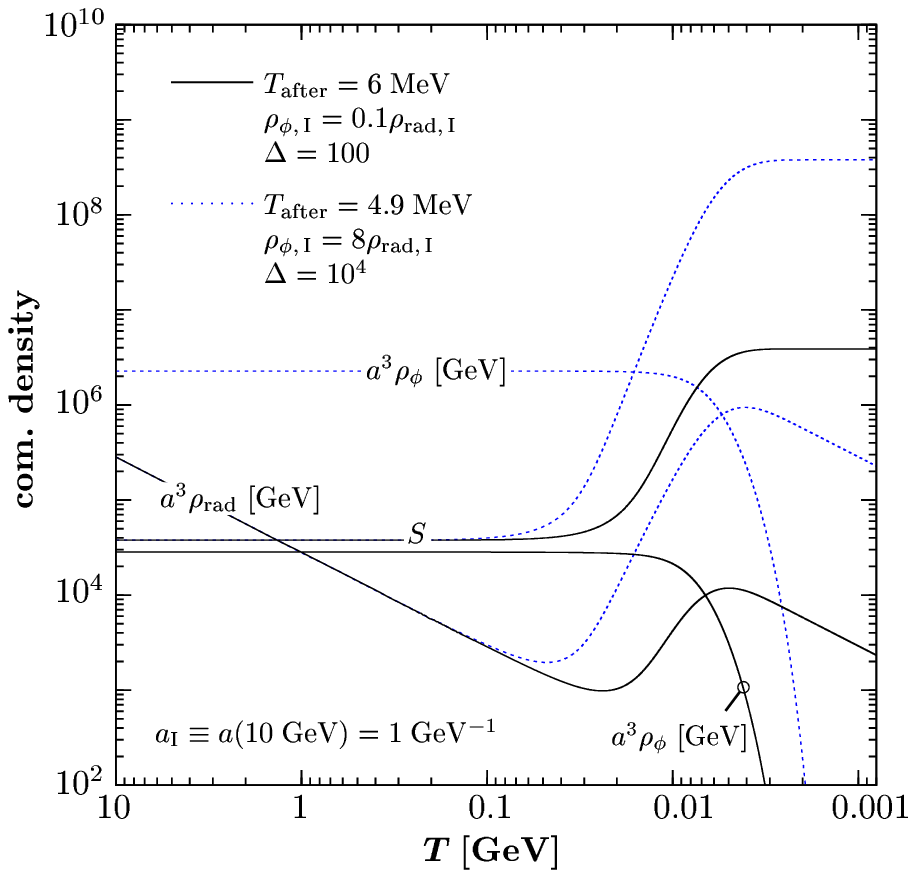}
\includegraphics[width=0.65\textwidth]{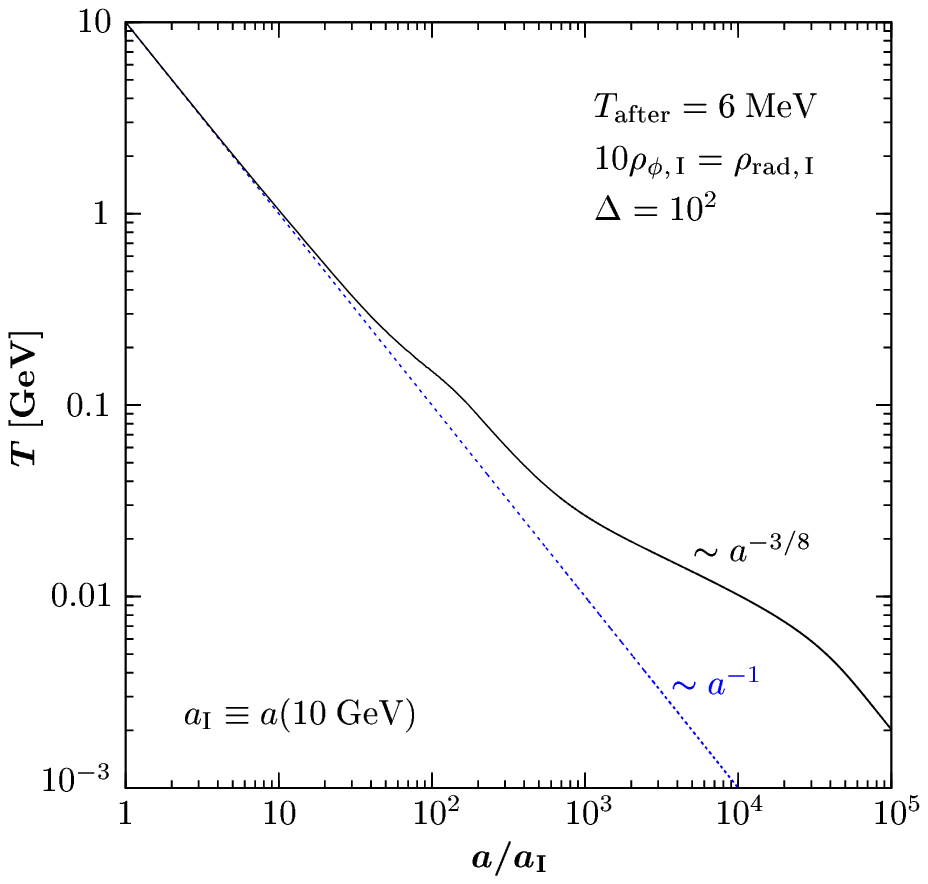}
\caption[Late time entropy production]{\textit{Top}: Evolution of
  $S$, $a^3\rho_{\mathrm{\phi}}$, and $a^3\rho_{\rad}$ as a function of
  $T$ for the normalization $a_{\mathrm{I}}\equiv
  a(10~\GeV)=1~\GeV^{-1}$.  The solid lines are obtained for
  $\rho_{\phi}(10~\GeV)=0.1\,\rho_{\rad}(10~\GeV)$ and
  $T_{\mathrm{after}}=6~\MeV$, the dotted (blue) lines for
  $\rho_{\phi}(10~\GeV)=8\,\rho_{\rad}(10~\GeV)$ and
  $T_{\mathrm{after}}=4.9~\MeV$. \textit{Bottom}: Evolution of the
  temperature for the scenario which is depicted by the solid lines in
  the left panel.}
\label{Fig:EntropyProduction}
\end{center}
\end{figure}

In the upper panel of Fig.~\ref{Fig:EntropyProduction} we show the
evolution of $S$, $a^3\rho_{\phi}$, and $a^3\rho_{\rad}$ for two
exemplary scenarios respecting~(\ref{Eq:Tafter_limit}). The scale
factor $a$ is normalized by%
\footnote{Giving $a$ the dimension of a length makes the radial
  coordinate in the Friedmann-Robertson-Walker metric dimensionless.}
$a_{\mathrm{I}}\equiv a(10~\GeV)=1~\GeV^{-1}$ and the temperature
dependence of $\heff$ is taken into account as determined
in~\cite{Gondolo:1990dk}.
For $\rho_{\phi}(10~\GeV)=0.1\,\rho_{\rad}(10~\GeV)$ and
$T_{\mathrm{after}}=6~\MeV$, $S$ increases by a factor of $\Delta=100$
as shown by the corresponding solid line.
For $\rho_{\phi}(10~\GeV)=8\,\rho_{\rad}(10~\GeV)$ and
$T_{\mathrm{after}}=4.9~\MeV$, $S$ increases by a factor of
$\Delta=10^4$ as shown by the corresponding dotted (blue) line.

The lower panel of Fig.~\ref{Fig:EntropyProduction} shows the
associated evolution of the photon temperature for the case
$\Delta=100$. As long as the comoving entropy is conserved, $T\propto
\heff^{-1/3} a^{-1}$. This causes $T$ to decrease slightly less slowly
than $a^{-1}$ during the quark-hadron transition at $T\sim 200\ \MeV$
which we have assumed to be adiabatic~\cite{Olive:1980wz}.  As already
encountered in Sec.~\ref{sec:reheating-phase}, the temperature scales
with $a^{-3/8}$ during entropy production, i.e., during reheating, the
temperature does not increase but rather drops less
slowly~\cite{Scherrer:1984fd}. Below $ T_{\mathrm{after}}$ the
Universe again expands as a radiation-dominated one.
\afterpage{\clearpage}

In the following, we restrict our study of entropy production at late
times, $T_{\mathrm{before}}\simeq\TL \ll \TR$, so that the thermal
production of gravitinos is not affected. We assume that the
production of gravitinos and NLSPs in the entropy producing event,
such as the direct production in decays of $\phi$, is negligible.%
\footnote{Note however, if kinematically allowed, a $\phi$ field will
  also typically decay into SUSY
  particles~\cite{Endo:2006zj,Nakamura:2006uc,Endo:2006tf}. Thus, the
  constraints discussed below shall therefore be considered as
  conservative/optimistic bounds.}
Moreover, in this section, we focus on scenarios in which the
decoupling of the NLSP is not or at most marginally affected by
entropy production, i.e., either $\TR \gg T_{\mathrm{after}}\gg
T_{\freezeout}^{\NLSP}$ or $\rho_{\rad}\gg\rho_{\phi}$ for $T\gtrsim
T_{\freezeout}^{\NLSP}$. Note that the latter condition excludes the
event shown in Fig.~\ref{Fig:EntropyProduction} with $\Delta=10^4$;
see, however, Sec.~\ref{sec:viab-therm-lept}.
Thus, the thermally produced gravitino yield and---in the case of
entropy production after NLSP decoupling---also the non-thermally
produced gravitino yield are diluted:
\begin{equation}
  Y_{\gravitino}(T_{\mathrm{after}}) 
  =  
  \frac{S({\TL})}{S(T_{\mathrm{after}})}\,
  Y_{\gravitino}(\TL)
  \ .
\end{equation}

In the case of late-time entropy production \textit{before} the
decoupling of the NLSP, 
we parameterize this by writing
\begin{equation}
Y_{\gravitino}^{\TP}(T_0) 
= 
\frac{1}{\delta}\,
Y_{\gravitino}^{\TP}(\TL)
\ .
\end{equation}
In this case, $Y_{\NLSP}(T_0)$ and thereby
$\Omega_{\widetilde{G}}^{\NTP}$ and the BBN constraints remain
unaffected.

Conversely, in the case of late-time entropy production \textit{after}
the decoupling of the NLSP (and before BBN)
both, $Y_{\gravitino}^{\TP}(T_0)$ and $Y_{\NLSP}(T_0)$, are reduced:
\begin{eqnarray}
 Y_{\gravitino}^{\TP}(T_0) 
 &=&  
 \frac{1}{\Delta}\,Y_{\gravitino}^{\TP}(\TL) ,
 \nonumber \\
 Y_{\NLSP}(T_0) 
 &=& 
 \frac{1}{\Delta} Y_{\NLSP}(\TL) .
\end{eqnarray}
Accordingly, $\Omega_{\widetilde{G}}^{\TP}$ and
$\Omega_{\widetilde{G}}^{\NTP}$ become smaller and the BBN constraints
can be relaxed. 

\begin{figure*}[t]
\begin{center}
\includegraphics[width=0.49\textwidth]{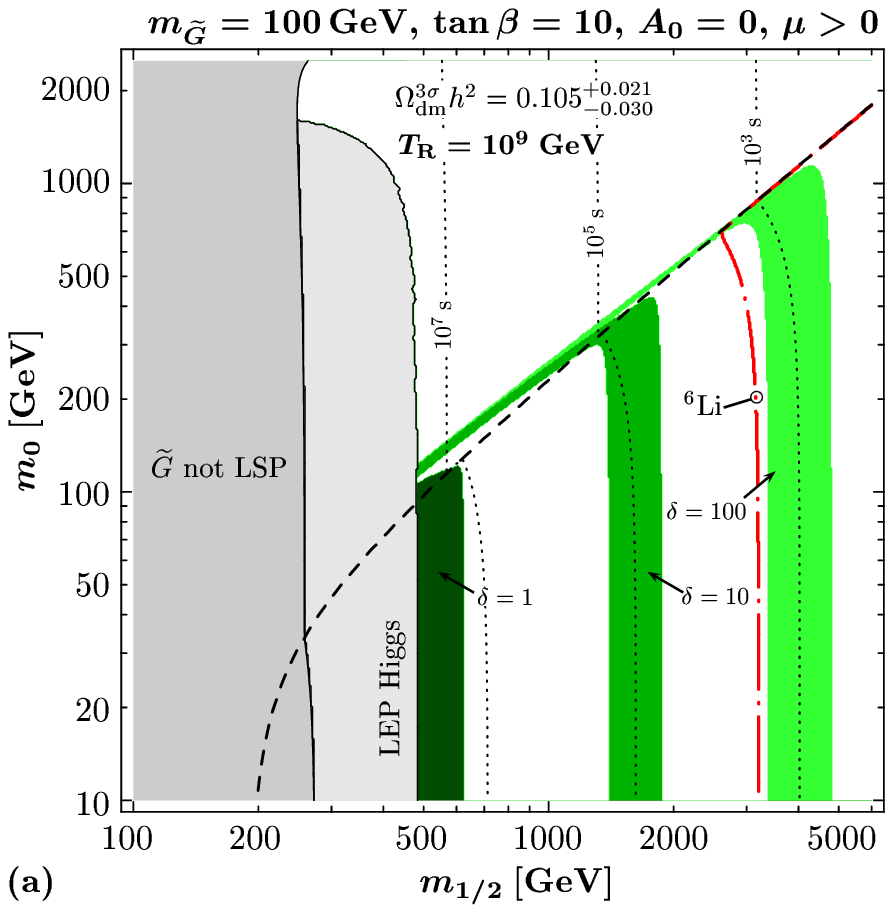} 
\includegraphics[width=0.49\textwidth]{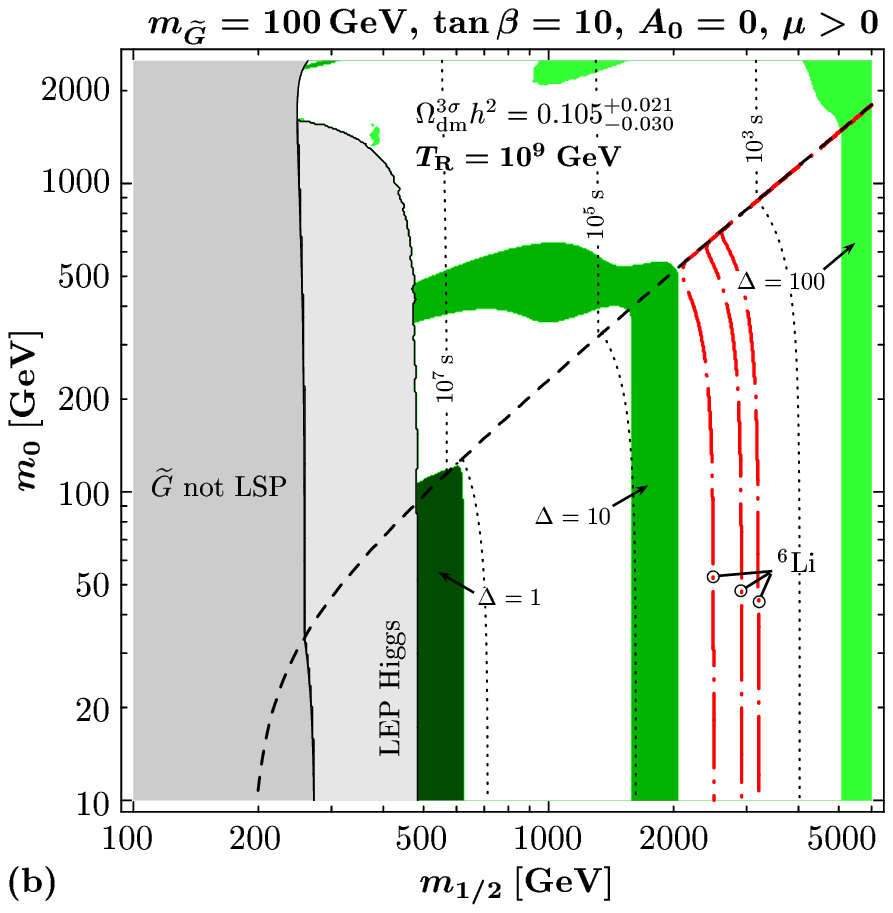} 
\includegraphics[width=0.49\textwidth]{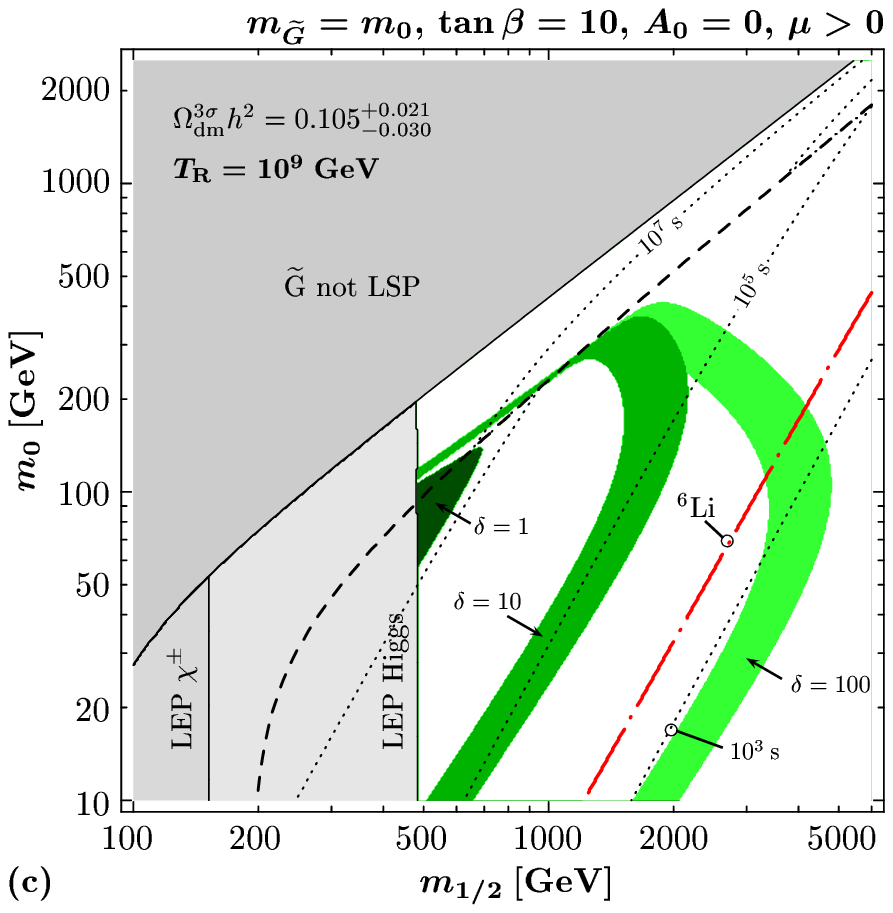} 
\includegraphics[width=0.49\textwidth]{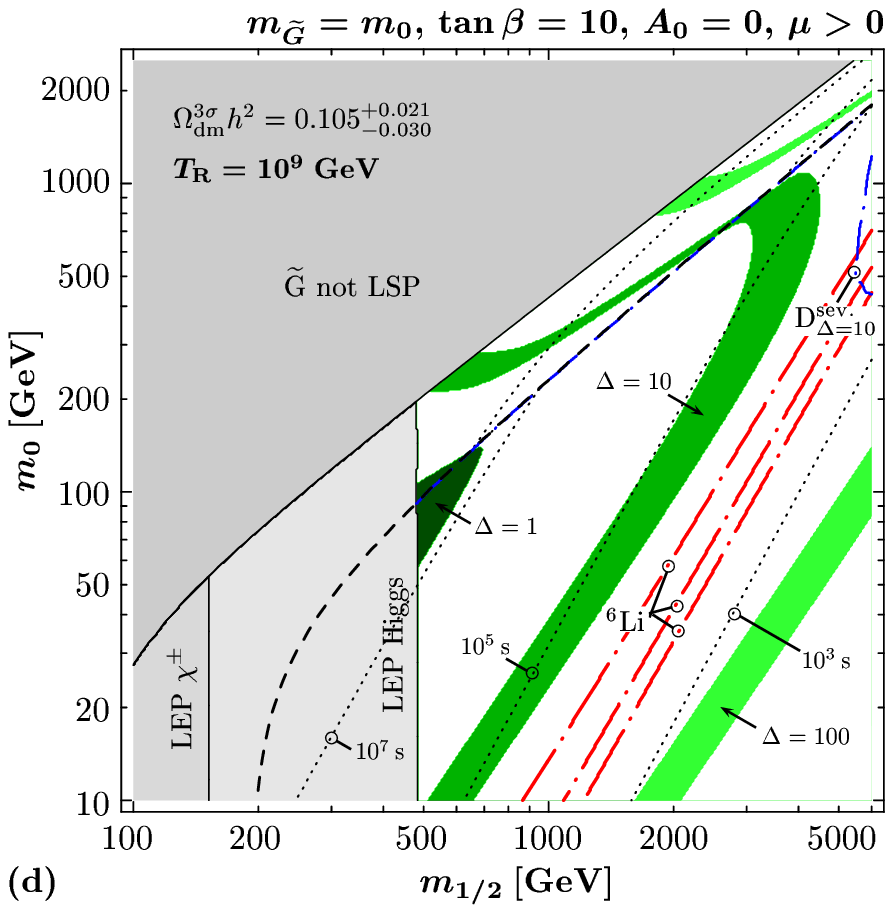} 
\caption[Entropy production before and after NLSP decoupling]{The
  effect of late-time entropy production before (left) and after
  (right) NLSP decoupling on regions in which
  $0.075\leq\Omega_{\gravitino}h^2\leq 0.126$ for $\TR=10^9~\GeV$.
  The $(m_{1/2},m_0)$ plane is shown for $\tan\beta=10$, $A_0=0$,
  $\mu>0$, $\mgr=100~\GeV$ (upper panels) and $\mgr=m_0$ (lower
  panels). The dark shaded (dark green) region is obtained without
  late-time entropy production $\delta=\Delta=1$. The medium and light
  shaded (medium and light green) bands are obtained with a dilution
  of $\Omega_{\gravitino}^{\TP}$
  ($\Omega_{\gravitino}^{\TP}+\Omega_{\gravitino}^{\NTP}$) by
  $\delta=10$ ($\Delta=10$) and $\delta=100$ ($\Delta=100$),
  respectively. The $\stau$ NLSP region to the right of the dot-dashed
  (red) line is cosmologically disfavored by the primordial $^6$Li
  abundance. Other curves and regions are identical to the ones in the
  corresponding panels of Fig.~\ref{Fig:CMSSMtB10}.  The severe D
  constraint for $\Delta=10$ appears only in panel~(d).}
\label{Fig:CMSSMEntropyProductionI}
\end{center}
\end{figure*}

In Fig.~\ref{Fig:CMSSMEntropyProductionI}
we show how late-time entropy production before (left) and after
(right) NLSP decoupling affects the $^6$Li constraint and the region
in which $0.075\leq\Omega_{\gravitino}h^2\leq 0.126$ for
$\TR=10^9~\GeV$.  The $(m_{1/2},m_0)$ planes are considered for
$\tan\beta=10$, $A_0=0$, $\mu>0$, $\mgr=100~\GeV$ (upper panels) and
$\mgr=m_0$ (lower panels).  The dark shaded (dark green) region is
obtained without late time entropy production, $\delta=\Delta=1$. The
medium and light shaded (medium and light green) bands are obtained
with a dilution of $\Omega_{\gravitino}^{\TP}$
($\Omega_{\gravitino}^{\TP}+\Omega_{\gravitino}^{\NTP}$) by
$\delta=10$ ($\Delta=10$) and $\delta=100$ ($\Delta=100$),
respectively. The dot-dashed (red) line illustrates
that the $^6$Li bound is independent of $\delta$, as shown in the
panels on the left-hand side, and becomes weaker (i.e., moves to the
left) with increasing $\Delta$, as shown in the panels on the
right-hand side. Other curves and regions are identical to the ones in
the corresponding panels of Fig.~\ref{Fig:CMSSMtB10}.
Note that we do not show the D constraint on late hadronic energy
injection since it is not sensitive to $\delta$ and vanishes already
for $\Delta=10$; an exception is the severe D constraint which still
appears for $\Delta=10$ in panel~(a).
BBN constraints on $\neutralino$ NLSP scenarios with entropy
production after NLSP decoupling will be studied elsewhere.

\afterpage{\clearpage} 

Comparing panels~(b) and~(d) of Fig.~\ref{Fig:CMSSMtB10} with
panels~(a) and~(c) in Fig.~\ref{Fig:CMSSMEntropyProductionI}, we find
that a dilution factor of $\delta=10$ ($100$) relaxes the $\TR$ bound
by a factor of 10 (100). Since the BBN constraints are unaffected by
$\delta$, the cosmologically disfavored range of NLSP masses cannot be
relaxed. With the dilution after NLSP decoupling, the relaxation of
the $\TR$ constraints is more pronounced.  Here also the
cosmologically disfavored range of NLSP masses can be
relaxed~\cite{Buchmuller:2006tt}. However, as can be seen in
panels~(b) and~(d) of Fig.~\ref{Fig:CMSSMEntropyProductionI}, the
$^6$Li bound is persistent. With a dilution factor of $\Delta=100$,
large regions of the $(m_{1/2},m_0)$ plane remain cosmologically
disfavored. For even larger factors of $\Delta$, however, the $^6$Li
bound can be evaded as will be shown explicitly below.

Figure~\ref{Fig:CMSSMEntropyProductionI} shows that inflation models
predicting, for example, $\TR=10^9~\GeV$ become allowed in the CMSSM
with gravitino dark matter for $\delta=\Delta\approx 100$. Here it is
not necessary to have late-time entropy production in the somewhat
narrow window between NLSP decoupling and BBN. This is different for
the viability of thermal leptogenesis in the considered scenarios
where $T^{\gravitino}_{\freezeout}>\TR$.

\subsection{Viability of thermal leptogenesis}
\label{sec:viab-therm-lept}

The constraint $\TR\lesssim 10^7~\GeV$ obtained in the considered
CMSSM scenarios for a standard cosmological history strongly disfavors
thermal leptogenesis.
However, if entropy is released after NLSP decoupling, a dilution
factor of $\Delta \gtrsim 10^3$ can render thermal leptogenesis viable
for $\TR \gtrsim 10^{12}~\GeV$.

Thermal leptogenesis---in its simplest form, with hierarchical heavy
right-handed Majorana neutrinos ---usually requires $\TR\gtrsim
10^9~\GeV$~\cite{Davidson:2002qv,Buchmuller:2004nz,Blanchet:2006be,Antusch:2006gy}.%
\footnote{Right-handed (heavy) neutrinos (and their
  superpartners) are not part of the (C)MSSM. Again, this section
  contains an updated discussion of the published
  work~\cite{Pradler:2006hh}; see Sec.~\ref{sec:exempl-param-scans}.}
However, late-time entropy production dilutes the baryon asymmetry
which is generated well before NLSP decoupling,
\begin{equation}
\etabaryon(T_{\mathrm{after}})
= 
\frac{1}{\Delta}\,
\etabaryon(T_{\mathrm{before}})
\ .
\end{equation}
Therefore, the baryon asymmetry before entropy production must be
larger by a factor of $\Delta$ in order to compensate for the
dilution. For $\Delta\sim 10^3$, this can be achieved in the case of
hierarchical neutrinos for
$M_{\mathrm{R}1}\sim\TR\gtrsim 10^{12}~\GeV$,
as can be seen in Fig.~7~(a) of Ref.~\cite{Buchmuller:2002rq} and in
Fig.~2 of Ref.~\cite{Buchmuller:2002jk}. Here, $M_{\mathrm{R}1}$ is the
mass of the lightest among the heavy right-handed Majorana neutrinos.

That it is indeed possible to produce such a large amount of entropy
in the narrow time window between NLSP decoupling an BBN, we show in
Fig.~\ref{Fig:EntropyProduction} a scenario in which a dilution factor
of even $\Delta=10^4$ is generated in the out-of-equilibrium decay of
a heavy particle $\phi$ [dotted (blue) lines].  However, because of
$\rho_{\phi}(10~\GeV)=8\,\rho_{\rad}(10~\GeV)$, the Hubble rate is
already enhanced already during the decoupling phase of the NLSP. This
leads to an increase of $T_{\freezeout}^{\NLSP}$ and
$Y_{\NLSP}(T_{\freezeout}^{\NLSP})$.
In the results shown below, we indeed account for this by using a
modified version of the \texttt{micrOMEGAs} code. Thereby, the
$Y_{\NLSP}$ contours shown in Fig.~\ref{Fig:YNLSP} do not apply in
this section.  After entropy production, the net effect is still a
significant reduction of $Y_{\NLSP}(T_0)$---provided that $\phi$ does
not decay into SUSY particles.

\begin{figure}[t]
\begin{center}
\includegraphics[width=0.49\textwidth]{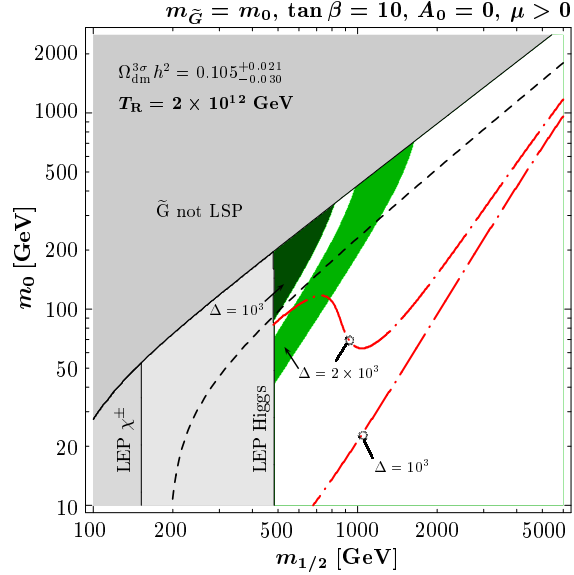} 
\caption[Entropy production and the viability of leptogenesis]{The
  effect of entropy production after NLSP decoupling for $\TR=2\times
  10^{12}~\GeV$ and $\Delta\geq 10^3$ in the $(m_{1/2},m_0)$ plane for
  $\tan\beta=10$, $A_0=0$, $\mu>0$, and $\mgr=m_0$.  The shaded
  (green) bands show the region in which
  $0.075\leq\Omega_{\gravitino}h^2\leq 0.126$ for $\Delta=10^3$ (dark)
  and $2\times 10^3$ (medium).  The dot-dashed (red) lines illustrate
  the corresponding evolution of the $^6$Li bound with the regions
  below being cosmologically allowed.}
\label{Fig:Leptogenesis}
\end{center}
\end{figure}
In Fig.~\ref{Fig:Leptogenesis} we consider now two concrete scenarios
with $\rho_{\phi}(10~\GeV)=8\,\rho_{\rad}(10~\GeV)$ as initial
condition and with $\Delta = 10^3$ and $ 2\times 10^3$ corresponding
to the respective reheating temperatures $T_{\mathrm{after}} = 48\
\MeV$ and $24\ \MeV $. We choose the $(\monetwo,\mzero)$-plane with
$\tan\beta=10$, $A_0=0$, $\mu>0$, and $\mgr=m_0$.  Here the shaded
(green) bands indicate the region in which
$0.075\leq\Omega_{\gravitino}h^2\leq 0.126$ for $\TR=2\times
10^{12}~\GeV$ and $\Delta=10^3$ (dark) and $2\times 10^3$ (medium). In
addition, the corresponding evolution of the $^6$Li bound is shown by
the dot-dashed (red) lines.  The regions in the \stauone-NLSP region
above the curves remain cosmologically disfavored.  The gray regions
are identical to the ones in Fig.~\ref{Fig:CMSSMtB30}.

We find that the $^6$Li bound can indeed be evaded for $\Delta=2\times
10^3$. Gravitino dark matter scenarios with successful thermal
leptogenesis in the $\stauone$ NLSP region are located on the
light-shaded (light green) band. 
As can be seen in Fig.~\ref{Fig:Leptogenesis}, the $\stauone$ NLSP region
with
$m_{1/2}\lesssim 700~\GeV$
where $m_{\stauone}\lesssim 250~\GeV$ (cf.~Fig.~\ref{Fig:YNLSP}), is no
longer disfavored by the $^6$Li bound provided $\Delta\gtrsim 10^3$.
Such scenarios are particularly promising since the long-lived $\stauone$
NLSP could provide striking signatures of gravitino dark matter at
future
colliders~\cite{Buchmuller:2004rq,Brandenburg:2005he,Steffen:2005cn,Martyn:2006as,Hamaguchi:2006vu}.

Finally, let us remark that the exact value of the actual
amount of entropy required is model dependent.
Though we have accounted for the fact that the presence of the energy
density $\rho_{\phi}$ during NLSP decoupling can affect $\Ydec$, we
have neglected a possible branching ratio of the $\phi$ decays into
$\stauone$ and/or gravitinos.
Moreover, such a scenario is---of course---fine-tuned since entropy
has to be released in a very narrow time window. However, with our
improved treatment of the CBBN yield of \lisx\ it becomes slightly
easier (when compared to \cite{Pradler:2006hh}) to circumvent the
stringent \lisx\ bound.%
\footnote{The value is now in the region provided by the more recent
  Ref.~\cite{Hamaguchi:2007mp} in which also our employed CBBN rate
  for \lisx\ production was obtained. The authors, however, only use
  the Saha approximation for the $(\hef{\stauone^-})$ bound state
  abundance; see Sec.~\ref{sec:recomb-phot-rates}.}
This is indicated by the fact that we were able to allow for an
increased NLSP decoupling yield by choosing
$\rho_{\phi}(10~\GeV)=8\,\rho_{\rad}(10~\GeV)$ which lead to the
generous choice of $T_{\mathrm{after}} >20\ \MeV$.
Since we have not provided explicitly a model for $\phi$ we understand the
results presented in this section as a proof-of-concept study that a
non-standard cosmological evolution can evade even the most stringent
of all BBN constraints while still allowing for successful
leptogenesis.


 \cleardoublepage
 \part{The long-lived stau as a thermal relic}
 \label{part:three}
 \chapter{Thermal relic stau abundances}
\label{cha:therm-relic-abund}

In Part~\ref{part:bbn} of this thesis we worked out the 
effects which 
a long-lived electrically charged massive particle species $\X^{\pm}$
has on the predictions of primordial nucleosynthesis.
In Part~\ref{part:two} we then considered concrete realizations of
$\X^{\pm}$ in gravitino dark matter scenarios in which
$\stauone$ is the lightest Standard Model superpartner. There, a
central parameter in the investigation is the abundance of $\stauone$
during/after BBN.
For a standard thermal history it is determined by $\taustau$ and by
its thermal freeze-out value which we shall call $Y_{\stau}\equiv
Y_{\stauone}^{\mathrm{dec}}$ in the following; recall that
$Y_{\stauone}^{\mathrm{dec}} = (n^{\mathrm{dec}}_{\stauone} +
n^{\mathrm{dec}}_{\stauone^*})/s$ is the total stau number density
normalized to the entropy density prior to its decay .

In particular, we saw that $\Ystau$ (i) governs the non-thermally
produced relic density~(\ref{Eq:GravitinoDensityNTP}) of gravitino
dark matter that originates from $\stauone$ decays, (ii) controls the
hadronic and electromagnetic BBN constraints by quantifying the
\textit{total} energy density $\mstauone s \Ystau$ which is
eventually be released in the decay and (iii) parameterizes the
abundance $\Ystau / 2 $ of recombination partners with the light
elements leading to the catalysis of BBN.

Importantly, the points (i)-(iii) have in common that they give rise
to upper limits on~$\Ystau$: (i)~For example, for
$m_{\gravitino}=50~\GeV$ and a thermally produced gravitino density
$\Omega_{\gravitino}^{\TP}=0.99\,\Omega_{\mathrm{dm}}$
($0.9\,\Omega_{\mathrm{dm}}$), one finds
from~(\ref{Eq:OgravitinoConstraint}) $\Ystau<10^{-13}$~ ($10^{-12}$);
see also Fig.~13 of~\cite{Steffen:2006hw}. (ii)~The BBN constraints on
hadronic and electromagnetic energy release can be as restrictive as
$\Ystau<10^{-14}$ ($10^{-15}$); cf.\ Fig.~12 of~\cite{Steffen:2006hw}
and Figs.~14 and~15 of~\cite{Kawasaki:2008qe}. (iii)~Catalyzed
production of $^9$Be (and $^6$Li) imposes restrictive upper limits of
$\Ystau\lesssim 2\times 10^{-15}$ ($2\times 10^{-15}$\,--\,$2\times
10^{-16})$ for $\tau_{\stauone}\gtrsim 10^5\,\seconds$;
see~Fig.~\ref{Fig:Ytau}.

In Chapter~\ref{cha:stau-nlsp} we have either made use of
representative values~(\ref{Eq:Yslepton}) of $\Ystau$ or performed
exemplary CMSSM parameter scans which gave us $\Ystau$, e.g., for
fixed values of $\tanb$ and  $A_{\tau}$.
In this chapter we now calculate $\Ystau$ by taking into account the
complete set of stau annihilation channels in the MSSM with real
parameters for SUSY spectra for which sparticle coannihilation is
negligible. Using our own code for the computation of the resulting
thermal relic stau abundance $\Ystau$, we examine explicitly (i)~the
effect of left--right mixing of the lighter stau, (ii)~the effect of
large stau--Higgs couplings, and (iii)~stau annihilation at the
resonance of the heavy CP-even Higgs boson $H^0$.  We consider both
the ``phenomenological MSSM'' (pMSSM) (see,
e.g.,~\cite{Djouadi:2002ze}) in which the soft SUSY breaking
parameters can be set at the weak scale, and the CMSSM.

Within the framework of the pMSSM, we show examples in which $\Ystau$
can be well below $10^{-15}$.  Even within the CMSSM, we encounter
regions with exceptionally small values of $\Ystau\lesssim 2\times
10^{-15}$. We stress that the results in the following are independent
on the nature of the LSP. However, we discuss the implications of
these findings for the gravitino-stau scenario.%
\footnote{For simplicity, we call the parameter region in which
  $m_{\neutralino}< \mstauone$ the ${\neutralino}$-NLSP region in the
  following.}
We also address the viability of a $\stauone$--$\stauone^*$ asymmetry.
The key quantities for the significant $\Ystau$ reduction could be
probed at both the LHC and the ILC.

The work presented in this chapter is based on the
publication~\cite{Pradler:2008qc}. A calculation of the thermal relic
abundance of long-lived staus has also been part of a detailed
study~\cite{Berger:2008ti} which focuses on gauge interactions and on
the effect of Sommerfeld enhancement. In contrast, the most striking
findings of our study---in which Sommerfeld enhancement is not taken
into account---are related to the Higgs sector of the MSSM.

We also remark that our work has some overlap with~\cite{Ratz:2008qh}
which appeared as~\cite{Pradler:2008qc} was being
finalized. Whereas~\cite{Ratz:2008qh} focuses on the potential
suppression in $\Ystau$ due to enhanced annihilation into the lighter
Higgs final state, our work provides an investigation of
stau decoupling based on a complete set of annihilation
channels. Thereby, we also study enhanced stau annihilation into
heavier Higgs final states and consider annihilation at the heavy
Higgs resonance.

The outline of this chapter is as follows.  In the next section we
review basic properties of the staus to introduce our notations and
conventions for the stau mixing angle.  Section~\ref{sec:prim-ann}
explains the way in which we calculate $\Ystau$ and provides the
complete list of stau annihilation channels.  In
Sec.~\ref{sec:dependence} we analyze the dependence of the most
relevant stau annihilation channels on the stau mixing angle.  Effects
of large stau--Higgs couplings and stau annihilation at the $H^0$
resonance are studied in Sects.~\ref{sec:enhanc-coupl-higgs}
and~\ref{sec:reson-annih}, respectively.  The viability of a
$\stauone$--$\stauone^*$ asymmetry is addressed in
Sec.~\ref{sec:comm-stau-stau}.  In Sec.~\ref{sec:annih-chann} we
present exemplary parameter scans within the CMSSM that exhibit
exceptionally small $\Ystau$ values.  Potential collider phenomenology
of the parameter regions associated with those exceptional relic
abundances and potential implications for gravitino dark matter
scenarios are discussed in Sects.~\ref{sec:collider}
and~\ref{sec:gravitino}, respectively.

\section{Stau mixing and mass eigenstates}
\label{sec:properties-stau}

In this section we review some basic properties of the stau to set the
notation. In absence of inter-generational mixing, the stau
mass-squared matrix in the basis of the gauge eigenstates $(\stauL,
\stauR)$ reads
\begin{equation}
    \label{eq:stau-mass-matrix}
    \stauMAT = 
    \begin{pmatrix}
    \mtau^2 + \mLL^2 & \mtau \Xtau^* \\
             \mtau \Xtau      & \mtau^2 + \mRR^2      
    \end{pmatrix} = 
    (\stauROT)^\dagger
    \begin{pmatrix}
    \mstauone^2 & 0\\
    0 & \mstautwo^2   
    \end{pmatrix}
    \stauROT
\end{equation}
with
\begin{eqnarray}
    \label{eq:M-mstau-entries}
    \mLL^2 & = & \mstauL^2 +  \left( - \frac{1}{2} + \sin^2{\theta_W}  \right) \mZ^2 \cos{2\beta}\\
    \mRR^2 & = & \mstauR^2 - \sin^2{\theta_W}  \mZ^2 \cos{2\beta} \\
    \Xtau & = &\Atau - \mu^* \tanb \ .
\end{eqnarray}
Here, \mstauL\ and \mstauR\ are the soft SUSY breaking masses, \Atau\
is the trilinear coupling, $\mu$ is the Higgs-higgsino mass parameter,
and $\tanb=v_2/v_1$ denotes the ratio of the two Higgs vacuum expectation
values. In this work we restrict ourselves to the MSSM with real
parameters. Then $\Xtau^* = \Xtau$ so that the mass eigenstates
\stauone\ and \stautwo\ are related to \stauL\ and \stauR\ by means of
an orthogonal transformation
\begin{equation}
    \label{eq:phys-stau-fields}
    \begin{pmatrix}
      \stauone \\ \stautwo
    \end{pmatrix}
    = \stauROT
    \begin{pmatrix}
      \stauL \\
      \stauR
    \end{pmatrix}
    \quad\textrm{with}\quad
      \stauROT =
    \begin{pmatrix}
      \cos{\thetastau} & \sin{\thetastau} \\
      -\sin{\thetastau} & \cos{\thetastau}
    \end{pmatrix}
\end{equation} 
with \thetastau\ denoting the stau mixing angle. Imposing the mass
ordering $\mstauone < \mstautwo$ and choosing $0\leq \thetastau < \pi
$, the mixing angle can be inferred from the elements of \stauMAT,
\begin{equation}
    \label{eq:thetastau}
    \tan{2 \thetastau}  =  \frac{ 2 \mtau \Xtau }
    {  \mLL^2 - \mRR^2 } = \frac{ 2 \mtau \Xtau }{\delta} \ , \qquad
    \sin{2\thetastau} = \frac{ 2 \mtau \Xtau}{\mstauone^2 - \mstautwo^2} \ ,
\end{equation}
where the sign of the second relation determines the quadrant of
$\thetastau $. In the first relation, we have introduced $\delta
\equiv \mLL^2 - \mRR^2$. In particular, $\thetastau=\pi/2$ corresponds
to a purely right-handed stau, $\stauone=\stauR$,
whereas 
maximal mixing occurs for $\thetastau=\pi/4$ and $3\pi/4$.  The
physical stau masses are then given by
\begin{equation}
   \label{eq:stau-masses}
   m_{\stau_{1,2}}^2 = \mtau^2 + \mRR^2 +  \frac{1}{2} \left[\delta \mp
      \sqrt{\delta^2 + 4\mtau^2\Xtau^2 } \right]
\end{equation}
from which we see that an increase of $|\Xtau|$ leads to a reduction
of \mstauone.

\section{Calculation of the thermal relic stau abundance}
\label{sec:prim-ann}

We have undertaken the effort to set up our own full-fledged relic
abundance calculation. Let us in the following give a description of
our approach to compute the stau yield $\Ystau$.  Throughout this work
we assume a standard cosmological history with a temperature $T$ of
the primordial plasma above the stau decoupling temperature $\Tf$ so
that the lighter stau $\stauone$ was once in thermal
equilibrium. Then, the total stau yield $\Ystau\equiv Y_{\stauone} +
Y_{\stauone^*} $ is found by solving the Boltzmann
equation~(\ref{eq:boltzmann-champ-in-Y}) with $Y_\X=\Ystau $
Using the Maxwell--Boltzmann approximation, the stau equilibrium yield
\Ystaueq\ is given by
\begin{align}
   \label{eq:stau-eq}
   \Ystaueq = \frac{ \mstauone^2 T}{\pi^2 s}\, K_{2}\left(\frac{\mstauone}{T}\right)
\end{align}
and the thermally averaged annihilation cross section by~\cite{Gondolo:1990dk}
\begin{align}
  \label{eq:thavg}
  \sigmav(T) =
  \frac{1}{2\mstauone^4 T [K_2(\mstauone/T)]^2}\,
  \int^\infty_{4\mstauone^2} ds \, 
  \sqrt{s} K_1\left(\frac{\sqrt{s}}{T}\right) \peff^2 {\boldsymbol \sigma}(s) \,
  ,
\end{align}
where $K_i$ is the modified Bessel function of order $i$ and
$\peff=\sqrt{s-4\mstauone^2}\Big/2$.

Note that \sigmav\ contains all the information from
the particle physics side. It is obtained by computing the total
stau-annihilation cross section,
\begin{align}
 \label{eq:sigmatot}
        {\boldsymbol \sigma} \equiv \frac{1}{2}\, \sigmatot
        \quad \mathrm{with} \quad
        \sigmatot =  \sigma_{\stauone\, \stauone
        \rightarrow \tau\tau } + \sum_X \sigma_{\stauone\, \stauone^*
        \rightarrow X }    
\ ,
\end{align}
where the sum for the annihilation of $\stauone\, \stauone^*$ pairs%
\footnote{Counting wise we distinguish between $\stauone\, \stauone^*
  \rightarrow X$ and the conjugate process $\stauone^*\, \stauone
  \rightarrow \overline{X}$. In absence of CP violation in the SUSY
  sector, their cross sections agree so that we can solve a single
  Boltzmann equation~(\ref{eq:boltzmann-champ-in-Y}) for obtaining
  $\Ystau$.}
has to be taken over all final states $X$. The factor $1/2$ is
convention but gives (\ref{eq:boltzmann-champ-in-Y}) its familiar form.
The complete list of annihilation processes in the MSSM with real
parameters---save for coannihilation processes---is given in
Table~\ref{tab:ann-channels}.%
\footnote{For a purely right-handed stau $\stauone=\stauR$, the stau
  annihilation channels and associated cross sections have already
  been presented in Ref.~\cite{Ellis:1999mm} in the context of
  $\neutralino$-$\stauone$ coannihilation.}
 In addition, this table shows all
possible particle exchanges, where $s$, $t$, and $u$ are the
Mandelstam variables which denote the respective channel. A number of
annihilation processes proceeds also via a four-point vertex. Those
are marked in the column named ``contact.'' Already by mere optical
inspection, we immediately see that the Higgs sector plays potentially
an important role in the determination of the stau yield $\Ystau$.

\begin{table}[t]
  \caption[Complete set of stau annihilation channels]{The complete set of stau annihilation channels in the MSSM with real parameters for scenarios in which sparticle coannihilations are negligible. The mass eigenstates of the Higgs fields are denoted by $\hhiggs$, $\Hhiggs$, $\Ahiggs$, and $\Hpmhiggs$ and the ones of the neutralinos, the charginos, and the tau sneutrino by $\widetilde{\chi}^0_{1,..,4}$, $\widetilde{\chi}^{\pm}_{1,2}$, and $\widetilde{\nu}_{\tau}$, respectively. Because of the absence of a $\stauone\stauone\Ahiggs$ coupling (cf.\ Sec.~\ref{sec:enhanc-coupl-higgs}), $s$-channel exchange of the CP-odd Higgs boson $\Ahiggs$ and also $\stauone\stauone^*\rightarrow\gamma\Ahiggs$ do not appear.} 
\label{tab:ann-channels}
\begin{center}
\begin{tabular}{llc@{\qquad}c@{\qquad}c}
\toprule 
$\stauone^{(*)}\,\stauone^{(*)} \rightarrow$& final state  &  $s$-channel &
$t(u)$-channel & contact\\
\midrule 
\vspace{-0.4cm} \\ 
   & $\tau\tau 
    \ (\overline{\tau}\overline{\tau})
   $ & --- & $\widetilde{\chi}^0_{1,..,4}$ &---\\
   \vspace{-0.4cm} \\
   \midrule
   $\stauone\stauone^* \rightarrow$& final state $X$${}^\dagger$ &  $s$-channel
   &
   $t(u)$-channel & contact\\ 
   \midrule
   \vspace{-0.4cm} \\ 
   & $\mu\overline{\mu}$, $e\overline{e}$ &
   $\hhiggs,\Hhiggs,\,\gamma,Z$ 
   & --- & ---\\\vspace{-0.4cm} \\ 
   & $\tau\overline{\tau}$ &
   $\hhiggs,\Hhiggs,\,\gamma,Z$ 
   & $\widetilde{\chi}^0_{1,..,4}$ & ---\\\vspace{-0.4cm} \\ 
   & $\nu_{e}\overline{\nu}_{e}$, $\nu_{\mu}\overline{\nu}_{\mu}$  & $Z$ & --- & ---\\\vspace{-0.4cm}\\
   & $\nu_{\tau}\overline{\nu}_{\tau}$  & $Z$ & $\widetilde{\chi}^{\pm}_{1,2}$ & ---\\\vspace{-0.4cm}\\
   & $q_k\overline{q}_k$ & $\hhiggs,\,\Hhiggs,\,\gamma,\, Z$ & ---
   & --- \\\vspace{-0.4cm}\\
   & $\gamma\gamma,\, \gamma Z$ & --- & $\stauone$ & $\checkmark$ \\\vspace{-0.4cm}\\
   & $ZZ$ & $\hhiggs,\Hhiggs$ & $\stau_{1,2}$ & $\checkmark$ \\\vspace{-0.4cm}\\
   & $W^+W^-$ & $\hhiggs,\,\Hhiggs,\,\gamma,\,Z$ &
   $\widetilde{\nu}_{\tau}$ & $\checkmark$ \\\vspace{-0.4cm}\\
   & $\gamma\hhiggs,\,\gamma\Hhiggs$ & --- & $\stauone$ & ---\\\vspace{-0.4cm}\\
   & $Z\hhiggs,\,Z\Hhiggs$ & $Z$  &  $\stau_{1,2}$ & ---\\\vspace{-0.4cm}\\
   & $Z\Ahiggs$ & $\hhiggs,\Hhiggs$  &  $\stau_{2}$ & ---\\\vspace{-0.4cm}\\
   & $W^{\mp}\Hpmhiggs$ & $\hhiggs,\Hhiggs$  &  $\widetilde{\nu}_{\tau}$ & ---\\\vspace{-0.4cm}\\
   & \parbox{2.5cm}{
     $\hhiggs\hhiggs , 
     \, \hhiggs\Hhiggs,$\\
     $\Hhiggs\Hhiggs$}
   & $\hhiggs, \Hhiggs$& $\stau_{1,2} $ & $\checkmark$
   \\\vspace{-0.4cm}\\\vspace{-0.4cm}\\
   & \parbox{2.5cm}{
     $\Ahiggs\Ahiggs$}
   & $\hhiggs, \Hhiggs$& $\stau_{2} $ & $\checkmark$
   \\\vspace{-0.4cm}\\\vspace{-0.4cm}\\
   &  $\hhiggs\Ahiggs,\,\Hhiggs\Ahiggs$ & $Z$ &
   $\stau_{2} $  & ---\\\vspace{-0.4cm}\\
   & $\Hphiggs\Hmhiggs$ & $\hhiggs,\,\Hhiggs,\,\gamma,\,Z$
   & $\widetilde{\nu}_{\tau}$& $\checkmark$\\\vspace{-0.4cm}\\
   \bottomrule
   \multicolumn{4}{l}{\footnotesize ${}^\dagger$  $k=u,\,d,\,c,\,s,\,t,\,b$}\\[-0.1cm]
\end{tabular}
\end{center}
\end{table}

\afterpage{\clearpage}

For all channels in Table~\ref{tab:ann-channels}, we generate
\texttt{Fortran} code for the squared matrix elements
$|\mathcal{M}_i|^2$ by using the computer algebra packages
\texttt{FeynArts~5.4}~\cite{Hahn:2000kx,Hahn:2001rv} and
\texttt{FormCalc~5.3}~\cite{Hahn:1998yk,Hahn:2006qw}. For a chosen
point in the SUSY parameter space, we then compute the radiatively
corrected superparticle spectrum by running the spectrum generator
\texttt{SuSpect 2.40}~\cite{Djouadi:2002ze}. Its output allows us to
set all SUSY parameters so that we can compute the total cross section
$\sigmatot(s)$ given by (\ref{eq:sigmatot}) and subsequently the
thermally averaged cross section~(\ref{eq:thavg}).
Numerically, the computation of (\ref{eq:thavg}) is the most demanding
part in the relic abundance calculation. In particular, we take special
care about the following cases:
\begin{itemize}
\item \Hhiggs -resonance: Resonant stau annihilation via $\Hhiggs$
  exchange is one of the central points in this part. In the
  generation of the matrix elements, we have therefore included the
  total \Hhiggs-width $\Gamma_{\Hhiggs}$ in the respective $s$-channel
  propagators.

\item Propagator poles: A diverging $t(u)$-channel propagator can be
  encountered when a production threshold is met. We overcome this
  problem by including a ``sparticle-width'' of $0.01\mstauone$ in the
  respective propagators in the vicinity of dangerous thresholds. A
  particularly interesting example with a diverging $t(u)$-channel
  propagator is given by the process $\stauone\,\stauone^*\rightarrow
  \gamma\Hhiggs$ if $\sqrt{s}=m_{\Hhiggs}$ is fulfilled since then the
  \Hhiggs-exchange in the $s$-channels of other processes is resonant
  simultaneously.

\item Bessel functions: The Bessel functions in (\ref{eq:stau-eq}) and
  (\ref{eq:thavg}) exhibit an exponential behavior for large
  arguments $x\gg 1$~\cite{abramowitz+stegun}
  \begin{align}
    \label{eq:bessel}
    K_n(x) \simeq \sqrt{\frac{\pi}{2\,x}} e^{-x} 
    \left( 
      1 + \frac{4n^2-1}{8x} + \dots 
    \right)\ .
  \end{align}
  For small temperatures $T$, the arguments of $K_1$ and $K_2$ in
  (\ref{eq:thavg}) become large simultaneously. Therefore, in order to
  ensure numerical stability, we expand the Bessel functions in
  (\ref{eq:thavg}) for $\mstauone/T>35$ as in (\ref{eq:bessel}) and
  cancel the exponents analytically.%
 \end{itemize}

\begin{figure}[h!t]
\begin{center}
\centerline{\includegraphics[totalheight=0.8\textheight]{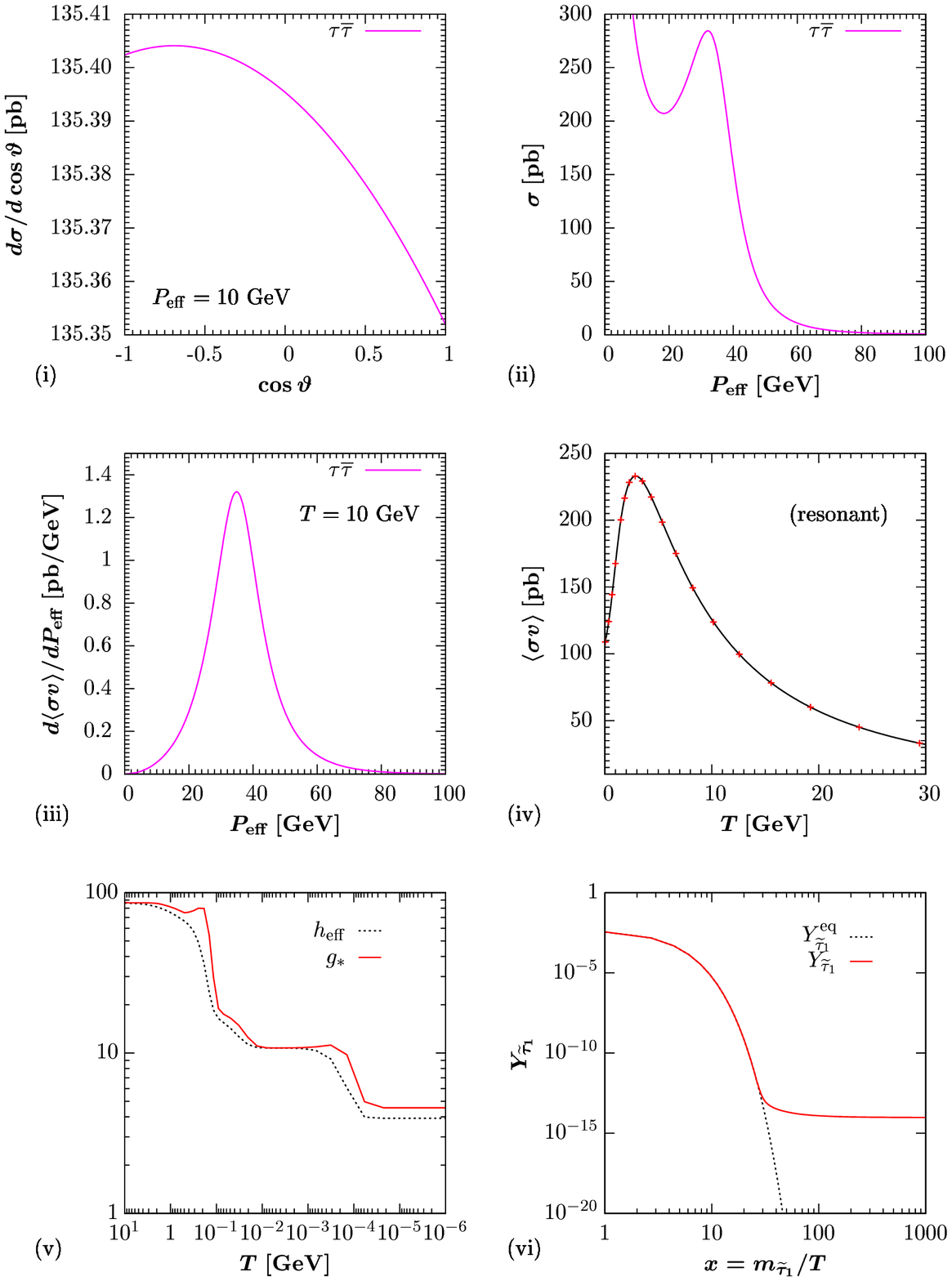}}
\caption[Schematic overview of the calculation of $\Ystau$]{Schematic
  overview of our approach to calculate \Ystau: (i) We generate matrix
  elements from which we obtain the differential cross section
  $d\sigma/d\cos{\vartheta}$; shown here is an example of the
  $\tau\overline{\tau}$ channel. (ii) Integration yields the invariant
  cross section~$\sigma$. The effect of ``thermal weighting'' of
  $\sigma$ is shown in (iii) by plotting~(\ref{eq:thavg}) before
  integration for the single $\tau\overline{\tau}$ channel (as a
  function of $\peff$.)  Summing up all annihilation channels and
  integration of (\ref{eq:thavg}) yields $\sigmav$ in (iv) at a grid
  of temperatures (crosses). (v)~Taking into account the temperature
  dependence of $g_*$ and $\heff$ and---upon cubic spline
  interpolation of $\sigmav$ in (iv)---we arrive at $\Ystau$ in (vi)
  from integration of~(\ref{eq:boltzmann-champ-in-Y}) . }
\label{Fig:control}
\end{center}
\end{figure}

\afterpage{\clearpage}

 We find the starting point for the numerical integration of
 (\ref{eq:boltzmann-champ-in-Y}) by solving~\cite{Belanger:2004yn}
 \begin{align}
   \label{eq:Tf1}
   \left. \frac{d\Ystaueq}{dT}\right|_{\Tfone} = \sqrt{\frac{8\pi^2
       g_{*}(T) }{45}} \MPl 
   \sigmav(\Ystaueq  )^2 \lambda(\lambda+2)
 \end{align} 
 where $g_{*}(T)$ is given by~(\ref{eq:gstar}).  $\Tfone$ marks the
 point at which the stau starts to decouple chemically from the
 background plasma, $\Ystau(\Tfone)-\Ystaueq(\Tfone)\simeq \lambda
 \Ystaueq(\Tfone)$ with $\lambda = 0.1$~\cite{Belanger:2004yn} chosen
 in our code.  Since we use a globally adaptive Gaussian integration
 routine to calculate (\ref{eq:thavg}), the computation of
 $\sigmav(T)$ is time-demanding.  Therefore, we evaluate
 (\ref{eq:thavg}) on a grid of different temperatures and use cubic
 spline interpolation to obtain values in between.
 We then solve the Boltzmann equation (\ref{eq:boltzmann-champ-in-Y}) by
 numerical integration from $\Tfone$ to zero.  There, we fully take
 into the account the temperature dependence of $g_{*}$ and \heff\ by
 interpolating the respective tabulated values provided as part of the
 relic density code \texttt{DarkSUSY~4.00}~\cite{Gondolo:2004sc}.  The
 freeze out temperature can then be defined by $\Tf\equiv
 (\Tfone+T_{\mathrm{f}2})/2$ where $T_{\mathrm{f}2}$ is given by
 $\Ystaueq(T_{\mathrm{f}2}) =
 \Ystau(T_{\mathrm{f}2})/10$~\cite{Belanger:2004yn}. For
 $T<T_{\mathrm{f}2}$, residual annihilations will further reduce
 $\Ystau$ so that we refer to the decoupling yield
 $\Ystau^{\mathrm{dec}}$ as the quantity at the endpoint of
 integration. As already pointed out in the introduction, for
 simplicity, we call this yield $\Ystau$. Moreover, we will quantify
 $T$ in terms of $x\equiv m_{\stauone}/T$ and in particular $\Tf$ in
 terms of $x_{\freezeout}\equiv m_{\stauone}/\Tf$.
 We have also schematically depicted the approach to $\Ystau$ in
 Fig.~\ref{Fig:control} for an exemplary scenario (which gives rise to
 resonant stau annihilation. See the figure caption for details.)

 Note that we have additionally modified the \texttt{FeynArts} MSSM
 model file for the generation of the matrix elements in two ways: The
 first version, which we use throughout
 Sects.~\ref{sec:dependence}--\ref{sec:reson-annih}, allows us to set
 all $q_k\overline{q}_k$--Higgs and all trilinear Higgs couplings by
 using the computer tool
 \texttt{FeynHiggs~2.6.3}~\cite{Heinemeyer:1998yj}; see also
 Sects.~\ref{sec:enhanc-coupl-higgs} and~\ref{sec:reson-annih}. The
 second version allows for a direct comparison with the existing
 computer code \texttt{micrOMEGAs 2.0.6}
 \cite{Belanger:2001fz,Belanger:2004yn,Belanger:2006is}. We have
 transcribed their routine~\cite{Pukhov:2004ca} for the computation of
 the running quark masses to \texttt{Fortran}, adopted all
 $q_k\overline{q}_k$--Higgs couplings, and modified all Higgs-self
 couplings of our matrix elements to match with their implemented
 version of the MSSM~\cite{Dubinin:1998nt}. Using this second version,
 we find perfect agreement between our codes.%
 \footnote{For our computation we use the Standard Model parameters
   $m_{\mathrm{t}}=172.5~\GeV$,
   $m_{\mathrm{b}}(m_{\mathrm{b}})^{\mathrm{\overline{MS}}} = 4.25\
   \GeV$, $\alpha_{\mathrm{s}}^{\mathrm{\overline{MS}}}(\mZ)=0.1172$,
   $\alpha_{\mathrm{em}}^{-1\mathrm{\overline{MS}}}(\mZ) = 127.932 $,
   and $\mZ=91.187\ \GeV$. Since \texttt{micrOMEGAs} has hard-coded
   $\sin{\theta_W}=0.481$ from which it computes $\mW$ using the
   on-shell relation with $\mZ$, we follow their convention to allow
   for a better comparison of our results with \texttt{micrOMEGAs}.}

 \section{Dependence of stau annihilation on the stau mixing angle}
\label{sec:dependence}

In order to isolate the distinct features of the different
annihilation processes we need to have full control over the
superparticle mass spectrum. Therefore, in the following, we will not
rely on any constrained model (such as the CMSSM) where the soft-SUSY
breaking parameters are subject to stringent boundary conditions at
some high scale (such as \mgut). In those models, the mass spectrum is
found only after renormalization group (RG) evolution from the high
scale down to the electroweak scale.  Instead, we choose to work in
the framework of the ``phenomenological MSSM'' (pMSSM), see,
e.g.,~\cite{Djouadi:2002ze}.  There, all soft-SUSY breaking parameters
can be set at the scale of electroweak symmetry breaking---a low
scale---which we fix to $\sim 2 \mstauone$. In particular, one can
also trade the Higgs mass-squared parameters $m_{H_u}^2$ and
$m_{H_d}^2$ against $\mu$ and the pseudoscalar Higgs boson mass
$m_{\Ahiggs}$.%
\footnote{Though the advocated procedure may require fine-tuning in
  the electroweak symmetry breaking conditions, it conveniently
  provides us with running parameters at the scale of stau
  annihilation.}
Choosing $\mu$ as an input parameter is very convenient for two
reasons: First, together with the specification of the gaugino masses
$M_{1,2}$ we have control over the gaugino/higgsino mixture of the
neutralinos $\widetilde{\chi}_i^0$. Second, $\mu$ enters directly into
the stau-Higgs couplings, whose importance will become clear in the
next section. Furthermore, in the following, we choose to set all
soft-SUSY breaking scalar masses (apart from \mstauL and \mstauR) to a
common value $M_S=1\ \TeV$. Thereby, we essentially decouple all
sfermions which are not of interest for us. This ensures also that we
never enter \emph{accidentally} any coannihilation regime.  Finally,
for simplicity, we set also all trilinear parameters to a common value
$A$. Given $\mu$, $\Atau=A$, and $\tanb$, and thereby $\Xtau$, we can
then fix $\mstauone$ and $\thetastau$ to arbitrary values by adjusting
$\mRR^2$ and $\delta$ in Eqs.~(\ref{eq:thetastau})
and~(\ref{eq:stau-masses}).

In the following, we will focus on two distinct regions of the SUSY
parameter space. In the beginning, we will choose $m_{\Ahiggs}$ to be
very large $m_{\Ahiggs}=1\ \TeV \gg \mZ$. This corresponds to the
decoupling limit of the MSSM where the following (tree-level)
relations hold~\cite{Gunion:2002zf}
\begin{align}
    \label{eq:DL-higgs-masses}
    \mh^2 &\simeq \mZ^2 \cos^{2}{2\beta}, \qquad \mH^2 \simeq \mA^2 +
    \mZ^2 \sin^{2}{2\beta},\\
    \label{eq:DL-cbma}   
    \mHpm^2 & = \mA^2 + \mW^2, \qquad \cos^{2}{(\beta-\alpha)} \simeq
    \frac{\mZ^4\sin^2{4\beta}}{4 \mA^4}.
\end{align}
Therefore, $\mA \simeq \mH \simeq \mHpm$ up to corrections
$\Orderof{\mZ^2/\mA}$ so that any of the stau annihilation channels
into heavy Higgs bosons is kinematically blocked. Furthermore,
$\cos{(\beta-\alpha)}=0$ up to corrections $\Orderof{\mZ^2/\mA^2}$
implies that the $\Hhiggs VV$ coupling ($V=Z,W$) becomes very small so
that we loose the $\Hhiggs$-exchanges in the stau annihilation
channels with a $VV$ final state. At the same time, the light Higgs
boson takes on its Standard Model value for the $\hhiggs VV$ coupling.
Complementary to that we will consider also regions of the SUSY
parameter space with smaller $\mA$, e.g., in the next section, where
we will put a stronger focus on the Higgs sector and its connection to
\Ystau.

\begin{figure}[t!]
\begin{center}
\centerline{\includegraphics[height=7.2cm]{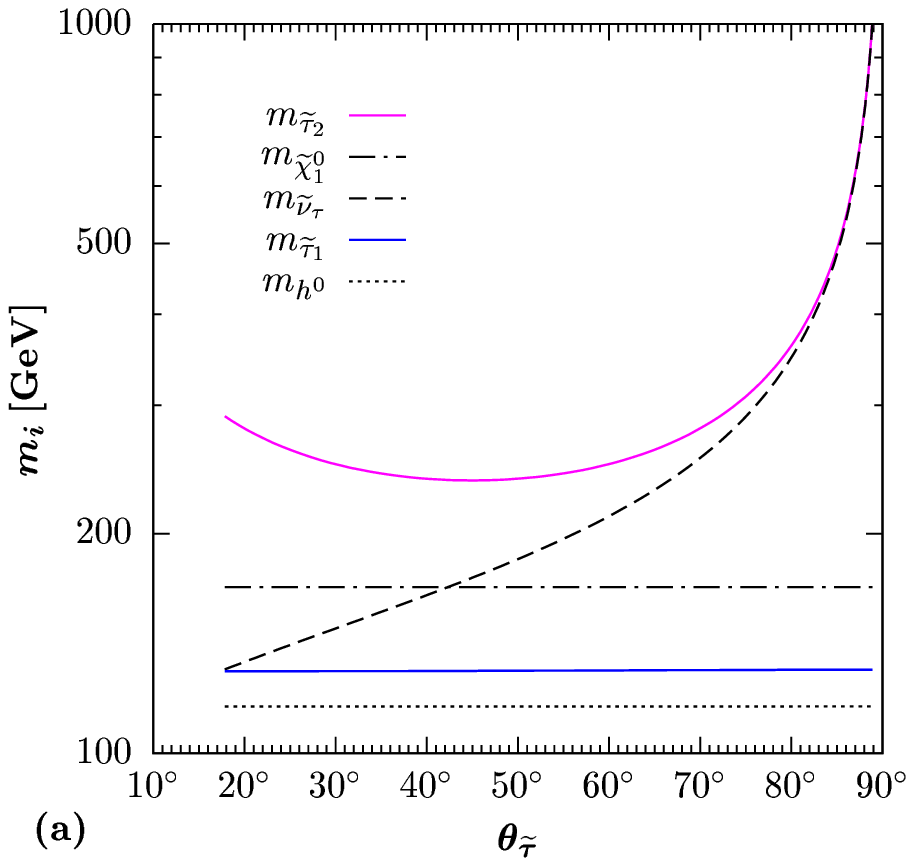}%
\includegraphics[height=7.2cm]{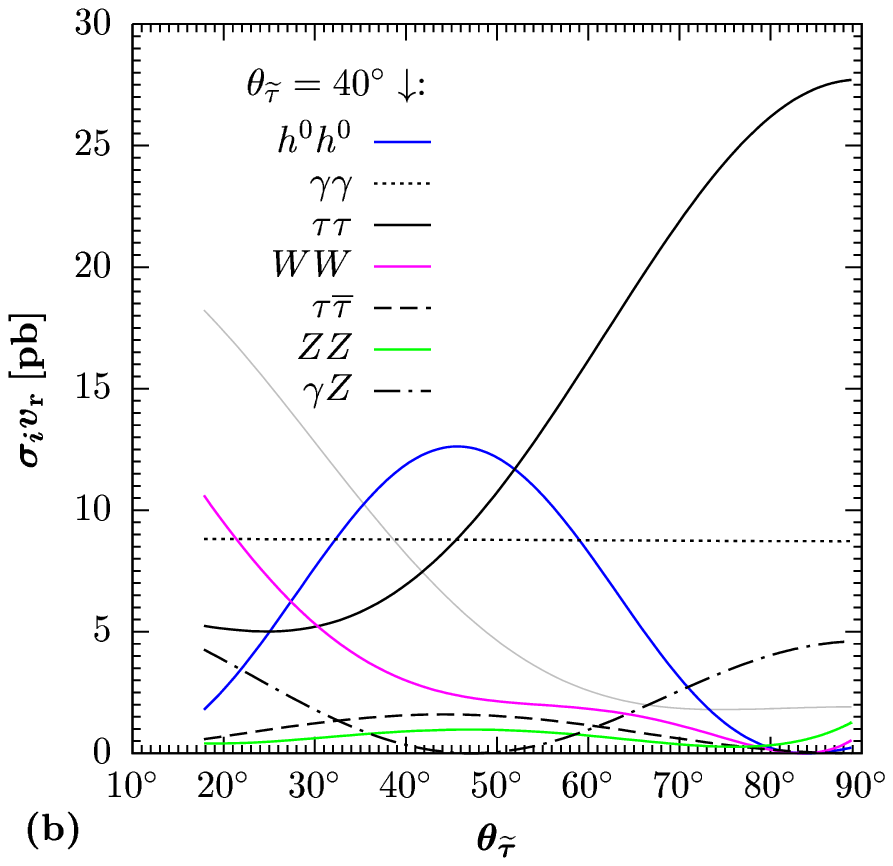}}
\caption[Dependencies of masses and annihilation cross sections on
\thetastau]{\small (a)~The dependence of $\mstautwo$ (curved solid
  line) and $m_{\widetilde{\nu}_{\tau}}$ (dashed line) on the stau
  mixing angle $\thetastau$ for the input parameters $\mstauone=130\
  \GeV$ (horizontal solid line), $\tanb=10$, $\mA=M_S=M_{3}=-A=1\
  \TeV$, and $6M_1=M_2=\mu=1\ \TeV$ (i.e.,
  $\neutralino\simeq\widetilde{B}$) for which $m_{\neutralino}=169\
  \GeV$ (dash-dotted line) and $m_{\hhiggs}=116\ \GeV$ (dotted
  line). (b)~Dominant stau annihilation cross sections times the
  relative velocity $v_{\mathrm{r}}$ of the incoming staus as a
  function of $\thetastau$ for $\peff=10\ \GeV$ and the same input
  parameters as in~(a). The curves show the channels with the
  following final states: $\hhiggs\hhiggs$, $\gamma\gamma$,
  $\tau\tau$, $WW$, $\tau\overline{\tau}$, $ZZ$, $\gamma Z$ (at
  $\thetastau=40^\circ$, from top to bottom).  In addition, we plot
  $\sigma_{\tau\tau}v_{\mathrm{r}}$ for the case of a wino-like
  neutralino, $\neutralino\simeq\widetilde{W}$, with
  $m_{\neutralino}=175\ \GeV$ as obtained with $M_{1}=6M_{2}=1\ \TeV$
  (thin gray line). No lines are shown for $\thetastau < 18^\circ$
  where $m_{\widetilde{\nu}_{\tau}}<\mstauone$.}
\label{Fig:mix}
\end{center}
\end{figure}

In Fig.~\ref{Fig:mix}a we show the $\thetastau$-dependence of the
masses of the heavier stau, $m_{\stautwo}$, (curved solid line) and
the tau-sneutrino, $m_{\widetilde{\nu}_{\tau}}$, (dashed line) for
fixed $\mstauone = 130\ \GeV$ and the input parameters $\tanb=10$,
$\mA=\mu=-A=1\ \TeV$, and $6 M_{1}=M_{2,3}=1\ \TeV$.  Because of SU(2)
gauge invariance, $\mstauL$ sets also the soft-breaking mass for the
tau-sneutrino hence approximately $m^2_{\widetilde{\nu}_{\tau}} \sim
\mstauR^2 + \delta$ so that $\widetilde{\nu}_{\tau}$ becomes lighter
than \stauone\ for $\thetastau \lesssim 18^\circ$ ($\delta$ is
negative in that region).  In addition, we plot the masses of the
lightest neutralino, $m_{\neutralino}=169\ \GeV$ (dash-dotted line),
the lighter stau, $\mstauone = 130\ \GeV$ (horizontal solid line), and
the lightest Higgs, $m_{\hhiggs}=116\ \GeV$ (dotted line). We note in
passing that $\mstauone$ may deviate slightly from its anticipated
input value due to radiative corrections. We then correct for this by
an adjustment of $\mstauR^2$ so that we indeed ensure $\mstauone$ to
be constant.

In Fig.~\ref{Fig:mix}b we plot the dominant stau annihilation cross
sections times the relative (non-relativistic) velocity in the
center-of-mass frame of the incoming staus, $v_{\mathrm{r}} =
2\peff/\mstauone$, for the same parameters as in Fig.~\ref{Fig:mix}a.
Owing to an (approximate) Maxwell-Boltzmann distribution of the stau
velocity, $\langle \peff \rangle|_{\Tf} \sim \sqrt{\mstauone\Tf}$, we
choose $\peff = 10\ \GeV$ as a representative value.%
\footnote{This value is actually at the somewhat lower end, given
  $\mstauone\gtrsim 100\ \GeV$ and $\Tf\simeq \mstauone/25$. However,
  $\sigma v_{\mathrm{r}}$ depends only weakly on $\peff$, and the
  thermally averaged $\sigmavof{i}$ will be shown in the upcoming
  figures.}
The curves show the annihilation channels with the following final
states: $\hhiggs\hhiggs$, $\gamma\gamma$, $\tau\tau$, $WW$,
$\tau\overline{\tau}$, $ZZ$, $\gamma Z$ (at $\thetastau=40^\circ$,
from top to bottom).
All channels except $\gamma\gamma$ show a strong dependence on
\thetastau. The $\hhiggs\hhiggs$ ($\tau\overline{\tau}$) channel peaks
at $\thetastau=\pi/4$---a feature which we will discuss in detail in
Sec.~\ref{sec:enhanc-coupl-higgs}. For the $\tau\tau$ channel, the
overall size of the cross section is governed by $m_{\neutralino}$
since this channel proceeds only via $t(u)$-channel exchanges of
neutralinos. Our chosen input values lead to a bino-like neutralino,
$\neutralino\simeq\widetilde{B}$, and $\sigma_{\tau\tau}$ drops for an
increasingly `left-handed' stau. (For comparison, the thin gray line
shows $\sigma_{\tau\tau}v_{\mathrm{r}}$ for the case of a wino-like
lightest neutralino, $\neutralino=\widetilde{W}$, of similar mass,
$m_{\neutralino}=175\ \GeV$, as obtained by changing the gaugino mass
input parameters to $M_{1}=6M_2=1\ \TeV$.)  The annihilation into a
$WW$ pair becomes important for an increasing $\stauL$ component in
$\stauone$, i.e., towards smaller $\thetastau$, since the
$t(u)$-channel exchange with the tau-sneutrino opens up; the
$\stauone\widetilde{\nu}_{\tau}W$ ($\stauone\stauone WW$) coupling is
proportional to $\cos{\thetastau}$ ($\cos^2{\thetastau}$).  The
modulation of the $\gamma Z$ channel can be understood by considering
the structure of the $\stauone\stauone Z$ coupling $\propto (1 -
4\sin^2{\theta_W} + \cos{2\thetastau})$. Note that the first two terms
practically cancel out. For stau annihilation into a $ZZ$ pair there
is an additional contribution from $\stautwo$-exchange with the
respective $\stauone\stautwo Z$ coupling $\propto \sin{2\thetastau}$.
Having discussed the dominant $\stauone$ annihilation channels in a
simple manner, we also warn the reader that interferences between the
different Feynman diagrams of a given channel may well lead to a
counter-intuitive behavior. In this regard, see
Ref.~\cite{Berger:2008ti} for a thorough discussion of
$\stauone\stauone^*$ annihilation into vector bosons. For the limiting
case of a purely `right-handed' stau, $\stauone\simeq\stauR$
($\thetastau\rightarrow \pi/2$), we recover the relative importance of
the annihilation cross sections into $\gamma\gamma$, $\gamma Z$, $Z
Z$, and $\tau\tau$ with bino $t(u)$-channel exchange found in
Ref.~\cite{Asaka:2000zh}.

\begin{figure}[t]
\begin{center}
\includegraphics[width=0.75\textwidth]{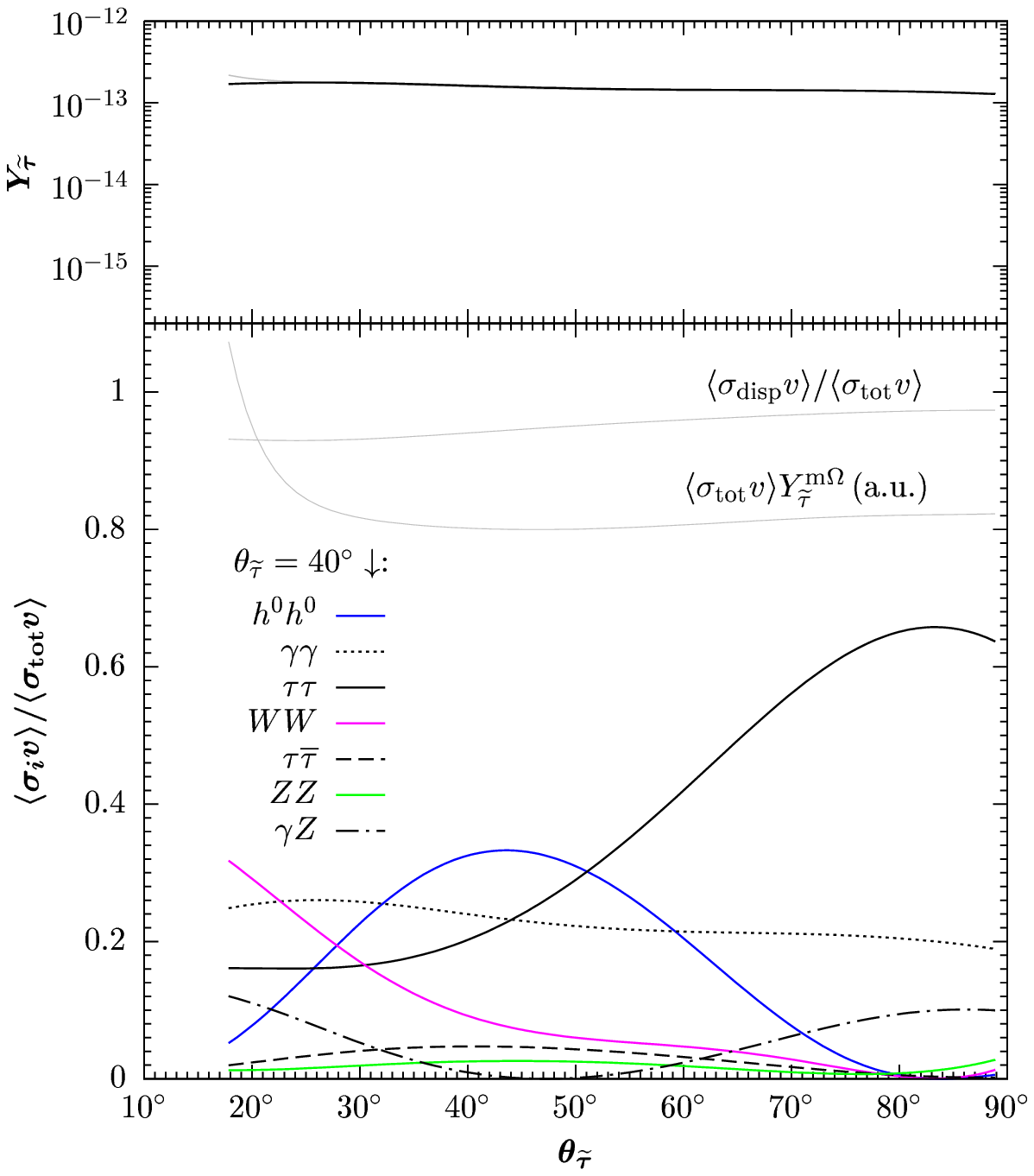}%
\caption[General dependence of $\Ystau$ on \thetastau]{\small Dependence
  of the stau yield $\Ystau$ (upper panel) and of the relative
  importance of the dominant thermally averaged cross sections,
  $\sigmavof{i}/\sigmavof{\mathrm{tot}}$, at $x=25$ (lower panel) on
  the stau-mixing angle $\thetastau$ for the same input parameters as
  in Fig.~\ref{Fig:mix}.  In the upper panel, the thick line shows the
  stau yield $\Ystau$ obtained with our relic abundance calculation
  and the thin gray line the one obtained with \texttt{micrOMEGAs} to
  which we refer as $\YstaumO$. In the lower panel, the line styles
  are associated with the same dominant annihilation channels as in
  Fig.~\ref{Fig:mix}b.  In addition, we show (as labeled) the relative
  importance of the sum of the displayed cross sections,
  $\sigmavof{\mathrm{disp}}/\sigmavof{\mathrm{tot}}$, and
  $\sigmavof{\mathrm{tot}}\,\YstaumO$ in arbitrary units (a.u.). No
  lines are shown for $\thetastau < 18^\circ$ where
  $m_{\widetilde{\nu}_{\tau}}<\mstauone$.}
\label{Fig:multi1}
\end{center}
\end{figure}
\afterpage{\clearpage}

Figure~\ref{Fig:multi1} shows the $\thetastau$-dependence of $\Ystau$
(upper panel) and of the relative importance of the dominant
\textit{thermally averaged} cross sections,
$\sigmavof{i}/\sigmavof{\mathrm{tot}}$, at $x=25$ (lower panel) for
the same input parameters as in Fig.~\ref{Fig:mix}.
The lines in the lower panel are associated with the same dominant
annihilation channels as in Fig.~\ref{Fig:mix}b.  In addition, the
relative importance of the sum of the displayed cross sections,
$\sigmavof{\mathrm{disp}}/\sigmavof{\mathrm{tot}}$, (thin line, as
labeled) is shown to demonstrate that the displayed channels
constitute indeed (up to at most about $10\%$) the dominant part of
$\sigmavof{\mathrm{tot}}$ for the chosen set of input parameters.
In the upper panel, the total stau decoupling yield obtained with our
own relic abundance calculation is shown by the thick line and the one
computed with \texttt{micrOMEGAs}, $\YstaumO$, by the thin gray line.
For $\thetastau\lesssim 25^\circ$, both curves start to deviate from
each other since one enters the $\widetilde{\nu}_{\tau}$--$\stauone$
coannihilation region in which the stau decoupling yield increases.
This coannihilation effect leads also to the rise of the thin gray
line that shows $\langle\sigma_{\mathrm{tot}}v\rangle\,\YstaumO$ in
arbitrary units (a.u.) in the lower panel. Note that the same line
illustrates $\Ystau\propto 1/\sigmavof{\mathrm{tot}}$ for
$\thetastau>25^\circ$, where the result of our relic abundance
calculation agrees with $\YstaumO$.  Interestingly, for the given
input parameters, $\Ystau$ is not overly affected by the variation in
$\thetastau$ in this region, which reflects the fact that
$\sigmavof{\mathrm{tot}}$ and thereby $\sigmav$ vary by less than a
factor of about $1.5$ at the relevant time of decoupling.
In the next sections, we will demonstrate that this picture changes
significantly for certain other choices of the input parameters.

\section{Effects of large stau-Higgs couplings}
\label{sec:enhanc-coupl-higgs}

Owing to the scalar nature of the stau, there exists a remarkable
difference between the standard neutralino decoupling and the scenario
in which the long-lived stau freezes out from the primordial plasma.
For the neutralino LSP, the $\mu$ parameter enters into the
annihilation cross sections only indirectly by influencing the
gaugino/higgsino mixture of $\neutralino$.  This stands in strong
contrast to the case in which a scalar particle is the lightest
Standard Model superpartner: the sfermions couple directly to
dimensionful parameters of the theory, namely, the trilinear couplings
$A$ and the Higgs-higgsino mass parameter $\mu$.  The corresponding
operators in the MSSM Lagrangian always contain a Higgs field. In
particular, the stau--Higgs couplings are given by
\begin{equation}
    \label{eq:L-stau-stau-Higgses}
    \Lagrangian_{\mathrm{MSSM}} \ni\frac{g}{\mW} \sum_{\alpha , \beta = \L , \R} \stau^*_{\alpha} 
    \couptriLR{\stau^*_{\alpha}}{\stau_{\beta}}{ \mathcal{H}}
    \stau_{\beta} \mathcal{H} 
\end{equation}
with $ \mathcal{H} = \hhiggs,\,\Hhiggs ,\,\Ahiggs$.  We have pulled
out the factor $g/\mW$ so that the `reduced' couplings
$\couptriLR{\stau^*_{\alpha}}{\stau_{\beta}}{ \mathcal{H}}$ among the
gauge eigenstates $\stauL$ and $\stauR$ are given
by~\cite{Haber:1997dt}
\begin{align}
    \label{eq:lighthiggs-stau-stau-couplings}
     \couptriLR{\stau^*}{\stau}{\hhiggs} & = 
    \begin{pmatrix}
      \displaystyle
      \left( -{\frac{1}{2}} + \ssqw \right) \mZ^2 \sapb + \mtau^2
      {\frac{\sa}{\cb}} 
      &
      \displaystyle
      \frac{\mtau}{2} \left( \Atau \frac{\sa}{\cb}
        + \mu \frac{\ca}{\cb} \right) 
      \\ 
      \displaystyle
      \frac{\mtau}{2} \left( \Atau
        \frac{\sa}{\cb} + \mu \frac{\ca}{\cb} \right) 
      & 
      \displaystyle
      - \ssqw \mZ^2
      \sapb + \mtau^2 {\frac{\sa}{\cb}}
    \end{pmatrix}\ ,
\\
    \label{eq:CPoddhiggs-stau-stau-couplings}
     \couptriLR{\stau^*}{\stau}{\Ahiggs}  &=
     \begin{pmatrix}
      0
      &
      \displaystyle
      +i \frac{\mtau}{2} \left(  \Atau \tanb + \mu\right) 
      \\ 
      \displaystyle
      -i \frac{\mtau}{2} \left(  \Atau \tanb + \mu\right) 
      & 
      0
    \end{pmatrix}\ ,
\end{align}
where $\couptriLR{\stau^*}{\stau}{\Hhiggs}$ can be obtained from
(\ref{eq:lighthiggs-stau-stau-couplings}) upon the replacement $\alpha
\rightarrow \alpha-\pi/2$.  Whenever convenient, we use the shorthand
notation $\ssqw=\sin^2{\theta_W}$, $c_{\gamma}=\cos{\gamma}$, and
$s_{\gamma}=\sin{\gamma}$. The parameters $\Atau$ and $\mu$ only
appear off-diagonal and they are multiplied with the associated
fermion mass, the tau mass $\mtau$.

Using $C = \stauROT \widetilde{C} \stauROT^\dagger$, one obtains the
couplings of the mass eigenstates $\stauone$ and $\stautwo$. In this
regard, it is important to note that the coupling of the CP-odd Higgs
boson to the lighter stau vanishes,
$\couptri{\stau^*_{1}}{\stau_{1}}{\Ahiggs}=0$. Therefore, we have not
listed the process $\stauone\stauone^*\rightarrow\gamma\Ahiggs$ in
Table~\ref{tab:ann-channels}. By the same token, there is also no
$s$-channel exchange of $\Ahiggs$ in any of the annihilation channels.
Note that this statement remains valid even after the inclusion of
radiative corrections: There is no induced mixing between
$\hhiggs(\Hhiggs)$ and $\Ahiggs$ in absence of CP-violating effects in
the SUSY sector.

Let us now turn to the probably most interesting couplings in the
context of $\stauone\stauone^*$ annihilation, namely, the ones of the
lighter stau to $\hhiggs$ and $\Hhiggs$. The `reduced'
$\stauone\stauone\hhiggs$ coupling reads
  \begin{align}
    \label{eq:higgses-stau-stau-couplings-PHYS}
     \couptri{\stau^*_{1}}{\stau_{1}}{\hhiggs}& =
      \displaystyle
      \left( -{\frac{1}{2}} \csqth + \ssqw \ctwoth \right) \mZ^2 \sapb + \mtau^2
      {\frac{\sa}{\cb}} + 
       \frac{\mtau}{2} \left( \Atau \frac{\sa}{\cb}
        + \mu \frac{\ca}{\cb} \right) \stwoth \ .
\end{align}
This is a complicated expression. However, if we choose $\mA$ to be
large, $\mA\gg\mZ$, we can
simplify~(\ref{eq:higgses-stau-stau-couplings-PHYS}) by using
$\cos{(\beta-\alpha)}=0$ [cf.~(\ref{eq:DL-cbma})],
\begin{align}
  \label{eq:coup-s1-s1-h-DL}
  \couptriDL{\stau^*_{1}}{\stau_{1}}{\hhiggs} & \simeq \displaystyle \left(
    {\frac{1}{2}} \csqth - \ssqw \ctwoth \right) \mZ^2 \ctwob -
  \mtau^2 - \frac{\mtau}{2} \Xtau \stwoth \ .
\end{align}
Thereby, we make an interesting observation: In the decoupling limit
(DL), the $\stauone\stauone\hhiggs$ coupling becomes proportional to
the left--right entry $\mtau \Xtau$ of the stau mass-squared matrix
(\ref{eq:stau-mass-matrix}) and to $\stwoth$. Therefore, it comes as
no surprise that the $\stauone\stauone^*$ annihilation cross section
into $\hhiggs\hhiggs$ peaks at $\thetastau = \pi/4$---the point of
maximal $\stauL$-$\stauR$ mixing---as can be seen, e.g., in
Fig.~\ref{Fig:mix}b. Analogously, one finds that the
$\stauone\stauone\Hhiggs$ coupling is proportional to
$\left(\Atau\tanb + \mu\right) \stwoth$ in the decoupling limit.
Complementary, the $\stauone\stautwo\hhiggs/\Hhiggs$ couplings exhibit
in this limit the same combination of $A$, $\mu$, and $\tanb$ as their
$\stauone\stauone$ counterparts but those terms are now multiplied by
$\ctwoth$ instead of $\stwoth$.

After the above discussion, it is clear that there exists the
possibility to enhance the total stau annihilation cross section
$\sigmatot$---and thereby to decrease $\Ystau\propto
1/\sigmavof{\mathrm{tot}}$---by choosing a proper combination of large
$A$, $\mu$, and $\tanb$. In the remainder of this section, we will
explore this possibility for two exemplary pMSSM scenarios.

Before proceeding let us make some technical comments. Large values of
the previously mentioned parameters may well lead to large radiative
corrections.%
\footnote{In this context, note that we introduce a large
  $m_\mathrm{t}$--$m_{\widetilde{\mathrm{t}}_{1,2}}$ splitting when
  choosing $M_S= 1\ \TeV$.}
In order to arrive at a proper loop-improved tree-level result, we
re-evaluate the entire Higgs sector using \texttt{FeynHiggs}.  In
particular, we have modified our generated matrix elements in a way
that allows us to set all trilinear Higgs couplings to their
loop-corrected values.%
\footnote{The author grateful to T.~Plehn and M.~Rauch for providing us,
  for cross-checking, with their implementation of a \texttt{Fortran}
  routine which calculates the Higgs self-couplings using the
  effective potential approach~\cite{Carena:1995bx}.}
Note that this goes well beyond a simple
$\alpha\rightarrow\alpha_{\mathrm{eff}}$ prescription. Only then, we
mostly find better agreement of our cross sections for stau
annihilation into two Higgses with the ones computed by
\texttt{micrOMEGAs}.  The latter program uses \texttt{CalcHEP}
\cite{Pukhov:2004ca} for the generation of the matrix elements. There,
the trilinear Higgs self-couplings have been expressed in terms of
$m_{\hhiggs}$, $m_{\Hhiggs}$, and $m_{\Ahiggs}$ which effectively
reabsorbs a bulk of the radiative corrections~\cite{Dubinin:1998nt}.
We therefore think that we do slightly better whenever we encounter
some disagreement between the mentioned cross sections. Though the
overall effect on $\Ystau$ is typically small, it can be at the level
of $20\%$ (see below).  Finally, it is well known that a large $A$
parameter may lead to charge/color breaking minima (CCB) in the scalar
MSSM potential; see, e.g., Ref.~\cite{Casas:1995pd}. \texttt{SuSpect}
performs some basic checks which we take into account to make sure
that we do not violate the constraints associated with CCB. We remark
that our pMSSM scenarios are chosen such as to allow us to extract the
important features of primordial stau annihilation in the most
transparent way---without emphasis on naturalness considerations.

\begin{figure}[t]
\begin{center}
\includegraphics[width=0.75\textwidth]{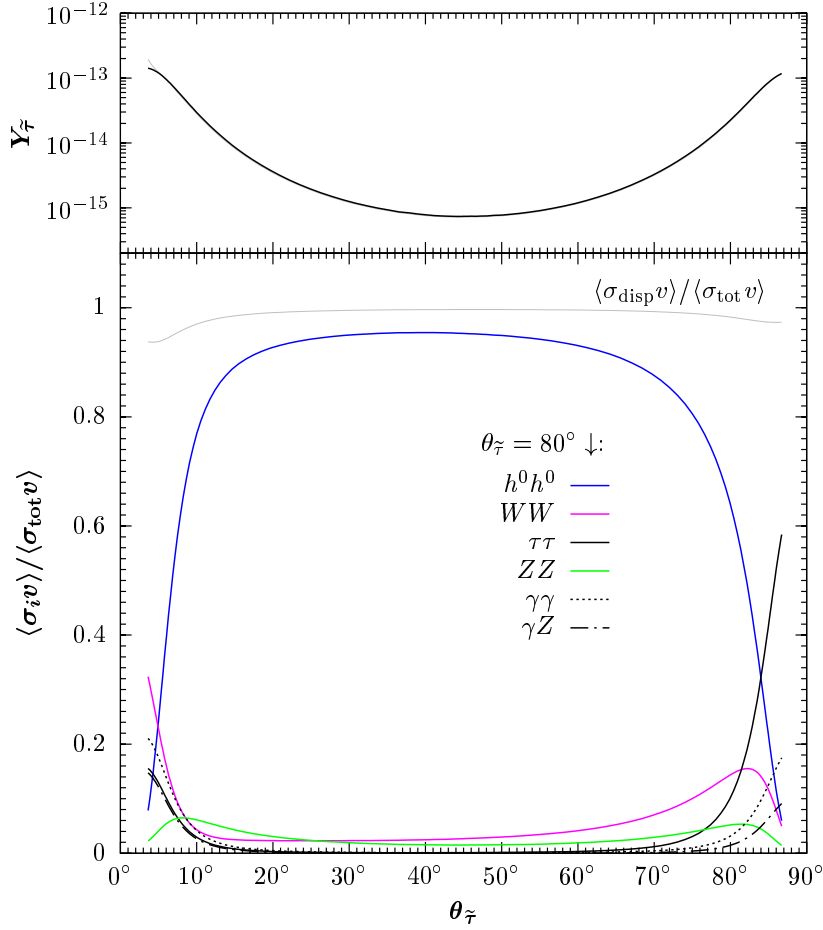} 
\caption[Enhanced annihilation into light Higgses]{\small Analogous to
  Fig.~\ref{Fig:multi1} but for the pMSSM scenario associated with
  $\mstauone=130\ \GeV$, $\tanb=50$, and
  $\mA=\mu=M_S=6M_1=M_{2,3}=-A=1\ \TeV$ and for $x=30$. The stau
  decoupling yield takes on its minimum value of $\Ystau = 7.4\times
  10^{-16}$ at $\thetastau=45^\circ$. The displayed stau annihilation
  channels are associated with the following final states:
  $\hhiggs\hhiggs$, $WW$, $\tau\tau$, $ZZ$, $\gamma\gamma$, and
  $\gamma Z$ (at $\thetastau=80^\circ$, from top to bottom). No lines
  are shown for $\thetastau < 4^\circ$ where
  $m_{\widetilde{\nu}_{\tau}}<\mstauone$.}
\label{Fig:multi2}
\end{center}
\end{figure}
\afterpage{\clearpage}

In Fig.~\ref{Fig:multi2} we demonstrate the effect associated with a
large $\stauone\stauone\hhiggs$ coupling by presenting the
$\thetastau$-dependence of $\Ystau$ (upper panel) and of the relative
importance of the dominant \textit{thermally averaged} cross sections,
$\sigmavof{i}/\sigmavof{\mathrm{tot}}$, at $x=30$ (lower panel) for
the pMSSM scenario associated with $\mstauone = 130\ \GeV$,
$\tanb=50$, $\mA=M_S=M_{3}=-A=1\ \TeV$, and $6M_1=M_2=\mu=1\ \TeV$.
In this scenario, $m_{\hhiggs}$ stays in the range $117-119\ \GeV$ and
the lightest neutralino is bino-like with a mass of
$m_{\neutralino}=169\ \GeV$.  Stau annihilation into heavy Higgses
remains kinematically forbidden.  The curves in the lower panel are
associated with stau annihilation into $\hhiggs\hhiggs$, $WW$,
$\tau\tau$, $ZZ$, $\gamma\gamma$, and $\gamma Z$ (at
$\thetastau=80^\circ$, from top to bottom).  As is evident, the
annihilation into $\hhiggs\hhiggs$ is enhanced already well before
$\thetastau=\pi/4$. At the peak position,
$\sigma_{\hhiggs\hhiggs}v_{\mathrm{r}}\simeq 8.8\times 10^3\
\mathrm{pb}$ for $\peff = 10\ \GeV$ (no thermal average), which is
still three orders of magnitude below the unitarity bound for
inelastic s-wave annihilation,
$\sigma_{\mathrm{u}}v_{\mathrm{r}}=8\pi/(\mstauone\peff)$
\cite{Griest:1989wd,Berger:2008ti}. Also the cross sections for stau
annihilation into $WW$ and $ZZ$ are strongly enhanced towards
$\thetastau=\pi/4$ since the $s$-channel contribution of
$\stauone\stauone\rightarrow {\hhiggs}^*\rightarrow VV$ becomes very
important. At their respective peak positions,
$\sigma_{WW}v_{\mathrm{r}}\simeq 250\ \mathrm{pb}$ and
$\sigma_{ZZ}v_{\mathrm{r}}\simeq 130\ \mathrm{pb}$ for $\peff = 10\
\GeV$. (Because of the dominance of the $\hhiggs\hhiggs$ channel, the
corresponding maxima do not show up in Fig.~\ref{Fig:multi2} where
$\sigmavof{i}/\sigmavof{\mathrm{tot}}$ is shown.) By the same token,
the cross sections of all (kinematically allowed) channels with a
fermion-antifermion final state (e.g.\ $\tau\overline{\tau}$)---which
are subdominant in the scenario considered in
Fig.~\ref{Fig:multi2}---experience an enhancement for
$\thetastau\rightarrow \pi/4$.  In total, there is an enhancement of
$\sigmavof{\mathrm{tot}}$ that delays the thermal freeze out of the
staus significantly, i.e., $x_{\mathrm f}\simeq 33$ for
$\thetastau\simeq\pi/4$.  As can be seen in the upper panel of
Fig.~\ref{Fig:multi2}, the decoupling yield is thereby reduced
dramatically down to a minimum value of $\Ystau=7.4\times 10^{-16}$
for maximal left--right mixing of the staus.%

\begin{figure}[t]
\begin{center}
\includegraphics[width=0.75\textwidth]{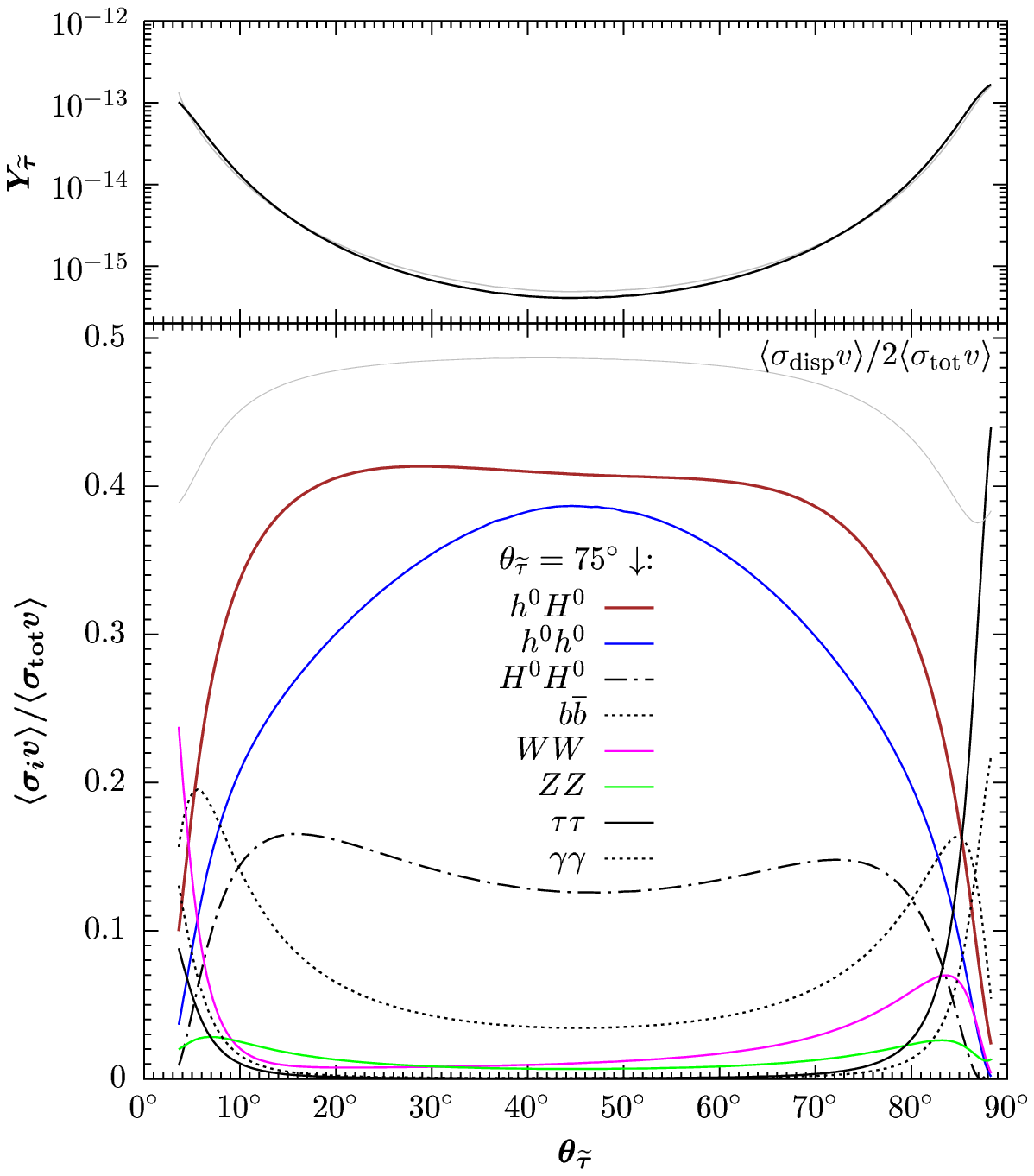} 
\caption[Enhanced annihilation into heavy Higgses]{\small Analogous to
  Fig.~\ref{Fig:multi2} but for the pMSSM scenario associated with
  $\mstauone=150\ \GeV$, $\tanb=50$, $\mA=130\ \GeV$, $M_S=M_{3}=-A=1\
  \TeV$, $3M_1=M_2=\mu=1\ \TeV$ and for $x=30$.  The stau decoupling
  yield reaches its minimum value of $\Ystau = 4.1\times 10^{-16}$ at
  $\thetastau=\pi/4$. The displayed stau annihilation channels are
  associated with the following final states: $\hhiggs\Hhiggs$,
  $\hhiggs\hhiggs$, $\Hhiggs\Hhiggs$, $b\overline{b}$, $WW$, $ZZ$,
  $\tau\tau$, and $\gamma\gamma$ (at $\thetastau=75^\circ$, from top
  to bottom). For an optimized presentation of those channels, the
  line indicating the relative importance of the sum of the displayed
  cross sections is scaled down by a factor of 1/2:
  $\sigmavof{\mathrm{disp}}/2\sigmavof{\mathrm{tot}}$.  No lines are
  shown for $\thetastau < 4^\circ$ where
  $m_{\widetilde{\nu}_{\tau}}<\mstauone$.}
\label{Fig:low-mA}
\end{center}
\end{figure}
\afterpage{\clearpage}

In the previous pMSSM examples, annihilation into final states
containing heavy Higgs bosons is kinematically forbidden. We can allow
for those channels by reducing the input value $\mA$. Indeed,
scenarios in which all Higgs bosons are very light in conjunction with
large $\tanb$ have been studied in the literature, see,
e.g.,~\cite{Boos:2002ze,Boos:2003jt} and references therein.  We thus
consider now the following pMSSM scenario: $\mA=130\ \GeV$,
$\mstauone=150\ \GeV$, $\tanb=50$, $M_S=M_{3}=-A=1\ \TeV$, and
$3M_1=M_2=\mu=1\ \TeV$.  In Fig.~\ref{Fig:low-mA}, the associated
$\thetastau$-dependence of $\Ystau$ and of
$\sigmavof{i}/\sigmavof{\mathrm{tot}}$ at $x=30$ for the now dominant
channels is shown in a similar way as in Fig.~\ref{Fig:multi2}; only
the relative importance of the sum of the displayed cross sections is
scaled down by a factor of 1/2,
$\sigmavof{\mathrm{disp}}/2\sigmavof{\mathrm{tot}}$, to allow for an
optimized presentation of the single dominant channels.  Throughout
the considered $\thetastau$ range, the masses of both CP-even Higgs
bosons are relatively light and remain rather constant:
$\mh=(118\pm1.5)\ \GeV$ and $\mH=(128.5\pm1)\ \GeV$.  Here the
dominant annihilation channels are associated with the following final
states: $\hhiggs\Hhiggs$, $\hhiggs\hhiggs$, $\Hhiggs\Hhiggs$,
$b\overline{b}$, $WW$, $ZZ$, $\tau\tau$, and $\gamma\gamma$ (at
$\thetastau = 75^\circ$, from top to bottom).  As can be seen, stau
annihilation into $\hhiggs\Hhiggs$ is now more dominant than the one
into $\hhiggs\hhiggs$ and also the $\Hhiggs\Hhiggs$ channel becomes
important, where each of those channels is indeed associated with an
(absolute) annihilation cross section $\sigmavof{i}$ that peaks at
$\thetastau=\pi/4$.  Also the annihilation into $b\overline{b}$ is
significant---a process which we will discuss in detail in the
following section. In this respect, one should stress that all
processes with $s$-channel \Hhiggs\ exchange are here less suppressed
by $\mH^2$ in the respective propagator than in the previously
considered scenarios.  Note that the asymmetry of
$\sigmavof{i}/\sigmavof{\mathrm{tot}}$ of those dominant channels
($\hhiggs\Hhiggs$, $\hhiggs\hhiggs$, $\Hhiggs\Hhiggs$,
$b\overline{b}$) with respect to a reflection at $\thetastau=\pi/4$ is
dominantly caused by the $\thetastau$-dependent modulation of the $WW$
channel.  As in the pMSSM scenario considered in
Fig.~\ref{Fig:multi2}, there is again an significant enhancement of
$\sigmavof{\mathrm{tot}}$ that delays the stau freeze out such that
$x_{\mathrm f}\simeq 33$ at $\thetastau\simeq\pi/4$.  Thereby, the
efficient annihilation into final state Higgses is accompanied by a
significant drop in \Ystau\ down to $\Ystau= 4.1\times 10^{-16}$ at
$\thetastau=\pi/4$ as can be seen in Fig.~\ref{Fig:low-mA}. At this
minimum, there is a $20\%$ disagreement between $\Ystau$ from our
calculation of stau decoupling (solid line) and the
\texttt{micrOMEGAs} result $\YstaumO$ (thin gray line) which is a
consequence of the different treatments of the Higgs sector described
above.

Let us finally remark that the Higgs couplings to fermions and vector
bosons as well as the Higgs self-couplings develop a strong dependence
on $\mA$ once we leave the decoupling regime ($\mA\lesssim 200\
\GeV$); for a comprehensive review see, e.g.,
Ref.~\cite{Djouadi:2005gj}.%
\footnote{The Higgs sector is also particularly sensitive to the
  mixing in the stop sector. In the considered pMSSM scenarios,
  $|X_t|\equiv |A_t-\mu\cot{\beta}|\sim M_S$ which corresponds to the
  ``typical-mixing scenario''~\cite{Carena:1999xa}.}
Changes in $\mA$ can therefore be accompanied by shifts in the
relative importance of the corresponding annihilation cross sections.
This underlines the fact that the details in the Higgs sector may very
well be crucial for the determination of the relic abundance of a
long-lived $\stauone$.

\section{Resonant stau annihilation}
\label{sec:reson-annih}
\begin{figure}[t]
\begin{center}
\includegraphics[width=0.75\textwidth]{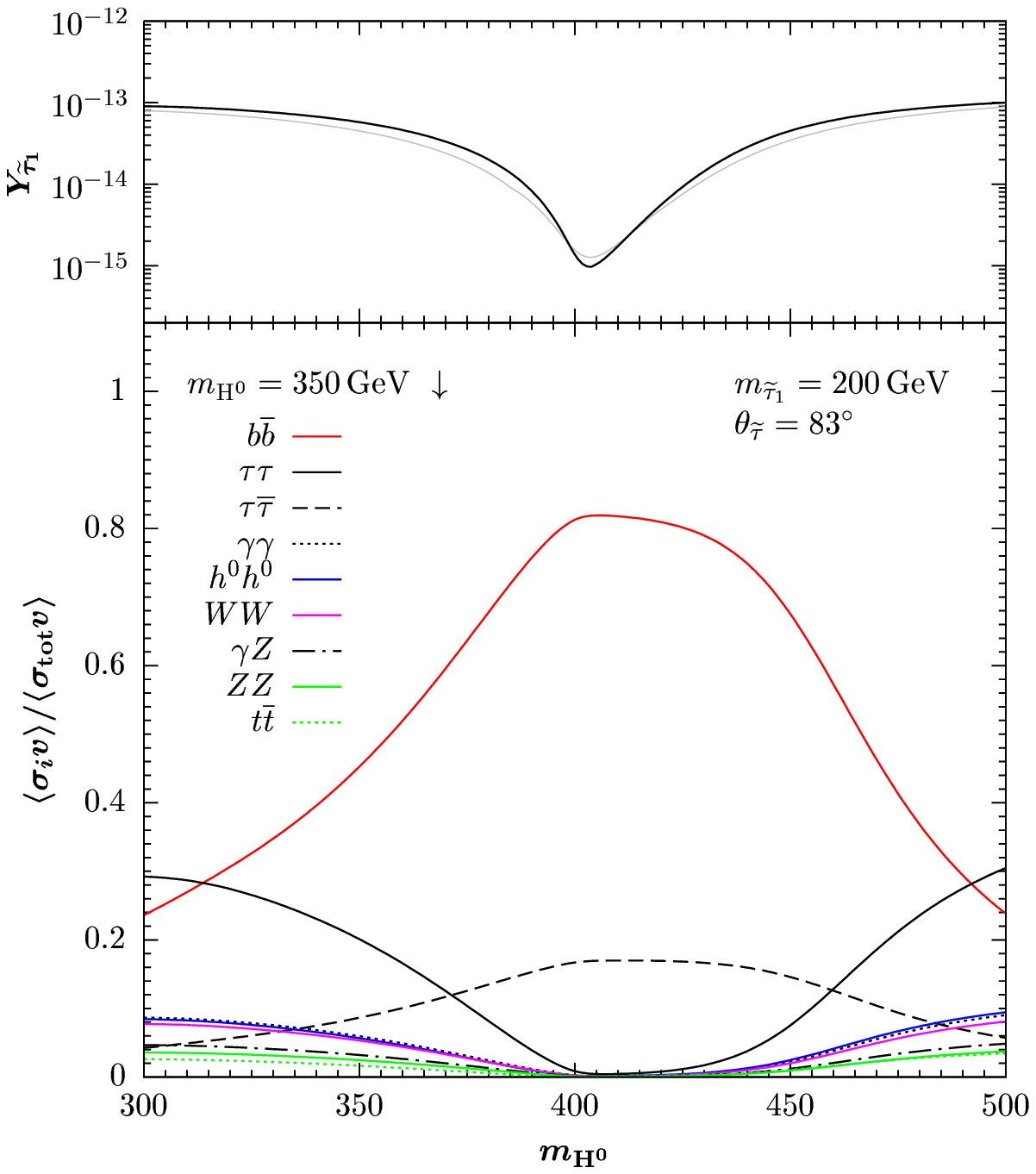} 
\caption[Resonant stau annihilation]{\small
  Dependence of $\Ystau$ (upper panel) and of
  $\sigmavof{i}/\sigmavof{\mathrm{tot}}$ at $x=25$ (lower panel) on
  $\mH$ for the pMSSM scenario associated with $\mstauone=200\ \GeV$,
  $\thetastau=83^\circ$, $\tanb=40$, and $-A=\mu=4M_1=M_{2,3}=M_S=1\
  \TeV$. In the upper panel, the dark line shows the stau yield
  $\Ystau$ obtained with our relic abundance calculation and the thin
  gray line the one obtained with \texttt{micrOMEGAs}. The stau
  decoupling yield takes on its minimum value of $\Ystau = 9.7\times
  10^{-16}$ at $\mH=404~\GeV$. In the lower panel, the displayed
  dominant stau annihilation channels are associated with the
  following final states: $b\overline{b}$, $\tau\tau$,
  $\tau\overline{\tau}$, $\gamma\gamma$, $\hhiggs\hhiggs$, $WW$,
  $\gamma Z$, $ZZ$, and $t\overline{t}$ (at $\mH=350\ \GeV$, from top
  to bottom).}
\label{Fig:res}
\end{center}
\end{figure}
\afterpage{\clearpage}

By inspection of Table~\ref{tab:ann-channels} it becomes clear that
primordial stau annihilation can also proceed resonantly via
$s$-channel exchange of the heavy CP-even Higgs boson~$\Hhiggs$ for
$\mH\simeq 2\mstauone$. While the LEP bound on the stau mass
$\mstauone \gtrsim 82\ \GeV$~\cite{Yao:2006px} forbids \hhiggs\ to
become on-shell ($\mh^{\mathrm{max}} \sim 140~\GeV$, e.g.,
\cite{Djouadi:2005gj}), the $s$-channel exchange of \Ahiggs\ is
absent%
\footnote{Even in absence of SUSY-induced CP violation, resonant
  annihilation via $\Ahiggs$-exchange may still proceed through
  $\stauone$-$\stautwo$ coannihilation. However, this scenario
  requires considerable fine-tuning in the stau mass-squared matrix
  since $\stauone$ and $\stautwo$ have to be nearly degenerate.}
because of \mbox{$C[\stauone,\stauone^*,\Ahiggs] = 0$} (see
Sec.~\ref{sec:enhanc-coupl-higgs}).  Again, our choice to work in the
framework of the pMSSM proves to be very helpful. Since the $\Hhiggs$
resonance occurs for $2\mstauone\simeq m_{\Hhiggs}$, one runs quickly
into the decoupling limit in which $\mH$ is governed by the input
parameter $\mA$ according to the simple
relation~(\ref{eq:DL-higgs-masses}). This allows us to scan through
the resonance easily.

Let us explore resonant stau annihilation by considering the exemplary
pMSSM scenario associated with $\mstauone=200\ \GeV$,
$\thetastau=83^\circ$ (i.e., a mostly `right-handed' $\stauone$),
$\tanb=40$, and $-A=\mu=4M_1=M_{2,3}=M_S=1\ \TeV$, for which we vary
$\mA$ (and thereby $\mH$) to scan through the resonance.
Figure~\ref{Fig:res} shows the resulting $\mH$-dependence of $\Ystau$
(upper panel) and of $\sigmavof{i}/\sigmavof{\mathrm{tot}}$ at $x=25$
for the dominant annihilation channels (lower panel). Those channels
are now associated with the following final states: $b\overline{b}$,
$\tau\tau$, $\tau\overline{\tau}$, $\gamma\gamma$, $\hhiggs\hhiggs$,
$WW$, $\gamma Z$, $ZZ$, and $t\overline{t}$ (at $\mH=350\ \GeV$, from
top to bottom).  In Table~\ref{tab:ann-channels} all resonant channels
can be identified.  Close to the resonance condition $2\mstauone\simeq
m_{\Hhiggs}$, the most important processes are stau annihilation into
$b\overline{b}$ and $\tau\overline{\tau}$. This is because the
couplings of those final state fermions to $\Hhiggs$ are $\tanb$
enhanced: for $\tanb\gg 1$, the \mbox{$f$$\overline{f}$$\Hhiggs$}
coupling $\sim m_{f} \sbma \tanb$ with $f=b,\tau$~\cite{Haber:1997dt}.
The (broad) peak associated with the resonance%
\footnote{Notice that we plot $\sigmavof{i}/\sigmavof{\mathrm{tot}}$
  so that the actual shape of the resonance looks somewhat different.}
already builds up for $\mH>2\mstauone=400\ \GeV$.  At zero relative
velocity, this would be a region in which the $\Hhiggs$ resonance
cannot occur. However, since $\stauone$ is in kinetic equilibrium at
the time of freeze out, resonant annihilation takes place already for
$2\mstauone<\mH$~\cite{Griest:1990kh}. For $\mH<2\mstauone=400\ \GeV$,
the processes containing $s$-channel $\Hhiggs$ exchange proceed with a
slightly faster rate (if kinematically allowed).  The impact of the
$\Hhiggs$ resonance on the thermal $\stauone$ freeze out and the
resulting $\Ystau$ is substantial.  Since the total width of $\Hhiggs$
is $\Gamma_{\Hhiggs} = (6-10)\ \GeV$ for $m_{\Hhiggs}= (300-500)\
\GeV$ in the considered pMSSM scenario, the reduction of $\Ystau$
extends over a relatively large $\mH$ range. In this regard, note that
$\Gamma_{\Hhiggs} $ could be substantially larger had we not
essentially decoupled all sfermions---except $\stauone$, $\stautwo$,
and $\widetilde{\nu}_{\tau}$---by choosing $M_S=1\ \TeV$.  For
$\mH\simeq 404~\GeV$, i.e., at the dip of the resonance, we find
$x_{\mathrm f}\simeq 33$ and a minimum stau decoupling yield of
$\Ystau=9.7\times 10^{-16}$ (dark line).  Thus, despite the (still)
moderate value of $\tanb=40$, a significant reduction of $\Ystau$ is
encountered. Indeed, $\Ystau$ can be even further suppressed for a
larger value of $\tanb$.  Let us remark that an accurate determination
of $\Ystau$ in the resonance region requires to take special care of
the $b\overline{b}\Hhiggs$ vertex.  This coupling is well known to
receive substantial radiative corrections for sizable values of
$\tanb$.  Therefore, we rely again on the computer tool
\texttt{FeynHiggs} to compute all quark--antiquark--Higgs couplings
and the total width $\Gamma_{\Hhiggs}$.  Also the \texttt{micrOMEGAs}
code takes special care of the $b\overline{b}\Hhiggs$ vertex.  We
therefore think that the difference between the yields shown in the
upper panel of Fig.~\ref{Fig:res} reflects the theoretical uncertainty
involved in the determination of $\Gamma_{\Hhiggs}$ as well as the
$b\overline{b}\Hhiggs$ vertex.

\section[On the viability of a $\stauone$-$\stauone^*$
asymmetry]{\texorpdfstring{On the viability of a
    \boldmath$\stauone$-$\stauone^*$ asymmetry}{On the viability of a
    stau-antistau asymmetry}}
\label{sec:comm-stau-stau}

Given the strong bounds on the abundance of negatively charged
$\stauone$ from bound-state effects during BBN, i.e., from CBBN of
$^6$Li and $^9$Be, it is natural to ask whether it is possible to have
an excess of positively charged $\stauone^*$'s over negatively charged
$\stauone$'s. The generation of a particle-antiparticle asymmetry
requires a departure from thermal equilibrium. Therefore, one might
think that a $\stauone$-$\stauone^*$ asymmetry can be produced at the
time of the stau freeze out if the (slepton number violating) process
$\stauone\stauone \rightarrow \tau\tau$ occurs at a different rate
than its conjugate counterpart.  Such a situation might indeed occur
if we allow for (CP-violating) complex values of the parameters
$\Atau$, $\mu$, and $M_{1,2}$ in the SUSY sector.  However, the staus
are still tightly coupled to Standard Model particles so that they
remain in kinetic equilibrium with the primordial plasma. Therefore,
any excess of $\stauone^*$ over $\stauone$ arising will be washed out
quickly by the inelastic scattering process
$\stauone^*\tau\leftrightarrow\stauone\overline{\tau}$.%
\footnote{Additional equilibrating processes are, e.g., $\stauone^*
  W^-\leftrightarrow \stauone W^+$ or
  $\stauone^*H^-\leftrightarrow\stauone H^+$, which are however
  Boltzmann-suppressed. Also note that a lepton asymmetry of the order
  of the baryon asymmetry is expected because of charge neutrality of
  the Universe; cf.~\cite{Caprini:2003gz} and references therein. }
Indeed, it is well-known~\cite{Griest:1990kh} that processes of the
latter type occur at much larger rates than the rates for the mutual
annihilation of the decoupling particle species. The same argument
given in~\cite{Griest:1990kh} can be adopted to our case. At the time
of freeze out, the reaction rates of interest can be estimated as
\begin{align}
  \label{eq:p1}
  \stauone\stauone \rightarrow \tau \tau\, :&\quad n_{\stauone}
  n_{\stauone} \sigma_{\stauone\stauone\rightarrow\tau \tau} \sim
  T^3 \mstauone^3 
  e^{-2\mstauone/T} \sigma_{\stauone\stauone\rightarrow\tau \tau} \, ,\\
  \label{eq:p2}
  \stauone^*\tau \rightarrow \stauone \overline{\tau}\, :& \quad
  n_{\stauone^*} n_{\tau} \sigma_{\stauone^* \tau \rightarrow \stauone
    \overline{\tau}} \sim T^{9/2} \mstauone^{3/2} e^{-\mstauone/T}
  \sigma_{\stauone^* \tau \rightarrow \stauone \overline{\tau}}\, ,
\end{align}
since $\stauone^{(*)}$ is approximately Boltzmann distributed. For
simplicity, we have treated the tau lepton $\tau$ as a (still)
relativistic species. By taking the ratio of~(\ref{eq:p2}) with
respect to~(\ref{eq:p1}),
\begin{align}
  \left( T/\mstauone \right)^{3/2} e^{\mstauone/T} \sim 10^9\quad
  \mathrm{for}\quad \mstauone/T \simeq 25\ ,
\end{align}
we find that the equilibrating process is by far more dominant. Here,
we have used that $\sigma_{\stauone\stauone\rightarrow\tau \tau}$ and
$\sigma_{\stauone^* \tau \rightarrow \stauone \overline{\tau}}$ are
not too different. In fact, both processes proceed at tree level
exclusively via $\widetilde{\chi}_i^0$ exchange so that one cannot
decouple (\ref{eq:p2}) from (\ref{eq:p1}) by a simple adjustment of
the neutralino mass spectrum.

\section{Exceptionally small abundances within the CMSSM}
\label{sec:annih-chann}

We have shown above that the total stau annihilation cross section can
be significantly enhanced. The thermal freeze out of $\stauone$'s is
thereby delayed such that their abundance prior to decay, \Ystau, is
suppressed.  In the following we focus on the CMSSM to see whether the
effects discussed in Sects.~\ref{sec:enhanc-coupl-higgs}
and~\ref{sec:reson-annih} do appear also in models in which the
pattern of soft-SUSY breaking parameters fulfills certain boundary
conditions at a high scale.  Note that we compute \Ystau\ with
\texttt{micrOMEGAs} in this section since coannihilation processes are
not included in our relic density code. In addition, we employ
\texttt{SPheno~2.2.3}~\cite{Porod:2003um} for the computation of the
mass spectrum and the low energy constraints associated with
$B(b\rightarrow s \gamma)$ and the anomalous magnetic moment of the
muon $a_{\mu}$.  Let us now proceed by discussing two exemplary CMSSM
parameter scans.

\begin{figure}[t]
\begin{center}
\includegraphics[width=0.65\textwidth]{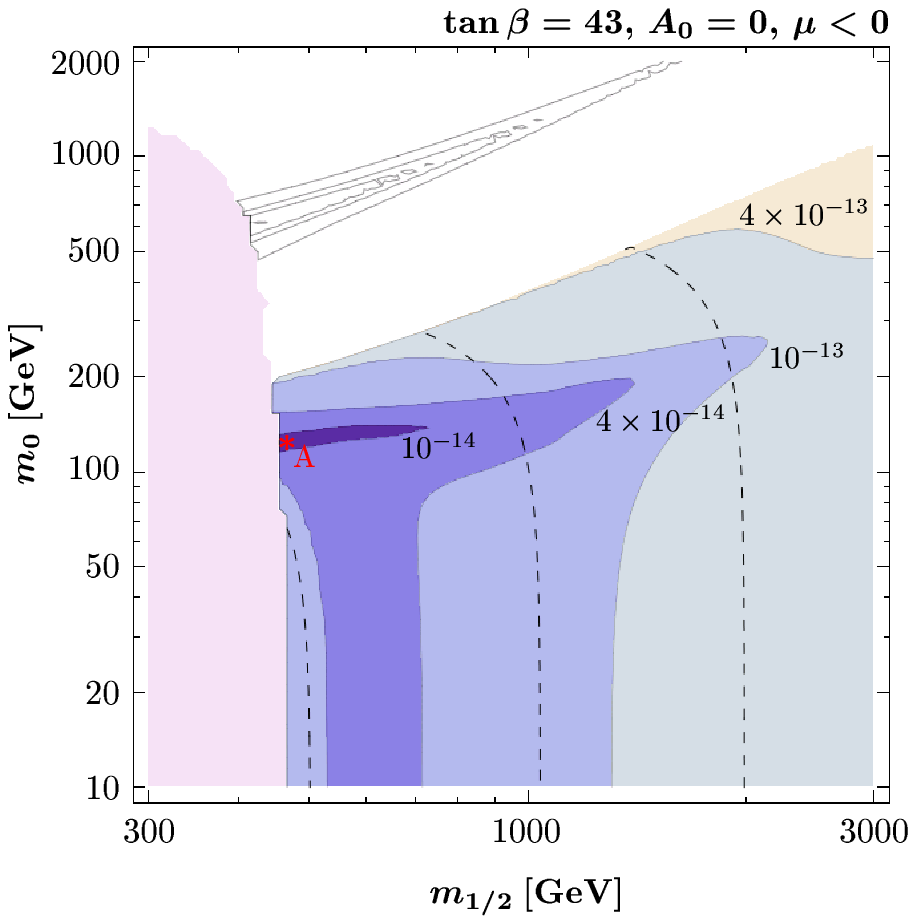}
\caption[Enhanced annihilation in the CMSSM for $\tanb=43$]{\small
  Contours of $\Ystau$ (as labeled) in the (\monetwo,\,\mzero) plane
  for $\tanb=43$, $A_0=0$, and $\mu<0$, where darker shadings imply
  smaller $\Ystau$ values.  The dashed lines are contours of
  $\mstauone=100$, $300$, and $600\ \GeV$ (from left to right).  The
  light-shaded region at $\monetwo \lesssim 450~\GeV$ is excluded by
  the LEP bound $\mh \le 114.4\ \GeV$~\cite{Yao:2006px}.  In the white
  area either $m_{\neutralino} < \mstauone$ or correct electroweak
  symmetry breaking is not established (in the very upper left
  corner), where the thin contours indicate the Higgs funnel in the
  $\neutralino$ NLSP region.  Table~\ref{tab:points} provides detailed
  information for the SUSY model represented by the point ``A'' that
  is indicated by the star.}
\label{Fig:cmssm1}
\end{center}
\end{figure}

Figure~\ref{Fig:cmssm1} shows contours of constant $\Ystau$ in the
(\monetwo,\,\mzero) plane for $\tanb =43$, $A_0 = 0$, and a negative
sign of the $\mu$ parameter.
The contour lines represent the values $\Ystau=10^{-14}$, $4\times
10^{-14}$, $10^{-13}$, and $4\times 10^{-13}$, where darker shadings
imply smaller values of $\Ystau$.
The dashed lines are contours of $\mstauone = 100$, $300$ and
$600~\GeV$ (from left to right). The light-shaded region at $\monetwo
\lesssim 450~\GeV$ is excluded by the mass bound $\mh \ge 114.4\ \GeV$
from Higgs searches at LEP~\cite{Yao:2006px}. The white area indicates
the region in which either correct electroweak symmetry breaking is
not established (in the very upper left corner) or in which
$m_{\neutralino}<\mstauone$. Since $\mu<0$, the plane is actually in
tension because of (negative) SUSY contributions
$a^{\mathrm{SUSY}}_\mu$ to the anomalous magnetic moment of the muon,
$a_\mu\equiv(g-2)_\mu/2$.

\begin{figure}[t]
\begin{center}
\includegraphics[width=0.65\textwidth]{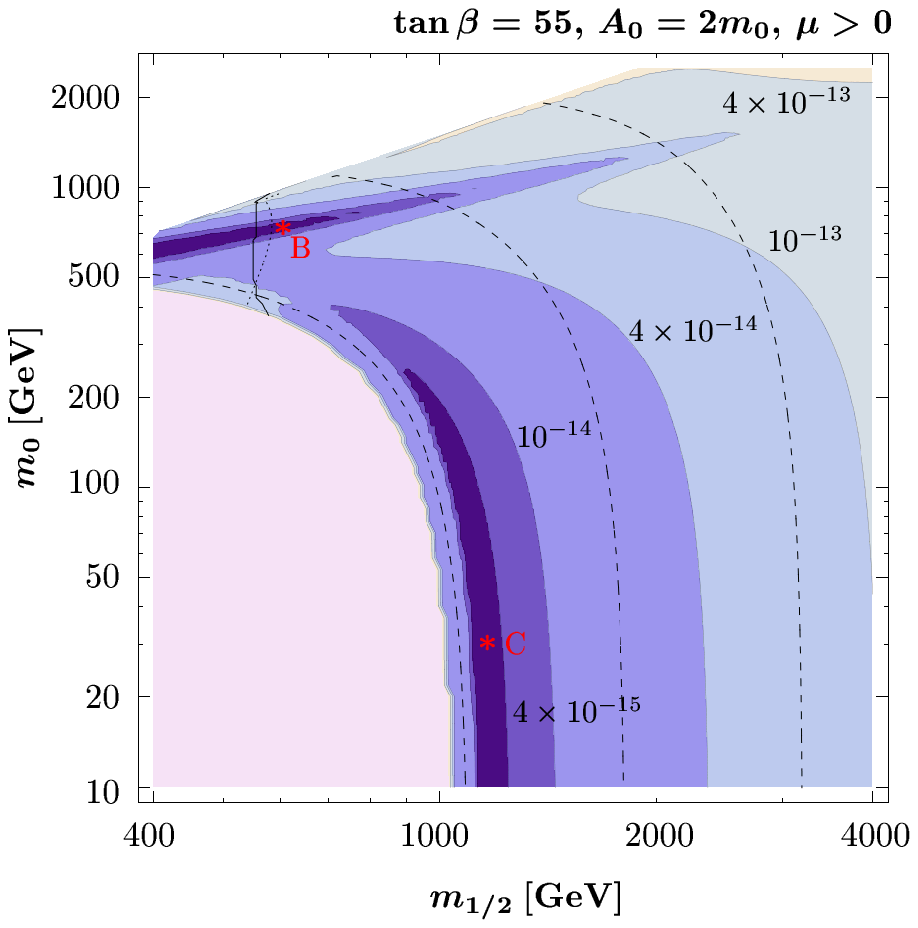}
\caption[Enhanced annihilation in the CMSSM for $\tanb=55$]{\small
  Contours of $\Ystau$ (as labeled) in the (\monetwo,\,\mzero) plane
  for $\tanb =55$, $A_0 = 2m_0$, and $\mu>0$, where darker shadings
  imply smaller $\Ystau$ values. The dashed lines are contours of
  $\mstauone = 100,\, 300,$ and $600\ \GeV$ (from left to right).  The
  large light-shaded region in the lower left corner is excluded by
  bounds from direct Higgs and SUSY searches (or by the appearance of
  a tachyonic spectrum). In the region to the left of the vertical
  solid and dotted lines, $\mh \le 114.4\ \GeV$~\cite{Yao:2006px} and
  $B(b \rightarrow s \gamma)\ge 4.84\times
  10^{-4}$~\cite{Mahmoudi:2007gd}, respectively.  In the white area,
  $m_{\neutralino} < \mstauone$. Table~\ref{tab:points} provides
  detailed information for the SUSY models represented by the stars
  ``B'' and ``C'' (as labeled).}
\label{Fig:cmssm2}
\end{center}
\end{figure}

Figure~\ref{Fig:cmssm2} presents a scan over the (\monetwo,\,\mzero)
plane for $\tanb =55$, $A_0 = 2m_0$, and $\mu>0$
with contours of $\Ystau=4\times 10^{-15}$, $10^{-14}$, $4\times
10^{-14}$, $10^{-13}$, and $4\times 10^{-13}$ (darker shadings
indicate smaller $\Ystau$ values) and $\mstauone=100$, $300$, and
$600~\GeV$ (dashed lines, from left to right). The large light-shaded
region in the lower left corner is excluded by the robust bound
$\mstauone\ge 82\ \GeV$~\cite{Yao:2006px} from collider searches of
charged sleptons (or by the appearance of a tachyonic spectrum).  The
LEP Higgs bound $\mh \le 114.4\ \GeV$~\cite{Yao:2006px} is situated
within this region in close vicinity to its boundary for
$\mzero\lesssim 400~\GeV$ and is indicated by the solid line for
$\mzero\gtrsim 400~\GeV$. In the region to the left of the dotted
line, $B(b\rightarrow s\gamma)\ge 4.84\times
10^{-4}$~\cite{Mahmoudi:2007gd}, which is in tension with the bounds
from inclusive $b\,\rightarrow\,s\gamma$ decays.

Let us now discuss some generic features of the stau yield within the
CMSSM on the basis of Figs.~\ref{Fig:cmssm1} and~\ref{Fig:cmssm2}.  We
note beforehand that our more general statements on the $\stauone$
NLSP region in the CMSSM are corroborated by a parameter scan over the
following range%
\footnote{Here, we disregard CMSSM parameter points in which \texttt{SPheno}
  flags an error in the spectrum calculation.}
\begin{alignat}{2}
  \label{eq:para-range}
  &\monetwo = (0.1 - 6)\ \TeV,\qquad &
  \tanb &= 2 - 60, \nonumber\\
   &-4 \mzero < A_{0} < 4 \mzero  ,
   &\mathrm{sgn\ \!}\mu &= \pm 1 .
\end{alignat}
In both figures an almost horizontal, narrow band of low $\Ystau$
appears in which $2\mstauone\simeq\mH$ holds so that stau annihilation
proceeds via resonant production of the heavy CP-even Higgs boson
\Hhiggs.
We have marked points the centers of the respective regions with ``A''
and ``B'' for which we provide detailed information in
Table~\ref{tab:points}.  Given a present uncertainty of $\sim 3\ \GeV$
in the determination of $m_{\hhiggs}$~\cite{Degrassi:2002fi}, we note
that the LEP Higgs bound has to be treated with some care. For
example, a (vertical) $m_{\hhiggs}=112\ \GeV$ contour would be
situated at $\monetwo\simeq 400\ \GeV$ in the resonance region of
Fig.~\ref{Fig:cmssm2}. Accordingly, one could consider the entire
resonance region shown to be compatible with direct Higgs searches.
However, due to the large value of $\tanb=55$, the bound on
$b\,\rightarrow\,s\gamma$ is very strong so that a large part of the
resonance region remains excluded by this constraint. In this regard,
it is interesting to see (Fig.~\ref{Fig:cmssm1}) that
$2\mstauone\simeq\mH$ also appears in the $\stauone$ NLSP region for
lower values of $\tanb$.  In the center of both resonance regions, the
yield becomes as low as $\Ystau= 4.2\times 10^{-15}$ (point A) and
$\Ystau=2.5\times 10^{-15}$ (point B). Despite the heavier mass of the
lighter stau (see Table~\ref{tab:points}), the suppression of $\Ystau$
is still more pronounced in Fig.~\ref{Fig:cmssm2} than in
Fig.~\ref{Fig:cmssm1}.  This is because the bottom Yukawa coupling
becomes larger with increasing $\tanb$, as discussed already in
Sec.~\ref{sec:reson-annih}. In fact, annihilation into
$b\overline{b}$ final states is in both cases by far the dominant
process with relative importance of $76\%$ (point A) and $87\%$
(point B). The extension of both resonance regions is due to the total
width of $\Hhiggs$ of respectively $\Gamma_{\Hhiggs}\simeq 9.6~\GeV$
(point A) and $\Gamma_{\Hhiggs}\simeq 22~\GeV$ (point B); note the
logarithmic scales in Figs~\ref{Fig:cmssm1} and~\ref{Fig:cmssm2}.  We
note in passing that the appearance of the $\Hhiggs$ resonance does
not imply the absence of the neutralino funnel region which is
indicated by the (unshaded) contour lines in the white area of
Fig.~\ref{Fig:cmssm1}

Of course, the question arises whether the appearance of the resonance
region is encountered more generically within the framework of the
CMSSM.
In principle, it is not easy to provide a simple quantitative
connection between $\mstauone$ and $\mH$ for arbitrary values of the
CMSSM parameters. However, without emphasis on an overall
applicability, a qualitative picture can be drawn. Let us start with
the mass of the CP-odd Higgs boson $\mA$ which can be written
as~\cite{Drees:1991mx,Drees:1995hj}
\begin{align}
  \label{eq:mA}
  \mA^2 \sim 1/\sin^{2}{\beta}\, (\mzero^2 + 0.52 \monetwo^2 + \mu^2  - \dots) .
\end{align}
Here, the ellipsis stand for contributions from the bottom and tau
Yukawa couplings. For $\tanb\gtrsim 20$, $\mA^2 \sim
\mzero^2+2.5\monetwo^2-\dots$, and the corrections from the bottom and
tau Yukawa couplings become important so that $\mA$ is driven towards
lower values;%
\footnote{The latter relation ignores contributions from $A$-terms
  which can be important but complicate the envisaged illustrative
  picture; for the derivation, we have used
  $m_{\mathrm{t}}(m_{\mathrm{t}})=163\ \GeV$ in Eq.~(2.25a) of
  Ref.~\cite{Drees:1995hj}.}
note that $\sin^{2}{\beta}\simeq1$ for $\tanb\gtrsim 20$. Indeed, this
property can be used to constrain $\tanb$ from above by confronting
$\mA$ with the lower bound from LEP, $\mA>93.4\
\GeV$~\cite{Yao:2006px}. On the other hand, for large $\monetwo$, one
also enters the decoupling limit of the MSSM so that $\mA$ and $\mH$
will be nearly degenerate in mass; cf.~(\ref{eq:DL-higgs-masses}).
This can be also seen from the exemplary points presented in
Table~\ref{tab:points}. Therefore, also $\mH$ will be driven towards
lower values for growing $\tanb$.  Now, left-right mixing of the
lighter stau for not too large values of $\tanb$ is small within the
CMSSM, $\stauone\simeq\stauR$, so that approximately $\mstauone^2 \sim
\mzero^2 + 0.15\monetwo^2$~\cite{Drees:1995hj}.  Therefore,
$2\mstauone <\mH$ is the relation that holds usually in the region in
which $\stauone$ is the lightest Standard Model superpartner.
However, for large $\tanb$, the contributions from the bottom Yukawa
coupling in~(\ref{eq:mA}) can become strong enough (growing with
$\mzero$~\cite{Drees:1995hj}) to overcome any additional decrease of
$\mstauone$ due to left-right mixing so that the resonance condition
$2\mstauone\simeq\mH$ can indeed be met. Nevertheless, from scanning
over the CMSSM parameter range (\ref{eq:para-range}) it seems to us
that the resonance condition $2\mstauone\simeq\mH$ is not easily
realized in the part of the $\stauone$ NLSP region in which
$\stauone$-$\neutralino$ coannihilations are negligible.  Conversely,
it is clear that relaxing the universality conditions for the soft-SUSY
breaking masses at $\mgut$ will make it easier to find parameter
regions in which the resonance condition $2\mstauone\simeq\mH$ is
satisfied. Of particular interest in this respect is the model with
non-universal Higgs masses (NUHM) with $m_{H_1}\neq m_{H_2} \neq m_0$
at $\mgut$. There, one can adjust the input parameters in order to
realize resonant stau annihilation.  Indeed, this model is
qualitatively the same as the class of pMSSM scenarios considered in
the previous sections, where $m_{H_1}$ and $m_{H_2}$ are traded (at
the low-scale) against $\mA$ and $\mu$ by using the electroweak
symmetry breaking conditions.

Low $\Ystau$ values are also realized in the narrow vertical region
around $\monetwo\sim 1.1\ \TeV$ in Fig.~\ref{Fig:cmssm2}. At the
representative point ``C'' of that region, $\Ystau = 2.2\times
10^{-15}$ and the main stau annihilation channels are the ones into
$\hhiggs\hhiggs$ ($90\%$) and $WW$ ($6\%$); see
Table~\ref{tab:points}.  For larger values of $\monetwo$, \Ystau\
exhibits its well known behavior and grows with $\mstauone$. To the
left of the $\Ystau=4\times 10^{-15}$ contour, the yield increases
quickly since the annihilation into $\hhiggs\hhiggs$ becomes
kinematically forbidden. Indeed, regions of low $\Ystau$ which are due
to the aforementioned annihilation channels are a commonplace
appearance in the CMSSM parameter space.  They are found slightly
above the lowest feasible values of $\monetwo$, i.e., close to the
boundary of the region which is excluded by direct Higgs and SUSY
searches and where $\mstauone>\mh$ still holds.  This is because
$\stauone$ is light in that region since the SUSY particle spectrum
scales with $\monetwo$ (typically, $m_0\ll\monetwo$ for $\stauone$
NLSP).  Moreover, we find that the LEP Higgs bound drops hardly below
$\monetwo\simeq 450\ \GeV$ for $\tanb\gtrsim 40$ and $\mzero\lesssim
100\ \GeV$.%
\footnote{The position of the LEP Higgs bound (which appears as a near
  to vertical line for low \mzero) is very sensitive to the value of
  $\mtop$. Lowering $\mtop$ shifts the bound towards larger values of
  \monetwo.}
Due to a strong correlation between the gaugino mass parameter
$\monetwo$ and the size of the $\mu$ parameter, $\mu^2\sim (1- 3)\,
\monetwo^2$~\cite{Carena:1994bv}, the value of $\mu$ in the
experimentally allowed region is large.  Recall from
Sec.~\ref{sec:enhanc-coupl-higgs} that the $\stauone\stauone\hhiggs$
coupling is $\sim\sin{2\thetastau}\Xtau$ ($\mA\gg \mZ$) so that
$|\Xtau|=|\Atau-\mu\tanb|$ will become sizeable by increasing $\tanb$.
This leads then to efficient stau annihilation into $\hhiggs\hhiggs$
final states.
Indeed, in those CMSSM regions, also $|\sin{2\thetastau}|$ is
maximized so that $\Ystau$ already starts to drop below the estimate
(\ref{Eq:Yslepton})  for $\tanb\gtrsim 40$.
Note, however, that the left-right mixing of $\stauone$ within the
CMSSM is somewhat constrained. Neglecting $\tau$-Yukawa contributions,
the RG-evolution induced splitting reads $\mstauL^2-\mstauR^2\sim
0.37\monetwo^2$~\cite{Drees:1995hj} and indeed $\stauone$ remains
mainly right-handed: By scanning over the parameter space, we
typically find $65^\circ\lesssim\thetastau\lesssim 115^\circ$ and thus
$|\sin{2\thetastau}|\lesssim 0.75$ in the $\stauone$ NLSP region in
which $\mstauone>\mh$ and $\mh>114.4\ \GeV$ holds.

\begin{table}[tb] 
  \caption[Parameters of exemplary CMSSM parameter points]{\small Exemplary CMSSM points A, B, and C shown in Figs.~\ref{Fig:cmssm1} and \ref{Fig:cmssm2}. In addition to the quantities explained in the main text, values of the gluino mass $m_{\widetilde{g}}$ and of the mass of the lighter stop $m_{\widetilde{t}_1}$ are given together with the relative importance of the dominant stau annihilation channels, $x_{\mathrm{f}}=\mstauone/\Tf$, and the decoupling yield $\Ystau$. For each point, we list gravitino dark matter scenarios with $\mgr = 100\ (50)\ \GeV$ and associated values of the stau lifetime $\taustau$, the non-thermally produced gravitino density $\Omega_{\gravitino}^{\NTP}h^2$, and the maximum reheating temperature $\TR^{\mathrm{max}}$.}
\label{tab:points} 
\begin{center}
\begin{tabular}{l@{}l@{\hspace*{0.5cm}}ccc}
\toprule
Point & & A & B & C  \\
\midrule
\monetwo & $[\GeV]$           &   456     &   600   &  1138   \\ 
\mzero &$[\GeV]$              &   124     &   748   &  30     \\
\tanb  &                      &   43      &   55    &  55     \\
\midrule
\mstauone &$[\GeV]$           &   130     &   197   &  127    \\
\mstautwo &$[\GeV]$           &   352     &   673   &  739    \\
\thetastau &                  &   114     &   80    &  75     \\
\midrule
$m_{\hhiggs}$ & $[\GeV]$        &   114.6     &   115   &  117.9   \\ 
$m_{\Hhiggs\! , \Ahiggs}$ &  $[\GeV]$ &   265     &   390   &  799     \\
$\Gamma_{\Hhiggs}$& $[\GeV]$    &   9.6      &   22    &  41     \\
$\mu$& $[\GeV]$               &   -565     &   666   &  1262    \\
\Atau & $[\GeV]$              &   -63     &   473   &  -164    \\
\midrule
$m_{\widetilde{g}}$& $[\GeV]$     &   1052       &   1375  &  2446   \\ 
$m_{\widetilde{t}_1}$ & $[\GeV]$   &   740       &   1091  &  1757   \\
\midrule
$b\overline{b}$   &   $[\%]$   &   76    &   87   &   $<1$   \\ 
$\hhiggs\hhiggs$  &   $[\%]$  &   10    &   $<1$   &   90   \\ 
$\tau\overline{\tau}$ & $[\%]$  &   9    &    11   &   $<1$   \\ 
$WW$   &   $[\%]$             &   2    &   $<1$   &   6   \\ 
$x_{\mathrm{f}}$ &            &   30      &   30    &  32   \\ 
\Ystau & $[10^{-15}]$         &   4.2    &   2.5   &   2.2   \\
\midrule
\mgr & $[\GeV]$               &   100    &   100    &   100    \\
     &                        &   (50)    &   (50)    &  (50)    \\[0.1cm]
\taustau  & $[\mathrm{s}]$ &  $5.7\times 10^9$  &  $6.5\times 10^7$  &  $8.5\times 10^9$  \\
          &                & $(7.5\times 10^7)$ & $(6.4\times 10^6)$ & $(8.7\times 10^7)$ \\[0.1cm]
$\Omega^{\mathrm{NTP}}_{\widetilde{G}}h^2$ & $[10^{-4}]$ & $1.2$ &  $0.7$ & $0.64$ \\
                                        &&   ($0.58$)  & ($0.35$) &  ($0.32$) \\[0.1cm]
$\TR^{\mathrm{max}}$ & $[\GeV]$ & $1.9\times 10^9$ &  $1.1\times 10^9$          &  $3.1\times 10^8$ \\
                     &          & ($9.5\times 10^8$)       & ($5.5\times 10^8$) & ($1.5\times 10^8$)    \\[0.1cm]
\bottomrule

\end{tabular}
\end{center}
\end{table}

\afterpage{\clearpage}

\section{Prospects for collider phenomenology}
\label{sec:collider}

If a SUSY model with a long-lived $\stauone$ of $\mstauone<0.7~\TeV$
is realized in nature, the $\stauone$ discovery potential will be
promising at the LHC with a luminosity of
100~fb$^{-1}$~\cite{Feng:2004mt}.
For $\mstauone<0.25~\TeV$ (0.5~TeV), $\stauone$'s can also be examined
in precision studies at the ILC with a c.m.\ range up to
$\sqrt{s}=0.5~\TeV$ (1~\TeV).
Once long-lived $\stauone$'s are produced, one should be able to
distinguish them from muons by considering the associated highly
ionizing tracks and with time-of-flight measurements.
One should then also be able to infer $\mstauone$ from measurements of
the $\stauone$ velocity and its momentum~\cite{Ambrosanio:2000ik} and
complementary from (threshold) studies of the process
$\positron\electron\to\stauone\stauone^*$ at the ILC.

Both mechanisms leading to exceptionally small $\Ystau$ values come
with testable predictions: certain ranges of the stau-mixing angle
$\thetastau$ together with large values of $\tanb$, $|\mu|$, and/or
$|\Atau|$ and, in the case of resonant stau annihilation, also
$\mH\simeq 2\mstauone$.
In particular, the large stau-Higgs couplings lead to an enhanced
production of light Higgs bosons in association with staus via
$\positron\electron\to\stauone\stauone^*\hhiggs$ and
$\gamma\gamma\to\stauone\stauone^*\hhiggs$.  The associated cross
sections can then be relatively large at the ILC with a sufficiently
high c.m.\ energy~\cite{Datta:2001sh}.  In addition, the above
reactions with $\Hhiggs$ instead of $\hhiggs$ in the final state can
have also relatively large cross sections if $\Hhiggs$ and $\stauone$
are sufficiently light.  These reactions will then allow for an
experimental determination of the stau-Higgs couplings and clarify
whether its values are compatible with an extremely small value of
$\Ystau$~\cite{Datta:2001sh}.  Moreover, a measurement of $\mH$
pointing to $\mH\simeq 2\mstauone$ could be an experimental hint for
resonant stau annihilation in the early Universe.

Indeed, the scenarios considered could allow for a determination of
both $\mh$ and $\mH$ already at the LHC.  Because of the large values
of $\tanb$, the dominant production mechanism for $\hhiggs/\Hhiggs$
will be the associated production of the neutral Higgs bosons with
bottom quark pairs, $\proton\proton\to
\bquark\antibquark\hhiggs/\Hhiggs$; see,
e.g.,~\cite{Dittmaier:2003ej,Dawson:2003kb,Harlander:2003ai} and
references therein.  In fact, associated
$\bquark\antibquark\hhiggs/\Hhiggs$ production with
$\hhiggs/\Hhiggs\to\mu^+\mu^-$ is considered as one of the most
promising processes for measurements of $\mH$ at the LHC despite the
relatively small $\hhiggs/\Hhiggs\to\mu^+\mu^-$ branching
ratio~\cite{Ball:2007zza}.  In SUSY scenarios with a sufficiently
light long-lived $\stauone$ NLSP, these processes will be complemented
by associated $\bquark\antibquark\hhiggs/\Hhiggs$ production with
$\hhiggs/\Hhiggs\to\stauone\stauone^*$, where measurements of the
invariant mass of the $\stauone\stauone^*$ pair could potentially
provide a unique way to infer $\mh$ and $\mH$ at the LHC. In fact,
$\hhiggs/\Hhiggs\to\stauone\stauone^*$ will occur most prominently
exactly in the regions associated with the exceptional $\Ystau$ values
due to the enhanced stau--Higgs couplings. Having outlined these
proposals, we leave a dedicated study for future work.

Table~\ref{tab:points} illustrates that the kinematical reach of both
the LHC and the ILC could be sufficiently large to allow for the
studies mentioned above.  In none of the given points does $\mstauone$
exceed $200~\GeV$ so that $\stauone\stauone^*$ pair production would
already be possible at the ILC with $\sqrt{s}\leq 0.5~\TeV$.  There,
one could also produce $\stauone\stauone^*\hhiggs$ final states in
scenarios A and C.  Even the condition $\mH\simeq 2\mstauone$ could be
probed in both scenarios A and B that allow for resonant stau
annihilation.

\section{Implications for gravitino dark matter scenarios}
\label{sec:gravitino}

We have seen in this thesis that $\Ystau$ is subject to stringent
cosmological constraints. Indeed, to decide on the cosmological
viability of a SUSY model, one has to confront the associated $\Ystau$
values with those constraints. For gravitino LSP scenarios with
unbroken R-parity, we have obtained restrictive cosmological
constraints in Part~\ref{part:two}.  In particular, in
Sects.~\ref{sec:typical-stau-abundance},
\ref{sec:lower-limit-monetwo}, and \ref{sec:upper-bound-TR} we have
derived constraints and implications thereof under the assumption that
$\Ystau$ can be described by~(\ref{eq:Ystau7}).  
However, while (\ref{eq:Ystau7}) is quite reliable for
$\stauone\simeq\stauR$~\cite{Asaka:2000zh,Fujii:2003nr,Pradler:2006hh,Berger:2008ti},
we have shown in the previous sections that $\Ystau$ (for a given
$\mstauone$) can be about two orders of magnitude smaller
than~(\ref{eq:Ystau7}). 

Generally speaking, in this chapter we have shown that islands exist
in which $\Ystau$ can be significantly below~(\ref{eq:Ystau7}) even
within the CMSSM and for a standard cosmological history.  Thus, in
gravitino dark matter scenarios with such exceptionally small $\Ystau$
values, our understanding of the cosmological constraints and the
associated implications could change significantly.
To demonstrate this point, let us indicate for which $\Ystau$ values
the existing cosmological constraints are respected:
\begin{itemize}
\item For $\Ystau<10^{-14}$, the upper limit on $\Ystau$ imposed by
  the non-thermal production of gravitinos in $\stauone$ decays,
  $\Omega_{\gravitino}^{\NTP}\leq f\,\Omega_{\mathrm{dm}}$---given
  explicitly in~(22) of Ref.~\cite{Steffen:2006hw}---is respected for
  $\mgr\lesssim 500~\GeV$ even if only a small fraction $f=0.01$ of
  dark matter is assumed to originate from $\stauone$ decays; cf.\
  Fig.~13 of Ref.~\cite{Steffen:2006hw}.
  This applies equally to other scenarios with an extremely weakly
  interacting LSP---such as the axino
  LSP~\cite{Covi:1999ty,Brandenburg:2005he,Freitas:2009fb}---originating
  from $\stauone$ decays.
\item For $\Ystau\lesssim 10^{-13}$, the BBN constraints associated
  with effects of hadronic energy release on the primordial D
  abundance can be respected for $\stauone\simeq\stauR$ and
  $\mstauone$ up to $10~\TeV$ independent of the $\stauone$ lifetime;
  cf.~Fig.~11 of Ref.~\cite{Steffen:2006hw}. For a sizable admixture
  of $\stauL$ in $\stauone$, this $\Ystau$ constraint can become more
  restrictive in particular with the enhanced stau--Higgs couplings
  allowing for exceptionally small $\Ystau$ values. Nevertheless,
  these exceptional values are typically associated with $\mstauone<
  300~\GeV$ where the $\Ystau$ limit is significantly more relaxed:
  $\Ystau\lesssim 10^{-11}$ for $\stauone\simeq\stauR$.
  A tightening to $\Ystau\lesssim 10^{-13}$ ($10^{-15}$) will then
  require an increase of (hadronic) $E_{\mathrm{vis}}$ by a factor of
  $10^2$ ($10^4$).
  On the other hand, sufficiently degenerate $\mgr$ and $\mstauone$
  will always be associated with small values of
  $E_{\mathrm{vis}}$ and thereby with relaxed $\Ystau$ limits
  from energy release, even in the case of strongly enhanced
  stau--Higgs couplings.
\item For $\Ystau\lesssim 10^{-14}~(10^{-15})$, the BBN constraints
  associated with effects of electromagnetic energy release on the
  primordial D ($^3$He) abundance can be respected independent of the
  $\stauone$ lifetime; cf.~upper panels of Fig.~12
  ($100~\GeV\leq\mstauone\leq 10~\TeV$) of Ref.~\cite{Steffen:2006hw}
  and Figs.~14 ($\mstauone=100~\GeV$) and~15 ($\mstauone=300~\GeV$) of
  Ref.~\cite{Kawasaki:2008qe}.
\item For $\Ystau\lesssim 2\times 10^{-15}~(2\times 10^{-16}\div2\times
  10^{-15})$, the BBN constraints associated with bound state effects
  allowing for CBBN of $^6$Li and $^9$Be can be respected even for
  $\taustau\gtrsim 10^5\,\seconds$; see Fig.~\ref{Fig:Ytau}.
  Recall, that these values correspond to upper limits on the
  primordial fractions of $^9$Be/H and $^6$Li/H of $2.1\times
  10^{-13}$ and to the generous range $10^{-11}\div 10^{-10}$,
  respectively.
\end{itemize}
Thus, the SUSY models which come with thermal relic stau abundances of
$\Ystau \lesssim 2\times 10^{-15}$ can respect each of those
cosmological constraints independently of the stau lifetime if a
primordial $^6$Li/H abundance of about $10^{-10}$ is viable. In
particular, the limit~(\ref{eq:lifetimebound}) of
$\tau_{\stauone}\lesssim 6\times 10^3\,\seconds$ and its implications
discussed in Chapter~\ref{cha:stau-nlsp} are then no longer valid even
for a standard cosmological history with primordial temperatures of
$T>\Tf$. Thereby, the regions with $\Ystau \lesssim 2\times 10^{-15}$
are associated with particularly attractive gravitino dark matter
scenarios:
\begin{itemize}
\item The gravitino mass can be within the range
  $0.1\lesssim\mgr<\mstauone$ for which its kinematical determination
  could be
  viable~\cite{Buchmuller:2004rq,Martyn:2006as,Hamaguchi:2006vu}.
  Together with measurements of $\mstauone$ and $\taustau$, a
  kinematically determined $\mgr$ would allow one to measure the
  Planck scale $\MPl$ at
  colliders~\cite{Buchmuller:2004rq,Martyn:2006as,Hamaguchi:2006vu}.
  Indeed, an agreement of the $\MPl$ value determined in collider
  experiments with the one inferred from Newton's constant $G_{\rm N}$
  would support the existence of supergravity in
  nature~\cite{Buchmuller:2004rq}.
\item For $\mgr$ sufficiently close to $\mstauone$, the spin-3/2
  character of the gravitino becomes relevant so that it could be
  probed in principle by analyzing the decays
  $\stauone\to\gravitino\tau\gamma$~\cite{Buchmuller:2004rq}.
\item With $\Ystau \lesssim 2\times 10^{-15}$,
  $\Omega_{\gravitino}^{\NTP}$ is negligible so that basically all of
  $\Omega_{\mathrm{dm}}$ can be provided by gravitinos from other
  sources such as thermal production.
  Indeed, if also gravitino production in decays of scalar fields such
  as the inflaton~\cite{Asaka:2006bv,Endo:2007sz} is negligible,
  reheating temperatures of $\TR\gtrsim 10^9\,\GeV$ could become
  viable for $\mgr\sim 100~\GeV$ and not too heavy gaugino masses;
  see, in particular, Sec.~\ref{sec:TR-bounds-from-TP}.
  This would mean that thermally produced gravitinos could provide the
  right amount of dark matter and that thermal leptogenesis (with
  $\TR\gtrsim 10^9\,\GeV$ as a benchmark
  value~\cite{Davidson:2002qv,Buchmuller:2004nz}) would be a viable
  explanation of the cosmic baryon asymmetry, i.e., there would be no
  gravitino problem.
\item With a kinematically determined $\mgr$, one would be able to
  probe the reheating temperature $\TR$ at colliders and thereby the
  viability of thermal leptogenesis~\cite{Pradler:2006qh}.
\item For $\taustau\gtrsim 10^4\,\seconds$, the small $\Ystau$ values
  could still allow for the primordial catalysis of $^6$Li and $^9$Be
  in agreement with existing astrophysical observations; see
  Sec.~\ref{Sec:9BeConstraints}.
\end{itemize}

Table~\ref{tab:points} illustrates that gravitino dark matter
scenarios of the type discussed above can even be accommodated within
the CMSSM.
For gravitino masses of $50~\GeV$ and $100~\GeV$, we list the
associated values of $\taustau$, of $\Omega_{\gravitino}^{\NTP}h^2$,
and of the maximum reheating temperature $\TR^{\mathrm{max}}$ under
the assumption that other gravitino sources can be neglected.  
The stau lifetime $\taustau$ is given in~(\ref{Eq:SleptonLifetime})
and the $\TR^{\mathrm{max}}$ values imposed by
$\Omega_{\gravitino}^{\TP}h^2\leq 0.126$ can be inferred
from~(\ref{eq:omega-tp}).
At each CMSSM point and for both $\mgr$ values, $\stauone$ is very
long lived, $\taustau> 10^6\,\seconds$, and gravitino production from
$\stauone$ decays is negligible,
$\Omega_{\gravitino}^{\NTP}h^2\lesssim 10^{-4}$.  In all cases, the
gravitino mass $\mgr=100~\GeV$ is sufficiently close to $\mstauone$ so
that the \mbox{spin-3/2} character of the gravitino can in principle be
probed~\cite{Buchmuller:2004rq}.  A reheating temperature of
$\TR\gtrsim 10^9\,\GeV$ is viable only for the points A and B with
$\monetwo$ significantly below $1~\TeV$, i.e., at the points at which
resonant stau annihilation leads to the reduction of $\Ystau$.
Because of $\taustau> 10^6\,\seconds$, the $\Ystau$ limit from CBBN of
$^9$Be is at $Y_{\mathrm{Be}}^{\mathrm{max}}\simeq 2\times 10^{-15}$
for each point as can be inferred from Fig.~\ref{Fig:Ytau}.  This
bound disfavors point A while the points B and C are associated with
$\Ystau$ values very close to this limit and thereby with
$^9\mathrm{Be/H}$ ($^6\mathrm{Li/H}$) values of about $2.1\times
10^{-13}$ ($10^{-10}$).

%
%
%
%

%
%
%
%
%
%
%
%
%

%
%
%
%
%
%
%
%
%
%
%
%
%
%
%
%

%
%
%
%
%
%
%
%
%
%
%
%
%
%
%
%
%
%

%
%
%
%
%
%
%
%
%
%
%
%
%
%
%

%
%
%
%
%
%
%
%

%
%
%
%
%
%
%
%
%
%
%
%
%
%

%
%
%
%
%
%
%
%
%
%


\cleardoublepage

\chapter*{Conclusions}
\label{cha:conclusions}
\addcontentsline{toc}{chapter}{Conclusions}
\pdfbookmark[-1]{Conclusions}{bla}

\setlength{\parindent}{20pt}

In this thesis we have worked out the cosmological implications of a
long-lived electromagnetically charged massive particle species
$\X^{\pm}$ also called CHAMP. Our working hypothesis has been that
$X$ possesses a weak-scale mass $\mx\gtrsim \Orderof{100\ \GeV}$ and 
typically a lifetime $\tauX\gtrsim 1\ \seconds$.
We have assumed that the temperature $T$
of the early Universe was high enough so
that \X\ has achieved chemical equilibrium with the primordial plasma.
Then, following a standard cosmological evolution, \X\ experienced a
thermal freeze-out once $T \lesssim \mx / 25$. This makes \X\ to what is
called a thermal relic (prior to its decay).

\section*{BBN with a long-lived CHAMP}
\label{sec:conclusions-part-i}

We have started our investigation with a brief introduction into the
framework of standard Big Bang Nucleosynthesis (SBBN) where we also
have given account to some of the latest measurements from which
primordial light element abundances are inferred.
In a simplified discussion of \X-decoupling we have argued that its
expected cosmological abundance prior to decay reads $10^{-18}\lesssim
Y_\X \lesssim 10^{-12} (\mx/100\GeV)$.
We have recalled that the long-lived CHAMP scenario is strongly
constrained by BBN limits on electromagnetic and hadronic energy
release in the \X-decay.
For a reliability check on hadronic BBN constraints we have worked out
the Coulomb stopping power of charged hadrons in the plasma.  In
particular, we have developed on a refined approach taking into
account peculiarities in the plasma-screening of the Coulomb
interaction and paying close attention to the velocity dependencies of
the cross sections.  We find reasonably good agreement with the
treatment used in Ref.~\cite{Kawasaki:2004qu} from which we have incorporated
the associated constraints.

Subsequently, the effects of \champ\ on BBN due to its binding onto
the light nuclei~\nuc{} have been considered.
Given that the \champ-catalysis of thermal nucleosynthesis reactions
had only been discovered recently~\cite{Pospelov:2006sc}, we have
laid out in detail the central points of CBBN.
Using the variational approach, we have obtained ground state energies
for bound states \BSx{\nuc{}}\ by taking into account the finite
nuclear charge radius of~$\nuc{}$. This leads to a reduction (in magnitude)
from the na\"ive point-like Coulomb values, e.g., for \BSx{\hef}\
by~$13\%$ and for \BSx{\ben}\ by~$60\%$ which has also been confirmed
upon numerical solution of the Schr\"odinger equation. For the
examples of \BSx{\hef}\ and \BSx{\beet}\ also the complete spectrum
for $n\leq 3$ has been computed. We have further obtained the
wave functions for the \nuc{}--\champ continuum. This has allowed us to
calculate the cross sections for \BSx{\nuc{}}\ photo-dissociation,
$\sigmavof{\mathrm{ph}}$, and radiative recombination,
$\sigmavof{\mathrm{rec}}$, including the finite charge radius
correction and taking into account recombinations into 1S as well as
2S states. Those rates (per particle pair) are important since they
control the fractional bound state abundance and thus the timing and
efficiency of CBBN. For example, for \BSx{\hef}\ and \BSx{\lisx}\ we
find a reduction of $\sigmavof{\mathrm{rec}}$ from the hydrogen-like
case by~$17\%$ and by~$74\%$, respectively.

From observations of beryllium in Population II halo stars at very low
metallicities, we have extracted a nominal upper limit on primordial
beryllium of $\ben/\Hyd \le 2.1\times 10^{-13}$.  This limit allows
one to set  constraints on models in which the primordial
$A=8$ divide is bridged by catalytic effects.  Considering the
primordial catalysis of $^9$Be~\cite{Pospelov:2007js}, we have derived
\tauX-dependent upper limits on the $X^-$-yield prior to decay,
\YXdec.
For a typical relic abundance $\YXdec\gtrsim 3 \times 10^{-14}$
($10^{-14}$), we find that this $\ben$ limit translates into an upper
limit on the $X^-$ lifetime of $\tauX\lesssim 6\times
10^{3}\,\seconds$ ($10^{4}\,\seconds$). Furthermore, we have also
worked out the catalytic production of \lisx\ which, depending on the
adopted upper limit on primordial \lisx, gives rise to similar bounds.

We have clarified that the presence of ($p$\xm) bound states cannot
relax the $\YXdec$ limits at long lifetimes $\tauX$ in any substantial
way. 
Indeed, we have shown explicitly by solving the associated full set of
Boltzmann equations that late-time effects of ($p$\xm) bound states
can affect the lithium and beryllium abundances synthesized at
$T\simeq8\ \keV$ by not more than a few percent.
Any substantial formation of ($p$\xm) at $T\simeq 0.7\ \keV$ is
immediately intercepted by the very efficient charge exchange reaction
of ($p$\xm) with \hef. This comes as no surprise given the large size
of the ($p$\xm) system $\sim 30~\fm$ and the fact that the proton
deconfinement probability approaches unity already for a \hef--\xm\
distance of $\sim 95~\fm$.  In particular, we find that the fractional
density of protons in bound states does not exceed the level of $\sim
10^{-6}$ for $\YX\lesssim Y_{\hef}$.  
By the same argument, the \ben\ yield also remains unaffected by
late-time catalysis.
Thus, we find that the possibility of allowed islands in the parameter
region with typical $\YXdec$ and large $\tauX$---which was advocated
in Ref.~\cite{Jedamzik:2007cp}---does not exist.

\section*{The gravitino-stau scenario}
\label{sec:conclusions-part-ii}

In the second part of this thesis we have considered cosmological
constraints and their implications for models in which the gravitino
is the LSP and the stau is the NLSP. We have first focused on
\gravitino\ as a dark matter candidate.
Using the full gauge-invariant result for the thermally produced
gravitino abundance $\Omega_{\gravitino}^{\TP}$ to leading order in
the Standard Model gauge
couplings~\cite{Pradler:2007ne,Pradler:2006qh}, we have studied bounds
on the reheating temperature $T_{\Reheating}$ from the constraint
$\Omega_{\gravitino}\leq \Omega_{\CDM}$. In particular, taking into
account the dependence of $\Omega_{\gravitino}^{\TP}$ on the masses of
the gauginos has allowed us to explore the dependence of the $\TR$
bounds on the gaugino-mass relation at the scale of grand unification
$M_{\GUT}$. We have explicitly studied the effect of \gravitino\
regeneration during a post-inflationary perturbative reheating
phase. Thereby, we have made contact between the notion of \TR\ as the
initial temperature of the radiation-dominated epoch in the analytical
expression (\ref{Eq:YgravitinoTP}) for the \gravitino\ abundance and
the definition of \TR\ in terms of the decay width $\Gamma_{\phi}$ of
the inflaton.

Applying the $\tauX$-dependent upper limits on $\YXdec$ derived from
the primordial catalysis of \ben\ and \lisx\ in Part~\ref{part:bbn},
we have analyzed the emerging constraints in the gravitino-stau
scenario, i.e., for $\stauone^-=X^-$.
For typical values~(\ref{eq:Ystau7})
 of the stau NLSP yield after
decoupling, the \ben\ and \lisx\ constraints have been found in close
vicinity to each other so that they lead to the same implications.
For example, for $\mgr = 10\ \GeV$, the CBBN constraints impose the
lower limit $\mstauone > 400\ \GeV$ with rising tendency for growing
\mgr. For \stauone\ being the lightest Standard Model superpartner
such a limit directly affects the testability of those SUSY scenarios
at future colliders.
Furthermore, for a primordial limit of $\lisx/\Hyd \lesssim 6\times
10^{-11}$ the calculated \lisx\ abundance drops below this
observational bound only for $\taustau \lesssim 6\times 10^{3}\
\seconds$ (likewise for \ben). Taken at face value, we find that this
constraint translates into a lower limit on the gaugino mass parameter
$\monetwo \geq 0.87\ \TeV\,(\mgr/10\ \GeV)^{2/5}$ in the entire
natural region of the CMSSM parameter space. This limit implies a
restrictive upper bound $\TR\lesssim 5\times 10^7\ \GeV\, (\mgr/10\
\GeV)^{1/5}$.

Using exemplary (\monetwo,\mzero) CMSSM planes where we explicitly
compute $Y_{\stauone}^{\mathrm{dec}}$ in every point, we have further
explored gravitino dark matter scenarios and the associated $\TR$
bounds for $\mgr \geq 10~\GeV$ and for temperatures as low as
$10^7~\GeV$. Taking into account the \lisx\ CBBN constraint as well as
the constraints on electromagnetic and hadronic energy injection from
\stauone-decays, we have illustrated that in the considered regions of the
CMSSM parameter space $\TR\lesssim 10^7~\GeV$ indeed is the highest
cosmologically viable temperature of the radiation-dominated epoch in
case of a standard thermal history of the Universe. Moreover, in the
\stauone\ NLSP region the lower bound on \monetwo\ typically implies a
very heavy superparticle mass spectrum where, e.g., $m_\gluino< 2.5\
\TeV$ can be well excluded and which makes such scenarios hard to
probe at the LHC.
The bound on $\TR$ imposes a serious constraint for inflation models.
Moreover, thermal leptogenesis seems to be strongly disfavored in the
considered regions of the CMSSM parameter space.

With late-time entropy release, the obtained limit $\TR\lesssim
10^7~\GeV$ can be relaxed.  For example, the dilution of the thermally
produced gravitino yield by a factor of $10$ relaxes the $\TR$ bound
by about one order of magnitude in regions where
$\Omega_{\gravitino}^{\TP}$ dominates $\Omega_{\gravitino}$.  In the
case of entropy production after NLSP decoupling, the yield of the
NLSP prior to its decay, $Y_{\NLSP}$, is reduced so that the BBN
constraints can be weakened. Although the $^6$Li bound is persistent,
we find that it disappears provided $Y_{\NLSP}$ is diluted by a factor
of $\Delta\gtrsim 10^3$.
We have discussed the viability of thermal leptogenesis in a
cosmological scenario with entropy production after NLSP decoupling.
We find that successful thermal leptogenesis can be revived in generic
regions of the CMSSM parameters space for 
$M_{\mathrm{R}1}\sim\TR \gtrsim 10^{12}~\GeV$ and $\Delta\gtrsim 10^3$,
where $M_{\mathrm{R}1}$ is the mass of the lightest among the heavy
right-handed Majorana neutrinos.  There, the collider-friendly
$\stau$ NLSP region with $m_{\stau}\lesssim 250~\GeV$ reopens as a
cosmologically allowed region in the CMSSM with the gravitino~LSP.

\section*{Thermal relic stau abundances}
\label{sec:conclusions-part-iii}

In the final part of this thesis we have carried out a thorough study
of primordial stau annihilation and the associated thermal freeze-out.
Taking into account the complete set of stau annihilation channels
within the MSSM with real parameters for cases with negligible
sparticle coannihilation,
the resulting thermal relic $\stauone$ yield $Y_{\stauone}^{\mathrm{dec}}$ has been
examined systematically.
While we have often (implicitly) focused on the $\stauone\simeq\stauR$
case in Part~\ref{part:two} by employing~(\ref{eq:Ystau7}), we have
investigated cases in Part~\ref{part:three} in which $\stauone$
contains a significant admixture of $\stauL$ including the maximal
mixing case as well as~$\stauone\simeq\stauL$.

We find that the variation of the stau mixing angle $\thetastau$
affects the relative importance of the different annihilation channels
significantly but not necessarily the resulting
$Y_{\stauone}^{\mathrm{dec}}$ value for relatively small values of $\tanb$.
By increasing $\tanb$, however, we encounter a dramatic change of this
picture for large absolute values of the Higgs-higgsino mass parameter
$\mu$ and/or of the trilinear coupling $\Atau$, which are the
dimensionful SUSY parameters that govern simultaneously stau
left-right mixing and the stau--Higgs couplings: Stau annihilation
into $\hhiggs\hhiggs$, $\hhiggs\Hhiggs$, and $\Hhiggs\Hhiggs$ can
become very efficient (if kinematically allowed) so that $Y_{\stauone}^{\mathrm{dec}}$ can
decrease to values well below $10^{-15}$. 
The scalar nature of $\stauone$ allows those parameters to enter
directly into the annihilation cross sections. This mechanism has no
analogue in calculations of the thermal relic density of the lightest
neutralino $\neutralino$.

The stau--Higgs couplings are crucial also for the second $Y_{\stauone}^{\mathrm{dec}}$
reduction mechanism identified in this work:
Even for moderate values of $\tanb$, we find that staus can annihilate
very efficiently into a $\bquark\antibquark$ pair via $s$-channel
exchange of the heavy CP-even Higgs boson $\Hhiggs$ provided the MSSM
spectrum exhibits the resonance condition $2\mstauone\simeq\mH$.
We have shown explicitly that the associated $Y_{\stauone}^{\mathrm{dec}}$ values can be
below $10^{-15}$ as well.
This mechanism is similar to the one that leads to the reduction of
the $\neutralino$ density in the Higgs funnel region in which
neutralino annihilation proceeds at the resonance of the CP-odd Higgs
boson $\Ahiggs$.

We have worked with an effective low energy version of the MSSM to
investigate the $\thetastau$-dependence of $Y_{\stauone}^{\mathrm{dec}}$ and the two
$Y_{\stauone}^{\mathrm{dec}}$-reduction mechanisms in a controlled way. In addition, we
have shown that the considered effects can be accommodated also with
restrictive assumptions on the soft-SUSY breaking sector at a high
scale. Within the CMSSM, we encounter both mechanisms each of which
leading to $Y_{\stauone}^{\mathrm{dec}}\simeq 2\times 10^{-15}$ in two distinct regions of
a single $(\monetwo,\,\mzero)$ plane.

We have discussed possibilities to probe the viability of the
presented $Y_{\stauone}^{\mathrm{dec}}$-reduction mechanisms at colliders.  While a $\mH$
measurement pointing to $\mH\simeq 2\mstauone$ would support resonant
primordial stau annihilation, studies of Higgs boson production in
association with staus, $\positron\electron\,(\gamma\gamma)
\to\stauone\stauone^*\hhiggs,\stauone\stauone^*\Hhiggs$ could allow
for an experimental determination of the relevant stau--Higgs
couplings, for example, at the ILC.
Moreover, we have outlined that associated
$\bquark\antibquark\hhiggs/\Hhiggs$ production with
$\hhiggs/\Hhiggs\to\stauone\stauone^*$ has the potential to allow for
a determination of both $\mh$ and $\mH$ at the LHC if a SUSY scenario
with large $\tanb$ and large stau--Higgs couplings is realized.

With the obtained small $Y_{\stauone}^{\mathrm{dec}}$ values, even the restrictive
constraints associated with CBBN could be respected so that attractive
gravitino dark matter scenarios could be revived to be cosmologically
viable even for a standard cosmological history. Within this class of
models, collider evidence for supergravity, for the gravitino being
the LSP, and for high values of the reheating temperatures of up to
$10^9\,\GeV$ is conceivable, which could thereby accommodate
simultaneously the explanation of the cosmic baryon asymmetry provided
by thermal leptogenesis and the hypothesis of thermally produced
gravitinos being the dark matter in our Universe.

%
%

%
%
%
%
%
%
%
%
%
%
%
%
%
%
%
%
%


\cleardoublepage

\phantomsection 
\addcontentsline{toc}{chapter}{References}
\bibliography{biblio}

\providecommand{\href}[2]{#2}\begingroup\raggedright\begin{thebibliography}{10%
0}

\bibitem{Pradler:2006hh}
J.~Pradler and F.~D. Steffen, {\it {Constraints on the reheating temperature in
  gravitino dark matter scenarios}},  {\em Phys. Lett.} {\bf B648} (2007)
  224--235 [\href{http://arXiv.org/abs/hep-ph/0612291}{{\tt hep-ph/0612291}}].

\bibitem{Pradler:2007is}
J.~Pradler and F.~D. Steffen, {\it {Implications of Catalyzed BBN in the CMSSM
  with Gravitino Dark Matter}},  {\em Phys. Lett.} {\bf B666} (2008) 181--184
  [\href{http://arXiv.org/abs/0710.2213}{{\tt 0710.2213}}].

\bibitem{Pradler:2007ar}
J.~Pradler and F.~D. Steffen, {\it {CBBN in the CMSSM}},  {\em Eur. Phys. J.}
  {\bf C56} (2008) 287--291 [\href{http://arXiv.org/abs/0710.4548}{{\tt
  0710.4548}}].

\bibitem{Pospelov:2008ta}
M.~Pospelov, J.~Pradler and F.~D. Steffen, {\it {Constraints on Supersymmetric
  Models from Catalytic Primordial Nucleosynthesis of Beryllium}},  {\em JCAP}
  {\bf 0811} (2008) 020 [\href{http://arXiv.org/abs/0807.4287}{{\tt
  0807.4287}}].

\bibitem{Pradler:2008qc}
J.~Pradler and F.~D. Steffen, {\it {Thermal relic abundances of long-lived
  staus}},  {\em Nucl. Phys.} {\bf B809} (2009) 318--346
  [\href{http://arXiv.org/abs/0808.2462}{{\tt 0808.2462}}].

\bibitem{Spergel:2003cb}
{\bf WMAP} Collaboration, D.~N. Spergel {\em et.~al.}, {\it {First Year
  Wilkinson Microwave Anisotropy Probe (WMAP) Observations: Determination of
  Cosmological Parameters}},  {\em Astrophys. J. Suppl.} {\bf 148} (2003)
  175--194 [\href{http://arXiv.org/abs/astro-ph/0302209}{{\tt
  astro-ph/0302209}}].

\bibitem{Spergel:2006hy}
D.~N. Spergel {\em et.~al.}, {\it Wilkinson microwave anisotropy probe {(WMAP)}
  three year results: Implications for cosmology},  {\em Astrophys. J. Suppl.}
  {\bf 170} (2007) 377 [\href{http://arXiv.org/abs/astro-ph/0603449}{{\tt
  astro-ph/0603449}}].

\bibitem{Komatsu:2008hk}
E.~Komatsu {\em et.~al.}, {\it {Five-Year Wilkinson Microwave Anisotropy Probe
  (WMAP) Observations: Cosmological Interpretation}},  {\em Astrophys. J.
  Suppl.} {\bf 180} (2009) 330--376 [\href{http://arXiv.org/abs/0803.0547}{{\tt
  0803.0547}}].

\bibitem{Dunkley:2008ie}
J.~Dunkley {\em et.~al.}, {\it {Five-Year Wilkinson Microwave Anisotropy Probe
  (WMAP) Observations: Likelihoods and Parameters from the WMAP data}},  {\em
  Astrophys. J. Suppl.} {\bf 180} (2009) 306--329
  [\href{http://arXiv.org/abs/0803.0586}{{\tt 0803.0586}}].

\bibitem{Steigman:2006nf}
G.~Steigman, {\it {The cosmological evolution of the average mass per baryon}},
   {\em JCAP} {\bf 0610} (2006) 016
  [\href{http://arXiv.org/abs/astro-ph/0606206}{{\tt astro-ph/0606206}}].

\bibitem{Pettini:2008mq}
M.~Pettini, B.~J. Zych, M.~T. Murphy, A.~Lewis and C.~C. Steidel, {\it
  {Deuterium Abundance in the Most Metal-Poor Damped Lyman alpha System:
  Converging on {$\Omega_{\mathrm{b},0}h^2$}}},
  \href{http://arXiv.org/abs/0805.0594}{{\tt 0805.0594}}.

\bibitem{Simha:2008mt}
V.~Simha and G.~Steigman, {\it {Constraining The Universal Lepton Asymmetry}},
  {\em JCAP} {\bf 0808} (2008) 011 [\href{http://arXiv.org/abs/0806.0179}{{\tt
  0806.0179}}].

\bibitem{Steigman:2007xt}
G.~Steigman, {\it {Primordial Nucleosynthesis in the Precision Cosmology Era}},
   {\em Ann. Rev. Nucl. Part. Sci.} {\bf 57} (2007) 463--491
  [\href{http://arXiv.org/abs/0712.1100}{{\tt 0712.1100}}].

\bibitem{Cyburt:2008zz}
R.~H. Cyburt, B.~D. Fields and K.~A. Olive, {\it {An update on the big bang
  nucleosynthesis prediction for Li-7: The problem worsens}},  {\em JCAP} {\bf
  0811} (2008) 012.

\bibitem{Spite:1982dd}
F.~Spite and M.~Spite, {\it {Abundance of lithium in unevolved halo stars and
  old disk stars: Interpretation and consequences}},  {\em Astron. Astrophys.}
  {\bf 115} (1982) 357--366.

\bibitem{2005ASPC..336...25A}
M.~{Asplund}, N.~{Grevesse} and A.~J. {Sauval}, {\em {The Solar Chemical
  Composition}}, vol.~336 of {\em Astronomical Society of the Pacific
  Conference Series}.
\newblock {Barnes}, III, T.~G. and {Bash}, F.~N., Sept., 2005.

\bibitem{Ryan:1999jq}
S.~G. Ryan, J.~E. Norris and T.~C. Beers, {\it {The Spite Lithium Plateau:
  Ultra-Thin but Post- Primordial}},  {\em Astrophys. J.} {\bf 523} (1999)
  654--677 [\href{http://arXiv.org/abs/astro-ph/9903059}{{\tt
  astro-ph/9903059}}].

\bibitem{Asplund:2005yt}
M.~Asplund, D.~L. Lambert, P.~E. Nissen, F.~Primas and V.~V. Smith, {\it
  {Lithium isotopic abundances in metal-poor halo stars}},  {\em Astrophys. J.}
  {\bf 644} (2006) 229--259 [\href{http://arXiv.org/abs/astro-ph/0510636}{{\tt
  astro-ph/0510636}}].

\bibitem{2007A&A...462..851B}
P.~{Bonifacio}, P.~{Molaro}, T.~{Sivarani}, R.~{Cayrel}, M.~{Spite},
  F.~{Spite}, B.~{Plez}, J.~{Andersen}, B.~{Barbuy}, T.~C. {Beers},
  E.~{Depagne}, V.~{Hill}, P.~{Fran{\c c}ois}, B.~{Nordstr{\"o}m} and
  F.~{Primas}, {\it {First stars VII - Lithium in extremely metal poor
  dwarfs}},  {\em \aap} {\bf 462} (Feb., 2007) 851--864
  [\href{http://arXiv.org/abs/astro-ph/0610245}{{\tt astro-ph/0610245}}].

\bibitem{Aoki:2009ce}
W.~Aoki {\em et.~al.}, {\it {Lithium Abundances of Extremely Metal-Poor
  Turn-off Stars}},  {\em Astrophys. J.} {\bf 698} (2009) 1803--1812
  [\href{http://arXiv.org/abs/0904.1448}{{\tt 0904.1448}}].

\bibitem{1993ApJ...408..262S}
V.~V. {Smith}, D.~L. {Lambert} and P.~E. {Nissen}, {\it {The 6Li/7Li ratio in
  the metal-poor halo dwarfs HD 19445 and HD 84937}},  {\em \apj} {\bf 408}
  (May, 1993) 262--276.

\bibitem{1994ApJ...428L..25H}
L.~M. {Hobbs} and J.~A. {Thorburn}, {\it {Lithium isotope ratios in six halo
  stars}},  {\em \apjl} {\bf 428} (June, 1994) L25--L28.

\bibitem{1998ApJ...506..405S}
V.~V. {Smith}, D.~L. {Lambert} and P.~E. {Nissen}, {\it {Isotopic Lithium
  Abundances in Nine Halo Stars}},  {\em \apj} {\bf 506} (Oct., 1998) 405--423.

\bibitem{1999A&A...343..923C}
R.~{Cayrel}, M.~{Spite}, F.~{Spite}, E.~{Vangioni-Flam}, M.~{Cass{\'e}} and
  J.~{Audouze}, {\it {New high S/N observations of the {6Li/7Li} blend in HD
  84937 and two other metal-poor stars}},  {\em \aap} {\bf 343} (Mar., 1999)
  923--932 [\href{http://arXiv.org/abs/astro-ph/9901205}{{\tt
  astro-ph/9901205}}].

\bibitem{1990ApJS...73...21D}
C.~P. {Deliyannis}, P.~{Demarque} and S.~D. {Kawaler}, {\it {Lithium in halo
  stars from standard stellar evolution}},  {\em \apjs} {\bf 73} (May, 1990)
  21--65.

\bibitem{1992ApJS...78..179P}
M.~H. {Pinsonneault}, C.~P. {Deliyannis} and P.~{Demarque}, {\it {Evolutionary
  models of halo stars with rotation. II - Effects of metallicity on lithium
  depletion, and possible implications for the primordial lithium abundance}},
  {\em \apjs} {\bf 78} (Jan., 1992) 179--203.

\bibitem{Takayama:2007du}
F.~Takayama, {\it {Extremely Long-Lived Charged Massive Particles as a Probe
  for Reheating of the Universe}},  {\em Phys. Rev.} {\bf D77} (2007) 116003
  [\href{http://arXiv.org/abs/0704.2785}{{\tt 0704.2785}}].

\bibitem{Gondolo:1990dk}
P.~Gondolo and G.~Gelmini, {\it Cosmic abundances of stable particles: Improved
  analysis},  {\em Nucl. Phys.} {\bf B360} (1991) 145--179.

\bibitem{Griest:1990kh}
K.~Griest and D.~Seckel, {\it {Three exceptions in the calculation of relic
  abundances}},  {\em Phys. Rev.} {\bf D43} (1991) 3191--3203.

\bibitem{Drees:2004jm}
M.~Drees, R.~Godbole and P.~Roy, {\it Theory and phenomenology of {Sparticles}:
  An account of four- dimensional {$N=1$} supersymmetry in high energy
  physics}, . Hackensack, USA: World Scientific (2004) 555 p.

\bibitem{Asaka:2000zh}
T.~Asaka, K.~Hamaguchi and K.~Suzuki, {\it Cosmological gravitino problem in
  gauge mediated supersymmetry breaking models},  {\em Phys. Lett.} {\bf B490}
  (2000) 136--146 [\href{http://arXiv.org/abs/hep-ph/0005136}{{\tt
  hep-ph/0005136}}].

\bibitem{Yao:2006px}
{\bf Particle Data Group} Collaboration, W.~M. Yao {\em et.~al.}, {\it Review
  of particle physics},  {\em J. Phys.} {\bf G33} (2006) 1--1232.

\bibitem{Gondolo:2004sc}
P.~Gondolo {\em et.~al.}, {\it {DarkSUSY: Computing supersymmetric dark matter
  properties numerically}},  {\em JCAP} {\bf 0407} (2004) 008
  [\href{http://arXiv.org/abs/astro-ph/0406204}{{\tt astro-ph/0406204}}].

\bibitem{nla.cat-vn2263194}
J.~M. Blatt and V.~F. Weisskopf, {\em Theoretical nuclear physics}.
\newblock John Wiley {\&} Sons Inc., New York, 1952.

\bibitem{Hemmick:1989ns}
T.~K. Hemmick {\em et.~al.}, {\it {A search for anomalously heavy isotopes of
  low {$Z$} nuclei}},  {\em Phys. Rev.} {\bf D41} (1990) 2074--2080.

\bibitem{Berger:2008ti}
C.~F. Berger, L.~Covi, S.~Kraml and F.~Palorini, {\it {The number density of a
  charged relic}},  {\em JCAP} {\bf 0810} (2008) 005
  [\href{http://arXiv.org/abs/0807.0211}{{\tt 0807.0211}}].

\bibitem{Reno:1987qw}
M.~H. Reno and D.~Seckel, {\it {Primordial Nucleosynthesis: The Effects of
  Injecting Hadrons}},  {\em Phys. Rev.} {\bf D37} (1988) 3441.

\bibitem{Dimopoulos:1988ue}
S.~Dimopoulos, R.~Esmailzadeh, L.~J. Hall and G.~D. Starkman, {\it {Limits on
  late decaying particles from nucleosynthesis}},  {\em Nucl. Phys.} {\bf B311}
  (1989) 699.

\bibitem{Cyburt:2002uv}
R.~H. Cyburt, J.~R. Ellis, B.~D. Fields and K.~A. Olive, {\it {Updated
  nucleosynthesis constraints on unstable relic particles}},  {\em Phys. Rev.}
  {\bf D67} (2003) 103521 [\href{http://arXiv.org/abs/astro-ph/0211258}{{\tt
  astro-ph/0211258}}].

\bibitem{Jedamzik:2004er}
K.~Jedamzik, {\it {Did something decay, evaporate, or annihilate during big
  bang nucleosynthesis?}},  {\em Phys. Rev.} {\bf D70} (2004) 063524
  [\href{http://arXiv.org/abs/astro-ph/0402344}{{\tt astro-ph/0402344}}].

\bibitem{Kawasaki:2004qu}
M.~Kawasaki, K.~Kohri and T.~Moroi, {\it {Big-bang nucleosynthesis and hadronic
  decay of long-lived massive particles}},  {\em Phys. Rev.} {\bf D71} (2005)
  083502 [\href{http://arXiv.org/abs/astro-ph/0408426}{{\tt
  astro-ph/0408426}}].

\bibitem{Bailly:2008yy}
S.~Bailly, K.~Jedamzik and G.~Moultaka, {\it {Gravitino Dark Matter and the
  Cosmic Lithium Abundances}},  \href{http://arXiv.org/abs/0812.0788}{{\tt
  0812.0788}}.

\bibitem{Kawasaki:1994sc}
M.~Kawasaki and T.~Moroi, {\it {Electromagnetic cascade in the early {U}niverse
  and its application to the big bang nucleosynthesis}},  {\em Astrophys. J.}
  {\bf 452} (1995) 506 [\href{http://arXiv.org/abs/astro-ph/9412055}{{\tt
  astro-ph/9412055}}].

\bibitem{Audi:2002rp}
G.~Audi, A.~H. Wapstra and C.~Thibault, {\it {The Ame2003 atomic mass
  evaluation (II). Tables, graphs and references}},  {\em Nucl. Phys.} {\bf
  A729} (2002) 337--676.

\bibitem{Jedamzik:1999di}
K.~Jedamzik, {\it {Lithium-6: A Probe of the Early Universe}},  {\em Phys. Rev.
  Lett.} {\bf 84} (2000) 3248
  [\href{http://arXiv.org/abs/astro-ph/9909445}{{\tt astro-ph/9909445}}].

\bibitem{Jedamzik:2004ip}
K.~Jedamzik, {\it {Neutralinos and Big Bang nucleosynthesis}},  {\em Phys.
  Rev.} {\bf D70} (2004) 083510
  [\href{http://arXiv.org/abs/astro-ph/0405583}{{\tt astro-ph/0405583}}].

\bibitem{Kawasaki:2004yh}
M.~Kawasaki, K.~Kohri and T.~Moroi, {\it {Hadronic decay of late-decaying
  particles and big-bang nucleosynthesis}},  {\em Phys. Lett.} {\bf B625}
  (2005) 7--12 [\href{http://arXiv.org/abs/astro-ph/0402490}{{\tt
  astro-ph/0402490}}].

\bibitem{Berestetsky:1982aq}
V.~B. Berestetsky, E.~M. Lifshitz and L.~P. Pitaevsky, {\em {Quantum
  Electrodynamics}}.
\newblock Oxford, Uk: Pergamon 652 P. ( Course Of Theoretical Physics, 4),
  1982.

\bibitem{Sachs:1962zzc}
R.~G. Sachs, {\it {High-Energy Behavior of Nucleon Electromagnetic Form
  Factors}},  {\em Phys. Rev.} {\bf 126} (1962) 2256--2260.

\bibitem{Rosenbluth:1950yq}
M.~N. Rosenbluth, {\it {High Energy Elastic Scattering of Electrons on
  Protons}},  {\em Phys. Rev.} {\bf 79} (1950) 615--619.

\bibitem{Walker:1993vj}
R.~C. Walker {\em et.~al.}, {\it {Measurements of the proton elastic
  form-factors for {$1~\mathrm{GeV}/c^2 \leq Q^2 \leq 3~\mathrm{GeV}/c^2 $} at
  SLAC}},  {\em Phys. Rev.} {\bf D49} (1994) 5671--5689.

\bibitem{Schiavilla:2001qe}
R.~Schiavilla and I.~Sick, {\it {Neutron charge form factor at large {$Q^2$}}},
   {\em Phys. Rev.} {\bf C64} (2001) 041002
  [\href{http://arXiv.org/abs/nucl-ex/0107004}{{\tt nucl-ex/0107004}}].

\bibitem{Raffelt:1985nk}
G.~G. Raffelt, {\it {Astrophysical axion bounds diminished by screening
  effects}},  {\em Phys. Rev.} {\bf D33} (1986) 897.

\bibitem{Raffelt:1996wa}
G.~G. Raffelt, {\em {Stars as laboratories for fundamental physics: The
  astrophysics of neutrinos, axions, and other weakly interacting particles}}.
\newblock Chicago, USA: Univ. Pr. 664 p, 1996.

\bibitem{Hahn:2004fe}
T.~Hahn, {\it Cuba: A library for multidimensional numerical integration},
  {\em Comput. Phys. Commun.} {\bf 168} (2005) 78--95
  [\href{http://arXiv.org/abs/hep-ph/0404043}{{\tt hep-ph/0404043}}].

\bibitem{Bernstein:1988bw}
J.~Bernstein, {\em {Kinetic theory in the expanding universe}}.
\newblock Cambridge, USA: Univ. Pr. 149p, 1988.

\bibitem{Bringmann:2006mu}
T.~Bringmann and S.~Hofmann, {\it {Thermal decoupling of WIMPs from first
  principles}},  {\em JCAP} {\bf 0407} (2007) 016
  [\href{http://arXiv.org/abs/hep-ph/0612238}{{\tt hep-ph/0612238}}].

\bibitem{1960ecm..book.....L}
E.~M. Lifshitz, L.~D. Landau and L.~P. Pitaevskii, {\em {Electrodynamics of
  continuous media}}.
\newblock Pergamon Press, 1984.

\bibitem{1975clel.book.....J}
J.~D. {Jackson}, {\em {Classical electrodynamics}}.
\newblock John Wiley {\&} Sons Inc., New York, 1975.

\bibitem{Braaten:1991jj}
E.~Braaten and M.~H. Thoma, {\it {Energy loss of a heavy fermion in a hot
  plasma}},  {\em Phys. Rev.} {\bf D44} (1991) 1298--1310.

\bibitem{2004ADNDT..87..185A}
I.~{Angeli}, {\it {A consistent set of nuclear rms charge radii: properties of
  the radius surface R(N,Z)}},  {\em Atomic Data and Nuclear Data Tables} {\bf
  87} (July, 2004) 185--206.

\bibitem{1997ADNDT..67..207V}
L.~{Visscher} and K.~G. {Dyall}, {\it {Dirac-Fock Atomic Electronic Structure
  Calculations Using Different Nuclear Charge Distributions}},  {\em Atomic
  Data and Nuclear Data Tables} {\bf 67} (1997) 207.

\bibitem{Pospelov:2007js}
M.~Pospelov, {\it {Bridging the primordial A=8 divide with Catalyzed Big Bang
  Nucleosynthesis}},  \href{http://arXiv.org/abs/0712.0647}{{\tt 0712.0647}}.

\bibitem{Kamimura:2008fx}
M.~Kamimura, Y.~Kino and E.~Hiyama, {\it {Big-Bang Nucleosynthesis Reactions
  Catalyzed by a Long- Lived Negatively-Charged Leptonic Particle}},
  \href{http://arXiv.org/abs/0809.4772}{{\tt 0809.4772}}.

\bibitem{Hiyama:1997ub}
E.~Hiyama, M.~Kamimura, T.~Motoba, T.~Yamada and Y.~Yamamoto, {\it {Three- and
  four-body cluster models of hypernuclei using the G-matrix $\Lambda$N
  interaction: ${}_{\Lambda}^9Be$, ${}_{\Lambda}^{13}C$,
  ${}_{\Lambda\Lambda}^6He$ and $^{10}_{\Lambda\Lambda}Be$}},  {\em Prog.
  Theor. Phys.} {\bf 97} (1997) 881--899.

\bibitem{Yost:1936zz}
F.~L. Yost, J.~A. Wheeler and G.~Breit, {\it {Coulomb Wave Functions in
  Repulsive Fields}},  {\em Phys. Rev.} {\bf 49} (1936) 174--189.

\bibitem{1964coth.book.....G}
M.~L. {Goldberger} and K.~M. {Watson}, {\em {Collision theory}}.
\newblock New York : Wiley, c1964.~Structure of matter series, 1964.

\bibitem{abramowitz+stegun}
M.~Abramowitz and I.~A. Stegun, {\em Handbook of Mathematical Functions with
  Formulas, Graphs, and Mathematical Tables}.
\newblock Dover, New York, tenth~ed., 1964.

\bibitem{1963cma..book.....W}
E.~T. {Whittaker} and G.~N. {Watson}, {\em {A course of modern analysis}}.
\newblock Cambridge: University Press, 4th ed., 1963.

\bibitem{1960MNRAS.120..121B}
A.~{Burgess} and M.~J. {Seaton}, {\it {A general formula for the calculation of
  atomic photo-ionization cross-sections}},  {\em \mnras} {\bf 120} (1960) 121.

\bibitem{1958MNRAS.118..504S}
M.~J. {Seaton}, {\it {The Quantum Defect Method}},  {\em \mnras} {\bf 118}
  (1958) 504.

\bibitem{PhysRev.67.11}
J.~G. Beckerley, {\it Expansion of positive energy {Coulomb} wave functions in
  powers of the energy},  {\em Phys. Rev.} {\bf 67} (Jan, 1945) 11--14.

\bibitem{Kohri:2006cn}
K.~Kohri and F.~Takayama, {\it {Big Bang Nucleosynthesis with Long Lived
  Charged Massive Particles}},  {\em Phys. Rev.} {\bf D76} (2007) 063507
  [\href{http://arXiv.org/abs/hep-ph/0605243}{{\tt hep-ph/0605243}}].

\bibitem{1965qume.book.....L}
L.~D. {Landau} and E.~M. {Lifshitz}, {\em {Quantum mechanics}}.
\newblock Course of theoretical physics, Oxford: Pergamon Press, 1965, 1965.

\bibitem{Bird:2007ge}
C.~Bird, K.~Koopmans and M.~Pospelov, {\it {Primordial Lithium Abundance in
  Catalyzed Big Bang Nucleosynthesis}},  {\em Phys. Rev.} {\bf D78} (2008)
  083010 [\href{http://arXiv.org/abs/hep-ph/0703096}{{\tt hep-ph/0703096}}].

\bibitem{Dimopoulos:1989hk}
S.~Dimopoulos, D.~Eichler, R.~Esmailzadeh and G.~D. Starkman, {\it {Getting a
  charge out of dark matter}},  {\em Phys. Rev.} {\bf D41} (1990) 2388.

\bibitem{DeRujula:1989fe}
A.~De~Rujula, S.~L. Glashow and U.~Sarid, {\it {Charged Dark Matter}},  {\em
  Nucl. Phys.} {\bf B333} (1990) 173.

\bibitem{Rafelski:1989pz}
J.~Rafelski, M.~Sawicki, M.~Gajda and D.~Harley, {\it {Reactions of charged
  massive particle in a deuterium environment}},  {\em Phys. Rev.} {\bf A44}
  (1991) 4345.

\bibitem{Pospelov:2006sc}
M.~Pospelov, {\it Particle physics catalysis of thermal big bang
  nucleosynthesis},  {\em Phys. Rev. Lett.} {\bf 98} (2007) 231301
  [\href{http://arXiv.org/abs/hep-ph/0605215}{{\tt hep-ph/0605215}}].

\bibitem{Angulo:1999zz}
C.~Angulo {\em et.~al.}, {\it {A compilation of charged-particle induced
  thermonuclear reaction rates}},  {\em Nucl. Phys.} {\bf A656} (1999) 3--183.

\bibitem{Cyburt:2008up}
R.~H. Cyburt and B.~Davids, {\it {Evaluation of Modern
  {${}^3\mathrm{He}(\alpha,\gamma){}^7$Be} Data}},  {\em Phys. Rev.} {\bf C78}
  (2008) 064614 [\href{http://arXiv.org/abs/0809.3240}{{\tt 0809.3240}}].

\bibitem{Nollett:2000ch}
K.~M. Nollett, R.~B. Wiringa and R.~Schiavilla, {\it A six-body calculation of
  the alpha-deuteron radiative capture cross section},  {\em Phys. Rev.} {\bf
  C63} (2001) 024003 [\href{http://arXiv.org/abs/nucl-th/0006064}{{\tt
  nucl-th/0006064}}].

\bibitem{Pisanti:2007hk}
O.~Pisanti {\em et.~al.}, {\it {PArthENoPE: Public Algorithm Evaluating the
  Nucleosynthesis of Primordial Elements}},  {\em Comp. Phys. Commun.} {\bf
  178} (2008) 956 [\href{http://arXiv.org/abs/0705.0290}{{\tt 0705.0290}}].

\bibitem{Hamaguchi:2007mp}
K.~Hamaguchi, T.~Hatsuda, M.~Kamimura, Y.~Kino and T.~T. Yanagida, {\it
  {Stau-catalyzed Li-6 production in big-bang nucleosynthesis}},  {\em Phys.
  Lett.} {\bf B650} (2007) 268--274
  [\href{http://arXiv.org/abs/hep-ph/0702274}{{\tt hep-ph/0702274}}].

\bibitem{Mohr:2005zz}
P.~J. Mohr and B.~N. Taylor, {\it {CODATA recommended values of the fundamental
  physical constants: 2002}},  {\em Rev. Mod. Phys.} {\bf 77} (2005) 1--107.

\bibitem{Caughlan:1987qf}
G.~R. Caughlan and W.~A. Fowler, {\it {Thermonuclear reaction rates. 5}},  {\em
  Atom. Data Nucl. Data Tabl.} {\bf 40} (1988) 283--334.

\bibitem{Thomas:1992tq}
D.~Thomas, D.~N. Schramm, K.~A. Olive and B.~D. Fields, {\it {Primordial
  nucleosynthesis and the abundance of beryllium and boron}},  {\em Astrophys.
  J.} {\bf 406} (1993) 569--579
  [\href{http://arXiv.org/abs/astro-ph/9206002}{{\tt astro-ph/9206002}}].

\bibitem{Mukhanov:2003xs}
V.~F. Mukhanov, {\it {Nucleosynthesis Without a Computer}},  {\em Int. J.
  Theor. Phys.} {\bf 43} (2004) 669--693
  [\href{http://arXiv.org/abs/astro-ph/0303073}{{\tt astro-ph/0303073}}].

\bibitem{Kawano:1992ua}
L.~Kawano, {\it {Let's go: Early {U}niverse. 2. Primordial nucleosynthesis: The
  Computer way}}, . FERMILAB-PUB-92-004-A.

\bibitem{Jedamzik:2007cp}
K.~Jedamzik, {\it {The cosmic 6Li and 7Li problems and BBN with long-lived
  charged massive particles}},  {\em Phys. Rev.} {\bf D77} (2008) 063524
  [\href{http://arXiv.org/abs/0707.2070}{{\tt 0707.2070}}].

\bibitem{1990ApJ...349..415S}
R.~{Svensson} and A.~{Zdziarski}, {\it {Photon-photon scattering of gamma rays
  at cosmological distances}},  {\em \apj} {\bf 349} (Feb., 1990) 415--428.

\bibitem{1976tper.book.....J}
J.~M. {Jauch} and F.~{Rohrlich}, {\em {The theory of photons and electrons. The
  relativistic quantum field theory of charged particles with spin one-half}}.
\newblock Texts and Monographs in Physics, New York: Springer, 2nd ed., 1976.

\bibitem{1970Natur.226..727R}
H.~{Reeves}, W.~A. {Fowler} and F.~{Hoyle}, {\it {Galactic Cosmic Ray Origin of
  Li, Be and B in Stars}},  {\em \nat} {\bf 226} (May, 1970) 727.

\bibitem{1971A&A....15..337M}
M.~{Meneguzzi}, J.~{Audouze} and H.~{Reeves}, {\it {The production of the
  elements Li, Be, B by galactic cosmic rays in space and its relation with
  stellar observations.}},  {\em \aap} {\bf 15} (1971) 337--359.

\bibitem{1990ApJ...364..568V}
E.~{Vangioni-Flam}, J.~{Audouze}, Y.~{Oberto} and M.~{Casse}, {\it {The
  evolution of Be-9}},  {\em \apj} {\bf 364} (Dec., 1990) 568--572.

\bibitem{Fields:1999ib}
B.~D. Fields, K.~A. Olive, E.~Vangioni-Flam and M.~Casse, {\it {Testing
  Spallation Processes With Beryllium and Boron}},  {\em Astrophys. J.} {\bf
  540} (2000) 930--945 [\href{http://arXiv.org/abs/astro-ph/9911320}{{\tt
  astro-ph/9911320}}].

\bibitem{Primas:2000gc}
F.~Primas, M.~Asplund, P.~E. Nissen and V.~Hill, {\it {The beryllium abundance
  in the very metal-poor halo star G 64-12 from VLT/UVES observations}},
  \href{http://arXiv.org/abs/astro-ph/0009482}{{\tt astro-ph/0009482}}.

\bibitem{Boesgaard:2005pf}
A.~M. Boesgaard and M.~C. Novicki, {\it {Beryllium in Disk and Halo Stars --
  Evidence for a Beryllium Dispersion in Old Stars}},  {\em Astrophys. J.} {\bf
  641} (2006) 1122--1130 [\href{http://arXiv.org/abs/astro-ph/0512317}{{\tt
  astro-ph/0512317}}].

\bibitem{Fields:2004ug}
B.~D. Fields, K.~A. Olive and E.~Vangioni-Flam, {\it {Implications of a new
  temperature scale for halo dwarfs on LiBeB and chemical evolution}},  {\em
  Astrophys. J.} {\bf 623} (2005) 1083--1091
  [\href{http://arXiv.org/abs/astro-ph/0411728}{{\tt astro-ph/0411728}}].

\bibitem{Takayama:2000uz}
F.~Takayama and M.~Yamaguchi, {\it {Gravitino dark matter without R-parity}},
  {\em Phys. Lett.} {\bf B485} (2000) 388--392
  [\href{http://arXiv.org/abs/hep-ph/0005214}{{\tt hep-ph/0005214}}].

\bibitem{Buchmuller:2007ui}
W.~Buchm{\"u}ller, L.~Covi, K.~Hamaguchi, A.~Ibarra and T.~Yanagida, {\it
  {Gravitino dark matter in R-parity breaking vacua}},  {\em JHEP} {\bf 03}
  (2007) 037 [\href{http://arXiv.org/abs/hep-ph/0702184}{{\tt
  hep-ph/0702184}}].

\bibitem{Ibarra:2008qg}
A.~Ibarra and D.~Tran, {\it {Antimatter Signatures of Gravitino Dark Matter
  Decay}},  {\em JCAP} {\bf 0807} (2008) 002
  [\href{http://arXiv.org/abs/0804.4596}{{\tt 0804.4596}}].

\bibitem{Hamaguchi:2009sz}
K.~Hamaguchi, F.~Takahashi and T.~T. Yanagida, {\it {Decaying gravitino dark
  matter and an upper bound on the gluino mass}},  {\em Phys. Lett.} {\bf B677}
  (2009) 59--61 [\href{http://arXiv.org/abs/0901.2168}{{\tt 0901.2168}}].

\bibitem{Coleman:1967ad}
S.~R. Coleman and J.~Mandula, {\it {All possible symmetries of the S-matrix}},
  {\em Phys. Rev.} {\bf 159} (1967) 1251--1256.

\bibitem{Haber:1984rc}
H.~E. Haber and G.~L. Kane, {\it The search for supersymmetry: Probing physics
  beyond the standard model},  {\em Phys. Rept.} {\bf 117} (1985) 75--263.

\bibitem{Martin:1997ns}
S.~P. Martin, {\it {A supersymmetry primer}},
  \href{http://arXiv.org/abs/hep-ph/9709356}{{\tt hep-ph/9709356}}.

\bibitem{Rarita:1941mf}
W.~Rarita and J.~S. Schwinger, {\it On a theory of particles with half-integral
  spin},  {\em Phys. Rev.} {\bf 60} (1941) 61.

\bibitem{Pradler:2007ne}
J.~Pradler, {\it {Electroweak Contributions to Thermal Gravitino Production}},
  \href{http://arXiv.org/abs/0708.2786}{{\tt 0708.2786}}. Diploma Thesis,
  MPP-2006-257.

\bibitem{wess:1992cp}
J.~Wess and J.~Bagger, {\em Supersymmetry and supergravity}.
\newblock {Princeton}, USA: Univ. Pr. 259 p, 1992.

\bibitem{Lee:1998aw}
T.~Lee and G.-H. Wu, {\it Interactions of a single goldstino},  {\em Phys.
  Lett.} {\bf B447} (1999) 83--88
  [\href{http://arXiv.org/abs/hep-ph/9805512}{{\tt hep-ph/9805512}}].

\bibitem{Linde:2005ht}
A.~D. Linde, {\it {Particle Physics and Inflationary Cosmology}},
  \href{http://arXiv.org/abs/hep-th/0503203}{{\tt hep-th/0503203}}.

\bibitem{Kolb:1990vq}
E.~W. {Kolb} and M.~S. {Turner}, {\em {The early {U}niverse}}.
\newblock Frontiers in Physics, Reading, MA: Addison-Wesley, 1988, 1990, 1990.

\bibitem{Linde:1991km}
A.~D. Linde, {\it Axions in inflationary cosmology},  {\em Phys. Lett.} {\bf
  B259} (1991) 38--47.

\bibitem{Khlopov:1984pf}
M.~Y. Khlopov and A.~D. Linde, {\it Is it easy to save the gravitino?},  {\em
  Phys. Lett.} {\bf B138} (1984) 265--268.

\bibitem{Ellis:1984eq}
J.~R. Ellis, J.~E. Kim and D.~V. Nanopoulos, {\it Cosmological gravitino
  regeneration and decay},  {\em Phys. Lett.} {\bf B145} (1984) 181.

\bibitem{Moroi:1993mb}
T.~Moroi, H.~Murayama and M.~Yamaguchi, {\it Cosmological constraints on the
  light stable gravitino},  {\em Phys. Lett.} {\bf B303} (1993) 289--294.

\bibitem{Bolz:1998ek}
M.~Bolz, W.~Buchm{\"u}ller and M.~Pl{\"u}macher, {\it {Baryon asymmetry and
  dark matter}},  {\em Phys. Lett.} {\bf B443} (1998) 209--213
  [\href{http://arXiv.org/abs/hep-ph/9809381}{{\tt hep-ph/9809381}}].

\bibitem{Bolz:2000fu}
M.~Bolz, A.~Brandenburg and W.~Buchm{\"u}ller, {\it Thermal production of
  gravitinos},  {\em Nucl. Phys.} {\bf B606} (2001) 518--544
  [\href{http://arXiv.org/abs/hep-ph/0012052}{{\tt hep-ph/0012052}}].

\bibitem{Pradler:2006qh}
J.~Pradler and F.~D. Steffen, {\it Thermal gravitino production and collider
  tests of leptogenesis},  {\em Phys. Rev.} {\bf D75} (2007) 023509
  [\href{http://arXiv.org/abs/hep-ph/0608344}{{\tt hep-ph/0608344}}].

\bibitem{Rychkov:2007uq}
V.~S. Rychkov and A.~Strumia, {\it {Thermal production of gravitinos}},  {\em
  Phys. Rev.} {\bf D75} (2007) 075011
  [\href{http://arXiv.org/abs/hep-ph/0701104}{{\tt hep-ph/0701104}}].

\bibitem{Asaka:2006bv}
T.~Asaka, S.~Nakamura and M.~Yamaguchi, {\it {Gravitinos from heavy scalar
  decay}},  {\em Phys. Rev.} {\bf D74} (2006) 023520
  [\href{http://arXiv.org/abs/hep-ph/0604132}{{\tt hep-ph/0604132}}].

\bibitem{Endo:2006tf}
M.~Endo, K.~Hamaguchi and F.~Takahashi, {\it {Moduli / inflaton mixing with
  supersymmetry breaking field}},  {\em Phys. Rev.} {\bf D74} (2006) 023531
  [\href{http://arXiv.org/abs/hep-ph/0605091}{{\tt hep-ph/0605091}}].

\bibitem{Endo:2007sz}
M.~Endo, F.~Takahashi and T.~T. Yanagida, {\it {Inflaton Decay in
  Supergravity}},  {\em Phys. Rev.} {\bf D76} (2007) 083509
  [\href{http://arXiv.org/abs/0706.0986}{{\tt 0706.0986}}].

\bibitem{Braaten:1991dd}
E.~Braaten and T.~C. Yuan, {\it Calculation of screening in a hot plasma},
  {\em Phys. Rev. Lett.} {\bf 66} (1991) 2183--2186.

\bibitem{Braaten:1989mz}
E.~Braaten and R.~D. Pisarski, {\it Soft amplitudes in hot gauge theories: A
  general analysis},  {\em Nucl. Phys.} {\bf B337} (1990) 569.

\bibitem{Brandenburg:2004du}
A.~Brandenburg and F.~D. Steffen, {\it {Axino dark matter from thermal
  production}},  {\em JCAP} {\bf 0408} (2004) 008
  [\href{http://arXiv.org/abs/hep-ph/0405158}{{\tt hep-ph/0405158}}].

\bibitem{Buchmuller:2008vw}
W.~Buchm{\"u}ller, M.~Endo and T.~Shindou, {\it {Superparticle Mass Window from
  Leptogenesis and Decaying Gravitino Dark Matter}},  {\em JHEP} {\bf 11}
  (2008) 079 [\href{http://arXiv.org/abs/0809.4667}{{\tt 0809.4667}}].

\bibitem{Kolb:2003ke}
E.~W. Kolb, A.~Notari and A.~Riotto, {\it {On the reheating stage after
  inflation}},  {\em Phys. Rev.} {\bf D68} (2003) 123505
  [\href{http://arXiv.org/abs/hep-ph/0307241}{{\tt hep-ph/0307241}}].

\bibitem{Kohri:2005wn}
K.~Kohri, T.~Moroi and A.~Yotsuyanagi, {\it {Big-bang nucleosynthesis with
  unstable gravitino and upper bound on the reheating temperature}},  {\em
  Phys. Rev.} {\bf D73} (2006) 123511
  [\href{http://arXiv.org/abs/hep-ph/0507245}{{\tt hep-ph/0507245}}].

\bibitem{Kawasaki:2008qe}
M.~Kawasaki, K.~Kohri, T.~Moroi and A.~Yotsuyanagi, {\it {Big-Bang
  Nucleosynthesis and Gravitino}},  {\em Phys. Rev.} {\bf D78} (2008) 065011
  [\href{http://arXiv.org/abs/0804.3745}{{\tt 0804.3745}}].

\bibitem{Rangarajan:2008zb}
R.~Rangarajan and N.~Sahu, {\it {Perturbative Reheating and Gravitino
  Production in Inflationary Models}},
  \href{http://arXiv.org/abs/0811.1866}{{\tt 0811.1866}}.

\bibitem{Bolz2008336}
M.~Bolz, A.~Brandenburg and W.~Buchm{\"u}ller, {\it {Erratum to: "Thermal
  production of gravitinos" [Nucl. Phys. B 606 (2001) 518-544]}},  {\em Nucl.
  Phys.} {\bf B790} (2008) 336 -- 337.

\bibitem{Fukugita:1986hr}
M.~Fukugita and T.~Yanagida, {\it Baryogenesis without grand unification},
  {\em Phys. Lett.} {\bf B174} (1986) 45.

\bibitem{Davidson:2002qv}
S.~Davidson and A.~Ibarra, {\it {A lower bound on the right-handed neutrino
  mass from leptogenesis}},  {\em Phys. Lett.} {\bf B535} (2002) 25--32
  [\href{http://arXiv.org/abs/hep-ph/0202239}{{\tt hep-ph/0202239}}].

\bibitem{Buchmuller:2004nz}
W.~Buchm{\"u}ller, P.~Di~Bari and M.~Pl{\"u}macher, {\it {Leptogenesis for
  pedestrians}},  {\em Ann. Phys.} {\bf 315} (2005) 305--351
  [\href{http://arXiv.org/abs/hep-ph/0401240}{{\tt hep-ph/0401240}}].

\bibitem{Anderson:1996bg}
G.~Anderson {\em et.~al.}, {\it {Motivations for and implications of
  non-universal GUT- scale boundary conditions for soft SUSY-breaking
  parameters}},  \href{http://arXiv.org/abs/hep-ph/9609457}{{\tt
  hep-ph/9609457}}.

\bibitem{Cerdeno:2005eu}
D.~G. Cerdeno, K.-Y. Choi, K.~Jedamzik, L.~Roszkowski and R.~Ruiz~de Austri,
  {\it {Gravitino dark matter in the CMSSM with improved constraints from
  BBN}},  {\em JCAP} {\bf 0606} (2006) 005
  [\href{http://arXiv.org/abs/hep-ph/0509275}{{\tt hep-ph/0509275}}].

\bibitem{Fujii:2003nr}
M.~Fujii, M.~Ibe and T.~Yanagida, {\it Upper bound on gluino mass from thermal
  leptogenesis},  {\em Phys. Lett.} {\bf B579} (2004) 6--12
  [\href{http://arXiv.org/abs/hep-ph/0310142}{{\tt hep-ph/0310142}}].

\bibitem{Steffen:2006hw}
F.~D. Steffen, {\it {Gravitino dark matter and cosmological constraints}},
  {\em JCAP} {\bf 0609} (2006) 001
  [\href{http://arXiv.org/abs/hep-ph/0605306}{{\tt hep-ph/0605306}}].

\bibitem{Steffen:2008bt}
F.~D. Steffen, {\it {Probing the Reheating Temperature at Colliders and with
  Primordial Nucleosynthesis}},  {\em Phys. Lett.} {\bf B669} (2008) 74--80
  [\href{http://arXiv.org/abs/0806.3266}{{\tt 0806.3266}}].

\bibitem{Feng:2004mt}
J.~L. Feng, S.~Su and F.~Takayama, {\it {Supergravity with a gravitino LSP}},
  {\em Phys. Rev.} {\bf D70} (2004) 075019
  [\href{http://arXiv.org/abs/hep-ph/0404231}{{\tt hep-ph/0404231}}].

\bibitem{Steffen:2006wx}
F.~D. Steffen, {\it {Constraints on gravitino dark matter scenarios with long-
  lived charged sleptons}},  {\em AIP Conf. Proc.} {\bf 903} (2007) 595--598
  [\href{http://arXiv.org/abs/hep-ph/0611027}{{\tt hep-ph/0611027}}].

\bibitem{Sigl:1995kk}
G.~Sigl, K.~Jedamzik, D.~N. Schramm and V.~S. Berezinsky, {\it {Helium
  photodisintegration and nucleosynthesis: Implications for topological
  defects, high-energy cosmic rays, and massive black holes}},  {\em Phys.
  Rev.} {\bf D52} (1995) 6682--6693
  [\href{http://arXiv.org/abs/astro-ph/9503094}{{\tt astro-ph/9503094}}].

\bibitem{Jedamzik:2006xz}
K.~Jedamzik, {\it {Big bang nucleosynthesis constraints on hadronically and
  electromagnetically decaying relic neutral particles}},  {\em Phys. Rev.}
  {\bf D74} (2006) 103509 [\href{http://arXiv.org/abs/hep-ph/0604251}{{\tt
  hep-ph/0604251}}].

\bibitem{Cyburt:2006uv}
R.~H. Cyburt, J.~R. Ellis, B.~D. Fields, K.~A. Olive and V.~C. Spanos, {\it
  {Bound-state effects on light-element abundances in gravitino dark matter
  scenarios}},  {\em JCAP} {\bf 0611} (2006) 014
  [\href{http://arXiv.org/abs/astro-ph/0608562}{{\tt astro-ph/0608562}}].

\bibitem{Jedamzik:2007qk}
K.~Jedamzik, {\it {Bounds on long-lived charged massive particles from Big Bang
  nucleosynthesis}},  {\em JCAP} {\bf 0803} (2008) 008
  [\href{http://arXiv.org/abs/0710.5153}{{\tt 0710.5153}}].

\bibitem{Cayrel:2007te}
R.~Cayrel {\em et.~al.}, {\it {Line shift, line asymmetry, and the 6Li/7Li
  isotopic ratio determination}},  \href{http://arXiv.org/abs/0708.3819}{{\tt
  0708.3819}}.

\bibitem{Cayrel:2008hk}
R.~Cayrel, M.~Steffen, P.~Bonifacio, H.-G. Ludwig and E.~Caffau, {\it {Overview
  of the lithium problem in metal-poor stars and new results on 6Li}},
  \href{http://arXiv.org/abs/0810.4290}{{\tt 0810.4290}}.

\bibitem{Korn:2006tv}
A.~J. Korn {\em et.~al.}, {\it {A probable stellar solution to the cosmological
  lithium discrepancy}},  {\em Nature} {\bf 442} (2006) 657--659
  [\href{http://arXiv.org/abs/astro-ph/0608201}{{\tt astro-ph/0608201}}].

\bibitem{Primas:2000ee}
F.~Primas, P.~Molaro, P.~Bonifacio and V.~Hill, {\it {First UVES observations
  of beryllium in very metal-poor stars}},
  \href{http://arXiv.org/abs/astro-ph/0008402}{{\tt astro-ph/0008402}}.

\bibitem{Boesgaard:2005jf}
A.~M. Boesgaard and M.~C. Novicki, {\it {Beryllium in the
  Ultra-Lithium-Deficient,Metal-Poor Halo Dwarf, G186-26}},  {\em Astrophys.
  J.} {\bf 633} (2005) L125--L128
  [\href{http://arXiv.org/abs/astro-ph/0509483}{{\tt astro-ph/0509483}}].

\bibitem{PhysRevC.63.018801}
H.~Utsunomiya, Y.~Yonezawa, H.~Akimune, T.~Yamagata, M.~Ohta, M.~Fujishiro,
  H.~Toyokawa and H.~Ohgaki, {\it Photodisintegration of $9be$ with
  laser-induced compton backscattered $\gamma{}$ rays},  {\em Phys. Rev. C}
  {\bf 63} (Dec, 2000) 018801.

\bibitem{Sumiyoshi2002467}
K.~Sumiyoshi, H.~Utsunomiya, S.~Goko and T.~Kajino, {\it Astrophysical reaction
  rate for {$\alpha(\alpha n,\gamma)^9$}{Be} by photodisintegration},  {\em
  Nucl. Phys.} {\bf B709} (2002) 467 -- 486.

\bibitem{Buchmuller:2004rq}
W.~Buchm{\"u}ller, K.~Hamaguchi, M.~Ratz and T.~Yanagida, {\it {Supergravity at
  colliders}},  {\em Phys. Lett.} {\bf B588} (2004) 90--98
  [\href{http://arXiv.org/abs/hep-ph/0402179}{{\tt hep-ph/0402179}}].

\bibitem{DiazCruz:2007fc}
J.~L. Diaz-Cruz, J.~R. Ellis, K.~A. Olive and Y.~Santoso, {\it {On the
  feasibility of a stop NLSP in gravitino dark matter scenarios}},  {\em JHEP}
  {\bf 05} (2007) 003 [\href{http://arXiv.org/abs/hep-ph/0701229}{{\tt
  hep-ph/0701229}}].

\bibitem{Roszkowski:2004jd}
L.~Roszkowski, R.~Ruiz~de Austri and K.-Y. Choi, {\it {Gravitino dark matter in
  the CMSSM and implications for leptogenesis and the LHC}},  {\em JHEP} {\bf
  08} (2005) 080 [\href{http://arXiv.org/abs/hep-ph/0408227}{{\tt
  hep-ph/0408227}}].

\bibitem{Porod:2003um}
W.~Porod, {\it {SPheno, a program for calculating supersymmetric spectra, SUSY
  particle decays and SUSY particle production at {$e^+ e^-$} colliders}},
  {\em Comput. Phys. Commun.} {\bf 153} (2003) 275--315
  [\href{http://arXiv.org/abs/hep-ph/0301101}{{\tt hep-ph/0301101}}].

\bibitem{Martin:1993ft}
S.~P. Martin and P.~Ramond, {\it {Sparticle spectrum constraints}},  {\em Phys.
  Rev.} {\bf D48} (1993) 5365--5375
  [\href{http://arXiv.org/abs/hep-ph/9306314}{{\tt hep-ph/9306314}}].

\bibitem{Kersten:2007ab}
J.~Kersten and K.~Schmidt-Hoberg, {\it {The Gravitino-Stau Scenario after
  Catalyzed BBN}},  {\em JCAP} {\bf 0801} (2008) 011
  [\href{http://arXiv.org/abs/0710.4528}{{\tt 0710.4528}}].

\bibitem{Bailly:2009pe}
S.~Bailly, K.-Y. Choi, K.~Jedamzik and L.~Roszkowski, {\it {A Re-analysis of
  Gravitino Dark Matter in the Constrained MSSM}},  {\em JHEP} {\bf 05} (2009)
  103 [\href{http://arXiv.org/abs/0903.3974}{{\tt 0903.3974}}].

\bibitem{Djouadi:2002ze}
A.~Djouadi, J.-L. Kneur and G.~Moultaka, {\it Suspect: A fortran code for the
  supersymmetric and higgs particle spectrum in the mssm},  {\em Comput. Phys.
  Commun.} {\bf 176} (2007) 426--455
  [\href{http://arXiv.org/abs/hep-ph/0211331}{{\tt hep-ph/0211331}}].

\bibitem{Belanger:2001fz}
G.~Belanger, F.~Boudjema, A.~Pukhov and A.~Semenov, {\it {micrOMEGAs: A program
  for calculating the relic density in the MSSM}},  {\em Comput. Phys. Commun.}
  {\bf 149} (2002) 103--120 [\href{http://arXiv.org/abs/hep-ph/0112278}{{\tt
  hep-ph/0112278}}].

\bibitem{Belanger:2004yn}
G.~Belanger, F.~Boudjema, A.~Pukhov and A.~Semenov, {\it {MicrOMEGAs: Version
  1.3}},  {\em Comput. Phys. Commun.} {\bf 174} (2006) 577--604
  [\href{http://arXiv.org/abs/hep-ph/0405253}{{\tt hep-ph/0405253}}].

\bibitem{Allahverdi:2006iq}
R.~Allahverdi, K.~Enqvist, J.~Garcia-Bellido and A.~Mazumdar, {\it {Gauge
  invariant MSSM inflaton}},  {\em Phys. Rev. Lett.} {\bf 97} (2006) 191304
  [\href{http://arXiv.org/abs/hep-ph/0605035}{{\tt hep-ph/0605035}}].

\bibitem{GonzalezGarcia:2007ib}
M.~C. Gonzalez-Garcia and M.~Maltoni, {\it {Phenomenology with Massive
  Neutrinos}},  {\em Phys. Rept.} {\bf 460} (2008) 1--129
  [\href{http://arXiv.org/abs/0704.1800}{{\tt 0704.1800}}].

\bibitem{Coughlan:1983ci}
G.~D. Coughlan, W.~Fischler, E.~W. Kolb, S.~Raby and G.~G. Ross, {\it
  {Cosmological Problems for the Polonyi Potential}},  {\em Phys. Lett.} {\bf
  B131} (1983) 59.

\bibitem{Banks:1993en}
T.~Banks, D.~B. Kaplan and A.~E. Nelson, {\it {Cosmological Implications of
  Dynamical Supersymmetry Breaking}},  {\em Phys. Rev.} {\bf D49} (1994)
  779--787 [\href{http://arXiv.org/abs/hep-ph/9308292}{{\tt hep-ph/9308292}}].

\bibitem{deCarlos:1993jw}
B.~de~Carlos, J.~A. Casas, F.~Quevedo and E.~Roulet, {\it {Model independent
  properties and cosmological implications of the dilaton and moduli sectors of
  4-d strings}},  {\em Phys. Lett.} {\bf B318} (1993) 447--456
  [\href{http://arXiv.org/abs/hep-ph/9308325}{{\tt hep-ph/9308325}}].

\bibitem{Kasuya:2007cy}
S.~Kasuya and F.~Takahashi, {\it {Entropy production by Q-ball decay for
  diluting long-lived charged particles}},  {\em JCAP} {\bf 0711} (2007) 019
  [\href{http://arXiv.org/abs/0709.2634}{{\tt 0709.2634}}].

\bibitem{Baltz:2001rq}
E.~A. Baltz and H.~Murayama, {\it {Gravitino warm dark matter with entropy
  production}},  {\em JHEP} {\bf 05} (2003) 067
  [\href{http://arXiv.org/abs/astro-ph/0108172}{{\tt astro-ph/0108172}}].

\bibitem{Fujii:2002fv}
M.~Fujii and T.~Yanagida, {\it {Natural gravitino dark matter and thermal
  leptogenesis in gauge-mediated supersymmetry-breaking models}},  {\em Phys.
  Lett.} {\bf B549} (2002) 273--283
  [\href{http://arXiv.org/abs/hep-ph/0208191}{{\tt hep-ph/0208191}}].

\bibitem{Fujii:2003iw}
M.~Fujii, M.~Ibe and T.~Yanagida, {\it {Thermal leptogenesis and gauge
  mediation}},  {\em Phys. Rev.} {\bf D69} (2004) 015006
  [\href{http://arXiv.org/abs/hep-ph/0309064}{{\tt hep-ph/0309064}}].

\bibitem{Jedamzik:2005ir}
K.~Jedamzik, M.~Lemoine and G.~Moultaka, {\it {Gravitino dark matter in gauge
  mediated supersymmetry breaking}},  {\em Phys. Rev.} {\bf D73} (2006) 043514
  [\href{http://arXiv.org/abs/hep-ph/0506129}{{\tt hep-ph/0506129}}].

\bibitem{Lemoine:2005hu}
M.~Lemoine, G.~Moultaka and K.~Jedamzik, {\it {Natural gravitino dark matter in
  SO(10) gauge mediated supersymmetry breaking}},  {\em Phys. Lett.} {\bf B645}
  (2007) 222--227 [\href{http://arXiv.org/abs/hep-ph/0504021}{{\tt
  hep-ph/0504021}}].

\bibitem{Buchmuller:2006tt}
W.~Buchm{\"u}ller, K.~Hamaguchi, M.~Ibe and T.~T. Yanagida, {\it {Eluding the
  BBN constraints on the stable gravitino}},  {\em Phys. Lett.} {\bf B643}
  (2006) 124--126 [\href{http://arXiv.org/abs/hep-ph/0605164}{{\tt
  hep-ph/0605164}}].

\bibitem{Scherrer:1984fd}
R.~J. Scherrer and M.~S. Turner, {\it Decaying particles do not heat up the
  universe},  {\em Phys. Rev.} {\bf D31} (1985) 681.

\bibitem{Kawasaki:1999na}
M.~Kawasaki, K.~Kohri and N.~Sugiyama, {\it {Cosmological Constraints on
  Late-time Entropy Production}},  {\em Phys. Rev. Lett.} {\bf 82} (1999) 4168
  [\href{http://arXiv.org/abs/astro-ph/9811437}{{\tt astro-ph/9811437}}].

\bibitem{Kawasaki:2000en}
M.~Kawasaki, K.~Kohri and N.~Sugiyama, {\it Mev-scale reheating temperature and
  thermalization of neutrino background},  {\em Phys. Rev.} {\bf D62} (2000)
  023506 [\href{http://arXiv.org/abs/astro-ph/0002127}{{\tt
  astro-ph/0002127}}].

\bibitem{Hannestad:2004px}
S.~Hannestad, {\it What is the lowest possible reheating temperature?},  {\em
  Phys. Rev.} {\bf D70} (2004) 043506
  [\href{http://arXiv.org/abs/astro-ph/0403291}{{\tt astro-ph/0403291}}].

\bibitem{Ichikawa:2005vw}
K.~Ichikawa, M.~Kawasaki and F.~Takahashi, {\it The oscillation effects on
  thermalization of the neutrinos in the universe with low reheating
  temperature},  {\em Phys. Rev.} {\bf D72} (2005) 043522
  [\href{http://arXiv.org/abs/astro-ph/0505395}{{\tt astro-ph/0505395}}].

\bibitem{Olive:1980wz}
K.~A. Olive, D.~N. Schramm and G.~Steigman, {\it {Limits on New Superweakly
  Interacting Particles from Primordial Nucleosynthesis}},  {\em Nucl. Phys.}
  {\bf B180} (1981) 497.

\bibitem{Endo:2006zj}
M.~Endo, K.~Hamaguchi and F.~Takahashi, {\it Moduli-induced gravitino problem},
   {\em Phys. Rev. Lett.} {\bf 96} (2006) 211301
  [\href{http://arXiv.org/abs/hep-ph/0602061}{{\tt hep-ph/0602061}}].

\bibitem{Nakamura:2006uc}
S.~Nakamura and M.~Yamaguchi, {\it Gravitino production from heavy moduli decay
  and cosmological moduli problem revived},  {\em Phys. Lett.} {\bf B638}
  (2006) 389--395 [\href{http://arXiv.org/abs/hep-ph/0602081}{{\tt
  hep-ph/0602081}}].

\bibitem{Blanchet:2006be}
S.~Blanchet and P.~Di~Bari, {\it {Flavor effects on leptogenesis predictions}},
   {\em JCAP} {\bf 0703} (2007) 018
  [\href{http://arXiv.org/abs/hep-ph/0607330}{{\tt hep-ph/0607330}}].

\bibitem{Antusch:2006gy}
S.~Antusch and A.~M. Teixeira, {\it Towards constraints on the susy seesaw from
  flavour- dependent leptogenesis},  {\em JCAP} {\bf 0702} (2007) 024
  [\href{http://arXiv.org/abs/hep-ph/0611232}{{\tt hep-ph/0611232}}].

\bibitem{Buchmuller:2002rq}
W.~Buchm{\"u}ller, P.~Di~Bari and M.~Pl{\"u}macher, {\it {Cosmic microwave
  background, matter-antimatter asymmetry and neutrino masses}},  {\em Nucl.
  Phys.} {\bf B643} (2002) 367--390
  [\href{http://arXiv.org/abs/hep-ph/0205349}{{\tt hep-ph/0205349}}].

\bibitem{Buchmuller:2002jk}
W.~Buchm{\"u}ller, P.~Di~Bari and M.~Pl{\"u}macher, {\it {A bound on neutrino
  masses from baryogenesis}},  {\em Phys. Lett.} {\bf B547} (2002) 128--132
  [\href{http://arXiv.org/abs/hep-ph/0209301}{{\tt hep-ph/0209301}}].

\bibitem{Brandenburg:2005he}
A.~Brandenburg, L.~Covi, K.~Hamaguchi, L.~Roszkowski and F.~D. Steffen, {\it
  {Signatures of axinos and gravitinos at colliders}},  {\em Phys. Lett.} {\bf
  B617} (2005) 99--111 [\href{http://arXiv.org/abs/hep-ph/0501287}{{\tt
  hep-ph/0501287}}].

\bibitem{Steffen:2005cn}
F.~D. Steffen, {\it {Collider signatures of axino and gravitino dark matter}},
  \href{http://arXiv.org/abs/hep-ph/0507003}{{\tt hep-ph/0507003}}.

\bibitem{Martyn:2006as}
H.~U. Martyn, {\it {Detecting metastable staus and gravitinos at the ILC}},
  {\em Eur. Phys. J.} {\bf C48} (2006) 15--24
  [\href{http://arXiv.org/abs/hep-ph/0605257}{{\tt hep-ph/0605257}}].

\bibitem{Hamaguchi:2006vu}
K.~Hamaguchi, M.~M. Nojiri and A.~de~Roeck, {\it {Prospects to study a
  long-lived charged next lightest supersymmetric particle at the LHC}},  {\em
  JHEP} {\bf 03} (2007) 046 [\href{http://arXiv.org/abs/hep-ph/0612060}{{\tt
  hep-ph/0612060}}].

\bibitem{Ratz:2008qh}
M.~Ratz, K.~Schmidt-Hoberg and M.~W. Winkler, {\it {A note on the primordial
  abundance of stau NLSPs}},  {\em JCAP} {\bf 0810} (2008) 026
  [\href{http://arXiv.org/abs/0808.0829}{{\tt 0808.0829}}].

\bibitem{Ellis:1999mm}
J.~R. Ellis, T.~Falk, K.~A. Olive and M.~Srednicki, {\it {Calculations of
  neutralino stau coannihilation channels and the cosmologically relevant
  region of MSSM parameter space}},  {\em Astropart. Phys.} {\bf 13} (2000)
  181--213 [\href{http://arXiv.org/abs/hep-ph/9905481}{{\tt hep-ph/9905481}}].

\bibitem{Hahn:2000kx}
T.~Hahn, {\it {Generating Feynman diagrams and amplitudes with FeynArts 3}},
  {\em Comput. Phys. Commun.} {\bf 140} (2001) 418--431
  [\href{http://arXiv.org/abs/hep-ph/0012260}{{\tt hep-ph/0012260}}].

\bibitem{Hahn:2001rv}
T.~Hahn and C.~Schappacher, {\it {The implementation of the minimal
  supersymmetric standard model in FeynArts and FormCalc}},  {\em Comput. Phys.
  Commun.} {\bf 143} (2002) 54--68
  [\href{http://arXiv.org/abs/hep-ph/0105349}{{\tt hep-ph/0105349}}].

\bibitem{Hahn:1998yk}
T.~Hahn and M.~Perez-Victoria, {\it {Automatized one-loop calculations in four
  and D dimensions}},  {\em Comput. Phys. Commun.} {\bf 118} (1999) 153--165
  [\href{http://arXiv.org/abs/hep-ph/9807565}{{\tt hep-ph/9807565}}].

\bibitem{Hahn:2006qw}
T.~Hahn and M.~Rauch, {\it {News from FormCalc and LoopTools}},  {\em Nucl.
  Phys. Proc. Suppl.} {\bf 157} (2006) 236--240
  [\href{http://arXiv.org/abs/hep-ph/0601248}{{\tt hep-ph/0601248}}].

\bibitem{Heinemeyer:1998yj}
S.~Heinemeyer, W.~Hollik and G.~Weiglein, {\it {FeynHiggs: A program for the
  calculation of the masses of the neutral CP-even Higgs bosons in the MSSM}},
  {\em Comput. Phys. Commun.} {\bf 124} (2000) 76--89
  [\href{http://arXiv.org/abs/hep-ph/9812320}{{\tt hep-ph/9812320}}].

\bibitem{Belanger:2006is}
G.~Belanger, F.~Boudjema, A.~Pukhov and A.~Semenov, {\it {micrOMEGAs2.0: A
  program to calculate the relic density of dark matter in a generic model}},
  {\em Comput. Phys. Commun.} {\bf 176} (2007) 367--382
  [\href{http://arXiv.org/abs/hep-ph/0607059}{{\tt hep-ph/0607059}}].

\bibitem{Pukhov:2004ca}
A.~Pukhov, {\it {CalcHEP 3.2: MSSM, structure functions, event generation,
  batchs, and generation of matrix elements for other packages}},
  \href{http://arXiv.org/abs/hep-ph/0412191}{{\tt hep-ph/0412191}}.

\bibitem{Dubinin:1998nt}
M.~N. Dubinin and A.~V. Semenov, {\it {Triple and quartic interactions of Higgs
  bosons in the general two-Higgs-doublet model}},
  \href{http://arXiv.org/abs/hep-ph/9812246}{{\tt hep-ph/9812246}}.

\bibitem{Gunion:2002zf}
J.~F. Gunion and H.~E. Haber, {\it {The CP-conserving two-Higgs-doublet model:
  The approach to the decoupling limit}},  {\em Phys. Rev.} {\bf D67} (2003)
  075019 [\href{http://arXiv.org/abs/hep-ph/0207010}{{\tt hep-ph/0207010}}].

\bibitem{Haber:1997dt}
H.~E. Haber, {\it {Higgs boson masses and couplings in the minimal
  supersymmetric model}},  \href{http://arXiv.org/abs/hep-ph/9707213}{{\tt
  hep-ph/9707213}}.

\bibitem{Carena:1995bx}
M.~S. Carena, J.~R. Espinosa, M.~Quiros and C.~E.~M. Wagner, {\it {Analytical
  expressions for radiatively corrected Higgs masses and couplings in the
  MSSM}},  {\em Phys. Lett.} {\bf B355} (1995) 209--221
  [\href{http://arXiv.org/abs/hep-ph/9504316}{{\tt hep-ph/9504316}}].

\bibitem{Casas:1995pd}
J.~A. Casas, A.~Lleyda and C.~Munoz, {\it {Strong constraints on the parameter
  space of the MSSM from charge and color breaking minima}},  {\em Nucl. Phys.}
  {\bf B471} (1996) 3--58 [\href{http://arXiv.org/abs/hep-ph/9507294}{{\tt
  hep-ph/9507294}}].

\bibitem{Griest:1989wd}
K.~Griest and M.~Kamionkowski, {\it {Unitarity Limits on the Mass and Radius of
  Dark Matter Particles}},  {\em Phys. Rev. Lett.} {\bf 64} (1990) 615.

\bibitem{Boos:2002ze}
E.~Boos, A.~Djouadi, M.~M{\"u}hlleitner and A.~Vologdin, {\it {The MSSM Higgs
  bosons in the intense-coupling regime}},  {\em Phys. Rev.} {\bf D66} (2002)
  055004 [\href{http://arXiv.org/abs/hep-ph/0205160}{{\tt hep-ph/0205160}}].

\bibitem{Boos:2003jt}
E.~Boos, A.~Djouadi and A.~Nikitenko, {\it {Detection of the neutral MSSM Higgs
  bosons in the intense- coupling regime at the LHC}},  {\em Phys. Lett.} {\bf
  B578} (2004) 384--393 [\href{http://arXiv.org/abs/hep-ph/0307079}{{\tt
  hep-ph/0307079}}].

\bibitem{Djouadi:2005gj}
A.~Djouadi, {\it {The anatomy of electro-weak symmetry breaking. II: The Higgs
  bosons in the minimal supersymmetric model}},  {\em Phys. Rept.} {\bf 459}
  (2008) 1--241 [\href{http://arXiv.org/abs/hep-ph/0503173}{{\tt
  hep-ph/0503173}}].

\bibitem{Carena:1999xa}
M.~S. Carena, S.~Heinemeyer, C.~E.~M. Wagner and G.~Weiglein, {\it {Suggestions
  for improved benchmark scenarios for Higgs- boson searches at LEP2}},
  \href{http://arXiv.org/abs/hep-ph/9912223}{{\tt hep-ph/9912223}}.

\bibitem{Caprini:2003gz}
C.~Caprini, S.~Biller and P.~G. Ferreira, {\it {Constraints on the electrical
  charge asymmetry of the universe}},  {\em JCAP} {\bf 0502} (2005) 006
  [\href{http://arXiv.org/abs/hep-ph/0310066}{{\tt hep-ph/0310066}}].

\bibitem{Mahmoudi:2007gd}
F.~Mahmoudi, {\it {New constraints on supersymmetric models from {$b\to s$}
  gamma}},  {\em JHEP} {\bf 12} (2007) 026
  [\href{http://arXiv.org/abs/0710.3791}{{\tt 0710.3791}}].

\bibitem{Degrassi:2002fi}
G.~Degrassi, S.~Heinemeyer, W.~Hollik, P.~Slavich and G.~Weiglein, {\it
  {Towards high-precision predictions for the MSSM Higgs sector}},  {\em Eur.
  Phys. J.} {\bf C28} (2003) 133--143
  [\href{http://arXiv.org/abs/hep-ph/0212020}{{\tt hep-ph/0212020}}].

\bibitem{Drees:1991mx}
M.~Drees and M.~M. Nojiri, {\it {One loop corrections to the Higgs sector in
  minimal supergravity models}},  {\em Phys. Rev.} {\bf D45} (1992) 2482--2492.

\bibitem{Drees:1995hj}
M.~Drees and S.~P. Martin, {\it {Implications of SUSY model building}},
  \href{http://arXiv.org/abs/hep-ph/9504324}{{\tt hep-ph/9504324}}.

\bibitem{Carena:1994bv}
M.~S. Carena, M.~Olechowski, S.~Pokorski and C.~E.~M. Wagner, {\it {Electroweak
  symmetry breaking and bottom - top Yukawa unification}},  {\em Nucl. Phys.}
  {\bf B426} (1994) 269--300 [\href{http://arXiv.org/abs/hep-ph/9402253}{{\tt
  hep-ph/9402253}}].

\bibitem{Ambrosanio:2000ik}
S.~Ambrosanio, B.~Mele, S.~Petrarca, G.~Polesello and A.~Rimoldi, {\it
  {Measuring the SUSY breaking scale at the LHC in the slepton NLSP scenario of
  GMSB models}},  {\em JHEP} {\bf 01} (2001) 014
  [\href{http://arXiv.org/abs/hep-ph/0010081}{{\tt hep-ph/0010081}}].

\bibitem{Datta:2001sh}
A.~Datta, A.~Djouadi and J.-L. Kneur, {\it {Probing the SUSY Higgs boson
  couplings to scalar leptons at high-energy {$e^+ e^-$} colliders}},  {\em
  Phys. Lett.} {\bf B509} (2001) 299--306
  [\href{http://arXiv.org/abs/hep-ph/0101353}{{\tt hep-ph/0101353}}].

\bibitem{Dittmaier:2003ej}
S.~Dittmaier, M.~Kramer and M.~Spira, {\it {Higgs radiation off bottom quarks
  at the Tevatron and the LHC}},  {\em Phys. Rev.} {\bf D70} (2004) 074010
  [\href{http://arXiv.org/abs/hep-ph/0309204}{{\tt hep-ph/0309204}}].

\bibitem{Dawson:2003kb}
S.~Dawson, C.~B. Jackson, L.~Reina and D.~Wackeroth, {\it {Exclusive Higgs
  boson production with bottom quarks at hadron colliders}},  {\em Phys. Rev.}
  {\bf D69} (2004) 074027 [\href{http://arXiv.org/abs/hep-ph/0311067}{{\tt
  hep-ph/0311067}}].

\bibitem{Harlander:2003ai}
R.~V. Harlander and W.~B. Kilgore, {\it {Higgs boson production in bottom quark
  fusion at next-to- next-to-leading order}},  {\em Phys. Rev.} {\bf D68}
  (2003) 013001 [\href{http://arXiv.org/abs/hep-ph/0304035}{{\tt
  hep-ph/0304035}}].

\bibitem{Ball:2007zza}
{\bf CMS} Collaboration, G.~L. Bayatian {\em et.~al.}, {\it {CMS technical
  design report, volume II: Physics performance}},  {\em J. Phys.} {\bf G34}
  (2007) 995--1579.

\bibitem{Covi:1999ty}
L.~Covi, J.~E. Kim and L.~Roszkowski, {\it {Axinos as cold dark matter}},  {\em
  Phys. Rev. Lett.} {\bf 82} (1999) 4180--4183
  [\href{http://arXiv.org/abs/hep-ph/9905212}{{\tt hep-ph/9905212}}].

\bibitem{Freitas:2009fb}
A.~Freitas, F.~D. Steffen, N.~Tajuddin and D.~Wyler, {\it {Upper Limits on the
  Peccei-Quinn Scale and on the Reheating Temperature in Axino Dark Matter
  Scenarios}},  \href{http://arXiv.org/abs/0904.3218}{{\tt 0904.3218}}.

\end{thebibliography}\endgroup

\end{document}